\documentclass[sn-mathphys,Numbered]{sn-jnl-mod2}

\usepackage{graphicx}%
\usepackage{multirow}%
\usepackage{amsmath,amssymb,amsfonts}%
\usepackage{amsthm}%
\usepackage{mathrsfs}%
\usepackage[title]{appendix}%
\usepackage{xcolor}%
\usepackage{textcomp}%
\usepackage{manyfoot}%
\usepackage{booktabs}%
\usepackage{algorithm}%
\usepackage{algorithmicx}%
\usepackage{algpseudocode}%
\usepackage{listings}%
\usepackage{multicol}
\usepackage{lmodern}

\usepackage{chemformula} 
\usepackage[free-standing-units=true]{siunitx}
\DeclareSIUnit{\revolution}{\text{rev}}
\DeclareSIUnit{\lightyear}{\text{ly}}
\DeclareSIUnit{\Torr}{\text{Torr}}
\DeclareSIUnit{\sccm}{\text{sccm}}
\DeclareSIUnit{\groove}{\text{gr}}
\DeclareSIUnit{\pixel}{\text{px}}
\DeclareSIUnit{\electronvolt}{\text{eV}}
\usepackage{xspace} 
\usepackage{xcolor}

\newcommand{\dC}{\textdegree C\xspace} 
\newcommand{\ie}{\textit{i.e.}} 
\newcommand{\eg}{\textit{e.g.}} 
\newcommand{\etc}{\textit{etc.}} 
\newcommand{\etal}{\textit{et~al.}} 
\newcommand{\via}{\textit{via}\xspace} 
\newcommand{\insitu}{\textit{in~situ}\xspace}
\newcommand{\ji}{\mathrm{i}} 
\newcommand{\overbar}[1]{\mkern 1.5mu\overline{\mkern-1.5mu#1\mkern-1.5mu}\mkern 1.5mu}
\newcommand{\app}{{\raise.17ex\hbox{$\scriptstyle\sim$}}} 

\usepackage{float} 

\newlength \figWidthCol
\newlength \figWidthFull
\newlength \figWidthFullExtra
\usepackage[switch,columnwise]{lineno}

\usepackage{tikz} 
\usepackage{pgfplots}
\usetikzlibrary{calc,decorations.markings,decorations.text,patterns,arrows.meta,intersections,pgfplots.fillbetween}

\usepackage{longtable,lscape} 

\usepackage{mathtools}
\DeclarePairedDelimiter\abs{\lvert}{\rvert}%
\begin{document}

\title[Experimental demonstration of corrugated nanolaminate films as reflective light sails]{Experimental demonstration of corrugated nanolaminate films as reflective light sails}

\author[1]{\fnm{Matthew F.} \sur{Campbell}}
\equalcont{These authors contributed equally to this work.}

\author[2,3]{\fnm{Pawan} \sur{Kumar}}
\equalcont{These authors contributed equally to this work.}

\author[2]{\fnm{Jason} \sur{Lynch}}

\author[4]{\fnm{Ramon} \sur{Gao}}

\author[2]{\fnm{Adam} \sur{Alfieri}}

\author[5,6]{\fnm{John} \sur{Brewer}}

\author[1,7]{\fnm{Thomas J.} \sur{Celenza}}

\author[1,8,9]{\fnm{Mohsen} \sur{Azadi}}

\author[4]{\fnm{Michael} \sur{Kelzenberg}}

\author[10]{\fnm{Eric} \sur{Stach}}

\author[5]{\fnm{Aaswath P.} \sur{Raman}}

\author[4]{\fnm{Harry A.} \sur{Atwater, Jr.}}

\author*[1]{\fnm{Igor} \sur{Bargatin}}\email{bargatin@seas.upenn.edu} 

\author*[2]{\fnm{Deep} \sur{Jariwala}}\email{dmj@seas.upenn.edu} 

\affil[1]{\orgdiv{Department of Mechanical Engineering and Applied Mechanics}, \orgname{University of Pennsylvania}, \orgaddress{\city{Philadelphia}, \state{PA}, \country{USA}, \postcode{19104}}}

\affil[2]{\orgdiv{Department of Electrical and Systems Engineering}, \orgname{University of Pennsylvania}, \orgaddress{\city{Philadelphia}, \state{PA}, \country{USA}, \postcode{19104}}}

\affil[3]{\orgname{Now at: Interuniversity Microelectronics Centre (IMEC)}, \orgaddress{\city{Leuven}, \country{Belgium}}}

\affil[4]{\orgdiv{Department of Applied Physics}, \orgname{California Institute of Technology}, \orgaddress{\city{Pasadena}, \state{CA}, \country{USA}, \postcode{91125}}}

\affil[5]{\orgdiv{Department of Materials Science and Engineering}, \orgname{University of California at Los Angeles}, \orgaddress{\city{Los Angeles}, \state{CA}, \country{USA}, \postcode{90024}}}

\affil[6]{\orgname{Now at: Leonardo DRS Daylight Solutions}, \orgaddress{\city{San Diego}, \state{CA}, \country{USA}, \postcode{92127}}}

\affil[7]{\orgname{Now at: Exponent, Inc.}, \orgaddress{\city{New York}, \state{NY}, \country{USA}, \postcode{10017}}}

\affil[8]{\orgdiv{Singh Center for Nanotechnology}, \orgname{University of Pennsylvania}, \orgaddress{\city{Philadelphia}, \state{PA}, \country{USA}, \postcode{19104}}}

\affil[9]{\orgname{Now at: Databuoy Corp.}, \orgaddress{\city{Vienna}, \state{VA}, \country{USA}, \postcode{22182}}}

\affil[10]{\orgdiv{Department of Materials Science and Engineering}, \orgname{University of Pennsylvania}, \orgaddress{\city{Philadelphia}, \state{PA}, \country{USA}, \postcode{19104}}}



\maketitle

\textbf{Abstract.} Achieving laser-driven, reflective, relativistic light sails would represent a tremendous breakthrough for humankind, allowing us to advance our understanding of the solar system and deep space far beyond what we know from space probes, telescopes, and objects passing near Earth. Numerous sail film designs have been proposed, but none have been demonstrated that satisfy all of the stringent optical, mechanical, and mass budget constraints. Here we overcome this challenge by experimentally demonstrating a novel class of optically-optimized nanolaminate sails with strong and flexible hexagonally-corrugated microstructures. Our prototypes, fabricated from alumina and molybdenum disulfide using scalable semiconductor processing techniques, feature ultra-low areal densities of $<\!1$~\si{\gram\per\meter\squared} and achieve experimentally-measured reflectivities of $>\!50\%$ and absorptivities of $<\!4\%$ within the Doppler-shifted laser wavelength range corresponding to accelerating to a fifth the speed of light. Moreover, we analyze reflectivity, strength, and mass constraints to show that our sails have the potential to achieve greater maximum velocities than other sail designs in the literature. Broadly, our films mark a significant leap forward toward plausible relativistic interstellar propulsion for intragalactic exploration. 

\begin{multicols}{2}

\begin{figure*}[hbt!]%
\centering
\includegraphics[width=0.9\linewidth]{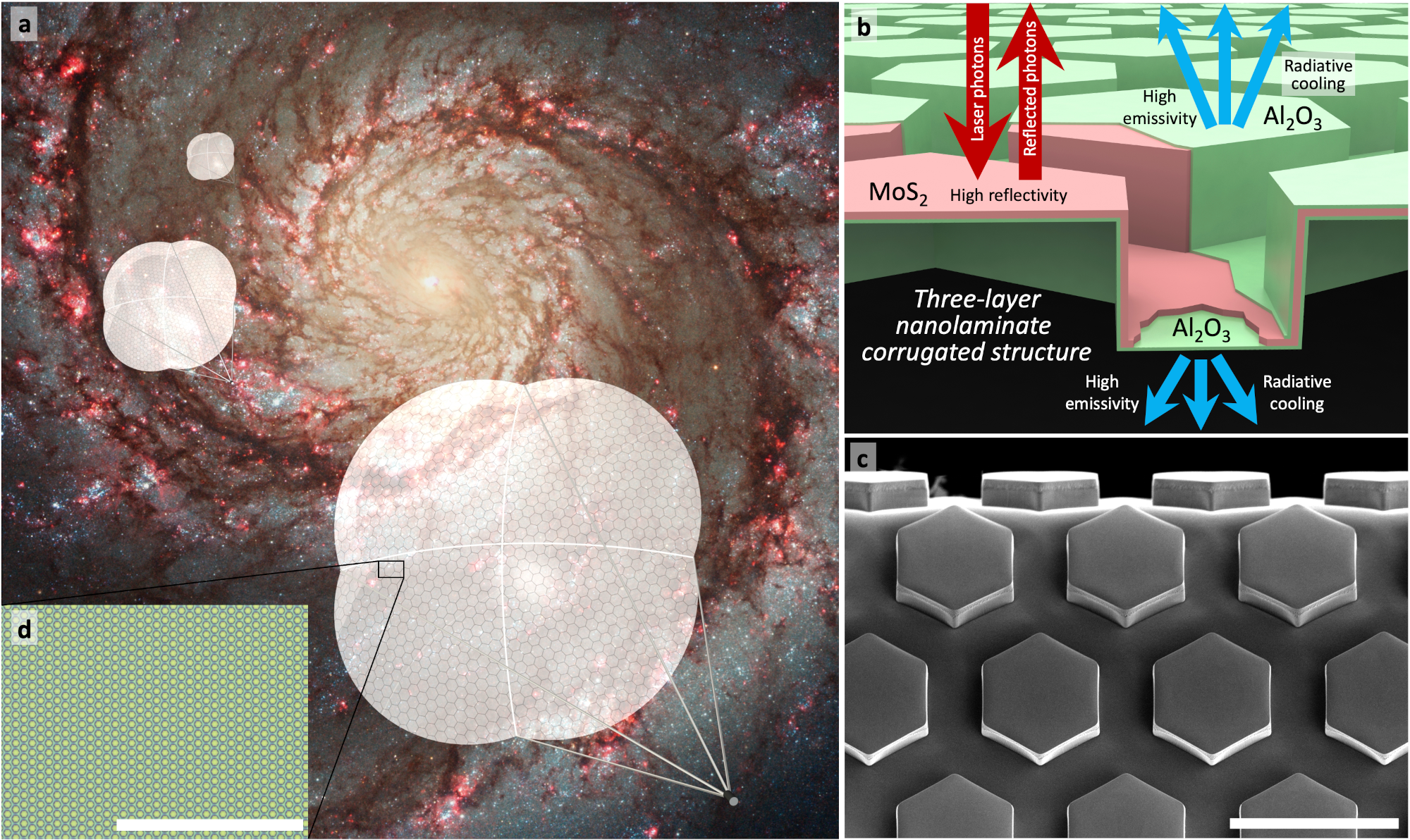}
\caption{\textbf{\textbar~ Overview of laser-accelerated light sails.} (\textbf{a}) Light sails carry gram-scale payloads at relativistic velocities for interstellar travel.  (\textbf{b}) Computer cut-away rendering of sail film showing three-layer nanolaminate corrugated structure. The \ch{MoS2} provides high reflectivity for acceleration and the flanking \ch{Al2O3} layers yield high emissivity for radiative cooling. The hexagonally corrugated microstructure increases the sail's bending rigidity, preventing wrinkles and allowing it to maintain its shape as it accelerates. (\textbf{c}) Scanning electron micrograph (SEM) showing prototype film bent away from the camera to emphasize its three-dimensional structure. (\textbf{d}) Micrograph of fully suspended prototype film (\ch{Si} substrate removed). Scale bars: (\textbf{c}) 50~\si{\micro\meter},  (\textbf{d}) 1~\si{\milli\meter}. Galaxy image obtained from NASA~\cite{NASA2011-photo}.}
\label{F:overview}
\end{figure*}

One hundred years ago, Friedrich Zander and Konstantin Tsiolkovsky proposed reflecting photons by thin mirror-like sheets as a means of space propulsion~\cite{Tsander1964-book, Tsander1967-TTF541}. Since that time, significant effort has been devoted to materializing their vision in the form of laser-driven light sails: highly reflective, ultra-lightweight membranes that accelerate miniature chip satellites to relativistic velocities~\cite{Marx1966-22, Lubin2016-40, Lubin2020-9, Atwater2018-861, Parkin2018-370,Lantin2022-261, Parkin2023-arXiv, Gao2024-4203, Michaeli2025-inPress, Lin2025-arxiv} (Figure~\ref{F:overview}). This design challenge is formidable. Light sails must be highly reflective at both the laser wavelength and at slightly longer wavelengths due to the Doppler shift as the sail's velocity increases~\cite{Ilic2018-5583}. They must also exhibit essentially zero absorptivity $(\alpha \lesssim 10^{-9})$ in the Doppler-shifted laser wavelength range and emit efficiently at longer infrared wavelengths $(\varepsilon_e \gtrsim 10^{-3})$ to avoid overheating~\cite{Brewer2022-594}. Furthermore, light sails must be lightweight yet mechanically robust, with ultra-low areal densities ($\rho_{a,target}\approx0.1$~\si{\gram\per\meter\squared})~\cite{Atwater2018-861} and high bending stiffnesses to avoid wrinkles~\cite{Campbell2022-90}. Previous designs, though maximizing reflectivity or minimizing areal density, struggled to address all of these requirements simultaneously~\cite{Ilic2018-5583, Myilswamy2020-8223, Siegel2019-2032, Jin2020-2350, Salary2020-1900311, Gieseler2021-21562, Brewer2022-594, Campbell2022-90, Gao2022-1965, Kudyshev2022-190, Lien2022-3032, Santi2022-16, Taghavi2022-20034, Tung2022-1108, Chang2024-6689, Gao2024-4203, Norder2025-2753}. 

Therefore, we have developed and tested a novel class of corrugated nanolaminate sail films by considering optical/thermal, mechanical, and mass budget constraints concurrently. Our films consist of a high-refractive-index multilayer molybdenum disulfide (\ch{MoS2}) core flanked by emissive alumina (\ch{Al2O3}) face sheets for thermal management. Below, we show that this low-areal-density combination provides high reflectivity and low absorptivity near the proposed laser wavelength of $\lambda_l=1.2$~\si{\micro\meter}. Moreover, our sails feature a hexagonally-patterned out-of-plane architected structure, which enhances their bending stiffness to prevent them from wrinkling. 

\ch{MoS2} is an advantageous material for laser-driven light sails because it has a high real index of refraction $n$ near $\lambda_l=1.2$~\si{\micro\meter} and can be fabricated in atomically-perfect sheets that exhibit near-zero infrared absorptivity~\cite{Atwater2018-861, Ermolaev2020-21}.  In order for the high $n$-values to beget high reflectivity, however, the \ch{MoS2} films must be grown as atomically-smooth sheets with multi-nanometer-scale thicknesses. Achieving such thick smooth films has historically been challenging~\cite{Akcay2021-1452}. 
We accomplished this using a two-step fabrication approach, in which we first sputtered \ch{Mo} onto a substrate, being careful to maintain precise control of the grain size and roughness, and subsequently sulfurized it into high-quality \ch{MoS2} multilayers in an elevated temperature \ch{H2}/\ch{H2S} environment~\cite{Altvater2024-2400463} (see Methods). 

\begin{figure*}[hbt!]%
\centering
\includegraphics[width=0.9\linewidth]{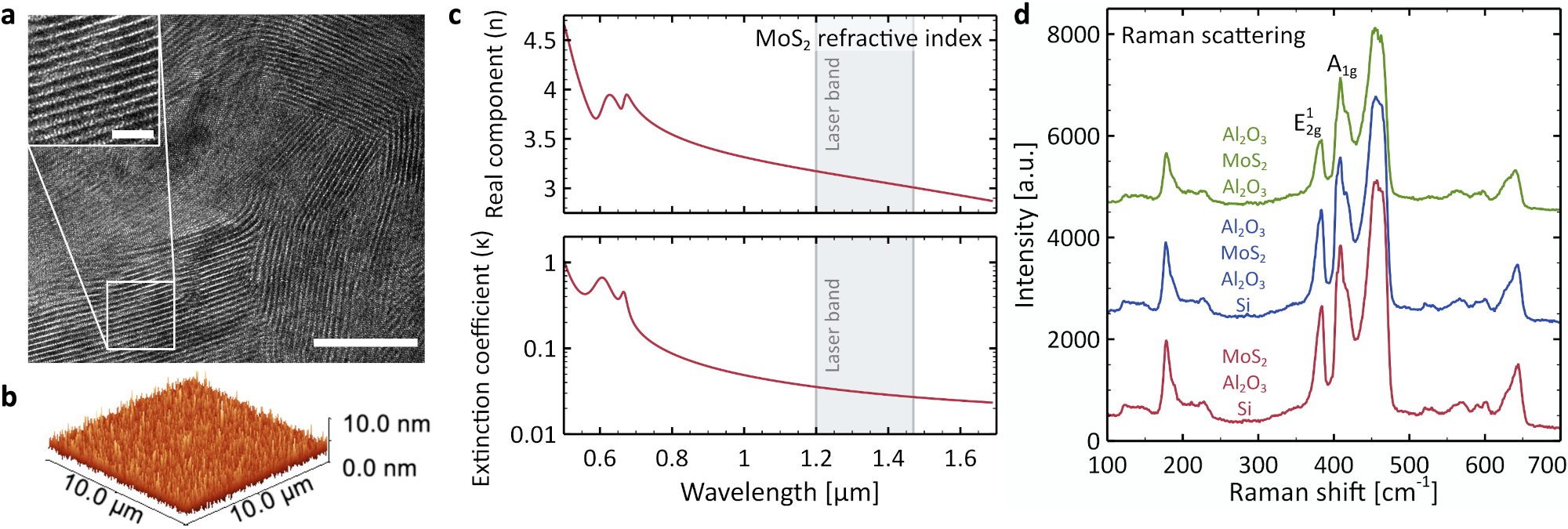}
\caption{\textbf{\textbar~ Characterization of thick \ch{MoS2} films.} (\textbf{a}) Transmission electron micrograph (TEM) of a roughly 75-\si{\nano\meter}-thick sample.  (\textbf{b}) Atomic force microscopy (AFM) scan of a 45-\si{\nano\meter}-thick sample (RMS roughness: 1~\si{\nano\meter}). (\textbf{c}) Refractive index information $\mathfrak{n}=n+\ji\kappa$ for a \ch{MoS2} film grown on \ch{Al2O3}, obtained \via spectroscopic ellipsometry. Here and elsewhere, the shaded gray range denotes the \emph{laser band}, or the Doppler-shifted wavelength range corresponding to a final relative velocity of $\beta_f=\frac{v_f}{c}=0.2$ ($v_f$ is the sail velocity and $c$ is the speed of light; $\lambda = \langle 1.2 , 1.4697 \rangle$~\si{\micro\meter}). (\textbf{d}) Raman scattering measurements of a \ch{MoS2} film on an \ch{Al2O3}-coated \ch{Si} substrate: uncovered (bottom record), with an \ch{Al2O3} top coating (middle record), and with the \ch{Si} substrate removed (top record).  Scale bars: (\textbf{a}) 10~\si{\nano\meter}, inset of (\textbf{a}): 2~\si{\nano\meter}. }
\label{F:opticalChar}
\end{figure*}

Using this approach, we produced thick \ch{MoS2} films on silicon, silicon dioxide, and alumina substrates and interrogated them to verify their quality (see Methods). Figure~\ref{F:opticalChar}(a) presents a planar-view transmission electron micrograph (TEM) of a \ch{MoS2} film transferred from a \ch{SiO2}/\ch{Si} substrate to a \ch{Cu}-TEM grid that clearly shows individual layers, indicating that the film has a vertically-oriented multilayer van der Waals structure (see also Figure~S7 in the Supplementary Information for cross-sectional TEM images). Panel (b) provides an atomic force microscopy (AFM) scan of a 45-\si{\nano\meter}-thick sample. The root-mean-square (RMS) roughness is just 1~\si{\nano\meter} over the 100~\si{\micro\meter\squared} area, indicating the film is morphologically smooth. 

Furthermore, we performed a spectroscopic ellipsometry (SE) analysis on a 60-\si{\nano\meter}-thick sample (Figure~\ref{F:opticalChar}(c)), finding that a graph of the complex index of refraction $\mathfrak{n}=n+\ji\kappa$ versus wavelength showed the same trends as measurements of bulk and monolayer samples in the literature~\cite{Song2019-1801250, Ermolaev2020-21, Islam2021-2000180, Munkhbat2022-2398} (see Figure~S16 in the Supplementary Information). The low extinction coefficient of $\kappa<0.1$ is consistent with complete transformation of the \ch{Mo} into \ch{MoS2}, and also indicates a low optical absorptivity. Finally, our Raman scattering measurements (panel (d)) confirm the formation of a crystalline multilayered \ch{MoS2} structure while negating any remaining oxide-based compounds.

\begin{figure*}[hbt!]%
\centering
\includegraphics[width=0.9\linewidth]{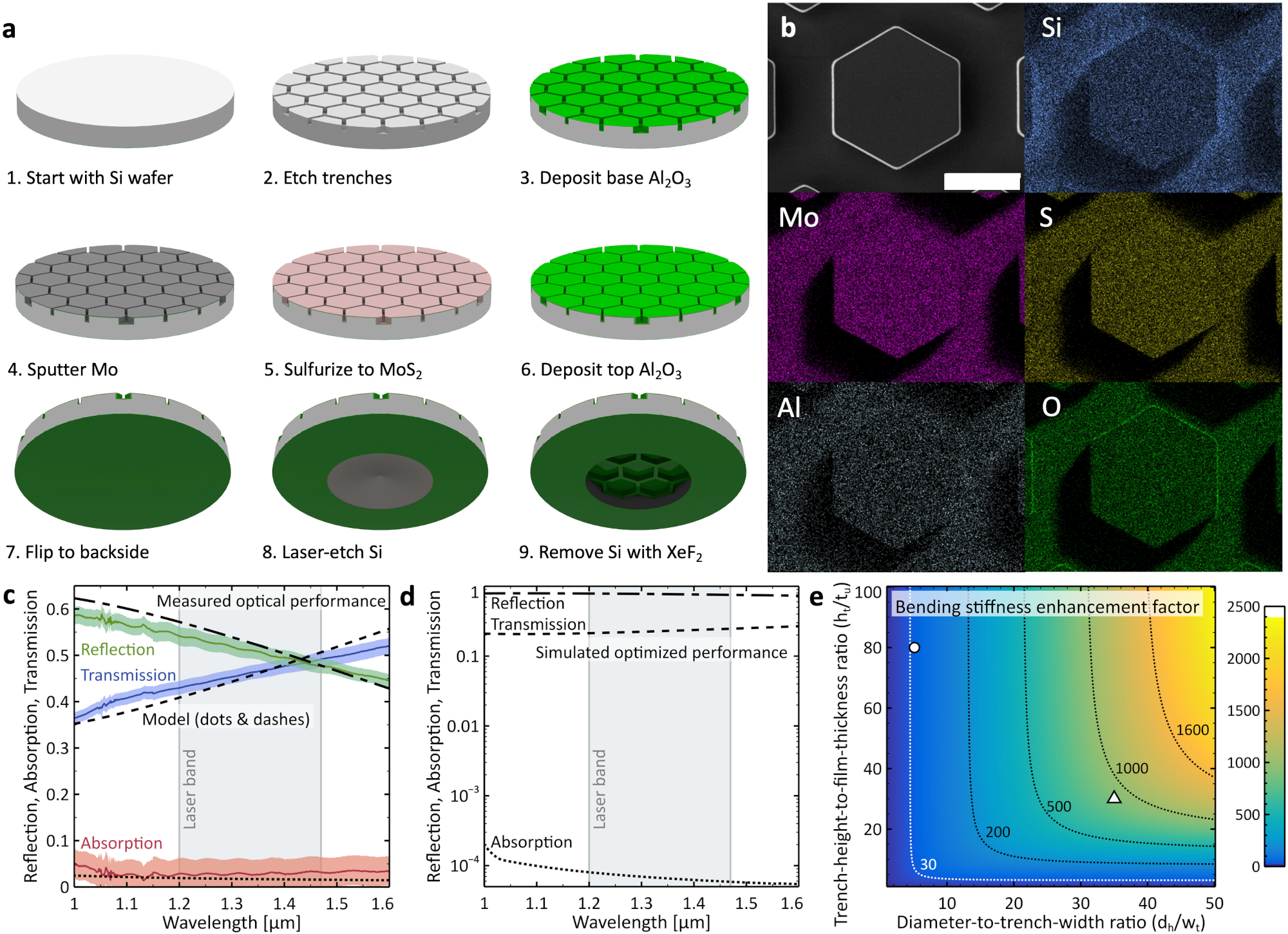}
\caption{\textbf{\textbar~ Fabrication and characterization of corrugated \ch{Al2O3}-\ch{MoS2}-\ch{Al2O3} films.} (\textbf{a}) Fabrication steps for sail film prototypes in the indented trench configuration. (\textbf{b}) Scanning electron micrograph (SEM) and energy dispersive X-ray spectroscopy (EDS) images of multilayer film on \ch{Si} substrate. The presence of color in the images corresponds to the presence of the indicated element. See Figure~S8 in the Supplementary Information for the associated spectrum. (\textbf{c}) Experimentally-measured normal-incidence reflectivity $(\varrho_{\lambda,\perp})$ and transmissivity $(\tau_{\lambda,\perp})$ spectra for fabricated prototype film, along with resulting absorptivity $(\alpha_{\lambda,\perp}=1-\varrho_{\lambda,\perp}-\tau_{\lambda,\perp})$ spectrum.  Shaded areas for $\varrho_{\lambda,\perp}$ and $\tau_{\lambda,\perp}$ reflect a 3\% uncertainty along with the maximum variation among three sets of measurements. Black dash/dot lines show simulated spectra obtained using the transfer-matrix method with complex indices of refraction obtained from ellipsometry measurements (see Methods and Section~S6 of the Supplementary Information). (\textbf{d}) Simulated spectrum for proposed optimized film, calculated using indices of refraction obtained from the literature~\cite{Munkhbat2022-2398, Kischkat2012-6789}. (\textbf{e}) Calculated bending stiffness enhancement factor of hexagonally corrugated structures relative to their planar (non-corrugated) counterparts~\cite{Davami2015-10019} as a function of the ratio of the hexagon diameter to trench width ($\frac{d_h}{w_t}$, ordinate) and the ratio of the trench height to the total composite film thickness ($\frac{h_t}{t_u}$, abscissa). Circle marks fabricated prototype film and triangle marks feasible proposed optimized film (see Section~S3 of the Supplementary Information). Prototype film dimensions: $d_h\approx77~\si{\micro\meter}$ (measured flat-to-flat), $w_t\approx15~\si{\micro\meter}$, $h_t\approx10~\si{\micro\meter}$, $t_{A,b}\approx 21~\si{\nano\meter}$ (bottom \ch{Al2O3} thickness), $t_{M}\approx 53~\si{\nano\meter}$ (\ch{MoS2}), $t_{A,t}\approx 51~\si{\nano\meter}$ (top \ch{Al2O3}), $\rho_{a,c}\approx0.7$~\si{\gram\per\meter\squared} (corrugated film areal density). Proposed optimized film dimensions: $d_h=70~\si{\micro\meter}$, $w_t=2~\si{\micro\meter}$, $h_t=3~\si{\micro\meter}$, $t_{A,b}=t_{A,t}\approx 19~\si{\nano\meter}$, $t_{M}\approx 63~\si{\nano\meter}$, $\rho_{a,c}\approx0.5$~\si{\gram\per\meter\squared}.  We define the total/unified film thickness to be $t_u=t_{A,b}+t_{M}+t_{A,t}$. Scale bar: (\textbf{b}) 20~\si{\micro\meter}. }
\label{F:compositeChar}
\end{figure*}

Our thick \ch{MoS2} films provide favorable optical constants for reflection but must be augmented to achieve sufficient thermal management and mechanical robustness. For thermal management, we sandwiched our multilayer \ch{MoS2} films between thin alumina layers, creating the film's characteristic nanolaminate structure (see Figure~\ref{F:compositeChar}(a)). The \ch{Al2O3} is minimally absorbing within the laser wavelength range but emits strongly at longer infrared wavelengths~\cite{Lingart1982-706, Querry1985-report, Kischkat2012-6789}, allowing the sail to reradiate the tiny fraction of incident laser photon energy it absorbs~\cite{Campbell2022-90}. For mechanical robustness, we imparted an out-of-plane corrugated microstructure pattern to our films by sequentially depositing the film layers on \ch{Si} substrate molds consisting of hexagons separated by either trenches cut into the substrate or ribs protruding from it (see Methods). The out-of-plane ``U''-shapes of this pattern greatly increase the bending stiffness of the architected films relative to their non-corrugated counterparts, much like the vertical section of an ``I''-beam increases its rigidity by increasing its bending moment of inertia~\cite{Davami2015-10019}.

After forming the corrugated molds and depositing the \ch{Al2O3} and \ch{MoS2} films thereon, we used a novel two-step process to suspend the sail prototypes for testing. First, we used laser ablation to remove a majority of the backside of the corrugated \ch{Si} mold. Second, we gently removed the remaining \ch{Si} using a gaseous \ch{XeF2}-etching process (see Methods). 

Having fabricated both chip-bound and freestanding film prototypes, we interrogated them to understand their composition and optical qualities. Figure~\ref{F:opticalChar}(d) provides a comparison of the Raman scattering signals for a \ch{MoS2} film on an \ch{Al2O3}-coated \ch{Si} substrate in three configurations: (1, bottom record) uncovered, (2, middle record) with an \ch{Al2O3} top coating, and (3, top record) with the \ch{Si} substrate etched away. The prominent and consistent $\text{E}^\text{1}_\text{2g}$ and $\text{A}_\text{1g}$ peaks indicate good crystallinity of the \ch{MoS2} throughout our fabrication process, and the difference in their positions (roughly 25~\si{\per\centi\meter}) is consistent throughout the fabrication process, which indicates good \ch{MoS2} uniformity~\cite{Song2019-1801250}. The absence of the characteristic peak near 520~\si{\per\centi\meter} in the top record reflects the fact that the \ch{Si} mold was completely etched away. Figure~\ref{F:compositeChar}(b) shows energy dispersive X-ray spectroscopy (EDS) images of a single hexagonal unit cell still attached to its \ch{Si} substrate, indicating the presence of all relevant elements. 

We conducted optical normal-incidence reflectivity $(\varrho_{\lambda,\perp})$ and transmissivity $(\tau_{\lambda,\perp})$ measurements of a suspended prototype film at wavelengths surrounding $\lambda_l=1.2$~\si{\micro\meter}, shown in Figure~\ref{F:compositeChar}(c) (see Methods and Section~S9 of the Supplementary Information). The data, shown with roughly 3\% uncertainty, are consistent with our thin film optical transfer-matrix method simulations (based on optical constants we measured for \ch{MoS2} and \ch{Al2O3} using spectroscopic ellipsometry; see Methods), and suggest low absorptivity ($\alpha_{\lambda,\perp}=1-\varrho_{\lambda,\perp}-\tau_{\lambda,\perp} < 0.04$ within our proposed laser band, $\lambda = \langle 1.2 , 1.4697 \rangle$~\si{\micro\meter}). To our knowledge, these are the lowest experimentally-measured absorptivity data for a prototype light sail.

In addition to being reflective and low-absorbing, our sails have low areal densities.  We obtained these values using the measured corrugation pattern dimensions and film thicknesses, together with published material density values~\cite{Groner2004-639, Ilic2010-044317, Graczykowski2017-7647}. Our corrugated films and their planar counterparts have areal densities of $\rho_{a,c}\approx0.7$~\si{\gram\per\meter\squared} and $\rho_{a,p}\approx0.5$~\si{\gram\per\meter\squared}, respectively, both of which fall within an order of magnitude of the Breakthrough Starshot Foundation's target~\cite{Parkin2018-370, Parkin2023-arXiv}. 

Although corrugated films weigh more than planar films, their structure brings critically important mechanical properties. First, corrugation reduces the films' contact areas and increases their bending stiffnesses (see Figure~\ref{F:compositeChar}(e)), such that they are less sticky, wrinkle less, and can be more readily assembled into working structures~\cite{Davami2015-10019, Jiao2019-034055}. Second, the corrugation can be tailored to increase the films' ``stretchiness'' to prevent tears at points of concentrated stress, such as those associated with sail support frames or cable tethers~\cite{Jiao2020-100599}. Third, as demonstrated by Davami~\etal~\cite{Davami2015-10019}, the corrugated structure prevents crack propagation by deflecting cracks at the vertical trench or rib walls, which is important given the likelihood of punctures by space dust~\cite{Weingartner2001-296, Early2015-205, Hoang2017-5}. Finally, our films' corrugation allows them to undergo and recover from extreme bending, which could allow fleets of light sails to be tightly folded, carried into low Earth orbit, and deployed~\cite{Jiao2019-034055}. 

A key figure of merit for light sails is their acceleration length $L$, that is, the distance they travel while being illuminated by laser-generated photons until reaching their target relative velocity, typically $\beta_f=\frac{v_f}{c}=0.2$, where $v_f$ is the final sail velocity and $c$ is the speed of light.  This length scales as $L\sim\frac{m_{tot} c v_f^2}{2 \Phi_l \varrho_a}$, where $m_{tot}$ is the total mass of the sail and payload, $\Phi_l$ is the laser output power, and $\varrho_a$ is the sail's average reflectivity near the laser wavelength~\cite{Campbell2022-90}. The acceleration length is of practical engineering importance because both the time duration that the laser array must produce photons and the laser array diameter on Earth required to perfectly focus the beam on the sail scale with it (see Section~S5 of the Supplementary Information).  We determined the acceleration trajectory for a sailcraft consisting of a $m_s=1$~\si{\gram} flat circular sail composed of the film we fabricated (using the experimentally-measured reflectivity of Figure~\ref{F:compositeChar}(c)) towing a $m_p=1$~\si{\gram} payload (tethers plus probe chip)  while illuminated by a constant $\Phi_l=100$~\si{\giga\watt} laser array with an output wavelength of $\lambda_l=1.2$~\si{\micro\meter}. Our calculations predict an acceleration length to $\beta=0.2$ of $L\approx15$~\si{\giga\meter}, a laser-on time of 7~\si{\minute}, and a required laser array diameter of 26~\si{\kilo\meter}.  These values indicate the practical feasibility of our films as light sails~\cite{Brewer2022-594, Campbell2022-90, Salary2020-1900311, Jin2020-2350}. 

\begin{figure*}[hbt!]%
\centering
\includegraphics[width=0.9\linewidth]{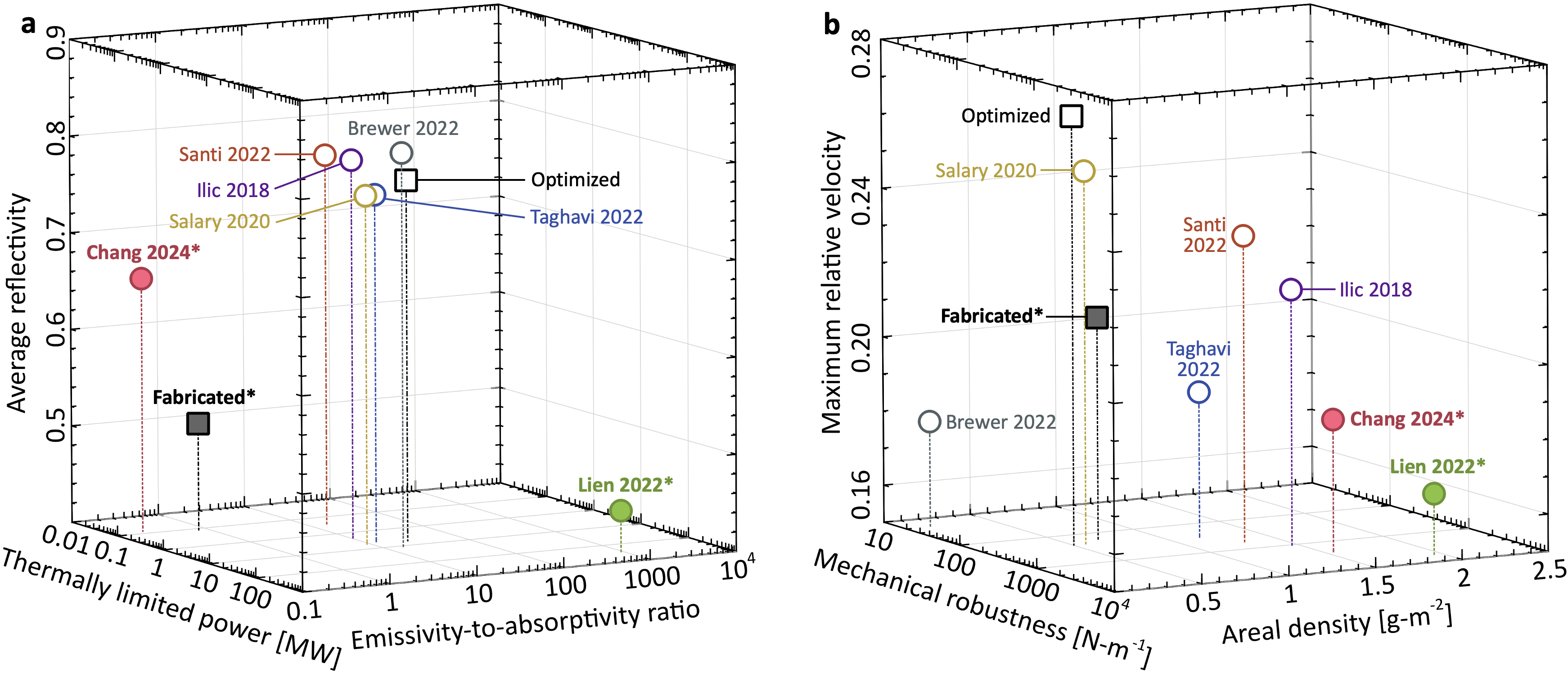}
\caption{\textbf{\textbar~ Optical, mechanical, and performance comparison of several light sail films.} Sails found in the literature are shown with circles, and our fabricated prototype and proposed optimized films are shown with squares. Solid shapes labeled with bold text and an asterisk symbol $(\ast)$ denote films that have at least some experimental measurements~\cite{Lien2022-3032, Chang2024-6689} and open shapes denote theoretical calculations~\cite{Ilic2018-5583, Taghavi2022-20034, Salary2020-1900311, Brewer2022-594, Santi2022-16}. (\textbf{a}) Comparison of laser-band average reflectivity $\overbar{\varrho_\perp}$, ratio of infrared effective emissivity (at $T=1000$~\si{\kelvin}, $\lambda=2-14$~\si{\micro\meter}) to laser-band average absorptivity $\frac{\varepsilon_e}{\overbar{\alpha_\perp}}$, and thermally limited power $\Phi_{s,max}$. (\textbf{b}) Comparison of maximum relative velocity $\beta_{max}$, areal density $\rho_a$, and membrane mechanical robustness $\wp$. The membrane mechanical robustness $\wp$ quantifies a sail's ability to resist photon pressure forces, and the maximum relative velocity $\beta_{max}$ estimates the fastest speed the sail can attain when constrained by mechanical forces and optical diffraction (see Methods). }
\label{F:benchmarking}
\end{figure*}

Our work marks an inflection point in the pursuit of interstellar travel \via unmanned probes. Never before has a macroscopic film been fabricated that comes within an order of magnitude of the required areal density while also exhibiting high broadband reflectivity and low experimentally-measured absorptivity. Figure~\ref{F:benchmarking}(a) compares our film to other sail prototypes in the literature in terms of their average reflectivity in the Doppler-shifted laser wavelength range (the \emph{laser band}), their infrared emissivity-to-laser band average absorptivity ratio for thermal management, and the maximum laser power they can tolerate without overheating (see Methods)~\cite{Ilic2018-5583, Lien2022-3032, Taghavi2022-20034, Salary2020-1900311, Brewer2022-594, Santi2022-16, Chang2024-6689}. Likewise, Figure~\ref{F:benchmarking}(b) examines the sails' acceleration performance, mechanical strength, and areal density values. Our measure of a sail's mechanical strength is a quantity that we call the \emph{membrane mechanical robustness}, defined here as the product of a film's thickness and tensile yield strength; it is indicative of a sail's resistance to failure/collapse upon incident laser photon pressure.  Similarly, our measure of accelerative performance is the maximum relative velocity $\beta_{max}$ to which a sail can be accelerated without tearing due to the photon pressure, under the assumption of zero optical absorptivity (\ie, no thermal limit, see Methods). 
Taken together, Figures~\ref{F:benchmarking}(a-b) illustrate the significant potential of our sail design for use in laser-propelled relativistic interstellar travel. Importantly, our model indicates that our film is the only one that could propel a sailcraft to the $\beta=0.2$ Starshot goal~\cite{Atwater2018-861, Parkin2018-370} based on an experimentally-measured, rather than a theoretically calculated, reflectivity spectrum. 

Our prototype sail is one embodiment among a large class of similarly-structured corrugated nanolaminate films. To demonstrate this design class's potential, under the assumption that future fabrication advances will allow thick, optically perfect, and mechanically strong \ch{MoS2} layers to be formed, we systematically changed the \ch{Al2O3} and \ch{MoS2} film thicknesses in order to maximize $\beta_{max}$ within the design space. Our models predict that the optimal film, whose simulated optical performance is shown in Figure~\ref{F:compositeChar}(d), can achieve $\beta_{max}\approx0.26$ (\ie, travel more than 25\% faster than the Starshot goal of $\beta=0.2$). It accomplishes this by decreasing the \ch{Al2O3} layer thicknesses (\app 19~\si{\nano\meter} each) to lower the areal density of the corrugated film (\app 0.5~\si{\gram\per\meter\squared}) while increasing the \ch{MoS2} thickness (\app 63~\si{\nano\meter}) to increase the reflectivity (laser band average: \app 0.77), all while maintaining the minimum required mechanical robustness to prevent tearing at the maximum laser power (see Methods). Even higher ultimate speeds could be achieved by throttling the laser power as the sail accelerates~\cite{Parkin2018-370, Campbell2022-90}. Our optimized prototype design achieves a higher $\beta_{max}$ value than all other designs in Figure~\ref{F:benchmarking}. Though the film proposed by Salary and Mosallaei~\cite{Salary2020-1900311} comes close, our optimized proposed prototype represents an experimentally-proven class of sails, suggesting the practical viability of our designs.

Although our films represent great strides in terms of reflectivity, mass, and mechanical integrity, they still need to be significantly improved in terms of their absorptivity and emissivity values. In fact, none of the sails represented in Figure~\ref{F:benchmarking} have a sufficiently high ratio of emissivity to absorptivity $(\frac{\varepsilon_e}{\alpha}\sim10^{6})$ to survive the strong laser photon fluxes required to achieve relativistic velocities within reasonable acceleration lengths (see Sections~S10 and~S14 of the Supplementary Information). 

With regard to our proposed optimized sail design, its absorptivity in the laser band will scale with the extinction coefficients of \ch{Al2O3} and \ch{MoS2}. To explore the possibility that future fabrication advances will allow minimally-absorbing films to be fabricated, we conducted a time-dependent energy balance to estimate the sail temperature during the acceleration phase (up to $\beta_f=0.2$) of our proposed optimized design using published index of refraction values for crystalline alumina~\cite{Lingart1982-706} and an extinction coefficient of $\kappa=10^{-8}$ for \ch{MoS2}, which approaches that measured for monolayers~\cite{Ermolaev2020-21} (see Section~S15 of the Supplementary Information). Even with a laser output power of 100~\si{\giga\watt}, our simulation indicated that, given these extinction coefficients, the sail temperature would not exceed the \ch{MoS2} high-vacuum sublimation point of about 1000~\si{\kelvin}~\cite{Cui2018-44}. This suggests that continuing improvements to film microfabrication processes will make our sail design viable for relativistic space flight.  Related questions remain; further research is needed to quantify the impact of temperature on thin-film absorption coefficients to predict thermal runaway scenarios~\cite{Holdman2022-2102835} and to explore the importance of two-photon absorption events for high photon fluxes~\cite{Dong2018-1558, Zhang2015-7142}. 

Our corrugated nanolaminate films have significant potential beyond relativistic interstellar travel. A more immediate application is intra-solar system transport for wafer-scale spacecraft~\cite{Tung2022-1108, Santi2023-19583}. For instance, thermal constraints aside, a 1~\si{\giga\watt} laser array with a diameter of 10~\si{\kilo\meter} could accelerate a 10-\si{gram} sailcraft employing a flat circular sail with our fabricated prototype's reflectivity (Figure~\ref{F:compositeChar}(c)), areal density, and mechanical robustness to 1\% of the speed of light $(\beta=0.01)$ in under 3~\si{\hour}, allowing it to reach Jupiter from Earth in less than three days. Such rapid transit would fundamentally alter humankind's approach to exploring our neighboring planets and sun, allowing us to take measurements with cutting-edge instruments and rapidly respond to phenomena such as sunspots, storms, and geological activity~\cite{Lainey2009-957, Chitta2023-867}. Similarly, fleets of sailcraft could be sent to pass through the focal point of the solar gravitational lens, located roughly 542~\si{AU} from the Sun, to sequentially image a distant planet with \si{\kilo\meter}-scale resolution~\cite{Einstein1936-506, Turyshev2020-044048, Turyshev2022-6122, Chang2024-6689}. 

We envision other uses for our design, as well. Our films could be employed as ultralightweight optical elements for orbiting laser-based deep space optical communication (DSOC) equipment~\cite{Deutsch2020-907}, offering high-bandwidth data transmission without the atmospheric interference and shadowing of ground-based telescopes. 
Similarly, arrays of our films could be placed at the inner Lagrangian point (L1) to deflect sunlight and cool the Earth, in the event of excessive warming and climate change~\cite{Angel2006-17184}. Furthermore, our high-quality multilayer \ch{MoS2} films, if coated with a thin metallic layer on one side, will be nearly perfectly absorbing, making them useful as components in ultra-thin metamaterial photovoltaics or photodiodes \cite{Jariwala2016-5482, Wong2017-7230, Zhang2020-3552, Kumar2022-182, Alfieri2023-2202011, Lin2024-13935, Alfieri2025-3020}. Finally, the relatively high emissivity of our designs could make them useful for passive radiative cooling of Earth-or-space-based equipment~\cite{Raman2014-540}. 
	
In summary, we created a new class of light sail films that makes interstellar travel plausible at relativistic speeds. These sails feature a nanolaminate structure consisting of high-quality multilayer \ch{MoS2} films flanked by thin \ch{Al2O3} face sheets. This combination provides high laser-band reflectivity ($>50\%$) and low absorptivity at a laser output wavelength of 1.2~\si{\micro\meter}, while offering strong emission at longer infrared wavelengths for thermal management. Our films also feature microscale out-of-plane corrugation, which makes them less prone to wrinkling. Importantly, these films' areal densities are within an order of magnitude of the target areal density of 0.1~\si{\gram\per\meter\squared}, positioning them for practical relativistic spaceflight. Moreover, we anticipate that future fabrication advances will lead to completely in-plane-aligned, near-epitaxial quality multilayer \ch{MoS2} films, which will increase the reflectivity of these sails and reduce their mass. Our reflective, strong, and lightweight nanolaminate corrugated films represent a giant leap for humankind toward Zander and Tsiolkovsky's vision of photon-driven relativistic interstellar space travel.  

\section*{Methods}


\textbf{Fabrication.}

\textbf{\emph{\ch{MoS2} films.}} We fabricated our \ch{MoS2} films \via a two-step process, in which we first deposited molybdenum onto a substrate and subsequently sulfurized it into high-quality \ch{MoS2}. We slowly deposited the \ch{Mo} onto silicon, silicon dioxide, or sapphire substrates using direct current sputtering (Denton Explorer 14, 25~\si{\watt}, 3~\si{\milli\Torr} pressure, rate $\approx 1$~\si{\nano\meter\per\minute}) in film thicknesses of roughly 15~\si{\nano\meter}. Importantly, our low deposition pressure helped to ensure highly directional sputtering, allowing uniform and conformal coverage on the vertical corrugated surfaces of our molds (trench or rib walls; see Figure~\ref{F:compositeChar})~\cite{Ji2007-1227}.  We used atomic force microscopy (see below) to measure the thicknesses of the deposited films by examining the step height created \via shadowing by the polyimide tape used to secure the substrates in the sputtering tool. For the subsequent sulfurization process, we used a horizontal tube furnace chemical vapor deposition system (Structured Materials Industries, Inc.). After placing the samples in the furnace at room temperature, we heated the tube to 750~\dC in the span of 15~\si{\minute} in a \ch{H2}/\ch{Ar} environment (volumetric flow rates: 10~\si{sccm}/100~\si{sccm}, total pressure: 10~\si{\Torr}). Upon the tube reaching 750~\dC, we introduced \ch{H2S} gas (volumetric flow rate: 25~\si{sccm}) while maintaining the same \ch{H2} and \ch{Ar} flow rates and the same total pressure of 10~\si{\Torr}. We maintained the temperature, pressure, and gas flow rate for 6~\si{\hour}, at which point we turned off the heater power and stopped the flows of \ch{H2} and \ch{H2S} (\ch{Ar} continued to flow at 100~\si{\sccm} at a total pressure of 10~\si{\Torr} during the cooling step.) The resulting films had thicknesses of roughly 60-70~\si{\nano\meter}, as measured and confirmed \via atomic force microscopy and ellipsometry. 

\textbf{\emph{Corrugated nanolaminate films.}} Our prototype fabrication process can be divided into three parts: (1) etching the shape of the corrugated \ch{Si} mold; (2) depositing the films to form the nanolaminate stack; and (3) removing the \ch{Si} mold to create a free-standing film.  These steps are summarized in Figure~\ref{F:compositeChar}(a) and in Figure~S1 in the Supplementary Information. 

We formed the corrugated molds out of 200-um-thick double-side-polished \ch{Si} wafers (p(\ch{B})-doped, resistivity 1-10~\si{\ohm\centi\meter}, University Wafer). We spin-coated an adhesion layer of SurPass~3000 (3000~\si{\revolution\per\minute} for 1~\si{\minute}) followed by a 2.5~\si{\micro\meter} layer of S1818 photoresist (3000~\si{\revolution\per\minute} for 1~\si{\minute}), and performed a soft bake of the wafer on a hot plate (115\dC for 1~\si{\minute}). We then exposed a hexagonal pattern (70~\si{\milli\joule\per\centi\meter\squared}, vacuum contact, SUSS MicroTec MA6 Gen3 Mask Aligner), developed the photoresist (AZ~300~MIF for 1.5~\si{\minute}, with a brief deionized \ch{H2O} rinse and \ch{N2} dry), and baked the wafer on a hot plate again (115\dC for 10~\si{\minute}). We used both darkfield and brightfield photomasks, which, combined with our positive-tone photoresist, allowed us to produce \ch{Si} molds with indented trenches and protruding ribs, respectively. The hexagon diameter $d_h$ (flat-to-flat) and trench/rib width ($w_t$ or $w_r$) are important design parameters. For the purposes of this study, to avoid tears and folds that are possible when handling in air, we selected low diameter-to-trench/rib-width ratios. There is a practical lower limit to this ratio, however; designs should at least select $d_h\ge3w_t$ (the same holds for $w_r$) to prevent film creases at vertical hexagon walls~\cite{Jiao2019-034055}. An example of this can be seen in Figure~\ref{F:overview}(c), where we selected $d_h\approx36~\si{\micro\meter}$ and $w_t\approx15~\si{\micro\meter}$ to allow the film to bend and show its three-dimensional architecture. For future designs to be handled in vacuum environments, significant improvements in stiffness could be realized by using thinner trenches/ribs (lower $w_t$ or $w_r$) or larger diameter hexagons (higher $d_h$); see Figure~\ref{F:compositeChar}(e).  

Continuing in our fabrication process, we used deep reactive ion etching (DRIE) to remove the exposed \ch{Si} areas (hexagon edges to form indented trenches or hexagon areas to form protruding ribs) to a depth of roughly 10~\si{\micro\meter} (as verified using stylus profilometry with a KLA Tencor P7 2D machine). We then stripped the photoresist using sonication (70\dC in KL Remover~1000 for 10~\si{\minute}) and \ch{O2} plasma ashing (1~\si{\hour}, 300~\si{\watt}, 100~sccm \ch{O2}, Anatech SCE-108 Barrel Asher). In some cases, we found that, due to their high aspect ratios, some of the protruding ribs cracked during the sonication process. A simple workaround was to replace the sonication step with a longer quiescent soak. The wafers designed with trenches were not subject to cracking, however. Finally, we cleaved the wafer into 1.5~\si{\centi\meter} squares. 

We deposited the film layers sequentially as follows. First, we used atomic layer deposition (ALD) at 250\dC to coat the chips with alumina (\ch{Al2O3}), using precursors water and trimethylaluminium (Cambridge Nanotech Savannah S-200 reactor). ALD is a conformal technique, such that the vertical sides of the trenches/ribs of the \ch{Si} mold, as well as the horizontal areas, were coated. Importantly, we positioned the chips on standoffs in the chamber, such that both their front and back sides were coated, completely encapsulating the \ch{Si}. This alumina coating typically had a thickness of 20-50~\si{\nano\meter}, as controlled by the number of deposition cycles and measured using a spectroscopic ellipsometer (either Woollam V-VASE or Filmetrics F40) on a \ch{Si} test chip placed in the chamber during the deposition. Next, we formed the \ch{MoS2} layer using the process outlined above, which involved first conformally sputtering a \ch{Mo} film (Denton Explorer14 Magnetron Sputterer) and subsequently sulfurizing it in a high-temperature tube furnace. Thirdly, we used ALD a second time to conformally coat the front and backsides of the chips with \ch{Al2O3}, again at 250\dC using the precursors and reactor listed above.  

Lastly, we removed the \ch{Si} substrate in the area of interest, usually a 0.5-3~\si{\milli\meter} diameter circle in the center of the chip, to reveal a fully-suspended corrugated nanolaminate film. This process involved two steps.  First, we rastered a 532-\si{\nano\meter} laser (IPG IX-280-DXF) over the backside of the chips to ablate away the \ch{Al2O3} coating and a majority of the \ch{Si} substrate. We carefully controlled the power and cycle time such that the etch depth was deepest (more material removed) in the center of the area and shallower at the perimeter. This promoted etching of the area from the center towards the perimeter in the subsequent \ch{XeF2} step, which protected the suspended film against tears. We also used a \ch{N2} purge to partially suppress the thermal formation of \ch{SiO2} dust in the ablation region. Next, we used gaseous \ch{XeF2} etching to remove the remaining backside \ch{Si} in the area of interest (SPTS Xactix). We performed etching in cycles to prevent overheating, where each cycle consisted of a 75~\si{\second} long etch in 3~\si{\Torr} undiluted \ch{XeF2} followed by a 3~\si{\second} vacuum rest. Importantly, we designed and 3D-printed a miniature fixture (FormLabs Form3 printer, standard resin, highest resolution) to fully enclose each chip as it was etched (see Figure~S6 in the Supplementary Information). The fixture held the chips in a vertical orientation and contained small (750~\si{\micro\meter} diameter) holes to expose both sides of the chip to gas at equal pressures, which we found to be important in preventing the thin released films from tearing when venting the chamber back to atmospheric pressure after the etching process was complete. \ch{XeF2} etching is highly selective to \ch{Si} over \ch{Al2O3}, allowing us to easily remove the \ch{Si} substrate in the area where the alumina had been laser-etched away.  Some lateral \ch{Si} etching also occurred during the \ch{XeF2} process, which slightly increased the area of the suspended film over that opened in the backside by the laser-ablation step. We also remark that the pinhole-free nature of ALD is important in preserving the \ch{MoS2} during the \ch{XeF2} process (\ch{XeF2} etches \ch{MoS2}), because the \ch{MoS2} was completely and conformally encapsulated by the \ch{Al2O3}. 

Finally, we note that the above process enabled us to produce corrugated films suspended on \ch{Si} substrate frames, which were ideal for optical characterization purposes.  However, the above process would be unsuitable for producing the large-area films required to construct a light sail, for multiple reasons including the fact that commercial \ch{Si} wafers are currently limited to roughly 300~\si{\milli\meter} diameter sizes~\cite{Chang2024-6689, Norder2025-2753}. We envision that roll-to-roll processing techniques conducted directly in the vacuum of low-Earth orbit would enable significant advances in terms of both areal dimensions and optical quality~\cite{Gupta2022-2109105}.

\textbf{Characterization.}

\textbf{\emph{Atomic force microscopy.}} We measured the thicknesses of our \ch{Mo} and \ch{MoS2} films and characterized the roughness of our \ch{MoS2} films using an atomic force microscope in tapping mode (AIST-NT Co.). We processed the images using Gwyddion software~\cite{Necas2012-181}. 

\textbf{\emph{Spectroscopic ellipsometry.}} We performed spectroscopic ellipsometry measurements using a J.\ A.\ Woollam W-VASE ellipsometer at wavelengths of roughly~0.4 to~1.7~\si{\micro\meter}. We used the CompleteEase software package to fit the data using a multi-Lorentz oscillator method to minimize the root mean square error~\cite{Fujiwara2007-book}. 

\textbf{\emph{Raman scattering.}} We used a confocal micro-Raman spectroscopy system (LabRAM HR Evolution, HORIBA Instruments, Inc.) to measure all the Raman scattering characteristics of our multi-layered \ch{MoS2} films. We used a 633~\si{\nano\meter} laser to excite the sample while keeping a 1800~\si{\groove\per\milli\meter} grating to obtain high spectral width resolution. We used a 3.2\% neutral density filter to keep the laser excitation power within a safe range to prevent local heating/modulation during the sample measurement. We used the same measurement conditions for Raman imaging, achieving a very high spatial resolution through the use of 200~\si{\nano\meter\per\pixel} steps (governed by a M\"{a}rzh\"{a}user Wetzlar X-Y stage). 

\textbf{\emph{Transmission electron micrography.}} We used a JEOL F200 system to acquire high-resolution transmission electron micrographs (TEM).  The instrument was equipped with a 200~\si{\kilo\volt} cold-field emission electron source and two detectors (Gatan, Inc.), and we used Gatan DigitalMicrograph 3.5 software to acquire images. We prepared our samples according to two different methods. For planar views, we used the wet-chemical transfer (``scooping'') technique~\cite{Kumar2022-182} to port films grown on \ch{Si}/\ch{SiO2} substrates directly to TEM grids.  For cross-sectional images, we prepared samples using the \ch{Xe} plasma focused ion beam approach~\cite{Vitale2022-646} and subsequently transferred them to half-grids using the \insitu liftoff technique~\cite{Kim2023-1044}. 

\textbf{\emph{Scanning electron micrography and energy dispersive X-ray spectroscopy.}} We acquired our scanning electron micrographs (SEM) and energy dispersive X-ray images using a TESCAN S8000X machine.  In SEM mode we used an electron beam energy of 5~\si{\kilo\electronvolt} and an in-lens detector to capture high-resolution pictures. 

\textbf{\emph{Laser reflection and transmission.}} Briefly, we focused tunable, chopped, monochromatic, linearly polarized laser light onto our fully suspended nanolaminate film samples (10-\si{\micro\meter}-diameter spot size, centered within a single hexagon unit cell) using a near-infrared objective ($20\times$, numerical aperture $NA = 0.4$, Mitutoyo). We collected the reflected and transmitted light using \ch{Ge} photodetectors, whose signals we amplified and measured using lock-in amplifiers. We normalized the reflectivity measurements using the reflection from a template-stripped, flat \ch{Au} sample on \ch{Si}, whose wavelength-dependent reflectivity we calculated based on literature-obtained indices of refraction~\cite{Olmon2012-235147, Schinke2015-067168}. To account for source intensity fluctuations, we normalized all experiments by a simultaneous reference measurement of the source beam. Additional information, including a schematic diagram and data processing methods, is provided in Section~S9 of the Supplementary Information. 

\textbf{Simulations.}

\textbf{\emph{Optical comparison of films.}} We simulated the optical performance of our sails, as well as that of other published sails, using the transfer-matrix method~\cite{Macleod2017-book, Campbell2022-90} with optical constants that we either measured directly, extrapolated from our direct measurements (in the case of \ch{MoS2} in the infrared), or obtained from the literature \cite{Querry1985-report, Poruba2000-148, Kischkat2012-6789, RodriguezdeMarcos2016-3622, Franta2017-405, Munkhbat2022-2398}. 
Additional details, including the optical data that we measured, are included in Section~S8 of the Supplementary Information. 

\textbf{\emph{Thermally limited power comparison of films.}} We estimated the maximum power that each film could tolerate up to its thermal limit according to $\Phi_{s,max} = 2 \frac{\varepsilon_e}{\overbar{\alpha_\perp}} A_\perp \sigma T_{max}^4$. Here, $\varepsilon_e$ is the effective emissivity at $T=1000$~\si{\kelvin} (calculated in the wavelength range $\lambda=2-14$~\si{\micro\meter}, see Section~S10 of the Supplementary Information), $\overbar{\alpha_\perp}$ is the laser-band average absorptivity, $A_\perp$ is the area of the sail perpendicular to the laser beam, $\sigma$ is the Stefan-Boltzmann constant, and $T_{max}$ is the limiting temperature of the sail~\cite{Schneider1967-317, Cui2018-44, Nannichi1963-586, Batha1973-365, Liehr1987-1559, Mizuno2002-1716}. We estimated the absorptivity for each sail using $\alpha_{\lambda,\perp} = \sum_{i=1}^{n} \! \frac{4 \pi \kappa_{\lambda,i} t_{f,i} F_i}{\lambda}$, where $\kappa_{\lambda,i}$ is the extinction coefficient of the $i$\textsuperscript{th} layer ($n$ total layers) at wavelength $\lambda$, $t_{f,i}$ is the $i$\textsuperscript{th} layer's thickness, and $F_i$ is the layer's material fill factor, estimated using the given geometric pattern of the sail. In our calculations, we used a consistent temperature of $T=1000$~\si{\kelvin} to calculate the effective emissivity, which is a reasonable simplification given the other uncertainties in this calculation, such as the temperature dependence of the indices of refraction of the sail materials. 

\textbf{\emph{Mechanical comparison of films.}} We used the areal density values provided in each article~\cite{Ilic2018-5583, Lien2022-3032, Taghavi2022-20034, Salary2020-1900311, Brewer2022-594, Santi2022-16, Chang2024-6689}, if available, or calculated them from the specified film thicknesses using material densities available in the literature~\cite{Groner2004-639, Ilic2010-044317, Graczykowski2017-7647, Petersen1982-420, Yen2003-1895, Kawase2009-101401, Saari2022-15357}. The membrane mechanical robustness, here denoted $\wp$, quantifies the ability of a sail film to resist the membrane stresses that occur due to the reflected photon pressure as it accelerates. We calculated these values according to $\wp = \sum_{i=1}^{n} G_i t_{f,i} \sigma_{y,i}$, where $t_{f,i}$ is the $i$\textsuperscript{th} layer's thickness ($n$ total layers), $\sigma_{y,i}$ is the tensile yield stress of the material's $i$\textsuperscript{th} layer~\cite{Sato1998-148, Sharpe1999-162, Tsuchiya2000-286, Yoshioka2000-291, Tsuchiya2005-665, Tavares2008-1434, Tsuchiya2010-1, Miller2010-58, Jen2011-084305, Bertolazzi2011-9703, Borgese2012-2459, Graczykowski2017-7647, Sledzinska2020-1169}, and $G_i$ are geometrical parameters that account for the presence of holes or other architected features in the film layers (see Section~S11 of the Supplementary Information). Our measure reflects the fact that a sail will tear if the stress it experiences is greater than its yield stress, and that the stress the film experiences is inversely proportional to its thickness~\cite{Timoshenko1959-book, Campbell2022-90}. For our designs, we conservatively set $G=1$ to examine the robustness of the film in a planar (non-corrugated) state, although the hexagonal structure can be designed to increase compliance to in-plane stresses, thereby increasing the practical membrane mechanical robustness~\cite{Jiao2020-100599}. Values for the films' membrane mechanical robustness range from about 10~\si{\newton\per\meter} to 10~\si{\kilo\newton\per\meter}, and that which we calculated for our fabricated prototype film is 184~\si{\newton\per\meter}. 

\textbf{\emph{Accelerative comparison of films.}} The maximum relative velocity, denoted $\beta_{max}$, achievable by a light sail is a benchmark for its practical accelerative performance. In our simulations, we allowed $\beta_{max}$ to depend on each sail's reflectivity, areal density, and mechanical robustness, but did not consider thermal constraints due to the relatively high absorptivity values of the films. To calculate the $\beta_{max}$ values, we simulated the acceleration of a $m_p=1$~\si{\gram} payload (tethers plus probe) towed by a $m_s=1$~\si{\gram} spherically-curved circular sail with equal radius of curvature and diameter $(s_s=d_s)$ having the areal density, mechanical robustness, and spectral reflectivity provided in or derived from each corresponding article~\cite{Ilic2018-5583, Lien2022-3032, Taghavi2022-20034, Salary2020-1900311, Brewer2022-594, Santi2022-16, Chang2024-6689}. Although we allowed the reflectivity to change with the Doppler-shifted wavelength, we did not consider the impact of the sail curvature on reflectivity and instead assumed that all laser photons were reflected normal to the sail surface. For each simulation, we iteratively determined the maximum constant-output laser power that the sail could sustain without tearing due to the photon pressure, up to a maximum of $\Phi_l^{max}=100$~\si{\giga\watt}~\cite{Campbell2022-90}.  We allowed each sail to accelerate until the distance it traveled $L$ reached the diffraction-limited distance to which the laser output light from a $d_{l,E}=30$~\si{\kilo\meter} diameter laser array on Earth could be focused; this distance scales as $L \sim \frac{d_{l,E}d_s}{2\lambda_l}$, where $\lambda_l$ is the laser output wavelength and $d_s$ is the sail diameter~\cite{Campbell2022-90}.  Additional details can be found in Sections~S12 and~S14 of the Supplementary Information.

\textbf{\emph{Optimization of corrugated nanolaminate design.}} To determine the film thicknesses that maximized $\beta_{max}$ within our design space, we numerically simulated a range of combinations of \ch{Al2O3} and \ch{MoS2} thicknesses ($t_A$ and $t_M$, respectively, with identical top and bottom alumina thicknesses). Anticipating future improvements in microfabrication, we used literature data for the indices of refraction of \ch{MoS2} and \ch{Al2O3}~\cite{Munkhbat2022-2398, Kischkat2012-6789} rather than the optical properties that we measured in-house, and we adopted a tensile yield strength for \ch{MoS2} that is 10\% of the single-layer crystalline value~\cite{Bertolazzi2011-9703}; our estimate is reasonable given comparable data for a similar material, \ch{MoSe2}~\cite{Babacic2021-2008614}. For simplicity, we selected a single hexagonal corrugation pattern for this optimization, with hexagon diameter $d_h=70~\si{\micro\meter}$, trench width $w_t=2~\si{\micro\meter}$, and trench height $h_t=3~\si{\micro\meter}$. More details are available in Section~S13 of the Supplementary Information. 

\textbf{\emph{Thermal performance of sail.}} We simulated the thermal performance of our proposed optimized film by solving the energy equation for laser photon energy absorbed and thermal energy radiated away. We did this by stepping through the acceleration phase from $\beta=0$ to $\beta=0.2$ and balancing the energy with a steady-state assumption, which is reasonable given the small thermal inertia (\ie, low mass) of the sail compared to the enormous incoming laser power. We obtained the required optical properties from Munkhbat~\etal~\cite{Munkhbat2022-2398}, Lingart, Petrov, and Tikhonova~\cite{Lingart1982-706}, and Querry~\cite{Querry1985-report}, except, as mentioned in the text, we specified the extinction coefficient of \ch{MoS2} to be $\kappa=10^{-8}$ at wavelengths equal to and longer than the laser wavelength. Additional details can be found in Section~S15 of the Supplementary Information. 


\section*{Data availability}
The data that support the findings of this study are available within
the Article and its Supplementary Information.

\section*{Acknowledgments}

We thank the University of Pennsylvania Libraries' Holman Biotech Commons for 3D-printing our \ch{XeF2} etching fixtures. 
We further thank Miranda Stern for assistance in fabricating the fixtures used to ship prototype films from Philadelphia, PA to Pasadena, CA for optical testing. 
This work was supported by the Breakthrough Initiatives, a division of the Breakthrough Prize Foundation. 
It was carried out in part at the Singh Center for Nanotechnology, which is supported by the NSF National Nanotechnology Coordinated Infrastructure Program under grant NNCI-2025608.
It was also funded in part by a National Science Foundation CAREER award under grant CBET-1845933. 
M.F.C is partially supported by the National Institutes of Health under grant number K25-AI-166040-01. 
During a portion of this work, J.B. was supported by a National Science Foundation Graduate Research Fellowship under grants DGE-1650605 and DGE-2034835. 
T.J.C.\ was supported by a NASA Space Technology Graduate Research Opportunities fellowship under grant 80NSSC20K1191.

\section*{Author contributions}
M.F.C., P.K., E.S., A.P.R., H.A.A., I.B., and D.J.\ conceptualized the study.
M.F.C., P.K., J.L., R.G., A.A., J.B., M.K., I.B., and D.J.\ conceived the methodology. 
M.F.C., P.K., J.L., T.J.C., and M.A.\ performed the fabrication.
P.K., J.L., and A.A.\ performed the film characterization.
R.G.\ and M.K.\ performed the optical measurements.
M.F.C.\ performed the numerical modeling. 
M.F.C.\ and P.K.\ made the figures. 
M.F.C., P.K., I.B., and D.J.\ wrote the manuscript. 

\section*{Competing interests}
The authors declare no competing interests.


\bibliography{sailProtoSourcesNatMater}


\begin{thebibliography}{173}
\ifx \bisbn   \undefined \def \bisbn  #1{ISBN #1}\fi
\ifx \binits  \undefined \def \binits#1{#1}\fi
\ifx \bauthor  \undefined \def \bauthor#1{#1}\fi
\ifx \batitle  \undefined \def \batitle#1{#1}\fi
\ifx \bjtitle  \undefined \def \bjtitle#1{#1}\fi
\ifx \bvolume  \undefined \def \bvolume#1{\textbf{#1}}\fi
\ifx \byear  \undefined \def \byear#1{#1}\fi
\ifx \bissue  \undefined \def \bissue#1{#1}\fi
\ifx \bfpage  \undefined \def \bfpage#1{#1}\fi
\ifx \blpage  \undefined \def \blpage #1{#1}\fi
\ifx \burl  \undefined \def \burl#1{\textsf{#1}}\fi
\ifx \doiurl  \undefined \def \doiurl#1{\url{https://doi.org/#1}}\fi
\ifx \betal  \undefined \def \betal{\textit{et al.}}\fi
\ifx \binstitute  \undefined \def \binstitute#1{#1}\fi
\ifx \binstitutionaled  \undefined \def \binstitutionaled#1{#1}\fi
\ifx \bctitle  \undefined \def \bctitle#1{#1}\fi
\ifx \beditor  \undefined \def \beditor#1{#1}\fi
\ifx \bpublisher  \undefined \def \bpublisher#1{#1}\fi
\ifx \bbtitle  \undefined \def \bbtitle#1{#1}\fi
\ifx \bedition  \undefined \def \bedition#1{#1}\fi
\ifx \bseriesno  \undefined \def \bseriesno#1{#1}\fi
\ifx \blocation  \undefined \def \blocation#1{#1}\fi
\ifx \bsertitle  \undefined \def \bsertitle#1{#1}\fi
\ifx \bsnm \undefined \def \bsnm#1{#1}\fi
\ifx \bsuffix \undefined \def \bsuffix#1{#1}\fi
\ifx \bparticle \undefined \def \bparticle#1{#1}\fi
\ifx \barticle \undefined \def \barticle#1{#1}\fi
\bibcommenthead
\ifx \bconfdate \undefined \def \bconfdate #1{#1}\fi
\ifx \botherref \undefined \def \botherref #1{#1}\fi
\ifx \url \undefined \def \url#1{\textsf{#1}}\fi
\ifx \bchapter \undefined \def \bchapter#1{#1}\fi
\ifx \bbook \undefined \def \bbook#1{#1}\fi
\ifx \bcomment \undefined \def \bcomment#1{#1}\fi
\ifx \oauthor \undefined \def \oauthor#1{#1}\fi
\ifx \citeauthoryear \undefined \def \citeauthoryear#1{#1}\fi
\ifx \endbibitem  \undefined \def \endbibitem {}\fi
\ifx \bconflocation  \undefined \def \bconflocation#1{#1}\fi
\ifx \arxivurl  \undefined \def \arxivurl#1{\textsf{#1}}\fi
\csname PreBibitemsHook\endcsname

\bibitem[\protect\citeauthoryear{Tsander}{1964}]{Tsander1964-book}
\begin{bbook}
\bauthor{\bsnm{Tsander}, \binits{F.A.}}:
\bbtitle{Problems of Flight by Jet Propulsion: {I}nterplanetary Flights},
\bedition{2$^{\text{nd}}$} edn.
\bpublisher{Israel Program for Scientific Translations},
\blocation{Jerusalem, {I}srael}
(\byear{1964}).
\bcomment{Translation of ``Problema Poleta Pri Pomoshchi Reaktivnykh Apparatov:
  {M}ezhplanetnye Polety''}
\end{bbook}
\endbibitem

\bibitem[\protect\citeauthoryear{Tsander}{1967}]{Tsander1967-TTF541}
\begin{botherref}
\oauthor{\bsnm{Tsander}, \binits{F.A.}}:
From a scientific heritage: {B}ibliography of works of {S}oviet engineer {F}.
  {A}. {T}sander.
Technical Report NASA-TT-F-541,
{NASA}
(1967).
Translation of ``Iz Nauchnogo Naslediya''
\end{botherref}
\endbibitem

\bibitem[\protect\citeauthoryear{Marx}{1966}]{Marx1966-22}
\begin{barticle}
\bauthor{\bsnm{Marx}, \binits{G.}}:
\batitle{Interstellar vehicle propelled by terrestrial laser beam}.
\bjtitle{Nature}
\bvolume{211}(\bissue{5044}),
\bfpage{22}--\blpage{23}
(\byear{1966})
\doiurl{10.1038/211022a0}
\end{barticle}
\endbibitem

\bibitem[\protect\citeauthoryear{Lubin}{2016}]{Lubin2016-40}
\begin{barticle}
\bauthor{\bsnm{Lubin}, \binits{P.}}:
\batitle{A roadmap to interstellar flight}.
\bjtitle{J. Br. Interplanet. Soc.}
\bvolume{69},
\bfpage{40}--\blpage{72}
(\byear{2016})
\end{barticle}
\endbibitem

\bibitem[\protect\citeauthoryear{Lubin and Hettel}{2020}]{Lubin2020-9}
\begin{barticle}
\bauthor{\bsnm{Lubin}, \binits{P.}},
\bauthor{\bsnm{Hettel}, \binits{W.}}:
\batitle{The path to interstellar flight}.
\bjtitle{Acta Futura}
\bvolume{12},
\bfpage{9}--\blpage{44}
(\byear{2020})
\doiurl{10.5281/zenodo.3747262}
\end{barticle}
\endbibitem

\bibitem[\protect\citeauthoryear{Atwater et~al.}{2018}]{Atwater2018-861}
\begin{barticle}
\bauthor{\bsnm{Atwater}, \binits{H.A.}},
\bauthor{\bsnm{Davoyan}, \binits{A.R.}},
\bauthor{\bsnm{Ilic}, \binits{O.}},
\bauthor{\bsnm{Jariwala}, \binits{D.}},
\bauthor{\bsnm{Sherrott}, \binits{M.C.}},
\bauthor{\bsnm{Went}, \binits{C.M.}},
\bauthor{\bsnm{Whitney}, \binits{W.S.}},
\bauthor{\bsnm{Wong}, \binits{J.}}:
\batitle{Materials challenges for the {S}tarshot lightsail}.
\bjtitle{Nat. Mater.}
\bvolume{17}(\bissue{10}),
\bfpage{861}--\blpage{867}
(\byear{2018})
\doiurl{10.1038/s41563-018-0075-8}
\end{barticle}
\endbibitem

\bibitem[\protect\citeauthoryear{Parkin}{2018}]{Parkin2018-370}
\begin{barticle}
\bauthor{\bsnm{Parkin}, \binits{K.L.G.}}:
\batitle{The {B}reakthrough {S}tarshot system model}.
\bjtitle{Acta Astronaut.}
\bvolume{152},
\bfpage{370}--\blpage{384}
(\byear{2018})
\doiurl{10.1016/j.actaastro.2018.08.035}
\end{barticle}
\endbibitem

\bibitem[\protect\citeauthoryear{Lantin et~al.}{2022}]{Lantin2022-261}
\begin{barticle}
\bauthor{\bsnm{Lantin}, \binits{S.}},
\bauthor{\bsnm{Mendell}, \binits{S.}},
\bauthor{\bsnm{Akkad}, \binits{G.}},
\bauthor{\bsnm{Cohen}, \binits{A.N.}},
\bauthor{\bsnm{Apicella}, \binits{X.}},
\bauthor{\bsnm{Mc{C}oy}, \binits{E.}},
\bauthor{\bsnm{Beltran-{P}ardo}, \binits{E.}},
\bauthor{\bsnm{Waltemathe}, \binits{M.}},
\bauthor{\bsnm{Srinivasan}, \binits{P.}},
\bauthor{\bsnm{Joshi}, \binits{P.M.}},
\bauthor{\bsnm{Rothman}, \binits{J.H.}},
\bauthor{\bsnm{Lubin}, \binits{P.}}:
\batitle{Interstellar space biology via {P}roject {S}tarlight}.
\bjtitle{Acta Astronautica}
\bvolume{190},
\bfpage{261}--\blpage{272}
(\byear{2022})
\doiurl{10.1016/j.actaastro.2021.10.009}
\end{barticle}
\endbibitem

\bibitem[\protect\citeauthoryear{Parkin}{2023}]{Parkin2023-arXiv}
\begin{botherref}
\oauthor{\bsnm{Parkin}, \binits{K.L.G.}}:
Cost-optimal laser-accelerated lightsails.
arXiv
(2023).
{URL}: \url{https://arxiv.org/abs/2205.13138}
\end{botherref}
\endbibitem

\bibitem[\protect\citeauthoryear{Gao et~al.}{2024}]{Gao2024-4203}
\begin{barticle}
\bauthor{\bsnm{Gao}, \binits{R.}},
\bauthor{\bsnm{Kelzenberg}, \binits{M.D.}},
\bauthor{\bsnm{Atwater}, \binits{H.A.}}:
\batitle{Dynamically stable radiation pressure propulsion of flexible
  lightsails for interstellar exploration}.
\bjtitle{Nature Communications}
\bvolume{15}(\bissue{1}),
\bfpage{4203}
(\byear{2024})
\doiurl{10.1038/s41467-024-47476-1}
\end{barticle}
\endbibitem

\bibitem[\protect\citeauthoryear{Michaeli et~al.}{2025}]{Michaeli2025-inPress}
\begin{botherref}
\oauthor{\bsnm{Michaeli}, \binits{L.}},
\oauthor{\bsnm{Gao}, \binits{R.}},
\oauthor{\bsnm{Kelzenberg}, \binits{M.D.}},
\oauthor{\bsnm{Hail}, \binits{C.U.}},
\oauthor{\bsnm{Merkt}, \binits{A.}},
\oauthor{\bsnm{Sader}, \binits{J.E.}},
\oauthor{\bsnm{Atwater}, \binits{H.A.}}:
Direct radiation pressure measurements for lightsail membranes.
Nature Photonics,
(2025)
\doiurl{10.1038/s41566-024-01605-w}
\end{botherref}
\endbibitem

\bibitem[\protect\citeauthoryear{Lin et~al.}{2025}]{Lin2025-arxiv}
\begin{botherref}
\oauthor{\bsnm{Lin}, \binits{J.Y.}},
\oauthor{\bsnm{{de Sterke}}, \binits{C.M.}},
\oauthor{\bsnm{Ilic}, \binits{O.}},
\oauthor{\bsnm{Kuhlmey}, \binits{B.T.}}:
Photonic lightsails: {F}ast and stable propulsion for interstellar travel.
arXiv
(2025).
{URL}: \url{https://arxiv.org/abs/2502.17828}
\end{botherref}
\endbibitem

\bibitem[\protect\citeauthoryear{Ilic et~al.}{2018}]{Ilic2018-5583}
\begin{barticle}
\bauthor{\bsnm{Ilic}, \binits{O.}},
\bauthor{\bsnm{Went}, \binits{C.M.}},
\bauthor{\bsnm{Atwater}, \binits{H.A.}}:
\batitle{Nanophotonic heterostructures for efficient propulsion and radiative
  cooling of relativistic light sails}.
\bjtitle{Nano Lett.}
\bvolume{18}(\bissue{9}),
\bfpage{5583}--\blpage{5589}
(\byear{2018})
\doiurl{10.1021/acs.nanolett.8b02035}
\end{barticle}
\endbibitem

\bibitem[\protect\citeauthoryear{Brewer et~al.}{2022}]{Brewer2022-594}
\begin{barticle}
\bauthor{\bsnm{Brewer}, \binits{J.}},
\bauthor{\bsnm{Campbell}, \binits{M.F.}},
\bauthor{\bsnm{Kumar}, \binits{P.}},
\bauthor{\bsnm{Kulkarni}, \binits{S.}},
\bauthor{\bsnm{Jariwala}, \binits{D.}},
\bauthor{\bsnm{Bargatin}, \binits{I.}},
\bauthor{\bsnm{Raman}, \binits{A.P.}}:
\batitle{Multiscale photonic emissivity engineering for relativistic lightsail
  thermal regulation}.
\bjtitle{Nano Letters}
\bvolume{22}(\bissue{2}),
\bfpage{594}--\blpage{601}
(\byear{2022})
\doiurl{10.1021/acs.nanolett.1c03273}
\end{barticle}
\endbibitem

\bibitem[\protect\citeauthoryear{Campbell et~al.}{2022}]{Campbell2022-90}
\begin{barticle}
\bauthor{\bsnm{Campbell}, \binits{M.F.}},
\bauthor{\bsnm{Brewer}, \binits{J.}},
\bauthor{\bsnm{Jariwala}, \binits{D.}},
\bauthor{\bsnm{Raman}, \binits{A.P.}},
\bauthor{\bsnm{Bargatin}, \binits{I.}}:
\batitle{Relativistic light sails need to billow}.
\bjtitle{Nano Letters}
\bvolume{22}(\bissue{1}),
\bfpage{90}--\blpage{96}
(\byear{2022})
\doiurl{10.1021/acs.nanolett.1c03272}
\end{barticle}
\endbibitem

\bibitem[\protect\citeauthoryear{Myilswamy et~al.}{2020}]{Myilswamy2020-8223}
\begin{barticle}
\bauthor{\bsnm{Myilswamy}, \binits{K.V.}},
\bauthor{\bsnm{Krishnan}, \binits{A.}},
\bauthor{\bsnm{Povinelli}, \binits{M.L.}}:
\batitle{Photonic crystal lightsail with nonlinear reflectivity for increased
  stability}.
\bjtitle{Opt. Express}
\bvolume{28}(\bissue{6}),
\bfpage{8223}--\blpage{8232}
(\byear{2020})
\doiurl{10.1364/OE.387687}
\end{barticle}
\endbibitem

\bibitem[\protect\citeauthoryear{Siegel et~al.}{2019}]{Siegel2019-2032}
\begin{barticle}
\bauthor{\bsnm{Siegel}, \binits{J.}},
\bauthor{\bsnm{Wang}, \binits{A.Y.}},
\bauthor{\bsnm{Menabde}, \binits{S.G.}},
\bauthor{\bsnm{Kats}, \binits{M.A.}},
\bauthor{\bsnm{Jang}, \binits{M.S.}},
\bauthor{\bsnm{Brar}, \binits{V.W.}}:
\batitle{Self-stabilizing laser sails based on optical metasurfaces}.
\bjtitle{ACS Photonics}
\bvolume{6}(\bissue{8}),
\bfpage{2032}--\blpage{2040}
(\byear{2019})
\doiurl{10.1021/acsphotonics.9b00484}
\end{barticle}
\endbibitem

\bibitem[\protect\citeauthoryear{Jin et~al.}{2020}]{Jin2020-2350}
\begin{barticle}
\bauthor{\bsnm{Jin}, \binits{W.}},
\bauthor{\bsnm{Li}, \binits{W.}},
\bauthor{\bsnm{Orenstein}, \binits{M.}},
\bauthor{\bsnm{Fan}, \binits{S.}}:
\batitle{Inverse design of lightweight broadband reflector for relativistic
  lightsail propulsion}.
\bjtitle{ACS Photonics}
\bvolume{7}(\bissue{9}),
\bfpage{2350}--\blpage{2355}
(\byear{2020})
\doiurl{10.1021/acsphotonics.0c00768}
\end{barticle}
\endbibitem

\bibitem[\protect\citeauthoryear{Salary and
  Mosallaei}{2020}]{Salary2020-1900311}
\begin{barticle}
\bauthor{\bsnm{Salary}, \binits{M.M.}},
\bauthor{\bsnm{Mosallaei}, \binits{H.}}:
\batitle{Photonic metasurfaces as relativistic light sails for
  {D}oppler-broadened stable beam-riding and radiative cooling}.
\bjtitle{Laser Photonics Rev.}
\bvolume{14}(\bissue{8}),
\bfpage{1900311}
(\byear{2020})
\doiurl{10.1002/lpor.201900311}
\end{barticle}
\endbibitem

\bibitem[\protect\citeauthoryear{Gieseler et~al.}{2021}]{Gieseler2021-21562}
\begin{barticle}
\bauthor{\bsnm{Gieseler}, \binits{N.}},
\bauthor{\bsnm{Rahimzadegan}, \binits{A.}},
\bauthor{\bsnm{Rockstuhl}, \binits{C.}}:
\batitle{Self-stabilizing curved metasurfaces as a sail for light-propelled
  spacecrafts}.
\bjtitle{Opt. Express}
\bvolume{29}(\bissue{14}),
\bfpage{21562}--\blpage{21575}
(\byear{2021})
\doiurl{10.1364/OE.420475}
\end{barticle}
\endbibitem

\bibitem[\protect\citeauthoryear{Gao et~al.}{2022}]{Gao2022-1965}
\begin{barticle}
\bauthor{\bsnm{Gao}, \binits{R.}},
\bauthor{\bsnm{Kelzenberg}, \binits{M.D.}},
\bauthor{\bsnm{Kim}, \binits{Y.}},
\bauthor{\bsnm{Ilic}, \binits{O.}},
\bauthor{\bsnm{Atwater}, \binits{H.A.}}:
\batitle{Optical characterization of silicon nitride metagrating-based
  lightsails for self-stabilization}.
\bjtitle{ACS Photonics}
\bvolume{9}(\bissue{6}),
\bfpage{1965}--\blpage{1972}
(\byear{2022})
\doiurl{10.1021/acsphotonics.1c02022}
\end{barticle}
\endbibitem

\bibitem[\protect\citeauthoryear{Kudyshev et~al.}{2022}]{Kudyshev2022-190}
\begin{barticle}
\bauthor{\bsnm{Kudyshev}, \binits{Z.A.}},
\bauthor{\bsnm{Kildishev}, \binits{A.V.}},
\bauthor{\bsnm{Shalaev}, \binits{V.M.}},
\bauthor{\bsnm{Boltasseva}, \binits{A.}}:
\batitle{Optimizing startshot lightsail design: {A} generative network-based
  approach}.
\bjtitle{ACS Photonics}
\bvolume{9}(\bissue{1}),
\bfpage{190}--\blpage{196}
(\byear{2022})
\doiurl{10.1021/acsphotonics.1c01352}
\end{barticle}
\endbibitem

\bibitem[\protect\citeauthoryear{Lien et~al.}{2022}]{Lien2022-3032}
\begin{barticle}
\bauthor{\bsnm{Lien}, \binits{M.R.}},
\bauthor{\bsnm{Meng}, \binits{D.}},
\bauthor{\bsnm{Liu}, \binits{Z.}},
\bauthor{\bsnm{Sakib}, \binits{M.A.}},
\bauthor{\bsnm{Tang}, \binits{Y.}},
\bauthor{\bsnm{Wu}, \binits{W.}},
\bauthor{\bsnm{Povinelli}, \binits{M.L.}}:
\batitle{Experimental characterization of a silicon nitride photonic crystal
  light sail}.
\bjtitle{Opt. Mater. Express}
\bvolume{12}(\bissue{8}),
\bfpage{3032}--\blpage{3042}
(\byear{2022})
\doiurl{10.1364/OME.464430}
\end{barticle}
\endbibitem

\bibitem[\protect\citeauthoryear{Santi et~al.}{2022}]{Santi2022-16}
\begin{barticle}
\bauthor{\bsnm{Santi}, \binits{G.}},
\bauthor{\bsnm{Favaro}, \binits{G.}},
\bauthor{\bsnm{Corso}, \binits{A.J.}},
\bauthor{\bsnm{Lubin}, \binits{P.}},
\bauthor{\bsnm{Bazzan}, \binits{M.}},
\bauthor{\bsnm{Ragazzoni}, \binits{R.}},
\bauthor{\bsnm{Garoli}, \binits{D.}},
\bauthor{\bsnm{Pelizzo}, \binits{M.G.}}:
\batitle{Multilayers for directed energy accelerated lightsails}.
\bjtitle{Communications Materials}
\bvolume{3}(\bissue{1}),
\bfpage{16}
(\byear{2022})
\doiurl{10.1038/s43246-022-00240-8}
\end{barticle}
\endbibitem

\bibitem[\protect\citeauthoryear{Taghavi and
  Mosallaei}{2022}]{Taghavi2022-20034}
\begin{barticle}
\bauthor{\bsnm{Taghavi}, \binits{M.}},
\bauthor{\bsnm{Mosallaei}, \binits{H.}}:
\batitle{Increasing the stability margins using multi-pattern metasails and
  multi-modal laser beams}.
\bjtitle{Scientific Reports}
\bvolume{12}(\bissue{1}),
\bfpage{20034}
(\byear{2022})
\doiurl{10.1038/s41598-022-24681-w}
\end{barticle}
\endbibitem

\bibitem[\protect\citeauthoryear{Tung and Davoyan}{2022}]{Tung2022-1108}
\begin{barticle}
\bauthor{\bsnm{Tung}, \binits{H.-T.}},
\bauthor{\bsnm{Davoyan}, \binits{A.R.}}:
\batitle{Low-power laser sailing for fast-transit space flight}.
\bjtitle{Nano Letters}
\bvolume{22}(\bissue{3}),
\bfpage{1108}--\blpage{1114}
(\byear{2022})
\doiurl{10.1021/acs.nanolett.1c04188}
\end{barticle}
\endbibitem

\bibitem[\protect\citeauthoryear{Chang et~al.}{2024}]{Chang2024-6689}
\begin{barticle}
\bauthor{\bsnm{Chang}, \binits{J.}},
\bauthor{\bsnm{Ji}, \binits{W.}},
\bauthor{\bsnm{Yao}, \binits{X.}},
\bauthor{\bsnm{Run}, \binits{A.J.}},
\bauthor{\bsnm{Gr{\"o}blacher}, \binits{S.}}:
\batitle{Broadband, high-reflectivity dielectric mirrors at wafer scale:
  {C}ombining photonic crystal and metasurface architectures for advanced
  lightsails}.
\bjtitle{Nano Letters}
\bvolume{24}(\bissue{22}),
\bfpage{6689}--\blpage{6695}
(\byear{2024})
\doiurl{10.1021/acs.nanolett.4c01374}
\end{barticle}
\endbibitem

\bibitem[\protect\citeauthoryear{Norder et~al.}{2025}]{Norder2025-2753}
\begin{barticle}
\bauthor{\bsnm{Norder}, \binits{L.}},
\bauthor{\bsnm{Yin}, \binits{S.}},
\bauthor{\bsnm{{de Jong}}, \binits{M.H.J.}},
\bauthor{\bsnm{Stallone}, \binits{F.}},
\bauthor{\bsnm{Aydogmus}, \binits{H.}},
\bauthor{\bsnm{Sberna}, \binits{P.M.}},
\bauthor{\bsnm{Bessa}, \binits{M.A.}},
\bauthor{\bsnm{Norte}, \binits{R.A.}}:
\batitle{Pentagonal photonic crystal mirrors: {S}calable lightsails with
  enhanced acceleration via neural topology optimization}.
\bjtitle{Nature Communications}
\bvolume{16}(\bissue{1}),
\bfpage{2753}
(\byear{2025})
\doiurl{10.1038/s41467-025-57749-y}
\end{barticle}
\endbibitem

\bibitem[\protect\citeauthoryear{{National Aeronautics and Space
  Administration}}{2011}]{NASA2011-photo}
\begin{botherref}
\oauthor{\bsnm{{National Aeronautics and Space Administration}}}:
The two-faced whirlpool galaxy.
Technical report,
{NASA}
(January 13 2011).
{URL}: \url{https://images.nasa.gov/details-GSFC_20171208_Archive_e001925}
\end{botherref}
\endbibitem

\bibitem[\protect\citeauthoryear{Ermolaev et~al.}{2020}]{Ermolaev2020-21}
\begin{barticle}
\bauthor{\bsnm{Ermolaev}, \binits{G.A.}},
\bauthor{\bsnm{Stebunov}, \binits{Y.V.}},
\bauthor{\bsnm{Vyshnevyy}, \binits{A.A.}},
\bauthor{\bsnm{Tatarkin}, \binits{D.E.}},
\bauthor{\bsnm{Yakubovsky}, \binits{D.I.}},
\bauthor{\bsnm{Novikov}, \binits{S.M.}},
\bauthor{\bsnm{Baranov}, \binits{D.G.}},
\bauthor{\bsnm{Shegai}, \binits{T.}},
\bauthor{\bsnm{Nikitin}, \binits{A.Y.}},
\bauthor{\bsnm{Arsenin}, \binits{A.V.}},
\bauthor{\bsnm{Volkov}, \binits{V.S.}}:
\batitle{Broadband optical properties of monolayer and bulk
  {M}o{S}$_\text{2}$}.
\bjtitle{{npj} 2{D} {M}ater. {A}ppl.}
\bvolume{4}(\bissue{1}),
\bfpage{21}
(\byear{2020})
\doiurl{10.1038/s41699-020-0155-x}
\end{barticle}
\endbibitem

\bibitem[\protect\citeauthoryear{Akcay et~al.}{2021}]{Akcay2021-1452}
\begin{barticle}
\bauthor{\bsnm{Akcay}, \binits{N.}},
\bauthor{\bsnm{Tivanov}, \binits{M.}},
\bauthor{\bsnm{Ozcelik}, \binits{S.}}:
\batitle{Mo{S}$_\text{2}$ thin films grown by sulfurization of {DC} sputtered
  {M}o thin films on {S}i/{S}i{O}$_\text{2}$ and {C}-plane sapphire
  substrates}.
\bjtitle{Journal of Electronic Materials}
\bvolume{50}(\bissue{3}),
\bfpage{1452}--\blpage{1466}
(\byear{2021})
\doiurl{10.1007/s11664-020-08687-6}
\end{barticle}
\endbibitem

\bibitem[\protect\citeauthoryear{Altvater et~al.}{2024}]{Altvater2024-2400463}
\begin{botherref}
\oauthor{\bsnm{Altvater}, \binits{M.}},
\oauthor{\bsnm{Muratore}, \binits{C.}},
\oauthor{\bsnm{Snure}, \binits{M.}},
\oauthor{\bsnm{Glavin}, \binits{N.R.}}:
Two-step conversion of metal and metal oxide precursor films to 2{D} transition
  metal dichalcogenides and heterostructures.
Small,
2400463
(2024)
\doiurl{10.1002/smll.202400463}
\end{botherref}
\endbibitem

\bibitem[\protect\citeauthoryear{Song et~al.}{2019}]{Song2019-1801250}
\begin{barticle}
\bauthor{\bsnm{Song}, \binits{B.}},
\bauthor{\bsnm{Gu}, \binits{H.}},
\bauthor{\bsnm{Fang}, \binits{M.}},
\bauthor{\bsnm{Chen}, \binits{X.}},
\bauthor{\bsnm{Jiang}, \binits{H.}},
\bauthor{\bsnm{Wang}, \binits{R.}},
\bauthor{\bsnm{Zhai}, \binits{T.}},
\bauthor{\bsnm{Ho}, \binits{Y.-T.}},
\bauthor{\bsnm{Liu}, \binits{S.}}:
\batitle{Layer-dependent dielectric function of wafer-scale 2{D}
  {M}o{S}$_\text{2}$}.
\bjtitle{Advanced Optical Materials}
\bvolume{7}(\bissue{2}),
\bfpage{1801250}
(\byear{2019})
\doiurl{10.1002/adom.201801250}
\end{barticle}
\endbibitem

\bibitem[\protect\citeauthoryear{Islam et~al.}{2021}]{Islam2021-2000180}
\begin{barticle}
\bauthor{\bsnm{Islam}, \binits{K.M.}},
\bauthor{\bsnm{Synowicki}, \binits{R.}},
\bauthor{\bsnm{Ismael}, \binits{T.}},
\bauthor{\bsnm{Oguntoye}, \binits{I.}},
\bauthor{\bsnm{Grinalds}, \binits{N.}},
\bauthor{\bsnm{Escarra}, \binits{M.D.}}:
\batitle{In-plane and out-of-plane optical properties of monolayer, few-layer,
  and thin-film {M}o{S}$_\text{2}$ from 190 to 1700 nm and their application in
  photonic device design}.
\bjtitle{Adv. Photon. Res.}
\bvolume{2}(\bissue{5}),
\bfpage{2000180}
(\byear{2021})
\doiurl{10.1002/adpr.202000180}
\end{barticle}
\endbibitem

\bibitem[\protect\citeauthoryear{Munkhbat et~al.}{2022}]{Munkhbat2022-2398}
\begin{barticle}
\bauthor{\bsnm{Munkhbat}, \binits{B.}},
\bauthor{\bsnm{Wr{\'o}bel}, \binits{P.}},
\bauthor{\bsnm{Antosiewicz}, \binits{T.J.}},
\bauthor{\bsnm{Shegai}, \binits{T.O.}}:
\batitle{Optical constants of several multilayer transition metal
  dichalcogenides measured by spectroscopic ellipsometry in the 300--1700 nm
  range: {H}igh index, anisotropy, and hyperbolicity}.
\bjtitle{ACS Photonics}
\bvolume{9}(\bissue{7}),
\bfpage{2398}--\blpage{2407}
(\byear{2022})
\doiurl{10.1021/acsphotonics.2c00433}
\end{barticle}
\endbibitem

\bibitem[\protect\citeauthoryear{Lingart et~al.}{1982}]{Lingart1982-706}
\begin{barticle}
\bauthor{\bsnm{Lingart}, \binits{Y.K.}},
\bauthor{\bsnm{Petrov}, \binits{V.A.}},
\bauthor{\bsnm{Tikhonova}, \binits{N.A.}}:
\batitle{Optical-properties of leucosapphire at high-temperatures. {I}.
  {T}ranslucent region}.
\bjtitle{High Temp.}
\bvolume{20}(\bissue{5}),
\bfpage{706}--\blpage{713}
(\byear{1982}).
\bcomment{{URL}: \url{http://mi.mathnet.ru/tvt6472}}
\end{barticle}
\endbibitem

\bibitem[\protect\citeauthoryear{Querry}{1985}]{Querry1985-report}
\begin{botherref}
\oauthor{\bsnm{Querry}, \binits{M.R.}}:
Optical constants.
Technical Report {CRDC}-{CR}-85034,
University of Missouri
(June 1985).
{URL}: \url{https://apps.dtic.mil/sti/pdfs/ADA158623.pdf}
\end{botherref}
\endbibitem

\bibitem[\protect\citeauthoryear{Kischkat et~al.}{2012}]{Kischkat2012-6789}
\begin{barticle}
\bauthor{\bsnm{Kischkat}, \binits{J.}},
\bauthor{\bsnm{Peters}, \binits{S.}},
\bauthor{\bsnm{Gruska}, \binits{B.}},
\bauthor{\bsnm{Semtsiv}, \binits{M.}},
\bauthor{\bsnm{Chashnikova}, \binits{M.}},
\bauthor{\bsnm{Klinkm{\"u}ller}, \binits{M.}},
\bauthor{\bsnm{Fedosenko}, \binits{O.}},
\bauthor{\bsnm{Machulik}, \binits{S.}},
\bauthor{\bsnm{Aleksandrova}, \binits{A.}},
\bauthor{\bsnm{Monastyrskyi}, \binits{G.}},
\bauthor{\bsnm{Flores}, \binits{Y.}},
\bauthor{\bsnm{Masselink}, \binits{W.T.}}:
\batitle{Mid-infrared optical properties of thin films of aluminum oxide,
  titanium dioxide, silicon dioxide, aluminum nitride, and silicon nitride}.
\bjtitle{Appl. Opt.}
\bvolume{51}(\bissue{28}),
\bfpage{6789}--\blpage{6798}
(\byear{2012})
\doiurl{10.1364/AO.51.006789}
\end{barticle}
\endbibitem

\bibitem[\protect\citeauthoryear{Davami et~al.}{2015}]{Davami2015-10019}
\begin{barticle}
\bauthor{\bsnm{Davami}, \binits{K.}},
\bauthor{\bsnm{Zhao}, \binits{L.}},
\bauthor{\bsnm{Lu}, \binits{E.}},
\bauthor{\bsnm{Cortes}, \binits{J.}},
\bauthor{\bsnm{Lin}, \binits{C.}},
\bauthor{\bsnm{Lilley}, \binits{D.E.}},
\bauthor{\bsnm{Purohit}, \binits{P.K.}},
\bauthor{\bsnm{Bargatin}, \binits{I.}}:
\batitle{Ultralight shape-recovering plate mechanical metamaterials}.
\bjtitle{Nature Communications}
\bvolume{6}(\bissue{1}),
\bfpage{10019}
(\byear{2015})
\doiurl{10.1038/ncomms10019}
\end{barticle}
\endbibitem

\bibitem[\protect\citeauthoryear{Groner et~al.}{2004}]{Groner2004-639}
\begin{barticle}
\bauthor{\bsnm{Groner}, \binits{M.D.}},
\bauthor{\bsnm{Fabreguette}, \binits{F.H.}},
\bauthor{\bsnm{Elam}, \binits{J.W.}},
\bauthor{\bsnm{George}, \binits{S.M.}}:
\batitle{Low-temperature {A}l$_\text{2}${O}$_\text{3}$ atomic layer
  deposition}.
\bjtitle{Chemistry of Materials}
\bvolume{16}(\bissue{4}),
\bfpage{639}--\blpage{645}
(\byear{2004})
\doiurl{10.1021/cm0304546}
\end{barticle}
\endbibitem

\bibitem[\protect\citeauthoryear{Ilic et~al.}{2010}]{Ilic2010-044317}
\begin{barticle}
\bauthor{\bsnm{Ilic}, \binits{B.}},
\bauthor{\bsnm{Krylov}, \binits{S.}},
\bauthor{\bsnm{Craighead}, \binits{H.G.}}:
\batitle{Young's modulus and density measurements of thin atomic layer
  deposited films using resonant nanomechanics}.
\bjtitle{Journal of Applied Physics}
\bvolume{108}(\bissue{4}),
\bfpage{044317}
(\byear{2010})
\doiurl{10.1063/1.3474987}
\end{barticle}
\endbibitem

\bibitem[\protect\citeauthoryear{Graczykowski
  et~al.}{2017}]{Graczykowski2017-7647}
\begin{barticle}
\bauthor{\bsnm{Graczykowski}, \binits{B.}},
\bauthor{\bsnm{Sledzinska}, \binits{M.}},
\bauthor{\bsnm{Placidi}, \binits{M.}},
\bauthor{\bsnm{Saleta~Reig}, \binits{D.}},
\bauthor{\bsnm{Kasprzak}, \binits{M.}},
\bauthor{\bsnm{Alzina}, \binits{F.}},
\bauthor{\bsnm{Sotomayor~Torres}, \binits{C.M.}}:
\batitle{Elastic properties of few nanometers thick polycrystalline
  {M}o{S}$_\text{2}$ membranes: {A} nondestructive study}.
\bjtitle{Nano Lett.}
\bvolume{17}(\bissue{12}),
\bfpage{7647}--\blpage{7651}
(\byear{2017})
\doiurl{10.1021/acs.nanolett.7b03669}
\end{barticle}
\endbibitem

\bibitem[\protect\citeauthoryear{Jiao et~al.}{2019}]{Jiao2019-034055}
\begin{barticle}
\bauthor{\bsnm{Jiao}, \binits{P.}},
\bauthor{\bsnm{Nicaise}, \binits{S.M.}},
\bauthor{\bsnm{Lin}, \binits{C.}},
\bauthor{\bsnm{Purohit}, \binits{P.K.}},
\bauthor{\bsnm{Bargatin}, \binits{I.}}:
\batitle{Extremely sharp bending and recoverability of nanoscale plates with
  honeycomb corrugation}.
\bjtitle{Phys. Rev. Appl.}
\bvolume{11}(\bissue{3}),
\bfpage{034055}
(\byear{2019})
\doiurl{10.1103/PhysRevApplied.11.034055}
\end{barticle}
\endbibitem

\bibitem[\protect\citeauthoryear{Jiao et~al.}{2020}]{Jiao2020-100599}
\begin{barticle}
\bauthor{\bsnm{Jiao}, \binits{P.}},
\bauthor{\bsnm{Nicaise}, \binits{S.M.}},
\bauthor{\bsnm{Azadi}, \binits{M.}},
\bauthor{\bsnm{Cortes}, \binits{J.}},
\bauthor{\bsnm{Lilley}, \binits{D.E.}},
\bauthor{\bsnm{Cha}, \binits{W.}},
\bauthor{\bsnm{Purohit}, \binits{P.K.}},
\bauthor{\bsnm{Bargatin}, \binits{I.}}:
\batitle{Tunable tensile response of honeycomb plates with nanoscale thickness:
  {T}esting and modeling}.
\bjtitle{Extreme Mechanics Letters}
\bvolume{34},
\bfpage{100599}
(\byear{2020})
\doiurl{10.1016/j.eml.2019.100599}
\end{barticle}
\endbibitem

\bibitem[\protect\citeauthoryear{Weingartner and
  Draine}{2001}]{Weingartner2001-296}
\begin{barticle}
\bauthor{\bsnm{Weingartner}, \binits{J.C.}},
\bauthor{\bsnm{Draine}, \binits{B.T.}}:
\batitle{Dust grain-size distributions and extinction in the {M}ilky {W}ay,
  {L}arge {M}agellanic {C}loud, and {S}mall {M}agellanic {C}loud}.
\bjtitle{The Astrophysical Journal}
\bvolume{548}(\bissue{1}),
\bfpage{296}--\blpage{309}
(\byear{2001})
\doiurl{10.1086/318651}
\end{barticle}
\endbibitem

\bibitem[\protect\citeauthoryear{Early and London}{2015}]{Early2015-205}
\begin{barticle}
\bauthor{\bsnm{Early}, \binits{J.T.}},
\bauthor{\bsnm{London}, \binits{R.A.}}:
\batitle{Dust grain damage to interstellar vehicles and lightsails}.
\bjtitle{J. Br. Interplanet. Soc.}
\bvolume{68},
\bfpage{205}--\blpage{210}
(\byear{2015})
\end{barticle}
\endbibitem

\bibitem[\protect\citeauthoryear{Hoang et~al.}{2017}]{Hoang2017-5}
\begin{barticle}
\bauthor{\bsnm{Hoang}, \binits{T.}},
\bauthor{\bsnm{Lazarian}, \binits{A.}},
\bauthor{\bsnm{Burkhart}, \binits{B.}},
\bauthor{\bsnm{Loeb}, \binits{A.}}:
\batitle{The interaction of relativistic spacecrafts with the interstellar
  medium}.
\bjtitle{Astrophys. J.}
\bvolume{837}(\bissue{1}),
\bfpage{5}
(\byear{2017})
\doiurl{10.3847/1538-4357/aa5da6}
\end{barticle}
\endbibitem

\bibitem[\protect\citeauthoryear{Cui et~al.}{2018}]{Cui2018-44}
\begin{barticle}
\bauthor{\bsnm{Cui}, \binits{S.}},
\bauthor{\bsnm{Hu}, \binits{B.}},
\bauthor{\bsnm{Ouyang}, \binits{B.}},
\bauthor{\bsnm{Zhao}, \binits{D.}}:
\batitle{Thermodynamic assessment of the {M}o-{S} system and its application in
  thermal decomposition of {M}o{S}$_\text{2}$}.
\bjtitle{Thermochim. Acta}
\bvolume{660},
\bfpage{44}--\blpage{55}
(\byear{2018})
\doiurl{10.1016/j.tca.2017.12.011}
\end{barticle}
\endbibitem

\bibitem[\protect\citeauthoryear{Holdman et~al.}{2022}]{Holdman2022-2102835}
\begin{barticle}
\bauthor{\bsnm{Holdman}, \binits{G.R.}},
\bauthor{\bsnm{Jaffe}, \binits{G.R.}},
\bauthor{\bsnm{Feng}, \binits{D.}},
\bauthor{\bsnm{Jang}, \binits{M.S.}},
\bauthor{\bsnm{Kats}, \binits{M.A.}},
\bauthor{\bsnm{Brar}, \binits{V.W.}}:
\batitle{Thermal runaway of silicon-based laser sails}.
\bjtitle{Advanced Optical Materials}
\bvolume{10}(\bissue{19}),
\bfpage{2102835}
(\byear{2022})
\doiurl{10.1002/adom.202102835}
\end{barticle}
\endbibitem

\bibitem[\protect\citeauthoryear{Dong et~al.}{2018}]{Dong2018-1558}
\begin{barticle}
\bauthor{\bsnm{Dong}, \binits{N.}},
\bauthor{\bsnm{Li}, \binits{Y.}},
\bauthor{\bsnm{Zhang}, \binits{S.}},
\bauthor{\bsnm{Mc{E}voy}, \binits{N.}},
\bauthor{\bsnm{Gatensby}, \binits{R.}},
\bauthor{\bsnm{Duesberg}, \binits{G.S.}},
\bauthor{\bsnm{Wang}, \binits{J.}}:
\batitle{Saturation of two-photon absorption in layered transition metal
  dichalcogenides: {E}xperiment and theory}.
\bjtitle{ACS Photonics}
\bvolume{5}(\bissue{4}),
\bfpage{1558}--\blpage{1565}
(\byear{2018})
\doiurl{10.1021/acsphotonics.8b00010}
\end{barticle}
\endbibitem

\bibitem[\protect\citeauthoryear{Zhang et~al.}{2015}]{Zhang2015-7142}
\begin{barticle}
\bauthor{\bsnm{Zhang}, \binits{S.}},
\bauthor{\bsnm{Dong}, \binits{N.}},
\bauthor{\bsnm{Mc{E}voy}, \binits{N.}},
\bauthor{\bsnm{O'{B}rien}, \binits{M.}},
\bauthor{\bsnm{Winters}, \binits{S.}},
\bauthor{\bsnm{Berner}, \binits{N.C.}},
\bauthor{\bsnm{Yim}, \binits{C.}},
\bauthor{\bsnm{Li}, \binits{Y.}},
\bauthor{\bsnm{Zhang}, \binits{X.}},
\bauthor{\bsnm{Chen}, \binits{Z.}},
\bauthor{\bsnm{Zhang}, \binits{L.}},
\bauthor{\bsnm{Duesberg}, \binits{G.S.}},
\bauthor{\bsnm{Wang}, \binits{J.}}:
\batitle{Direct observation of degenerate two-photon absorption and its
  saturation in {WS}$_\text{2}$ and {M}o{S}$_\text{2}$ monolayer and few-layer
  films}.
\bjtitle{ACS Nano}
\bvolume{9}(\bissue{7}),
\bfpage{7142}--\blpage{7150}
(\byear{2015})
\doiurl{10.1021/acsnano.5b03480}
\end{barticle}
\endbibitem

\bibitem[\protect\citeauthoryear{Santi et~al.}{2023}]{Santi2023-19583}
\begin{barticle}
\bauthor{\bsnm{Santi}, \binits{G.}},
\bauthor{\bsnm{Corso}, \binits{A.J.}},
\bauthor{\bsnm{Garoli}, \binits{D.}},
\bauthor{\bsnm{Lio}, \binits{G.E.}},
\bauthor{\bsnm{Manente}, \binits{M.}},
\bauthor{\bsnm{Favaro}, \binits{G.}},
\bauthor{\bsnm{Bazzan}, \binits{M.}},
\bauthor{\bsnm{Piotto}, \binits{G.}},
\bauthor{\bsnm{Andriolli}, \binits{N.}},
\bauthor{\bsnm{Strambini}, \binits{L.}},
\bauthor{\bsnm{Pavarin}, \binits{D.}},
\bauthor{\bsnm{Badia}, \binits{L.}},
\bauthor{\bsnm{Proietti~Zaccaria}, \binits{R.}},
\bauthor{\bsnm{Lubin}, \binits{P.}},
\bauthor{\bsnm{Ragazzoni}, \binits{R.}},
\bauthor{\bsnm{Pelizzo}, \binits{M.G.}}:
\batitle{Swarm of lightsail nanosatellites for {S}olar {S}ystem exploration}.
\bjtitle{Scientific Reports}
\bvolume{13}(\bissue{1}),
\bfpage{19583}
(\byear{2023})
\doiurl{10.1038/s41598-023-46101-3}
\end{barticle}
\endbibitem

\bibitem[\protect\citeauthoryear{Lainey et~al.}{2009}]{Lainey2009-957}
\begin{barticle}
\bauthor{\bsnm{Lainey}, \binits{V.}},
\bauthor{\bsnm{Arlot}, \binits{J.-E.}},
\bauthor{\bsnm{Karatekin}, \binits{{\"O}.}},
\bauthor{\bsnm{Van~{H}oolst}, \binits{T.}}:
\batitle{Strong tidal dissipation in {I}o and {J}upiter from astrometric
  observations}.
\bjtitle{Nature}
\bvolume{459}(\bissue{7249}),
\bfpage{957}--\blpage{959}
(\byear{2009})
\doiurl{10.1038/nature08108}
\end{barticle}
\endbibitem

\bibitem[\protect\citeauthoryear{Chitta et~al.}{2023}]{Chitta2023-867}
\begin{barticle}
\bauthor{\bsnm{Chitta}, \binits{L.P.}},
\bauthor{\bsnm{Zhukov}, \binits{A.N.}},
\bauthor{\bsnm{Berghmans}, \binits{D.}},
\bauthor{\bsnm{Peter}, \binits{H.}},
\bauthor{\bsnm{Parenti}, \binits{S.}},
\bauthor{\bsnm{Mandal}, \binits{S.}},
\bauthor{\bsnm{Cuadrado}, \binits{R.A.}},
\bauthor{\bsnm{Sch{\"u}hle}, \binits{U.}},
\bauthor{\bsnm{Teriaca}, \binits{L.}},
\bauthor{\bsnm{Auch{\'e}re}, \binits{F.}},
\bauthor{\bsnm{Barczynski}, \binits{K.}},
\bauthor{\bsnm{Buchlin}, \binits{{\'E}.}},
\bauthor{\bsnm{Harra}, \binits{L.}},
\bauthor{\bsnm{Kraaikamp}, \binits{E.}},
\bauthor{\bsnm{Long}, \binits{D.M.}},
\bauthor{\bsnm{Rodriguez}, \binits{L.}},
\bauthor{\bsnm{Schwanitz}, \binits{C.}},
\bauthor{\bsnm{Smith}, \binits{P.J.}},
\bauthor{\bsnm{Verbeeck}, \binits{C.}},
\bauthor{\bsnm{Seaton}, \binits{D.B.}}:
\batitle{Picoflare jets power the solar wind emerging from a coronal hole on
  the {S}un}.
\bjtitle{Science}
\bvolume{381}(\bissue{6660}),
\bfpage{867}--\blpage{872}
(\byear{2023})
\doiurl{10.1126/science.ade5801}
\end{barticle}
\endbibitem

\bibitem[\protect\citeauthoryear{Einstein}{1936}]{Einstein1936-506}
\begin{barticle}
\bauthor{\bsnm{Einstein}, \binits{A.}}:
\batitle{Lens-like action of a star by the deviation of light in the
  gravitational field}.
\bjtitle{Science}
\bvolume{84}(\bissue{2188}),
\bfpage{506}--\blpage{507}
(\byear{1936})
\doiurl{10.1126/science.84.2188.506}
\end{barticle}
\endbibitem

\bibitem[\protect\citeauthoryear{Turyshev and Toth}{2020}]{Turyshev2020-044048}
\begin{barticle}
\bauthor{\bsnm{Turyshev}, \binits{S.G.}},
\bauthor{\bsnm{Toth}, \binits{V.T.}}:
\batitle{Image formation process with the solar gravitational lens}.
\bjtitle{Phys. Rev. D}
\bvolume{101}(\bissue{4}),
\bfpage{044048}
(\byear{2020})
\doiurl{10.1103/PhysRevD.101.044048}
\end{barticle}
\endbibitem

\bibitem[\protect\citeauthoryear{Turyshev and Toth}{2022}]{Turyshev2022-6122}
\begin{barticle}
\bauthor{\bsnm{Turyshev}, \binits{S.G.}},
\bauthor{\bsnm{Toth}, \binits{V.T.}}:
\batitle{Resolved imaging of exoplanets with the solar gravitational lens}.
\bjtitle{Monthly Notices of the Royal Astronomical Society}
\bvolume{515}(\bissue{4}),
\bfpage{6122}--\blpage{6132}
(\byear{2022})
\doiurl{10.1093/mnras/stac2130}
\end{barticle}
\endbibitem

\bibitem[\protect\citeauthoryear{Deutsch}{2020}]{Deutsch2020-907}
\begin{barticle}
\bauthor{\bsnm{Deutsch}, \binits{L.J.}}:
\batitle{Towards deep space optical communications}.
\bjtitle{Nature Astronomy}
\bvolume{4}(\bissue{9}),
\bfpage{907}--\blpage{907}
(\byear{2020})
\doiurl{10.1038/s41550-020-1193-1}
\end{barticle}
\endbibitem

\bibitem[\protect\citeauthoryear{Angel}{2006}]{Angel2006-17184}
\begin{barticle}
\bauthor{\bsnm{Angel}, \binits{R.}}:
\batitle{Feasibility of cooling the {E}arth with a cloud of small spacecraft
  near the inner {L}agrange point ({L}1)}.
\bjtitle{Proceedings of the National Academy of Sciences of the United States
  of America}
\bvolume{103}(\bissue{46}),
\bfpage{17184}--\blpage{17189}
(\byear{2006})
\doiurl{10.1073/pnas.0608163103}
\end{barticle}
\endbibitem

\bibitem[\protect\citeauthoryear{Jariwala et~al.}{2016}]{Jariwala2016-5482}
\begin{barticle}
\bauthor{\bsnm{Jariwala}, \binits{D.}},
\bauthor{\bsnm{Davoyan}, \binits{A.R.}},
\bauthor{\bsnm{Tagliabue}, \binits{G.}},
\bauthor{\bsnm{Sherrott}, \binits{M.C.}},
\bauthor{\bsnm{Wong}, \binits{J.}},
\bauthor{\bsnm{Atwater}, \binits{H.A.}}:
\batitle{Near-unity absorption in van der {W}aals semiconductors for ultrathin
  optoelectronics}.
\bjtitle{Nano Letters}
\bvolume{16}(\bissue{9}),
\bfpage{5482}--\blpage{5487}
(\byear{2016})
\doiurl{10.1021/acs.nanolett.6b01914}
\end{barticle}
\endbibitem

\bibitem[\protect\citeauthoryear{Wong et~al.}{2017}]{Wong2017-7230}
\begin{barticle}
\bauthor{\bsnm{Wong}, \binits{J.}},
\bauthor{\bsnm{Jariwala}, \binits{D.}},
\bauthor{\bsnm{Tagliabue}, \binits{G.}},
\bauthor{\bsnm{Tat}, \binits{K.}},
\bauthor{\bsnm{Davoyan}, \binits{A.R.}},
\bauthor{\bsnm{Sherrott}, \binits{M.C.}},
\bauthor{\bsnm{Atwater}, \binits{H.A.}}:
\batitle{High photovoltaic quantum efficiency in ultrathin van der {W}aals
  heterostructures}.
\bjtitle{ACS Nano}
\bvolume{11}(\bissue{7}),
\bfpage{7230}--\blpage{7240}
(\byear{2017})
\doiurl{10.1021/acsnano.7b03148}
\end{barticle}
\endbibitem

\bibitem[\protect\citeauthoryear{Zhang et~al.}{2020}]{Zhang2020-3552}
\begin{barticle}
\bauthor{\bsnm{Zhang}, \binits{H.}},
\bauthor{\bsnm{Abhiraman}, \binits{B.}},
\bauthor{\bsnm{Zhang}, \binits{Q.}},
\bauthor{\bsnm{Miao}, \binits{J.}},
\bauthor{\bsnm{Jo}, \binits{K.}},
\bauthor{\bsnm{Roccasecca}, \binits{S.}},
\bauthor{\bsnm{Knight}, \binits{M.W.}},
\bauthor{\bsnm{Davoyan}, \binits{A.R.}},
\bauthor{\bsnm{Jariwala}, \binits{D.}}:
\batitle{Hybrid exciton-plasmon-polaritons in van der {W}aals semiconductor
  gratings}.
\bjtitle{Nature Communications}
\bvolume{11}(\bissue{1}),
\bfpage{3552}
(\byear{2020})
\doiurl{10.1038/s41467-020-17313-2}
\end{barticle}
\endbibitem

\bibitem[\protect\citeauthoryear{Kumar et~al.}{2022}]{Kumar2022-182}
\begin{barticle}
\bauthor{\bsnm{Kumar}, \binits{P.}},
\bauthor{\bsnm{Lynch}, \binits{J.}},
\bauthor{\bsnm{Song}, \binits{B.}},
\bauthor{\bsnm{Ling}, \binits{H.}},
\bauthor{\bsnm{Barrera}, \binits{F.}},
\bauthor{\bsnm{Kisslinger}, \binits{K.}},
\bauthor{\bsnm{Zhang}, \binits{H.}},
\bauthor{\bsnm{Anantharaman}, \binits{S.B.}},
\bauthor{\bsnm{Digani}, \binits{J.}},
\bauthor{\bsnm{Zhu}, \binits{H.}},
\bauthor{\bsnm{Choudhury}, \binits{T.H.}},
\bauthor{\bsnm{Mc{A}leese}, \binits{C.}},
\bauthor{\bsnm{Wang}, \binits{X.}},
\bauthor{\bsnm{Conran}, \binits{B.R.}},
\bauthor{\bsnm{Whear}, \binits{O.}},
\bauthor{\bsnm{Motala}, \binits{M.J.}},
\bauthor{\bsnm{Snure}, \binits{M.}},
\bauthor{\bsnm{Muratore}, \binits{C.}},
\bauthor{\bsnm{Redwing}, \binits{J.M.}},
\bauthor{\bsnm{Glavin}, \binits{N.R.}},
\bauthor{\bsnm{Stach}, \binits{E.A.}},
\bauthor{\bsnm{Davoyan}, \binits{A.R.}},
\bauthor{\bsnm{Jariwala}, \binits{D.}}:
\batitle{Light--matter coupling in large-area van der {W}aals superlattices}.
\bjtitle{Nature Nanotechnology}
\bvolume{17}(\bissue{2}),
\bfpage{182}--\blpage{189}
(\byear{2022})
\doiurl{10.1038/s41565-021-01023-x}
\end{barticle}
\endbibitem

\bibitem[\protect\citeauthoryear{Alfieri et~al.}{2023}]{Alfieri2023-2202011}
\begin{barticle}
\bauthor{\bsnm{Alfieri}, \binits{A.D.}},
\bauthor{\bsnm{Motala}, \binits{M.J.}},
\bauthor{\bsnm{Snure}, \binits{M.}},
\bauthor{\bsnm{Lynch}, \binits{J.}},
\bauthor{\bsnm{Kumar}, \binits{P.}},
\bauthor{\bsnm{Zhang}, \binits{H.}},
\bauthor{\bsnm{Post}, \binits{S.}},
\bauthor{\bsnm{Bowen}, \binits{T.}},
\bauthor{\bsnm{Muratore}, \binits{C.}},
\bauthor{\bsnm{Robinson}, \binits{J.A.}},
\bauthor{\bsnm{Hendrickson}, \binits{J.R.}},
\bauthor{\bsnm{Glavin}, \binits{N.R.}},
\bauthor{\bsnm{Jariwala}, \binits{D.}}:
\batitle{Ultrathin broadband metasurface superabsorbers from a van der {W}aals
  semimetal}.
\bjtitle{Advanced Optical Materials}
\bvolume{11}(\bissue{4}),
\bfpage{2202011}
(\byear{2023})
\doiurl{10.1002/adom.202202011}
\end{barticle}
\endbibitem

\bibitem[\protect\citeauthoryear{Lin et~al.}{2024}]{Lin2024-13935}
\begin{barticle}
\bauthor{\bsnm{Lin}, \binits{D.}},
\bauthor{\bsnm{Lynch}, \binits{J.}},
\bauthor{\bsnm{Wang}, \binits{S.}},
\bauthor{\bsnm{Hu}, \binits{Z.}},
\bauthor{\bsnm{Rai}, \binits{R.K.}},
\bauthor{\bsnm{Zhang}, \binits{H.}},
\bauthor{\bsnm{Chen}, \binits{C.}},
\bauthor{\bsnm{Kumari}, \binits{S.}},
\bauthor{\bsnm{Stach}, \binits{E.A.}},
\bauthor{\bsnm{Davydov}, \binits{A.V.}},
\bauthor{\bsnm{Redwing}, \binits{J.M.}},
\bauthor{\bsnm{Jariwala}, \binits{D.}}:
\batitle{Broadband light harvesting from scalable two-dimensional semiconductor
  multi-heterostructures}.
\bjtitle{Nano Letters}
\bvolume{24}(\bissue{44}),
\bfpage{13935}--\blpage{13944}
(\byear{2024})
\doiurl{10.1021/acs.nanolett.4c02963}
\end{barticle}
\endbibitem

\bibitem[\protect\citeauthoryear{Alfieri et~al.}{2025}]{Alfieri2025-3020}
\begin{barticle}
\bauthor{\bsnm{Alfieri}, \binits{A.D.}},
\bauthor{\bsnm{Ruth}, \binits{T.}},
\bauthor{\bsnm{Lim}, \binits{C.}},
\bauthor{\bsnm{Lynch}, \binits{J.}},
\bauthor{\bsnm{Jariwala}, \binits{D.}}:
\batitle{Effects of self-hybridized exciton-polaritons on {TMDC}
  photovoltaics}.
\bjtitle{Nano Letters}
\bvolume{25}(\bissue{7}),
\bfpage{3020}--\blpage{3026}
(\byear{2025})
\doiurl{10.1021/acs.nanolett.5c00399}
\end{barticle}
\endbibitem

\bibitem[\protect\citeauthoryear{Raman et~al.}{2014}]{Raman2014-540}
\begin{barticle}
\bauthor{\bsnm{Raman}, \binits{A.P.}},
\bauthor{\bsnm{Anoma}, \binits{M.A.}},
\bauthor{\bsnm{Zhu}, \binits{L.}},
\bauthor{\bsnm{Rephaeli}, \binits{E.}},
\bauthor{\bsnm{Fan}, \binits{S.}}:
\batitle{Passive radiative cooling below ambient air temperature under direct
  sunlight}.
\bjtitle{Nature}
\bvolume{515}(\bissue{7528}),
\bfpage{540}--\blpage{544}
(\byear{2014})
\doiurl{10.1038/nature13883}
\end{barticle}
\endbibitem

\bibitem[\protect\citeauthoryear{Ji et~al.}{2007}]{Ji2007-1227}
\begin{barticle}
\bauthor{\bsnm{Ji}, \binits{L.}},
\bauthor{\bsnm{Kim}, \binits{J.-K.}},
\bauthor{\bsnm{Ji}, \binits{Q.}},
\bauthor{\bsnm{Leung}, \binits{K.-N.}},
\bauthor{\bsnm{Chen}, \binits{Y.}},
\bauthor{\bsnm{Gough}, \binits{R.A.}}:
\batitle{Conformal metal thin-film coatings in high-aspect-ratio trenches using
  a self-sputtered {RF}-driven plasma source}.
\bjtitle{Journal of Vacuum Science \& Technology {B}: {M}icroelectronics and
  Nanometer Structures Processing, Measurement, and Phenomena}
\bvolume{25}(\bissue{4}),
\bfpage{1227}--\blpage{1230}
(\byear{2007})
\doiurl{10.1116/1.2749527}
\end{barticle}
\endbibitem

\bibitem[\protect\citeauthoryear{Gupta et~al.}{2022}]{Gupta2022-2109105}
\begin{barticle}
\bauthor{\bsnm{Gupta}, \binits{B.}},
\bauthor{\bsnm{Hossain}, \binits{M.A.}},
\bauthor{\bsnm{Riaz}, \binits{A.}},
\bauthor{\bsnm{Sharma}, \binits{A.}},
\bauthor{\bsnm{Zhang}, \binits{D.}},
\bauthor{\bsnm{Tan}, \binits{H.H.}},
\bauthor{\bsnm{Jagadish}, \binits{C.}},
\bauthor{\bsnm{Catchpole}, \binits{K.}},
\bauthor{\bsnm{Hoex}, \binits{B.}},
\bauthor{\bsnm{Karuturi}, \binits{S.}}:
\batitle{Recent advances in materials design using atomic layer deposition for
  energy applications}.
\bjtitle{Advanced Functional Materials}
\bvolume{32}(\bissue{3}),
\bfpage{2109105}
(\byear{2022})
\doiurl{10.1002/adfm.202109105}
\end{barticle}
\endbibitem

\bibitem[\protect\citeauthoryear{Ne\u{c}as and Klapetek}{2012}]{Necas2012-181}
\begin{barticle}
\bauthor{\bsnm{Ne\u{c}as}, \binits{D.}},
\bauthor{\bsnm{Klapetek}, \binits{P.}}:
\batitle{Gwyddion: {A}n open-source software for {SPM} data analysis}.
\bjtitle{Cent. Eur. J. Phys.}
\bvolume{10}(\bissue{1}),
\bfpage{181}--\blpage{188}
(\byear{2012})
\doiurl{10.2478/s11534-011-0096-2}
\end{barticle}
\endbibitem

\bibitem[\protect\citeauthoryear{Fujiwara}{2007}]{Fujiwara2007-book}
\begin{bbook}
\bauthor{\bsnm{Fujiwara}, \binits{H.}}:
\bbtitle{Spectroscopic Ellipsometry: Principles and Applications}.
\bpublisher{John Wiley \& Sons, Ltd.},
\blocation{Chichester, UK}
(\byear{2007}).
\doiurl{10.1002/9780470060193}
\end{bbook}
\endbibitem

\bibitem[\protect\citeauthoryear{Vitale and Sugar}{2022}]{Vitale2022-646}
\begin{barticle}
\bauthor{\bsnm{Vitale}, \binits{S.M.}},
\bauthor{\bsnm{Sugar}, \binits{J.D.}}:
\batitle{Using {X}e plasma {FIB} for high-quality {TEM} sample preparation}.
\bjtitle{Microscopy and Microanalysis}
\bvolume{28}(\bissue{3}),
\bfpage{646}--\blpage{658}
(\byear{2022})
\doiurl{10.1017/S1431927622000344}
\end{barticle}
\endbibitem

\bibitem[\protect\citeauthoryear{Kim et~al.}{2023}]{Kim2023-1044}
\begin{barticle}
\bauthor{\bsnm{Kim}, \binits{K.-H.}},
\bauthor{\bsnm{Oh}, \binits{S.}},
\bauthor{\bsnm{Fiagbenu}, \binits{M.M.A.}},
\bauthor{\bsnm{Zheng}, \binits{J.}},
\bauthor{\bsnm{Musavigharavi}, \binits{P.}},
\bauthor{\bsnm{Kumar}, \binits{P.}},
\bauthor{\bsnm{Trainor}, \binits{N.}},
\bauthor{\bsnm{Aljarb}, \binits{A.}},
\bauthor{\bsnm{Wan}, \binits{Y.}},
\bauthor{\bsnm{Kim}, \binits{H.M.}},
\bauthor{\bsnm{Katti}, \binits{K.}},
\bauthor{\bsnm{Song}, \binits{S.}},
\bauthor{\bsnm{Kim}, \binits{G.}},
\bauthor{\bsnm{Tang}, \binits{Z.}},
\bauthor{\bsnm{Fu}, \binits{J.-H.}},
\bauthor{\bsnm{Hakami}, \binits{M.}},
\bauthor{\bsnm{Tung}, \binits{V.}},
\bauthor{\bsnm{Redwing}, \binits{J.M.}},
\bauthor{\bsnm{Stach}, \binits{E.A.}},
\bauthor{\bsnm{Olsson}, \binits{R.H.} \bsuffix{III}},
\bauthor{\bsnm{Jariwala}, \binits{D.}}:
\batitle{Scalable {CMOS} back-end-of-line-compatible
  {A}l{S}c{N}/two-dimensional channel ferroelectric field-effect transistors}.
\bjtitle{Nature Nanotechnology}
\bvolume{18}(\bissue{9}),
\bfpage{1044}--\blpage{1050}
(\byear{2023})
\doiurl{10.1038/s41565-023-01399-y}
\end{barticle}
\endbibitem

\bibitem[\protect\citeauthoryear{Olmon et~al.}{2012}]{Olmon2012-235147}
\begin{barticle}
\bauthor{\bsnm{Olmon}, \binits{R.L.}},
\bauthor{\bsnm{Slovick}, \binits{B.}},
\bauthor{\bsnm{Johnson}, \binits{T.W.}},
\bauthor{\bsnm{Shelton}, \binits{D.}},
\bauthor{\bsnm{Oh}, \binits{S.-H.}},
\bauthor{\bsnm{Boreman}, \binits{G.D.}},
\bauthor{\bsnm{Raschke}, \binits{M.B.}}:
\batitle{Optical dielectric function of gold}.
\bjtitle{Phys. Rev. B}
\bvolume{86}(\bissue{23}),
\bfpage{235147}
(\byear{2012})
\doiurl{10.1103/PhysRevB.86.235147}
\end{barticle}
\endbibitem

\bibitem[\protect\citeauthoryear{Schinke et~al.}{2015}]{Schinke2015-067168}
\begin{barticle}
\bauthor{\bsnm{Schinke}, \binits{C.}},
\bauthor{\bsnm{Christian~Peest}, \binits{P.}},
\bauthor{\bsnm{Schmidt}, \binits{J.}},
\bauthor{\bsnm{Brendel}, \binits{R.}},
\bauthor{\bsnm{Bothe}, \binits{K.}},
\bauthor{\bsnm{Vogt}, \binits{M.R.}},
\bauthor{\bsnm{Kr{\"o}ger}, \binits{I.}},
\bauthor{\bsnm{Winter}, \binits{S.}},
\bauthor{\bsnm{Schirmacher}, \binits{A.}},
\bauthor{\bsnm{Lim}, \binits{S.}},
\bauthor{\bsnm{Nguyen}, \binits{H.T.}},
\bauthor{\bsnm{Mac{D}onald}, \binits{D.}}:
\batitle{Uncertainty analysis for the coefficient of band-to-band absorption of
  crystalline silicon}.
\bjtitle{AIP Advances}
\bvolume{5}(\bissue{6}),
\bfpage{067168}
(\byear{2015})
\doiurl{10.1063/1.4923379}
\end{barticle}
\endbibitem

\bibitem[\protect\citeauthoryear{Macleod}{2017}]{Macleod2017-book}
\begin{bbook}
\bauthor{\bsnm{Macleod}, \binits{H.A.}}:
\bbtitle{Thin-film Optical Filters},
\bedition{5$^\text{th}$} edn.
\bpublisher{CRC Press},
\blocation{New York, NY, USA}
(\byear{2017})
\end{bbook}
\endbibitem

\bibitem[\protect\citeauthoryear{Poruba et~al.}{2000}]{Poruba2000-148}
\begin{barticle}
\bauthor{\bsnm{Poruba}, \binits{A.}},
\bauthor{\bsnm{Fejfar}, \binits{A.}},
\bauthor{\bsnm{Reme{\u s}}, \binits{Z.}},
\bauthor{\bsnm{{\u S}pringer}, \binits{J.}},
\bauthor{\bsnm{Van{\u e}{\u c}ek}, \binits{M.}},
\bauthor{\bsnm{Ko{\u c}ka}, \binits{J.}},
\bauthor{\bsnm{Meier}, \binits{J.}},
\bauthor{\bsnm{Torres}, \binits{P.}},
\bauthor{\bsnm{Shah}, \binits{A.}}:
\batitle{Optical absorption and light scattering in microcrystalline silicon
  thin films and solar cells}.
\bjtitle{Journal of Applied Physics}
\bvolume{88}(\bissue{1}),
\bfpage{148}--\blpage{160}
(\byear{2000})
\doiurl{10.1063/1.373635}
\end{barticle}
\endbibitem

\bibitem[\protect\citeauthoryear{de~Marcos
  et~al.}{2016}]{RodriguezdeMarcos2016-3622}
\begin{barticle}
\bauthor{\bsnm{Marcos}, \binits{L.V.R.-d.}},
\bauthor{\bsnm{Larruquert}, \binits{J.I.}},
\bauthor{\bsnm{M{\'e}ndez}, \binits{J.A.}},
\bauthor{\bsnm{Azn{\'a}rez}, \binits{J.A.}}:
\batitle{Self-consistent optical constants of {S}i{O}$_\text{2}$ and
  {T}a$_\text{2}${O}$_\text{5}$ films}.
\bjtitle{Opt. Mater. Express}
\bvolume{6}(\bissue{11}),
\bfpage{3622}--\blpage{3637}
(\byear{2016})
\doiurl{10.1364/OME.6.003622}
\end{barticle}
\endbibitem

\bibitem[\protect\citeauthoryear{Franta et~al.}{2017}]{Franta2017-405}
\begin{barticle}
\bauthor{\bsnm{Franta}, \binits{D.}},
\bauthor{\bsnm{Dubroka}, \binits{A.}},
\bauthor{\bsnm{Wang}, \binits{C.}},
\bauthor{\bsnm{Giglia}, \binits{A.}},
\bauthor{\bsnm{Voh{\'a}nka}, \binits{J.}},
\bauthor{\bsnm{Franta}, \binits{P.}},
\bauthor{\bsnm{Ohl{\'\i}dal}, \binits{I.}}:
\batitle{Temperature-dependent dispersion model of float zone crystalline
  silicon}.
\bjtitle{Applied Surface Science}
\bvolume{421},
\bfpage{405}--\blpage{419}
(\byear{2017})
\doiurl{10.1016/j.apsusc.2017.02.021}
\end{barticle}
\endbibitem

\bibitem[\protect\citeauthoryear{Schneider and
  Mc{D}aniel}{1967}]{Schneider1967-317}
\begin{barticle}
\bauthor{\bsnm{Schneider}, \binits{S.J.}},
\bauthor{\bsnm{Mc{D}aniel}, \binits{C.L.}}:
\batitle{Effect of environment upon the melting point of
  {A}l$_\text{2}${O}$_\text{3}$}.
\bjtitle{J. Res. Natl. Bur. Stand. A Phys. Chem.}
\bvolume{71A}(\bissue{4}),
\bfpage{317}--\blpage{333}
(\byear{1967})
\doiurl{10.6028/jres.071A.038}
\end{barticle}
\endbibitem

\bibitem[\protect\citeauthoryear{Nannichi}{1963}]{Nannichi1963-586}
\begin{barticle}
\bauthor{\bsnm{Nannichi}, \binits{Y.}}:
\batitle{Sublimation rate of silicon in high vacuum}.
\bjtitle{Japanese Journal of Applied Physics}
\bvolume{2}(\bissue{9}),
\bfpage{586}--\blpage{587}
(\byear{1963})
\doiurl{10.1143/JJAP.2.586}
\end{barticle}
\endbibitem

\bibitem[\protect\citeauthoryear{Batha and Whitney}{1973}]{Batha1973-365}
\begin{barticle}
\bauthor{\bsnm{Batha}, \binits{H.D.}},
\bauthor{\bsnm{Whitney}, \binits{E.D.}}:
\batitle{Kinetics and mechanism of the thermal decomposition of
  {S}i$_\text{3}${N}$_\text{4}$}.
\bjtitle{Journal of the American Ceramic Society}
\bvolume{56}(\bissue{7}),
\bfpage{365}--\blpage{369}
(\byear{1973})
\doiurl{10.1111/j.1151-2916.1973.tb12687.x}
\end{barticle}
\endbibitem

\bibitem[\protect\citeauthoryear{Liehr et~al.}{1987}]{Liehr1987-1559}
\begin{barticle}
\bauthor{\bsnm{Liehr}, \binits{M.}},
\bauthor{\bsnm{Lewis}, \binits{J.E.}},
\bauthor{\bsnm{Rubloff}, \binits{G.W.}}:
\batitle{Kinetics of high‐temperature thermal decomposition of
  {S}i{O}$_\text{2}$ on {S}i(100)}.
\bjtitle{Journal of Vacuum Science \& Technology A}
\bvolume{5}(\bissue{4}),
\bfpage{1559}--\blpage{1562}
(\byear{1987})
\doiurl{10.1116/1.574564}
\end{barticle}
\endbibitem

\bibitem[\protect\citeauthoryear{Mizuno et~al.}{2002}]{Mizuno2002-1716}
\begin{barticle}
\bauthor{\bsnm{Mizuno}, \binits{Y.}},
\bauthor{\bsnm{King}, \binits{F.K.}},
\bauthor{\bsnm{Yamauchi}, \binits{Y.}},
\bauthor{\bsnm{Homma}, \binits{T.}},
\bauthor{\bsnm{Tanaka}, \binits{A.}},
\bauthor{\bsnm{Takakuwa}, \binits{Y.}},
\bauthor{\bsnm{Momose}, \binits{T.}}:
\batitle{Temperature dependence of oxide decomposition on titanium surfaces in
  ultrahigh vacuum}.
\bjtitle{Journal of Vacuum Science \& Technology A}
\bvolume{20}(\bissue{5}),
\bfpage{1716}--\blpage{1721}
(\byear{2002})
\doiurl{10.1116/1.1500746}
\end{barticle}
\endbibitem

\bibitem[\protect\citeauthoryear{Petersen}{1982}]{Petersen1982-420}
\begin{barticle}
\bauthor{\bsnm{Petersen}, \binits{K.E.}}:
\batitle{Silicon as a mechanical material}.
\bjtitle{Proceedings of the IEEE}
\bvolume{70}(\bissue{5}),
\bfpage{420}--\blpage{457}
(\byear{1982})
\doiurl{10.1109/PROC.1982.12331}
\end{barticle}
\endbibitem

\bibitem[\protect\citeauthoryear{Yen et~al.}{2003}]{Yen2003-1895}
\begin{barticle}
\bauthor{\bsnm{Yen}, \binits{B.K.}},
\bauthor{\bsnm{White}, \binits{R.L.}},
\bauthor{\bsnm{Waltman}, \binits{R.J.}},
\bauthor{\bsnm{Dai}, \binits{Q.}},
\bauthor{\bsnm{Miller}, \binits{D.C.}},
\bauthor{\bsnm{Kellock}, \binits{A.J.}},
\bauthor{\bsnm{Marchon}, \binits{B.}},
\bauthor{\bsnm{Kasai}, \binits{P.H.}},
\bauthor{\bsnm{Toney}, \binits{M.F.}},
\bauthor{\bsnm{York}, \binits{B.R.}},
\bauthor{\bsnm{Deng}, \binits{H.}},
\bauthor{\bsnm{Xiao}, \binits{Q.-F.}},
\bauthor{\bsnm{Raman}, \binits{V.}}:
\batitle{Microstructure and properties of ultrathin amorphous silicon nitride
  protective coating}.
\bjtitle{Journal of Vacuum Science \& Technology A}
\bvolume{21}(\bissue{6}),
\bfpage{1895}--\blpage{1904}
(\byear{2003})
\doiurl{10.1116/1.1615974}
\end{barticle}
\endbibitem

\bibitem[\protect\citeauthoryear{Kawase et~al.}{2009}]{Kawase2009-101401}
\begin{barticle}
\bauthor{\bsnm{Kawase}, \binits{K.}},
\bauthor{\bsnm{Noda}, \binits{S.}},
\bauthor{\bsnm{Nakai}, \binits{T.}},
\bauthor{\bsnm{Uehara}, \binits{Y.}}:
\batitle{Densification of chemical vapor deposition silicon dioxide film using
  ozone treatment}.
\bjtitle{Japanese Journal of Applied Physics}
\bvolume{48}(\bissue{10R}),
\bfpage{101401}
(\byear{2009})
\doiurl{10.1143/JJAP.48.101401}
\end{barticle}
\endbibitem

\bibitem[\protect\citeauthoryear{Saari et~al.}{2022}]{Saari2022-15357}
\begin{barticle}
\bauthor{\bsnm{Saari}, \binits{J.}},
\bauthor{\bsnm{Ali-{L}{\"o}ytty}, \binits{H.}},
\bauthor{\bsnm{Lahtonen}, \binits{K.}},
\bauthor{\bsnm{Hannula}, \binits{M.}},
\bauthor{\bsnm{Palmolahti}, \binits{L.}},
\bauthor{\bsnm{Tukiainen}, \binits{A.}},
\bauthor{\bsnm{Valden}, \binits{M.}}:
\batitle{Low-temperature route to direct amorphous to rutile crystallization of
  {T}i{O}$_\text{2}$ thin films grown by atomic layer deposition}.
\bjtitle{The Journal of Physical Chemistry C}
\bvolume{126}(\bissue{36}),
\bfpage{15357}--\blpage{15366}
(\byear{2022})
\doiurl{10.1021/acs.jpcc.2c04905}
\end{barticle}
\endbibitem

\bibitem[\protect\citeauthoryear{Sato et~al.}{1998}]{Sato1998-148}
\begin{barticle}
\bauthor{\bsnm{Sato}, \binits{K.}},
\bauthor{\bsnm{Yoshioka}, \binits{T.}},
\bauthor{\bsnm{Ando}, \binits{T.}},
\bauthor{\bsnm{Shikida}, \binits{M.}},
\bauthor{\bsnm{Kawabata}, \binits{T.}}:
\batitle{Tensile testing of silicon film having different crystallographic
  orientations carried out on a silicon chip}.
\bjtitle{Sensors and {A}ctuators {A}: {P}hysical}
\bvolume{70}(\bissue{1}),
\bfpage{148}--\blpage{152}
(\byear{1998})
\doiurl{10.1016/S0924-4247(98)00125-3}
\end{barticle}
\endbibitem

\bibitem[\protect\citeauthoryear{Sharpe et~al.}{1999}]{Sharpe1999-162}
\begin{barticle}
\bauthor{\bsnm{Sharpe}, \binits{W.N.}},
\bauthor{\bsnm{Turner}, \binits{K.T.}},
\bauthor{\bsnm{Edwards}, \binits{R.L.}}:
\batitle{Tensile testing of polysilicon}.
\bjtitle{Experimental Mechanics}
\bvolume{39}(\bissue{3}),
\bfpage{162}--\blpage{170}
(\byear{1999})
\doiurl{10.1007/BF02323548}
\end{barticle}
\endbibitem

\bibitem[\protect\citeauthoryear{Tsuchiya et~al.}{2000}]{Tsuchiya2000-286}
\begin{barticle}
\bauthor{\bsnm{Tsuchiya}, \binits{T.}},
\bauthor{\bsnm{Inoue}, \binits{A.}},
\bauthor{\bsnm{Sakata}, \binits{J.}}:
\batitle{Tensile testing of insulating thin films; humidity effect on tensile
  strength of {S}i{O}$_\text{2}$ films}.
\bjtitle{Sensors and {A}ctuators {A}: {P}hysical}
\bvolume{82}(\bissue{1}),
\bfpage{286}--\blpage{290}
(\byear{2000})
\doiurl{10.1016/S0924-4247(99)00363-5}
\end{barticle}
\endbibitem

\bibitem[\protect\citeauthoryear{Yoshioka et~al.}{2000}]{Yoshioka2000-291}
\begin{barticle}
\bauthor{\bsnm{Yoshioka}, \binits{T.}},
\bauthor{\bsnm{Ando}, \binits{T.}},
\bauthor{\bsnm{Shikida}, \binits{M.}},
\bauthor{\bsnm{Sato}, \binits{K.}}:
\batitle{Tensile testing of {S}i{O}$_\text{2}$ and
  {S}i$_\text{3}${N}$_\text{4}$ films carried out on a silicon chip}.
\bjtitle{Sensors and Actuators A: Physical}
\bvolume{82}(\bissue{1}),
\bfpage{291}--\blpage{296}
(\byear{2000})
\doiurl{10.1016/S0924-4247(99)00364-7}
\end{barticle}
\endbibitem

\bibitem[\protect\citeauthoryear{Tsuchiya}{2005}]{Tsuchiya2005-665}
\begin{barticle}
\bauthor{\bsnm{Tsuchiya}, \binits{T.}}:
\batitle{Tensile testing of silicon thin films}.
\bjtitle{Fatigue \& Fracture of Engineering Materials \& Structures}
\bvolume{28}(\bissue{8}),
\bfpage{665}--\blpage{674}
(\byear{2005})
\doiurl{10.1111/j.1460-2695.2005.00910.x}
\end{barticle}
\endbibitem

\bibitem[\protect\citeauthoryear{Tavares et~al.}{2008}]{Tavares2008-1434}
\begin{barticle}
\bauthor{\bsnm{Tavares}, \binits{C.J.}},
\bauthor{\bsnm{Marques}, \binits{S.M.}},
\bauthor{\bsnm{Lanceros-{M}{\'e}ndez}, \binits{S.}},
\bauthor{\bsnm{Sencadas}, \binits{V.}},
\bauthor{\bsnm{Teixeira}, \binits{V.}},
\bauthor{\bsnm{Carneiro}, \binits{J.O.}},
\bauthor{\bsnm{Martins}, \binits{A.J.}},
\bauthor{\bsnm{Fernandes}, \binits{A.J.}}:
\batitle{Strain analysis of photocatalytic {T}i{O}$_\text{2}$ thin films on
  polymer substrates}.
\bjtitle{Thin Solid Films}
\bvolume{516}(\bissue{7}),
\bfpage{1434}--\blpage{1438}
(\byear{2008})
\doiurl{10.1016/j.tsf.2007.03.134}
\end{barticle}
\endbibitem

\bibitem[\protect\citeauthoryear{Tsuchiya et~al.}{2010}]{Tsuchiya2010-1}
\begin{barticle}
\bauthor{\bsnm{Tsuchiya}, \binits{T.}},
\bauthor{\bsnm{Ikeda}, \binits{T.}},
\bauthor{\bsnm{Tsunematsu}, \binits{A.}},
\bauthor{\bsnm{Sugano}, \binits{K.}},
\bauthor{\bsnm{Tabata}, \binits{O.}}:
\batitle{Tensile testing of single-crystal silicon thin films at
  600${}^\circ${C} using infrared radiation heating}.
\bjtitle{Sensors and Materials}
\bvolume{22}(\bissue{1}),
\bfpage{1}--\blpage{11}
(\byear{2010})
\doiurl{10.18494/SAM.2010.619}
\end{barticle}
\endbibitem

\bibitem[\protect\citeauthoryear{Miller et~al.}{2010}]{Miller2010-58}
\begin{barticle}
\bauthor{\bsnm{Miller}, \binits{D.C.}},
\bauthor{\bsnm{Foster}, \binits{R.R.}},
\bauthor{\bsnm{Jen}, \binits{S.-H.}},
\bauthor{\bsnm{Bertrand}, \binits{J.A.}},
\bauthor{\bsnm{Cunningham}, \binits{S.J.}},
\bauthor{\bsnm{Morris}, \binits{A.S.}},
\bauthor{\bsnm{Lee}, \binits{Y.-C.}},
\bauthor{\bsnm{George}, \binits{S.M.}},
\bauthor{\bsnm{Dunn}, \binits{M.L.}}:
\batitle{Thermo-mechanical properties of alumina films created using the atomic
  layer deposition technique}.
\bjtitle{Sens. Actuators, A}
\bvolume{164}(\bissue{1}),
\bfpage{58}--\blpage{67}
(\byear{2010})
\doiurl{10.1016/j.sna.2010.09.018}
\end{barticle}
\endbibitem

\bibitem[\protect\citeauthoryear{Jen et~al.}{2011}]{Jen2011-084305}
\begin{barticle}
\bauthor{\bsnm{Jen}, \binits{S.-H.}},
\bauthor{\bsnm{Bertrand}, \binits{J.A.}},
\bauthor{\bsnm{George}, \binits{S.M.}}:
\batitle{Critical tensile and compressive strains for cracking of
  {A}l$_\text{2}${O}$_\text{3}$ films grown by atomic layer deposition}.
\bjtitle{J. Appl. Phys.}
\bvolume{109}(\bissue{8}),
\bfpage{084305}
(\byear{2011})
\doiurl{10.1063/1.3567912}
\end{barticle}
\endbibitem

\bibitem[\protect\citeauthoryear{Bertolazzi et~al.}{2011}]{Bertolazzi2011-9703}
\begin{barticle}
\bauthor{\bsnm{Bertolazzi}, \binits{S.}},
\bauthor{\bsnm{Brivio}, \binits{J.}},
\bauthor{\bsnm{Kis}, \binits{A.}}:
\batitle{Stretching and breaking of ultrathin {M}o{S}$_2$}.
\bjtitle{ACS Nano}
\bvolume{5}(\bissue{12}),
\bfpage{9703}--\blpage{9709}
(\byear{2011})
\doiurl{10.1021/nn203879f}
\end{barticle}
\endbibitem

\bibitem[\protect\citeauthoryear{Borgese et~al.}{2012}]{Borgese2012-2459}
\begin{barticle}
\bauthor{\bsnm{Borgese}, \binits{L.}},
\bauthor{\bsnm{Gelfi}, \binits{M.}},
\bauthor{\bsnm{Bontempi}, \binits{E.}},
\bauthor{\bsnm{Goudeau}, \binits{P.}},
\bauthor{\bsnm{Geandier}, \binits{G.}},
\bauthor{\bsnm{Thiaudi{\`e}re}, \binits{D.}},
\bauthor{\bsnm{Depero}, \binits{L.E.}}:
\batitle{{Y}oung modulus and {P}oisson ratio measurements of {T}i{O}$_\text{2}$
  thin films deposited with atomic layer deposition}.
\bjtitle{Surface and Coatings Technology}
\bvolume{206}(\bissue{8}),
\bfpage{2459}--\blpage{2463}
(\byear{2012})
\doiurl{10.1016/j.surfcoat.2011.10.050}
\end{barticle}
\endbibitem

\bibitem[\protect\citeauthoryear{Sledzinska et~al.}{2020}]{Sledzinska2020-1169}
\begin{barticle}
\bauthor{\bsnm{Sledzinska}, \binits{M.}},
\bauthor{\bsnm{Jumbert}, \binits{G.}},
\bauthor{\bsnm{Placidi}, \binits{M.}},
\bauthor{\bsnm{Arrighi}, \binits{A.}},
\bauthor{\bsnm{Xiao}, \binits{P.}},
\bauthor{\bsnm{Alzina}, \binits{F.}},
\bauthor{\bsnm{{Sotomayor Torres}}, \binits{C.M.}}:
\batitle{Fracturing of polycrystalline {M}o{S}$_\text{2}$ nanofilms}.
\bjtitle{ACS Appl. Electron. Mater.}
\bvolume{2}(\bissue{4}),
\bfpage{1169}--\blpage{1175}
(\byear{2020})
\doiurl{10.1021/acsaelm.0c00189}
\end{barticle}
\endbibitem

\bibitem[\protect\citeauthoryear{Timoshenko and
  Woinowsky-{K}rieger}{1959}]{Timoshenko1959-book}
\begin{bbook}
\bauthor{\bsnm{Timoshenko}, \binits{S.}},
\bauthor{\bsnm{Woinowsky-{K}rieger}, \binits{S.}}:
\bbtitle{Theory of Plates and Shells}.
\bpublisher{Mc{G}raw-{H}ill Book Company},
\blocation{New York, NY, USA}
(\byear{1959})
\end{bbook}
\endbibitem

\bibitem[\protect\citeauthoryear{Babacic et~al.}{2021}]{Babacic2021-2008614}
\begin{barticle}
\bauthor{\bsnm{Babacic}, \binits{V.}},
\bauthor{\bsnm{Saleta~Reig}, \binits{D.}},
\bauthor{\bsnm{Varghese}, \binits{S.}},
\bauthor{\bsnm{Vasileiadis}, \binits{T.}},
\bauthor{\bsnm{Coy}, \binits{E.}},
\bauthor{\bsnm{Tielrooij}, \binits{K.-J.}},
\bauthor{\bsnm{Graczykowski}, \binits{B.}}:
\batitle{Thickness-dependent elastic softening of few-layer free-standing
  {M}o{S}e$_\text{2}$}.
\bjtitle{Advanced Materials}
\bvolume{33}(\bissue{23}),
\bfpage{2008614}
(\byear{2021})
\doiurl{10.1002/adma.202008614}
\end{barticle}
\endbibitem

\bibitem[\protect\citeauthoryear{Fei et~al.}{2016}]{Fei2016-12206}
\begin{barticle}
\bauthor{\bsnm{Fei}, \binits{L.}},
\bauthor{\bsnm{Lei}, \binits{S.}},
\bauthor{\bsnm{Zhang}, \binits{W.-B.}},
\bauthor{\bsnm{Lu}, \binits{W.}},
\bauthor{\bsnm{Lin}, \binits{Z.}},
\bauthor{\bsnm{Lam}, \binits{C.H.}},
\bauthor{\bsnm{Chai}, \binits{Y.}},
\bauthor{\bsnm{Wang}, \binits{Y.}}:
\batitle{Direct {TEM} observations of growth mechanisms of two-dimensional
  mo{S}$_\text{2}$ flakes}.
\bjtitle{Nature Communications}
\bvolume{7}(\bissue{1}),
\bfpage{12206}
(\byear{2016})
\doiurl{10.1038/ncomms12206}
\end{barticle}
\endbibitem

\bibitem[\protect\citeauthoryear{Kumar and Viswanath}{2017}]{Kumar2017-5068}
\begin{barticle}
\bauthor{\bsnm{Kumar}, \binits{P.}},
\bauthor{\bsnm{Viswanath}, \binits{B.}}:
\batitle{Horizontally and vertically aligned growth of strained
  mo{S}$_\text{2}$ layers with dissimilar wetting and catalytic behaviors}.
\bjtitle{CrystEngComm}
\bvolume{19},
\bfpage{5068}--\blpage{5078}
(\byear{2017})
\doiurl{10.1039/C7CE01162H}
\end{barticle}
\endbibitem

\bibitem[\protect\citeauthoryear{Foral}{1979}]{Foral1979-200}
\begin{barticle}
\bauthor{\bsnm{Foral}, \binits{R.F.}}:
\batitle{Composite spherical pressure vessels with hardening metal liners}.
\bjtitle{J. Press. Vessel Technol.}
\bvolume{101}(\bissue{3}),
\bfpage{200}--\blpage{206}
(\byear{1979})
\doiurl{10.1115/1.3454623}
\end{barticle}
\endbibitem

\bibitem[\protect\citeauthoryear{Roy and Massard}{1992}]{Roy1992-479}
\begin{barticle}
\bauthor{\bsnm{Roy}, \binits{A.K.}},
\bauthor{\bsnm{Massard}, \binits{T.N.}}:
\batitle{A design study of thick multilayered composite spherical pressure
  vessels}.
\bjtitle{J. Reinf. Plast. Compos.}
\bvolume{11}(\bissue{5}),
\bfpage{479}--\blpage{493}
(\byear{1992})
\doiurl{10.1177/073168449201100502}
\end{barticle}
\endbibitem

\bibitem[\protect\citeauthoryear{Liu et~al.}{2009}]{Liu2009-2198}
\begin{barticle}
\bauthor{\bsnm{Liu}, \binits{B.}},
\bauthor{\bsnm{Feng}, \binits{X.}},
\bauthor{\bsnm{Zhang}, \binits{S.-M.}}:
\batitle{The effective {Y}oung's modulus of composites beyond the {V}oigt
  estimation due to the {P}oisson effect}.
\bjtitle{Composites Science and Technology}
\bvolume{69}(\bissue{13}),
\bfpage{2198}--\blpage{2204}
(\byear{2009})
\doiurl{10.1016/j.compscitech.2009.06.004}
\end{barticle}
\endbibitem

\bibitem[\protect\citeauthoryear{You et~al.}{2017}]{You2017-682}
\begin{barticle}
\bauthor{\bsnm{You}, \binits{Y.-J.}},
\bauthor{\bsnm{Kim}, \binits{J.-H.J.}},
\bauthor{\bsnm{Park}, \binits{K.-T.}},
\bauthor{\bsnm{Seo}, \binits{D.-W.}},
\bauthor{\bsnm{Lee}, \binits{T.-H.}}:
\batitle{Modification of rule of mixtures for tensile strength estimation of
  circular {GFRP} rebars}.
\bjtitle{Polymers}
\bvolume{9}(\bissue{12}),
\bfpage{682}
(\byear{2017})
\doiurl{10.3390/polym9120682}
\end{barticle}
\endbibitem

\bibitem[\protect\citeauthoryear{Raju et~al.}{2018}]{Raju2018-607}
\begin{barticle}
\bauthor{\bsnm{Raju}, \binits{B.}},
\bauthor{\bsnm{Hiremath}, \binits{S.R.}},
\bauthor{\bsnm{{Roy Mahapatra}}, \binits{D.}}:
\batitle{A review of micromechanics based models for effective elastic
  properties of reinforced polymer matrix composites}.
\bjtitle{Composite Structures}
\bvolume{204},
\bfpage{607}--\blpage{619}
(\byear{2018})
\doiurl{10.1016/j.compstruct.2018.07.125}
\end{barticle}
\endbibitem

\bibitem[\protect\citeauthoryear{Ylivaara et~al.}{2014}]{Ylivaara2014-124}
\begin{barticle}
\bauthor{\bsnm{Ylivaara}, \binits{O.M.E.}},
\bauthor{\bsnm{Liu}, \binits{X.}},
\bauthor{\bsnm{Kilpi}, \binits{L.}},
\bauthor{\bsnm{Lyytinen}, \binits{J.}},
\bauthor{\bsnm{Schneider}, \binits{D.}},
\bauthor{\bsnm{Laitinen}, \binits{M.}},
\bauthor{\bsnm{Julin}, \binits{J.}},
\bauthor{\bsnm{Ali}, \binits{S.}},
\bauthor{\bsnm{Sintonen}, \binits{S.}},
\bauthor{\bsnm{Berdova}, \binits{M.}},
\bauthor{\bsnm{Haimi}, \binits{E.}},
\bauthor{\bsnm{Sajavaara}, \binits{T.}},
\bauthor{\bsnm{Ronkainen}, \binits{H.}},
\bauthor{\bsnm{Lipsanen}, \binits{H.}},
\bauthor{\bsnm{Koskinen}, \binits{J.}},
\bauthor{\bsnm{Hannula}, \binits{S.-P.}},
\bauthor{\bsnm{Puurunen}, \binits{R.L.}}:
\batitle{Aluminum oxide from trimethylaluminum and water by atomic layer
  deposition: {T}he temperature dependence of residual stress, elastic modulus,
  hardness and adhesion}.
\bjtitle{Thin Solid Films}
\bvolume{552},
\bfpage{124}--\blpage{135}
(\byear{2014})
\doiurl{10.1016/j.tsf.2013.11.112}
\end{barticle}
\endbibitem

\bibitem[\protect\citeauthoryear{Cooper et~al.}{2013}]{Cooper2013-035423}
\begin{barticle}
\bauthor{\bsnm{Cooper}, \binits{R.C.}},
\bauthor{\bsnm{Lee}, \binits{C.}},
\bauthor{\bsnm{Marianetti}, \binits{C.A.}},
\bauthor{\bsnm{Wei}, \binits{X.}},
\bauthor{\bsnm{Hone}, \binits{J.}},
\bauthor{\bsnm{Kysar}, \binits{J.W.}}:
\batitle{Nonlinear elastic behavior of two-dimensional molybdenum disulfide}.
\bjtitle{Phys. Rev. B}
\bvolume{87}(\bissue{3}),
\bfpage{035423}
(\byear{2013})
\doiurl{10.1103/PhysRevB.87.035423}
\end{barticle}
\endbibitem

\bibitem[\protect\citeauthoryear{Peng and De}{2013}]{Peng2013-19427}
\begin{barticle}
\bauthor{\bsnm{Peng}, \binits{Q.}},
\bauthor{\bsnm{De}, \binits{S.}}:
\batitle{Outstanding mechanical properties of monolayer {M}o{S}$_\text{2}$ and
  its application in elastic energy storage}.
\bjtitle{Phys. Chem. Chem. Phys.}
\bvolume{15}(\bissue{44}),
\bfpage{19427}--\blpage{19437}
(\byear{2013})
\doiurl{10.1039/C3CP52879K}
\end{barticle}
\endbibitem

\bibitem[\protect\citeauthoryear{Woo et~al.}{2016}]{Woo2016-075420}
\begin{barticle}
\bauthor{\bsnm{Woo}, \binits{S.}},
\bauthor{\bsnm{Park}, \binits{H.C.}},
\bauthor{\bsnm{Son}, \binits{Y.-W.}}:
\batitle{Poisson's ratio in layered two-dimensional crystals}.
\bjtitle{Phys. Rev. B}
\bvolume{93}(\bissue{7}),
\bfpage{075420}
(\byear{2016})
\doiurl{10.1103/PhysRevB.93.075420}
\end{barticle}
\endbibitem

\bibitem[\protect\citeauthoryear{Tripp et~al.}{2006}]{Tripp2006-419}
\begin{barticle}
\bauthor{\bsnm{Tripp}, \binits{M.K.}},
\bauthor{\bsnm{Stampfer}, \binits{C.}},
\bauthor{\bsnm{Miller}, \binits{D.C.}},
\bauthor{\bsnm{Helbling}, \binits{T.}},
\bauthor{\bsnm{Herrmann}, \binits{C.F.}},
\bauthor{\bsnm{Hierold}, \binits{C.}},
\bauthor{\bsnm{Gall}, \binits{K.}},
\bauthor{\bsnm{George}, \binits{S.M.}},
\bauthor{\bsnm{Bright}, \binits{V.M.}}:
\batitle{The mechanical properties of atomic layer deposited alumina for use in
  micro- and nano-electromechanical systems}.
\bjtitle{Sens. Actuators, A}
\bvolume{130--131},
\bfpage{419}--\blpage{429}
(\byear{2006})
\doiurl{10.1016/j.sna.2006.01.029}
\end{barticle}
\endbibitem

\bibitem[\protect\citeauthoryear{Gayler}{1938}]{Gayler1938-478}
\begin{barticle}
\bauthor{\bsnm{Gayler}, \binits{M.L.V.}}:
\batitle{Melting point of high-purity silicon}.
\bjtitle{Nature}
\bvolume{142}(\bissue{3593}),
\bfpage{478}
(\byear{1938})
\doiurl{10.1038/142478a0}
\end{barticle}
\endbibitem

\bibitem[\protect\citeauthoryear{Rumble}{2024}]{CRCbook2024}
\begin{bbook}
\bauthor{\bsnm{Rumble}, \binits{J.R.}}:
\bbtitle{CRC Handbook of Chemistry and Physics},
\bedition{105$^\text{th}$} edn.
\bpublisher{CRC Press},
\blocation{Boca Raton, FL}
(\byear{2024})
\end{bbook}
\endbibitem

\bibitem[\protect\citeauthoryear{{St.\ Pierre}}{1952}]{StPierre1952-188}
\begin{barticle}
\bauthor{\bsnm{{St.\ Pierre}}, \binits{P.D.S.}}:
\batitle{A note on the melting point of titanium dioxide}.
\bjtitle{Journal of the American Ceramic Society}
\bvolume{35}(\bissue{7}),
\bfpage{188}--\blpage{188}
(\byear{1952})
\doiurl{10.1111/j.1151-2916.1952.tb13097.x}
\end{barticle}
\endbibitem

\bibitem[\protect\citeauthoryear{Kulkarni et~al.}{2016}]{Kulkarni2016-43}
\begin{bchapter}
\bauthor{\bsnm{Kulkarni}, \binits{N.}},
\bauthor{\bsnm{Lubin}, \binits{P.M.}},
\bauthor{\bsnm{Zhang}, \binits{Q.}}:
\bctitle{Relativistic solutions to directed energy}.
In: \beditor{\bsnm{Hughes}, \binits{G.B.}} (ed.)
\bbtitle{Planetary Defense and Space Environment Applications},
vol. \bseriesno{9981},
pp. \bfpage{43}--\blpage{52}.
\bpublisher{SPIE},
\blocation{Bellingham, WA, USA}
(\byear{2016}).
\doiurl{10.1117/12.2238094}
\end{bchapter}
\endbibitem

\bibitem[\protect\citeauthoryear{F\H{u}zfa et~al.}{2020}]{Fuzfa2020-043186}
\begin{barticle}
\bauthor{\bsnm{F\H{u}zfa}, \binits{A.}},
\bauthor{\bsnm{Dhelonga-Biarufu}, \binits{W.}},
\bauthor{\bsnm{Welcomme}, \binits{O.}}:
\batitle{Sailing towards the stars close to the speed of light}.
\bjtitle{Phys. Rev. Research}
\bvolume{2},
\bfpage{043186}
(\byear{2020})
\doiurl{10.1103/PhysRevResearch.2.043186}
\end{barticle}
\endbibitem

\bibitem[\protect\citeauthoryear{Pegoraro et~al.}{2021}]{Pegoraro2021-485}
\begin{barticle}
\bauthor{\bsnm{Pegoraro}, \binits{F.}},
\bauthor{\bsnm{Livi}, \binits{C.}},
\bauthor{\bsnm{Macchi}, \binits{A.}}:
\batitle{Light sail boosted by instantaneous radiation pressure}.
\bjtitle{Eur. Phys. J. Plus}
\bvolume{136}(\bissue{5}),
\bfpage{485}
(\byear{2021})
\doiurl{10.1140/epjp/s13360-021-01357-4}
\end{barticle}
\endbibitem

\bibitem[\protect\citeauthoryear{Kulkarni et~al.}{2018}]{Kulkarni2018-155}
\begin{barticle}
\bauthor{\bsnm{Kulkarni}, \binits{N.}},
\bauthor{\bsnm{Lubin}, \binits{P.}},
\bauthor{\bsnm{Zhang}, \binits{Q.}}:
\batitle{Relativistic spacecraft propelled by directed energy}.
\bjtitle{Astron. J.}
\bvolume{155}(\bissue{4}),
\bfpage{155}
(\byear{2018})
\doiurl{10.3847/1538-3881/aaafd2}
\end{barticle}
\endbibitem

\bibitem[\protect\citeauthoryear{Nichols and Hull}{1903}]{Nichols1903-315}
\begin{barticle}
\bauthor{\bsnm{Nichols}, \binits{E.F.}},
\bauthor{\bsnm{Hull}, \binits{G.F.}}:
\batitle{The pressure due to radiation}.
\bjtitle{Astrophys. J.}
\bvolume{17}(\bissue{5}),
\bfpage{315}--\blpage{351}
(\byear{1903})
\doiurl{10.1086/141035}
\end{barticle}
\endbibitem

\bibitem[\protect\citeauthoryear{Ohta and Ishida}{1990}]{Ohta1990-2466}
\begin{barticle}
\bauthor{\bsnm{Ohta}, \binits{K.}},
\bauthor{\bsnm{Ishida}, \binits{H.}}:
\batitle{Matrix formalism for calculation of the light beam intensity in
  stratified multilayered films, and its use in the analysis of emission
  spectra}.
\bjtitle{Appl. Opt.}
\bvolume{29}(\bissue{16}),
\bfpage{2466}--\blpage{2473}
(\byear{1990})
\doiurl{10.1364/AO.29.002466}
\end{barticle}
\endbibitem

\bibitem[\protect\citeauthoryear{Boyajian}{1923}]{Boyajian1923-155}
\begin{barticle}
\bauthor{\bsnm{Boyajian}, \binits{A.}}:
\batitle{Physical interpretation of complex angles and their functions}.
\bjtitle{J. Am. Inst. Electr. Eng.}
\bvolume{42}(\bissue{2}),
\bfpage{155}--\blpage{164}
(\byear{1923})
\doiurl{10.1109/JoAIEE.1923.6592034}
\end{barticle}
\endbibitem

\bibitem[\protect\citeauthoryear{Popova et~al.}{2016}]{Popova2016-1346}
\begin{barticle}
\bauthor{\bsnm{Popova}, \binits{E.}},
\bauthor{\bsnm{Efendiev}, \binits{M.}},
\bauthor{\bsnm{Gabitov}, \binits{I.}}:
\batitle{On the stability of a space vehicle riding on an intense laser beam}.
\bjtitle{Math. Meth. Appl. Sci.}
\bvolume{40},
\bfpage{1346}--\blpage{1354}
(\byear{2016})
\doiurl{10.1002/mma.4282}
\end{barticle}
\endbibitem

\bibitem[\protect\citeauthoryear{Mistrik et~al.}{2023}]{Jan2023-2911}
\begin{barticle}
\bauthor{\bsnm{Mistrik}, \binits{J.}},
\bauthor{\bsnm{Krbal}, \binits{M.}},
\bauthor{\bsnm{Prokop}, \binits{V.}},
\bauthor{\bsnm{Prikryl}, \binits{J.}}:
\batitle{Giant change of {M}o{S}$_\text{2}$ optical properties along
  amorphous--crystalline transition: {B}roadband spectroscopic study including
  the {NIR} therapeutic window}.
\bjtitle{Nanoscale Adv.}
\bvolume{5}(\bissue{11}),
\bfpage{2911}--\blpage{2920}
(\byear{2023})
\doiurl{10.1039/D3NA00111C}
\end{barticle}
\endbibitem

\bibitem[\protect\citeauthoryear{Beal and Hughes}{1979}]{Beal1979-881}
\begin{barticle}
\bauthor{\bsnm{Beal}, \binits{A.R.}},
\bauthor{\bsnm{Hughes}, \binits{H.P.}}:
\batitle{Kramers-{K}r{\"o}nig analysis of the reflectivity spectra of
  2{H}-{M}o{S}$_\text{2}$, 2{H}-{M}o{S}e$_\text{2}$, and
  2{H}-{M}o{T}e$_\text{2}$}.
\bjtitle{J. Phys. C: Solid State Phys.}
\bvolume{12}(\bissue{5}),
\bfpage{881}--\blpage{890}
(\byear{1979})
\doiurl{10.1088/0022-3719/12/5/017}
\end{barticle}
\endbibitem

\bibitem[\protect\citeauthoryear{Roxlo et~al.}{1987}]{Roxio1987-555}
\begin{barticle}
\bauthor{\bsnm{Roxlo}, \binits{C.B.}},
\bauthor{\bsnm{Chianelli}, \binits{R.R.}},
\bauthor{\bsnm{Deckman}, \binits{H.W.}},
\bauthor{\bsnm{Ruppert}, \binits{A.F.}},
\bauthor{\bsnm{Wong}, \binits{P.P.}}:
\batitle{Bulk and surface optical absorption in molybdenum disulfide}.
\bjtitle{Journal of Vacuum Science \& Technology A}
\bvolume{5}(\bissue{4}),
\bfpage{555}--\blpage{557}
(\byear{1987})
\doiurl{10.1116/1.574671}
\end{barticle}
\endbibitem

\bibitem[\protect\citeauthoryear{Yim et~al.}{2014}]{Yim2014-103114}
\begin{barticle}
\bauthor{\bsnm{Yim}, \binits{C.}},
\bauthor{\bsnm{{O'B}rien}, \binits{M.}},
\bauthor{\bsnm{Mc{E}voy}, \binits{N.}},
\bauthor{\bsnm{Winters}, \binits{S.}},
\bauthor{\bsnm{Mirza}, \binits{I.}},
\bauthor{\bsnm{Lunney}, \binits{J.G.}},
\bauthor{\bsnm{Duesberg}, \binits{G.S.}}:
\batitle{Investigation of the optical properties of {M}o{S}$_\text{2}$ thin
  films using spectroscopic ellipsometry}.
\bjtitle{Appl. Phys. Lett.}
\bvolume{104}(\bissue{10}),
\bfpage{103114}
(\byear{2014})
\doiurl{10.1063/1.4868108}
\end{barticle}
\endbibitem

\bibitem[\protect\citeauthoryear{Hsu et~al.}{2019}]{Hsu2019-1900239}
\begin{barticle}
\bauthor{\bsnm{Hsu}, \binits{C.}},
\bauthor{\bsnm{Frisenda}, \binits{R.}},
\bauthor{\bsnm{Schmidt}, \binits{R.}},
\bauthor{\bsnm{Arora}, \binits{A.}},
\bauthor{\bsnm{{de Vasconcellos}}, \binits{S.M.}},
\bauthor{\bsnm{Bratschitsch}, \binits{R.}},
\bauthor{\bsnm{{van der Zant}}, \binits{H.S.J.}},
\bauthor{\bsnm{Castellanos-{G}omez}, \binits{A.}}:
\batitle{Thickness-dependent refractive index of 1{L}, 2{L}, and 3{L}
  {M}o{S}$_\text{2}$, {M}o{S}e$_\text{2}$, {WS}$_\text{2}$, and
  {WS}e$_\text{2}$}.
\bjtitle{Advanced Optical Materials}
\bvolume{7}(\bissue{13}),
\bfpage{1900239}
(\byear{2019})
\doiurl{10.1002/adom.201900239}
\end{barticle}
\endbibitem

\bibitem[\protect\citeauthoryear{Liu et~al.}{2020}]{Liu2020-15282}
\begin{barticle}
\bauthor{\bsnm{Liu}, \binits{H.-L.}},
\bauthor{\bsnm{Yang}, \binits{T.}},
\bauthor{\bsnm{Chen}, \binits{J.-H.}},
\bauthor{\bsnm{Chen}, \binits{H.-W.}},
\bauthor{\bsnm{Guo}, \binits{H.}},
\bauthor{\bsnm{Saito}, \binits{R.}},
\bauthor{\bsnm{Li}, \binits{M.-Y.}},
\bauthor{\bsnm{Li}, \binits{L.-J.}}:
\batitle{Temperature-dependent optical constants of monolayer
  {M}o{S}$_\text{2}$, {M}o{Se}$_\text{2}$, {W}{S}$_\text{2}$, and
  {W}{Se}$_\text{2}$: {S}pectroscopic ellipsometry and first-principles
  calculations}.
\bjtitle{Sci. Rep.}
\bvolume{10}(\bissue{1}),
\bfpage{15282}
(\byear{2020})
\doiurl{10.1038/s41598-020-71808-y}
\end{barticle}
\endbibitem

\bibitem[\protect\citeauthoryear{Ermolaev et~al.}{2021}]{Ermolaev2021-854}
\begin{barticle}
\bauthor{\bsnm{Ermolaev}, \binits{G.A.}},
\bauthor{\bsnm{Grudinin}, \binits{D.V.}},
\bauthor{\bsnm{Stebunov}, \binits{Y.V.}},
\bauthor{\bsnm{Voronin}, \binits{K.V.}},
\bauthor{\bsnm{Kravets}, \binits{V.G.}},
\bauthor{\bsnm{Duan}, \binits{J.}},
\bauthor{\bsnm{Mazitov}, \binits{A.B.}},
\bauthor{\bsnm{Tselikov}, \binits{G.I.}},
\bauthor{\bsnm{Bylinkin}, \binits{A.}},
\bauthor{\bsnm{Yakubovsky}, \binits{D.I.}},
\bauthor{\bsnm{Novikov}, \binits{S.M.}},
\bauthor{\bsnm{Baranov}, \binits{D.G.}},
\bauthor{\bsnm{Nikitin}, \binits{A.Y.}},
\bauthor{\bsnm{Kruglov}, \binits{I.A.}},
\bauthor{\bsnm{Shegai}, \binits{T.}},
\bauthor{\bsnm{Alonso-{G}onz{\'a}lez}, \binits{P.}},
\bauthor{\bsnm{Grigorenko}, \binits{A.N.}},
\bauthor{\bsnm{Arsenin}, \binits{A.V.}},
\bauthor{\bsnm{Novoselov}, \binits{K.S.}},
\bauthor{\bsnm{Volkov}, \binits{V.S.}}:
\batitle{Giant optical anisotropy in transition metal dichalcogenides for
  next-generation photonics}.
\bjtitle{Nat. Commun.}
\bvolume{12}(\bissue{1}),
\bfpage{854}
(\byear{2021})
\doiurl{10.1038/s41467-021-21139-x}
\end{barticle}
\endbibitem

\bibitem[\protect\citeauthoryear{Zotev et~al.}{2023}]{Zotev2023-2200957}
\begin{barticle}
\bauthor{\bsnm{Zotev}, \binits{P.G.}},
\bauthor{\bsnm{Wang}, \binits{Y.}},
\bauthor{\bsnm{Andres-{P}enares}, \binits{D.}},
\bauthor{\bsnm{Severs-{M}illard}, \binits{T.}},
\bauthor{\bsnm{Randerson}, \binits{S.}},
\bauthor{\bsnm{Hu}, \binits{X.}},
\bauthor{\bsnm{Sortino}, \binits{L.}},
\bauthor{\bsnm{Louca}, \binits{C.}},
\bauthor{\bsnm{Brotons-Gisbert}, \binits{M.}},
\bauthor{\bsnm{Huq}, \binits{T.}},
\bauthor{\bsnm{Vezzoli}, \binits{S.}},
\bauthor{\bsnm{Sapienza}, \binits{R.}},
\bauthor{\bsnm{Krauss}, \binits{T.F.}},
\bauthor{\bsnm{Gerardot}, \binits{B.D.}},
\bauthor{\bsnm{Tartakovskii}, \binits{A.I.}}:
\batitle{Van der {W}aals materials for applications in nanophotonics}.
\bjtitle{Laser \& Photonics Reviews}
\bvolume{17}(\bissue{8}),
\bfpage{2200957}
(\byear{2023})
\doiurl{10.1002/lpor.202200957}
\end{barticle}
\endbibitem

\bibitem[\protect\citeauthoryear{Polyanskiy}{2024}]{Polyanskiy2024-94}
\begin{barticle}
\bauthor{\bsnm{Polyanskiy}, \binits{M.N.}}:
\batitle{Refractiveindex.info database of optical constants}.
\bjtitle{Scientific Data}
\bvolume{11}(\bissue{1}),
\bfpage{94}
(\byear{2024})
\doiurl{10.1038/s41597-023-02898-2}
\end{barticle}
\endbibitem

\bibitem[\protect\citeauthoryear{Lee and Kingery}{1960}]{Lee1960-594}
\begin{barticle}
\bauthor{\bsnm{Lee}, \binits{D.W.}},
\bauthor{\bsnm{Kingery}, \binits{W.D.}}:
\batitle{Radiation energy transfer and thermal conductivity of ceramic oxides}.
\bjtitle{J. Am. Ceram. Soc.}
\bvolume{43}(\bissue{11}),
\bfpage{594}--\blpage{607}
(\byear{1960})
\doiurl{10.1111/j.1151-2916.1960.tb13623.x}
\end{barticle}
\endbibitem

\bibitem[\protect\citeauthoryear{Oppenheim and Even}{1962}]{Oppenheim1962-1078}
\begin{barticle}
\bauthor{\bsnm{Oppenheim}, \binits{U.P.}},
\bauthor{\bsnm{Even}, \binits{U.}}:
\batitle{Infrared properties of sapphire at elevated temperatures}.
\bjtitle{J. Opt. Soc. Am.}
\bvolume{52}(\bissue{9}),
\bfpage{1078}--\blpage{1079}
(\byear{1962})
\doiurl{10.1364/JOSA.52.1078_1}
\end{barticle}
\endbibitem

\bibitem[\protect\citeauthoryear{Gillespie et~al.}{1965}]{Gillespie1965-1488}
\begin{barticle}
\bauthor{\bsnm{Gillespie}, \binits{D.T.}},
\bauthor{\bsnm{Olsen}, \binits{A.L.}},
\bauthor{\bsnm{Nichols}, \binits{L.W.}}:
\batitle{Transmittance of optical materials at high temperatures in the 1-$\mu$
  to 12-$\mu$ range}.
\bjtitle{Appl. Opt.}
\bvolume{4}(\bissue{11}),
\bfpage{1488}--\blpage{1493}
(\byear{1965})
\doiurl{10.1364/AO.4.001488}
\end{barticle}
\endbibitem

\bibitem[\protect\citeauthoryear{Gryvnak and Burch}{1965}]{Gryvnak1965-625}
\begin{barticle}
\bauthor{\bsnm{Gryvnak}, \binits{D.A.}},
\bauthor{\bsnm{Burch}, \binits{D.E.}}:
\batitle{Optical and infrared properties of {A}l$_\text{2}${O}$_\text{3}$ at
  elevated temperatures}.
\bjtitle{J. Opt. Soc. Am.}
\bvolume{55}(\bissue{6}),
\bfpage{625}--\blpage{629}
(\byear{1965})
\doiurl{10.1364/JOSA.55.000625}
\end{barticle}
\endbibitem

\bibitem[\protect\citeauthoryear{Billard and Piriou}{1974}]{Billard1974-943}
\begin{barticle}
\bauthor{\bsnm{Billard}, \binits{D.}},
\bauthor{\bsnm{Piriou}, \binits{B.}}:
\batitle{Absorption infrarouge du corindon de 77 a 2075 {K}}.
\bjtitle{Mater. Res. Bull.}
\bvolume{9}(\bissue{7}),
\bfpage{943}--\blpage{950}
(\byear{1974})
\doiurl{10.1016/0025-5408(74)90174-3}
\end{barticle}
\endbibitem

\bibitem[\protect\citeauthoryear{Billard et~al.}{1976}]{Billard1976-117}
\begin{barticle}
\bauthor{\bsnm{Billard}, \binits{D.}},
\bauthor{\bsnm{Gervais}, \binits{F.}},
\bauthor{\bsnm{Piriou}, \binits{B.}}:
\batitle{Analysis of multiphonon absorption in corundum}.
\bjtitle{Phys. Stat. Sol. (B)}
\bvolume{75}(\bissue{1}),
\bfpage{117}--\blpage{126}
(\byear{1976})
\doiurl{10.1002/pssb.2220750111}
\end{barticle}
\endbibitem

\bibitem[\protect\citeauthoryear{Malitson}{1962}]{Malitson1962-1377}
\begin{barticle}
\bauthor{\bsnm{Malitson}, \binits{I.H.}}:
\batitle{Refraction and dispersion of synthetic sapphire}.
\bjtitle{J. Opt. Soc. Am.}
\bvolume{52}(\bissue{12}),
\bfpage{1377}--\blpage{1379}
(\byear{1962})
\doiurl{10.1364/JOSA.52.001377}
\end{barticle}
\endbibitem

\bibitem[\protect\citeauthoryear{Hagemann et~al.}{1975}]{Hagemann1975-742}
\begin{barticle}
\bauthor{\bsnm{Hagemann}, \binits{H.-J.}},
\bauthor{\bsnm{Gudat}, \binits{W.}},
\bauthor{\bsnm{Kunz}, \binits{C.}}:
\batitle{Optical constants from the far infrared to the {X}-ray region: {M}g,
  {A}l, {C}u, {A}g, {A}u, {B}i, {C}, and {A}l$_\text{2}${O}$_\text{3}$}.
\bjtitle{J. Opt. Soc. Am.}
\bvolume{65}(\bissue{6}),
\bfpage{742}--\blpage{744}
(\byear{1975})
\doiurl{10.1364/JOSA.65.000742}
\end{barticle}
\endbibitem

\bibitem[\protect\citeauthoryear{Billard et~al.}{1980}]{Billard1980-641}
\begin{barticle}
\bauthor{\bsnm{Billard}, \binits{D.}},
\bauthor{\bsnm{Gervais}, \binits{F.}},
\bauthor{\bsnm{Piriou}, \binits{B.}}:
\batitle{Farinfrared absorption in {A}l$_\text{2}${O}$_\text{3}$ and {M}g{O}}.
\bjtitle{Int. J. Infrared and Millim. Waves}
\bvolume{1}(\bissue{4}),
\bfpage{641}--\blpage{647}
(\byear{1980})
\doiurl{10.1007/BF01013473}
\end{barticle}
\endbibitem

\bibitem[\protect\citeauthoryear{Cabannes and Billard}{1987}]{Cabannes1987-97}
\begin{barticle}
\bauthor{\bsnm{Cabannes}, \binits{F.}},
\bauthor{\bsnm{Billard}, \binits{D.}}:
\batitle{Measurement of infrared absorption of some oxides in connection with
  the radiative transfer in porous and fibrous materials}.
\bjtitle{Int. J. Thermophys.}
\bvolume{8}(\bissue{1}),
\bfpage{97}--\blpage{118}
(\byear{1987})
\doiurl{10.1007/BF00503227}
\end{barticle}
\endbibitem

\bibitem[\protect\citeauthoryear{Sarou-{K}anian
  et~al.}{2005}]{Sarou-Kanian2005-1263}
\begin{barticle}
\bauthor{\bsnm{Sarou-{K}anian}, \binits{V.}},
\bauthor{\bsnm{Rifflet}, \binits{J.C.}},
\bauthor{\bsnm{Millot}, \binits{F.}}:
\batitle{{IR} radiative properties of solid and liquid alumina: {E}ffects of
  temperature and gaseous environment}.
\bjtitle{Int. J. Thermophys.}
\bvolume{26}(\bissue{4}),
\bfpage{1263}--\blpage{1275}
(\byear{2005})
\doiurl{10.1007/s10765-005-6725-5}
\end{barticle}
\endbibitem

\bibitem[\protect\citeauthoryear{Lee et~al.}{2011}]{Lee2011-1448}
\begin{barticle}
\bauthor{\bsnm{Lee}, \binits{G.W.}},
\bauthor{\bsnm{Jeon}, \binits{S.}},
\bauthor{\bsnm{Park}, \binits{S.-N.}},
\bauthor{\bsnm{Yoo}, \binits{Y.S.}},
\bauthor{\bsnm{Park}, \binits{C.-W.}}:
\batitle{Temperature and thickness dependence of {IR} optical properties of
  sapphire at moderate temperature}.
\bjtitle{Int. J. Thermophys.}
\bvolume{32}(\bissue{7}),
\bfpage{1448}--\blpage{1456}
(\byear{2011})
\doiurl{10.1007/s10765-011-0990-2}
\end{barticle}
\endbibitem

\bibitem[\protect\citeauthoryear{Franta et~al.}{2015}]{Franta2015-96281U}
\begin{bchapter}
\bauthor{\bsnm{Franta}, \binits{D.}},
\bauthor{\bsnm{Ne{\u c}as}, \binits{D.}},
\bauthor{\bsnm{Ohl{\'\i}dal}, \binits{I.}},
\bauthor{\bsnm{Giglia}, \binits{A.}}:
\bctitle{Dispersion model for optical thin films applicable in wide spectral
  range}.
In: \beditor{\bsnm{Duparr{\'e}}, \binits{A.}},
\beditor{\bsnm{Geyl}, \binits{R.}} (eds.)
\bbtitle{Optical Systems Design 2015: {O}ptical Fabrication, Testing, and
  Metrology {V}},
vol. \bseriesno{9628},
p. \bfpage{96281}.
\bpublisher{SPIE},
\blocation{Bellingham, WA, USA}
(\byear{2015}).
\doiurl{10.1117/12.2190104}
\end{bchapter}
\endbibitem

\bibitem[\protect\citeauthoryear{Kalman et~al.}{2015}]{Kalman2015-74}
\begin{barticle}
\bauthor{\bsnm{Kalman}, \binits{J.}},
\bauthor{\bsnm{Allen}, \binits{D.}},
\bauthor{\bsnm{Glumac}, \binits{N.}},
\bauthor{\bsnm{Krier}, \binits{H.}}:
\batitle{Optical depth effects on aluminum oxide spectral emissivity}.
\bjtitle{J. Thermophys. Heat Trans.}
\bvolume{29}(\bissue{1}),
\bfpage{74}--\blpage{82}
(\byear{2015})
\doiurl{10.2514/1.T4260}
\end{barticle}
\endbibitem

\bibitem[\protect\citeauthoryear{Yang et~al.}{2016}]{Yang2016-111}
\begin{barticle}
\bauthor{\bsnm{Yang}, \binits{J.Y.}},
\bauthor{\bsnm{Xu}, \binits{M.}},
\bauthor{\bsnm{Liu}, \binits{L.H.}}:
\batitle{Infrared radiative properties of alumina up to the melting point: {A}
  first-principles study}.
\bjtitle{J. Quant. Spectrosc. Radiat. Transf.}
\bvolume{184},
\bfpage{111}--\blpage{117}
(\byear{2016})
\doiurl{10.1016/j.jqsrt.2016.07.006}
\end{barticle}
\endbibitem

\bibitem[\protect\citeauthoryear{Boidin et~al.}{2016}]{Boidin2016-1177}
\begin{barticle}
\bauthor{\bsnm{Boidin}, \binits{R.}},
\bauthor{\bsnm{Halenkovi\u{c}}, \binits{T.}},
\bauthor{\bsnm{Nazabal}, \binits{V.}},
\bauthor{\bsnm{Bene\u{s}}, \binits{L.}},
\bauthor{\bsnm{N\u{e}mec}, \binits{P.}}:
\batitle{Pulsed laser deposited alumina thin films}.
\bjtitle{Ceram.}
\bvolume{42}(\bissue{1, Part B}),
\bfpage{1177}--\blpage{1182}
(\byear{2016})
\doiurl{10.1016/j.ceramint.2015.09.048}
\end{barticle}
\endbibitem

\bibitem[\protect\citeauthoryear{Marr and Wilkin}{2012}]{Marr2012-399}
\begin{barticle}
\bauthor{\bsnm{Marr}, \binits{J.M.}},
\bauthor{\bsnm{Wilkin}, \binits{F.P.}}:
\batitle{A better presentation of {P}lanck's radiation law}.
\bjtitle{American Journal of Physics}
\bvolume{80}(\bissue{5}),
\bfpage{399}--\blpage{405}
(\byear{2012})
\doiurl{10.1119/1.3696974}
\end{barticle}
\endbibitem

\bibitem[\protect\citeauthoryear{Gordon}{1978}]{Gordon1978-book}
\begin{bbook}
\bauthor{\bsnm{Gordon}, \binits{J.E.}}:
\bbtitle{Structures or {W}hy Things Don't Fall Down}.
\bpublisher{Plenum Press},
\blocation{New York, NY, USA}
(\byear{1978}).
\bcomment{Chapter 6: {T}ension structures and pressure vessels - with some
  remarks on boilers, bats, and {C}hinese junks}
\end{bbook}
\endbibitem

\bibitem[\protect\citeauthoryear{Leff}{2002}]{Leff2002-792}
\begin{barticle}
\bauthor{\bsnm{Leff}, \binits{H.S.}}:
\batitle{Teaching the photon gas in introductory physics}.
\bjtitle{Am. J. Phys.}
\bvolume{70}(\bissue{8}),
\bfpage{792}--\blpage{797}
(\byear{2002})
\doiurl{10.1119/1.1479743}
\end{barticle}
\endbibitem

\bibitem[\protect\citeauthoryear{Sakamoto et~al.}{2007}]{Sakamoto2007-514}
\begin{barticle}
\bauthor{\bsnm{Sakamoto}, \binits{H.}},
\bauthor{\bsnm{Miyazaki}, \binits{Y.}},
\bauthor{\bsnm{Park}, \binits{K.C.}}:
\batitle{Finite element modeling of sail deformation under solar radiation
  pressure}.
\bjtitle{J. Spacecr. Rockets}
\bvolume{44}(\bissue{3}),
\bfpage{514}--\blpage{521}
(\byear{2007})
\doiurl{10.2514/1.23474}
\end{barticle}
\endbibitem

\bibitem[\protect\citeauthoryear{Lee et~al.}{2008}]{Lee2008-385}
\begin{barticle}
\bauthor{\bsnm{Lee}, \binits{C.}},
\bauthor{\bsnm{Wei}, \binits{X.}},
\bauthor{\bsnm{Kysar}, \binits{J.W.}},
\bauthor{\bsnm{Hone}, \binits{J.}}:
\batitle{Measurement of the elastic properties and intrinsic strength of
  monolayer graphene}.
\bjtitle{Science}
\bvolume{321}(\bissue{5887}),
\bfpage{385}--\blpage{388}
(\byear{2008})
\doiurl{10.1126/science.1157996}
\end{barticle}
\endbibitem

\bibitem[\protect\citeauthoryear{Yokoyama et~al.}{2007}]{Yokoyama2007-s68}
\begin{barticle}
\bauthor{\bsnm{Yokoyama}, \binits{T.}},
\bauthor{\bsnm{Nakai}, \binits{K.}},
\bauthor{\bsnm{Odamura}, \binits{T.}}:
\batitle{Tensile stress-strain properties of paper and paperboard and their
  constitutive equations}.
\bjtitle{Journal of the Japanese Society for Experimental Mechanics}
\bvolume{7},
\bfpage{68}--\blpage{73}
(\byear{2007})
\doiurl{10.11395/jjsem.7.s68}
\end{barticle}
\endbibitem

\bibitem[\protect\citeauthoryear{Whittam et~al.}{2025}]{Whittam2025-345}
\begin{barticle}
\bauthor{\bsnm{Whittam}, \binits{M.R.}},
\bauthor{\bsnm{Rebholz}, \binits{L.}},
\bauthor{\bsnm{Zerulla}, \binits{B.}},
\bauthor{\bsnm{Rockstuhl}, \binits{C.}}:
\batitle{Analyzing the acceleration time and reflectance of light sails made
  from homogeneous and core-shell spheres}.
\bjtitle{Opt. Mater. Express}
\bvolume{15}(\bissue{2}),
\bfpage{345}--\blpage{361}
(\byear{2025})
\doiurl{10.1364/OME.545481}
\end{barticle}
\endbibitem

\bibitem[\protect\citeauthoryear{Camino et~al.}{2001}]{Camino2001-2395}
\begin{barticle}
\bauthor{\bsnm{Camino}, \binits{G.}},
\bauthor{\bsnm{Lomakin}, \binits{S.M.}},
\bauthor{\bsnm{Lazzari}, \binits{M.}}:
\batitle{Polydimethylsiloxane thermal degradation. {P}art 1. {K}inetic
  aspects}.
\bjtitle{Polymer}
\bvolume{42}(\bissue{6}),
\bfpage{2395}--\blpage{2402}
(\byear{2001})
\doiurl{10.1016/S0032-3861(00)00652-2}
\end{barticle}
\endbibitem

\bibitem[\protect\citeauthoryear{Ariati et~al.}{2021}]{Ariati2021-4258}
\begin{barticle}
\bauthor{\bsnm{Ariati}, \binits{R.}},
\bauthor{\bsnm{Sales}, \binits{F.}},
\bauthor{\bsnm{Souza}, \binits{A.}},
\bauthor{\bsnm{Lima}, \binits{R.A.}},
\bauthor{\bsnm{Ribeiro}, \binits{J.}}:
\batitle{Polydimethylsiloxane composites characterization and its applications:
  {A} review}.
\bjtitle{Polymers}
\bvolume{13}(\bissue{23}),
\bfpage{4258}
(\byear{2021})
\doiurl{10.3390/polym13234258}
\end{barticle}
\endbibitem

\bibitem[\protect\citeauthoryear{El-Haija}{2003}]{El-Haija2003-2590}
\begin{barticle}
\bauthor{\bsnm{El-Haija}, \binits{A.J.A.}}:
\batitle{Effective medium approximation for the effective optical constants of
  a bilayer and a multilayer structure based on the characteristic matrix
  technique}.
\bjtitle{Journal of Applied Physics}
\bvolume{93}(\bissue{5}),
\bfpage{2590}--\blpage{2594}
(\byear{2003})
\doiurl{10.1063/1.1543229}
\end{barticle}
\endbibitem

\bibitem[\protect\citeauthoryear{Broas et~al.}{2017}]{Broas2017-3390}
\begin{barticle}
\bauthor{\bsnm{Broas}, \binits{M.}},
\bauthor{\bsnm{Kanninen}, \binits{O.}},
\bauthor{\bsnm{Vuorinen}, \binits{V.}},
\bauthor{\bsnm{Tilli}, \binits{M.}},
\bauthor{\bsnm{Paulasto-{K}r{\"o}ckel}, \binits{M.}}:
\batitle{Chemically stable atomic-layer-deposited {A}l$_\text{2}${O}$_\text{3}$
  films for processability}.
\bjtitle{ACS Omega}
\bvolume{2}(\bissue{7}),
\bfpage{3390}--\blpage{3398}
(\byear{2017})
\doiurl{10.1021/acsomega.7b00443}
\end{barticle}
\endbibitem

\bibitem[\protect\citeauthoryear{Broas et~al.}{2019}]{Broas2019-147}
\begin{barticle}
\bauthor{\bsnm{Broas}, \binits{M.}},
\bauthor{\bsnm{Lemettinen}, \binits{J.}},
\bauthor{\bsnm{Sajavaara}, \binits{T.}},
\bauthor{\bsnm{Tilli}, \binits{M.}},
\bauthor{\bsnm{Vuorinen}, \binits{V.}},
\bauthor{\bsnm{Suihkonen}, \binits{S.}},
\bauthor{\bsnm{Paulasto-{K}r{\"o}ckel}, \binits{M.}}:
\batitle{In-situ annealing characterization of atomic-layer-deposited
  {A}l$_\text{2}${O}$_\text{3}$ in {N}$_\text{2}$, {H}$_\text{2}$ and vacuum
  atmospheres}.
\bjtitle{Thin Solid Films}
\bvolume{682},
\bfpage{147}--\blpage{155}
(\byear{2019})
\doiurl{10.1016/j.tsf.2019.03.010}
\end{barticle}
\endbibitem

\bibitem[\protect\citeauthoryear{Zhang et~al.}{2007}]{Zhang2007-3707}
\begin{barticle}
\bauthor{\bsnm{Zhang}, \binits{L.}},
\bauthor{\bsnm{Jiang}, \binits{H.C.}},
\bauthor{\bsnm{Liu}, \binits{C.}},
\bauthor{\bsnm{Dong}, \binits{J.W.}},
\bauthor{\bsnm{Chow}, \binits{P.}}:
\batitle{Annealing of {A}l$_\text{2}${O}$_\text{3}$ thin films prepared by
  atomic layer deposition}.
\bjtitle{Journal of Physics {D}: {A}pplied Physics}
\bvolume{40}(\bissue{12}),
\bfpage{3707}
(\byear{2007})
\doiurl{10.1088/0022-3727/40/12/025}
\end{barticle}
\endbibitem

\bibitem[\protect\citeauthoryear{Wang et~al.}{2015}]{Wang2015-46}
\begin{barticle}
\bauthor{\bsnm{Wang}, \binits{Z.-Y.}},
\bauthor{\bsnm{Zhang}, \binits{R.-J.}},
\bauthor{\bsnm{Lu}, \binits{H.-L.}},
\bauthor{\bsnm{Chen}, \binits{X.}},
\bauthor{\bsnm{Sun}, \binits{Y.}},
\bauthor{\bsnm{Zhang}, \binits{Y.}},
\bauthor{\bsnm{Wei}, \binits{Y.-F.}},
\bauthor{\bsnm{Xu}, \binits{J.-P.}},
\bauthor{\bsnm{Wang}, \binits{S.-Y.}},
\bauthor{\bsnm{Zheng}, \binits{Y.-X.}},
\bauthor{\bsnm{Chen}, \binits{L.-Y.}}:
\batitle{The impact of thickness and thermal annealing on refractive index for
  aluminum oxide thin films deposited by atomic layer deposition}.
\bjtitle{Nanoscale Research Letters}
\bvolume{10}(\bissue{1}),
\bfpage{46}
(\byear{2015})
\doiurl{10.1186/s11671-015-0757-y}
\end{barticle}
\endbibitem

\bibitem[\protect\citeauthoryear{Ylivaara et~al.}{2022}]{Ylivaara2022-062424}
\begin{barticle}
\bauthor{\bsnm{Ylivaara}, \binits{O.M.E.}},
\bauthor{\bsnm{Langner}, \binits{A.}},
\bauthor{\bsnm{Ek}, \binits{S.}},
\bauthor{\bsnm{Malm}, \binits{J.}},
\bauthor{\bsnm{Julin}, \binits{J.}},
\bauthor{\bsnm{Laitinen}, \binits{M.}},
\bauthor{\bsnm{Ali}, \binits{S.}},
\bauthor{\bsnm{Sintonen}, \binits{S.}},
\bauthor{\bsnm{Lipsanen}, \binits{H.}},
\bauthor{\bsnm{Sajavaara}, \binits{T.}},
\bauthor{\bsnm{Puurunen}, \binits{R.L.}}:
\batitle{Thermomechanical properties of aluminum oxide thin films made by
  atomic layer deposition}.
\bjtitle{Journal of Vacuum Science \& Technology A}
\bvolume{40}(\bissue{6}),
\bfpage{062414}
(\byear{2022})
\doiurl{10.1116/6.0002095}
\end{barticle}
\endbibitem

\bibitem[\protect\citeauthoryear{Lu et~al.}{2013}]{Lu2013-8904}
\begin{barticle}
\bauthor{\bsnm{Lu}, \binits{X.}},
\bauthor{\bsnm{Utama}, \binits{M.I.B.}},
\bauthor{\bsnm{Zhang}, \binits{J.}},
\bauthor{\bsnm{Zhao}, \binits{Y.}},
\bauthor{\bsnm{Xiong}, \binits{Q.}}:
\batitle{Layer-by-layer thinning of {M}o{S}$_\text{2}$ by thermal annealing}.
\bjtitle{Nanoscale}
\bvolume{5}(\bissue{19}),
\bfpage{8904}--\blpage{8908}
(\byear{2013})
\doiurl{10.1039/C3NR03101B}
\end{barticle}
\endbibitem

\bibitem[\protect\citeauthoryear{Brenner}{1962}]{Brenner1962-33}
\begin{barticle}
\bauthor{\bsnm{Brenner}, \binits{S.S.}}:
\batitle{Mechanical behavior of sapphire whiskers at elevated temperatures}.
\bjtitle{Journal of Applied Physics}
\bvolume{33}(\bissue{1}),
\bfpage{33}--\blpage{39}
(\byear{1962})
\doiurl{10.1063/1.1728523}
\end{barticle}
\endbibitem

\bibitem[\protect\citeauthoryear{S\'{a}nchez-{G}onz\'{a}lez
  et~al.}{2007}]{Sanchez-Gonzalez2007-3345}
\begin{barticle}
\bauthor{\bsnm{S\'{a}nchez-{G}onz\'{a}lez}, \binits{E.}},
\bauthor{\bsnm{Miranda}, \binits{P.}},
\bauthor{\bsnm{Mel\'{e}ndez-{M}art\'{e}nez}, \binits{J.J.}},
\bauthor{\bsnm{Guiberteau}, \binits{F.}},
\bauthor{\bsnm{Pajares}, \binits{A.}}:
\batitle{Temperature dependence of mechanical properties of alumina up to the
  onset of creep}.
\bjtitle{J. Eur. Ceram. Soc.}
\bvolume{27}(\bissue{11}),
\bfpage{3345}--\blpage{3349}
(\byear{2007})
\doiurl{10.1016/j.jeurceramsoc.2007.02.191}
\end{barticle}
\endbibitem

\bibitem[\protect\citeauthoryear{Pham and Fang}{2022}]{Pham2022-7777}
\begin{barticle}
\bauthor{\bsnm{Pham}, \binits{V.-T.}},
\bauthor{\bsnm{Fang}, \binits{T.-H.}}:
\batitle{Thermal and mechanical characterization of nanoporous two-dimensional
  {M}o{S}$_\text{2}$ membranes}.
\bjtitle{Scientific Reports}
\bvolume{12}(\bissue{1}),
\bfpage{7777}
(\byear{2022})
\doiurl{10.1038/s41598-022-11883-5}
\end{barticle}
\endbibitem

\bibitem[\protect\citeauthoryear{Despont et~al.}{1996}]{Despont1996-219}
\begin{barticle}
\bauthor{\bsnm{Despont}, \binits{M.}},
\bauthor{\bsnm{Gross}, \binits{H.}},
\bauthor{\bsnm{Arrouy}, \binits{F.}},
\bauthor{\bsnm{Stebler}, \binits{C.}},
\bauthor{\bsnm{Staufer}, \binits{U.}}:
\batitle{Fabrication of a silicon-{P}yrex-silicon stack by {A}.{C}. anodic
  bonding}.
\bjtitle{Sens. Actuators, A}
\bvolume{55}(\bissue{2}),
\bfpage{219}--\blpage{224}
(\byear{1996})
\doiurl{10.1016/S0924-4247(97)80081-7}
\end{barticle}
\endbibitem

\bibitem[\protect\citeauthoryear{Hayashi et~al.}{2000}]{Hayashi2000-77}
\begin{barticle}
\bauthor{\bsnm{Hayashi}, \binits{H.}},
\bauthor{\bsnm{Watanabe}, \binits{M.}},
\bauthor{\bsnm{Inaba}, \binits{H.}}:
\batitle{Measurement of thermal expansion coefficient of
  {L}a{C}r{O}$_\text{3}$}.
\bjtitle{Thermochim. Acta}
\bvolume{359}(\bissue{1}),
\bfpage{77}--\blpage{85}
(\byear{2000})
\doiurl{10.1016/S0040-6031(00)00507-4}
\end{barticle}
\endbibitem

\bibitem[\protect\citeauthoryear{Huang et~al.}{2014}]{Huang2014-045409}
\begin{barticle}
\bauthor{\bsnm{Huang}, \binits{L.F.}},
\bauthor{\bsnm{Gong}, \binits{P.L.}},
\bauthor{\bsnm{Zeng}, \binits{Z.}}:
\batitle{Correlation between structure, phonon spectra, thermal expansion, and
  thermomechanics of single-layer {M}o{S}$_\text{2}$}.
\bjtitle{Phys. Rev. B}
\bvolume{90}(\bissue{4}),
\bfpage{045409}
(\byear{2014})
\doiurl{10.1103/PhysRevB.90.045409}
\end{barticle}
\endbibitem

\bibitem[\protect\citeauthoryear{Touloukian and
  Ho}{1977}]{Touloukian1977-report}
\begin{botherref}
\oauthor{\bsnm{Touloukian}, \binits{Y.S.}},
\oauthor{\bsnm{Ho}, \binits{C.Y.}}:
Thermophysical properties of matter: {T}he {TPRC} data series. {V}olume 13:
  {T}hermal expansion - {N}onmetallic solids.
Technical report,
Purdue University
(1977)
\end{botherref}
\endbibitem

\end{thebibliography}
\normalsize

\end{multicols}

\pagebreak
\begin{center}
\textbf{\huge \\ \textit{Supplementary information:} \\Experimental demonstration of corrugated nanolaminate films as
reflective light sails \\}
\end{center}
\setcounter{equation}{0}
\setcounter{figure}{0}
\setcounter{table}{0}
\makeatletter
\renewcommand{\theequation}{S\arabic{equation}}
\renewcommand{\thefigure}{S\arabic{figure}}
\renewcommand{\thetable}{S\arabic{table}}

\bigskip

\begin{multicols}{2}

\section{Additional information on film fabrication}\label{S:filmFab}

\begin{figure*}
\centering
\includegraphics[width=0.9\figWidthFull]{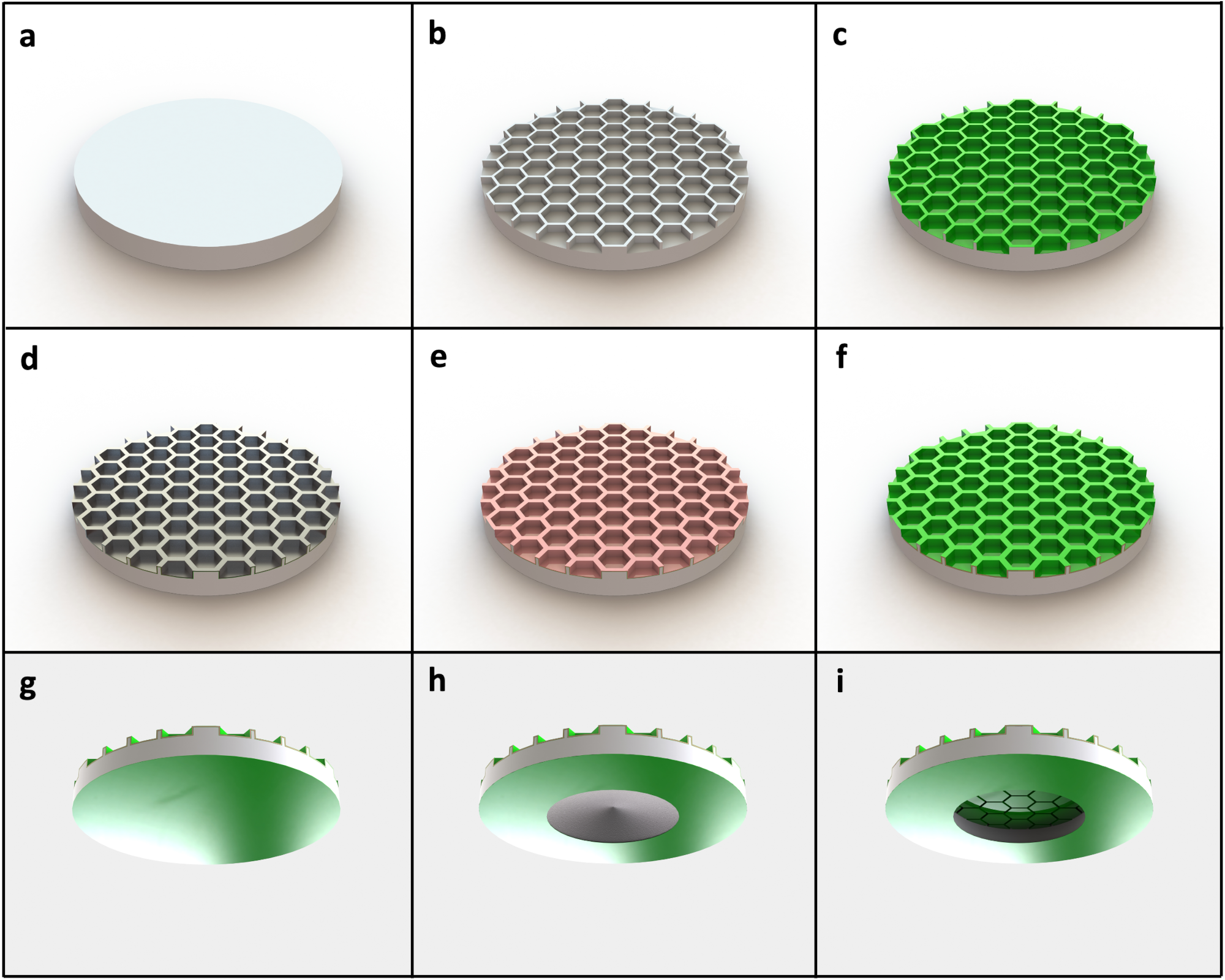}%
\caption{\textbf{\textbar~ Fabrication steps for sail film prototypes in the protruding rib configuration.} See also Figure~\ref{F:compositeChar}(a) in the main article. (\textbf{a}) Begin with a plain 200-\si{\micro\meter}-thick double-side-polished \ch{Si} wafer. (\textbf{b}) Form hexagon shapes/ribs using photolithography and deep reactive ion etching. (\textbf{c}) Use atomic layer deposition to produce the lower conformal \ch{Al2O3} film. (\textbf{d}) Conformally sputter \ch{Mo}. (\textbf{e}) Sulfurize the \ch{Mo} in a high-temperature tube furnace. (\textbf{f})  Use atomic layer deposition to produce the upper conformal \ch{Al2O3} film. (\textbf{g}) Invert the wafer. Note that the backside was coated in \ch{Al2O3} in steps (\textbf{c}) and (\textbf{f}). (\textbf{h}) Use laser micromachining to cut through the backside \ch{Al2O3} and remove a majority of the \ch{Si} substrate. (\textbf{i}) Use \ch{XeF2} gaseous etching to remove the rest of the \ch{Si} mold. The film is now suspended. }%
\label{F:fabDiagRibs}%
\end{figure*}

\begin{figure*}
\centering
\includegraphics[width=\figWidthCol]{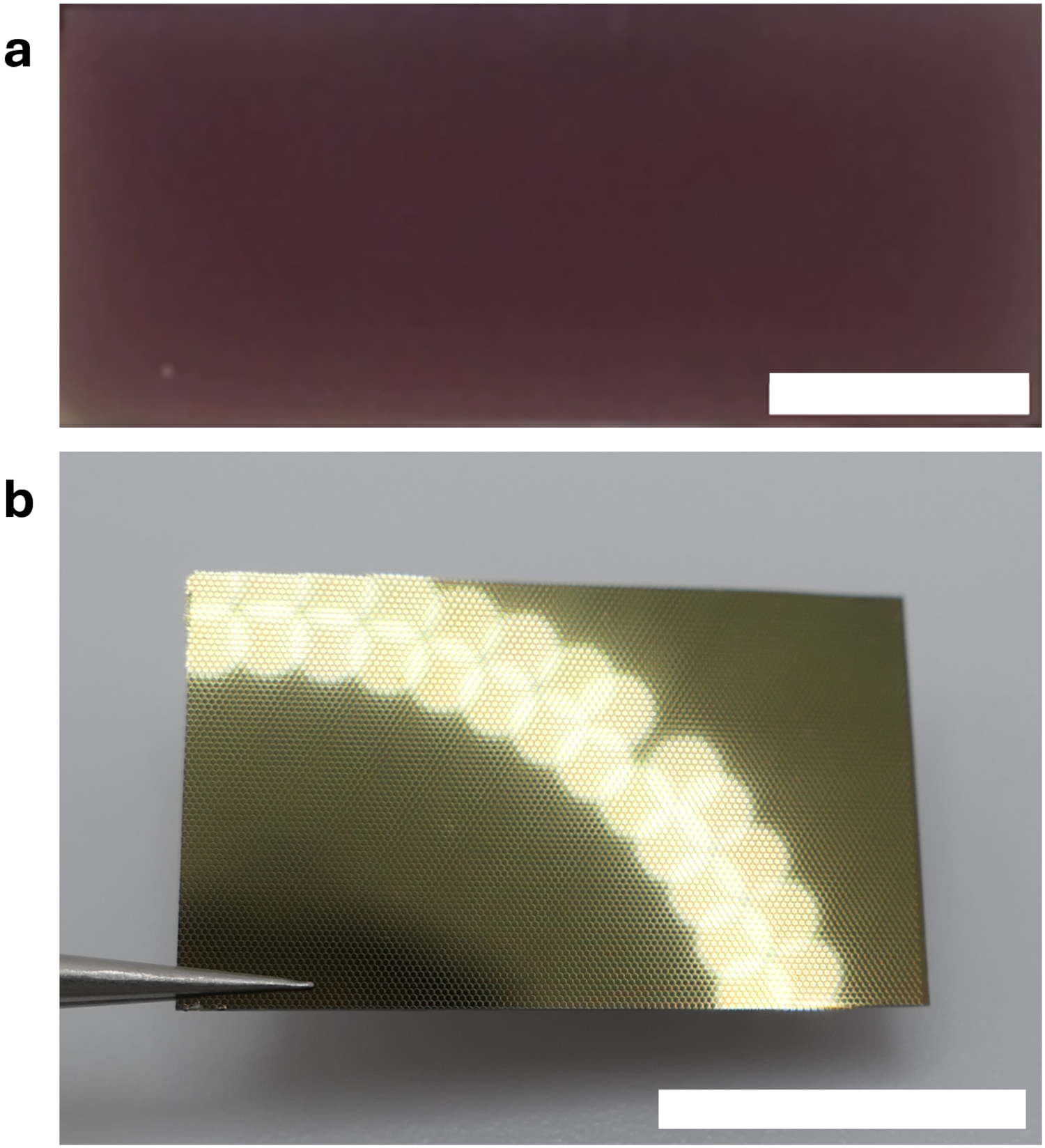}%
\caption{\textbf{\textbar~ Photographs of \ch{Si} chips with films attached.} (\textbf{a}) 45-\si{\nano\meter}-thick sample of \ch{MoS2} on flat (non-corrugated) substrate. (\textbf{b}) Prototype composite film on trench-corrugated \ch{Si} chip (held on the left by tweezers). The corrugation in this particular embodiment is faintly visible to the naked eye. Sample dimensions: $d_h\approx154~\si{\micro\meter}$, $w_t\approx30~\si{\micro\meter}$, $h_t\approx10~\si{\micro\meter}$, $t_{A,b}\approx 21~\si{\nano\meter}$, $t_{M}\approx 60~\si{\nano\meter}$, $t_{A,t}\approx 51~\si{\nano\meter}$. Scale bars: (\textbf{a}) 5~\si{\milli\meter}, (\textbf{b}) 10~\si{\milli\meter}.}%
\label{F:photosOfCompositeFilms}%
\end{figure*}

\begin{figure*}
\centering
\includegraphics[width=\figWidthFull]{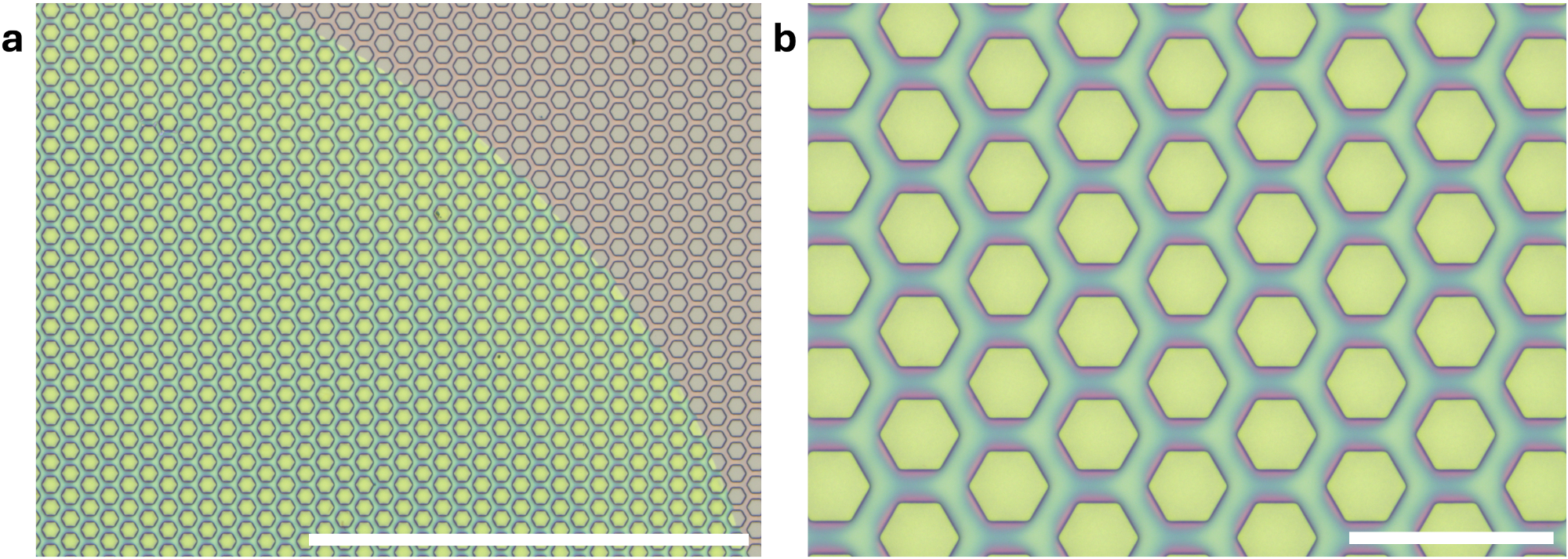}
\caption{\textbf{\textbar~ Additional micrographs of the fully suspended prototype film (indented trenches configuration) shown in Figure~\ref{F:overview}(b) of the main article.}  (\textbf{a}) Perimeter of suspended region. Red area is on \ch{Si} substrate and green area has substrate removed. (\textbf{b}) Enlarged view of suspended area. Sample dimensions: $d_h\approx36~\si{\micro\meter}$, $w_t\approx15~\si{\micro\meter}$, $h_t\approx10~\si{\micro\meter}$, $t_{A,b}\approx 15~\si{\nano\meter}$, $t_{M}\approx 75~\si{\nano\meter}$, $t_{A,t}\approx 53~\si{\nano\meter}$. Scale bars: (\textbf{a}) 1000~\si{\micro\meter} and (\textbf{b}) 100~\si{\micro\meter}. }%
\label{F:moreImagesBigSuspended}%
\end{figure*}

\begin{figure*}
\centering
\includegraphics[width=\figWidthFull]{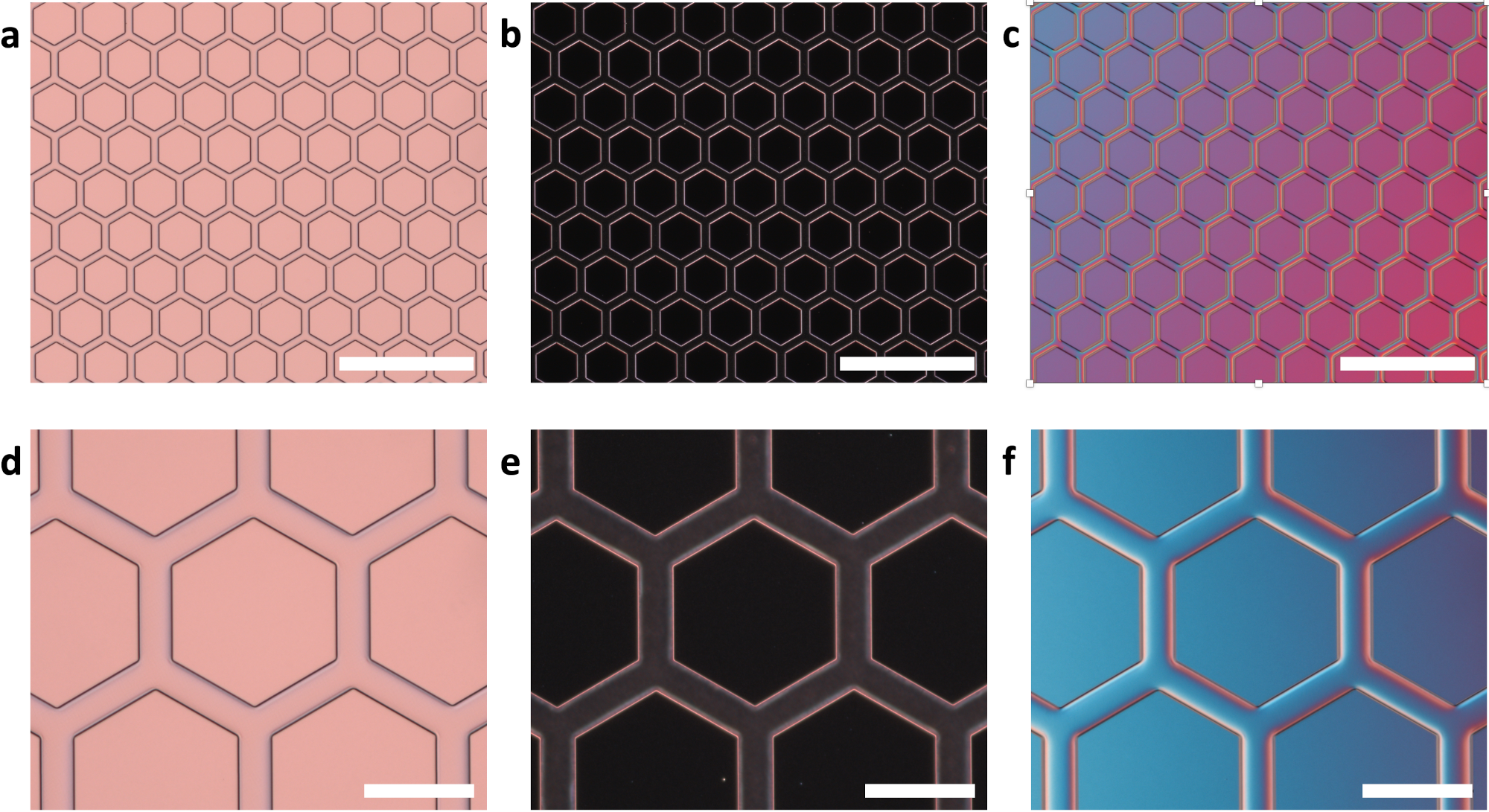}
\caption{\textbf{\textbar~ Micrographs of a corrugated composite film on a \ch{Si} substrate.} This is the same sample shown in Figure~\ref{F:photosOfCompositeFilms}(b).  The pattern features hexagonal trenches etched down into the \ch{Si}. (\textbf{a}, \textbf{d}): Brightfield images. (\textbf{b}, \textbf{e}): Darkfield images. (\textbf{c}, \textbf{f}): Reflected differential image contrast (DIC) images. Images obtained on a Zeiss Axio Imager M2m Microscope. Scale bars: (\textbf{a}, \textbf{b}, \textbf{c}) 500~\si{\micro\meter} and (\textbf{d}, \textbf{e}, \textbf{f}) 100~\si{\micro\meter}. }%
\label{F:filmImagesOnSi}%
\end{figure*}

\begin{figure*}
\centering
\includegraphics[width=\figWidthCol]{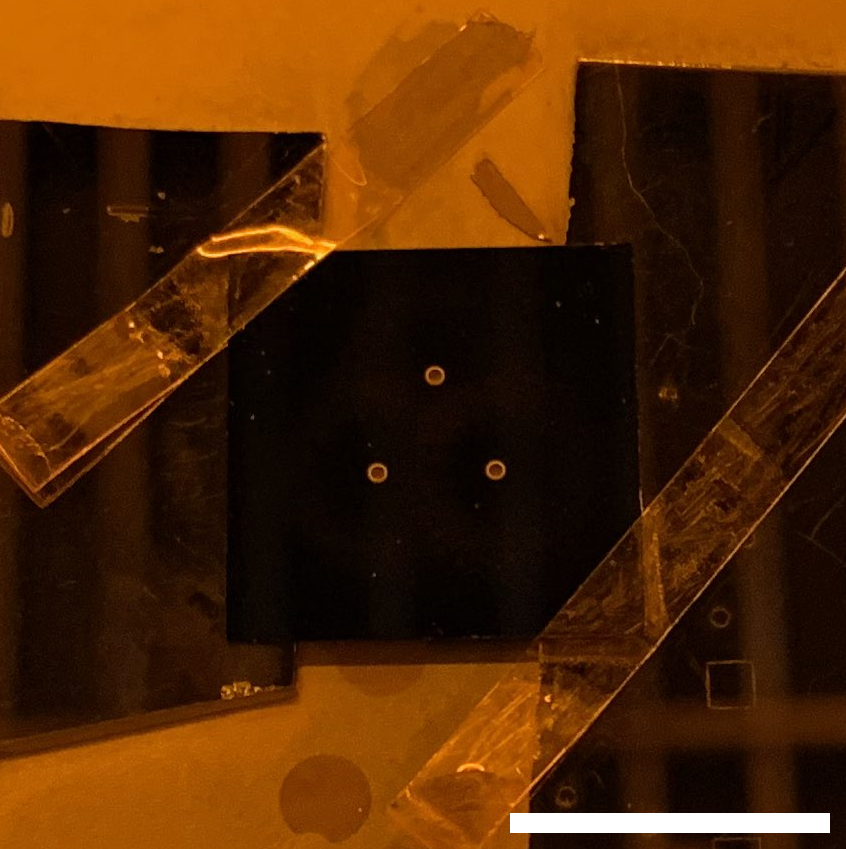}%
\caption{\textbf{\textbar~ Results of laser etching.} Photograph shows 12-\si{\milli\meter} square \ch{Si} chip with three 500~\si{\micro\meter} laser-drilled holes in its back side. Scale bar: 10~\si{\milli\meter}.}%
\label{F:laserDrilledHoles}%
\end{figure*}

\begin{figure*}
\centering
\includegraphics[width=\figWidthCol]{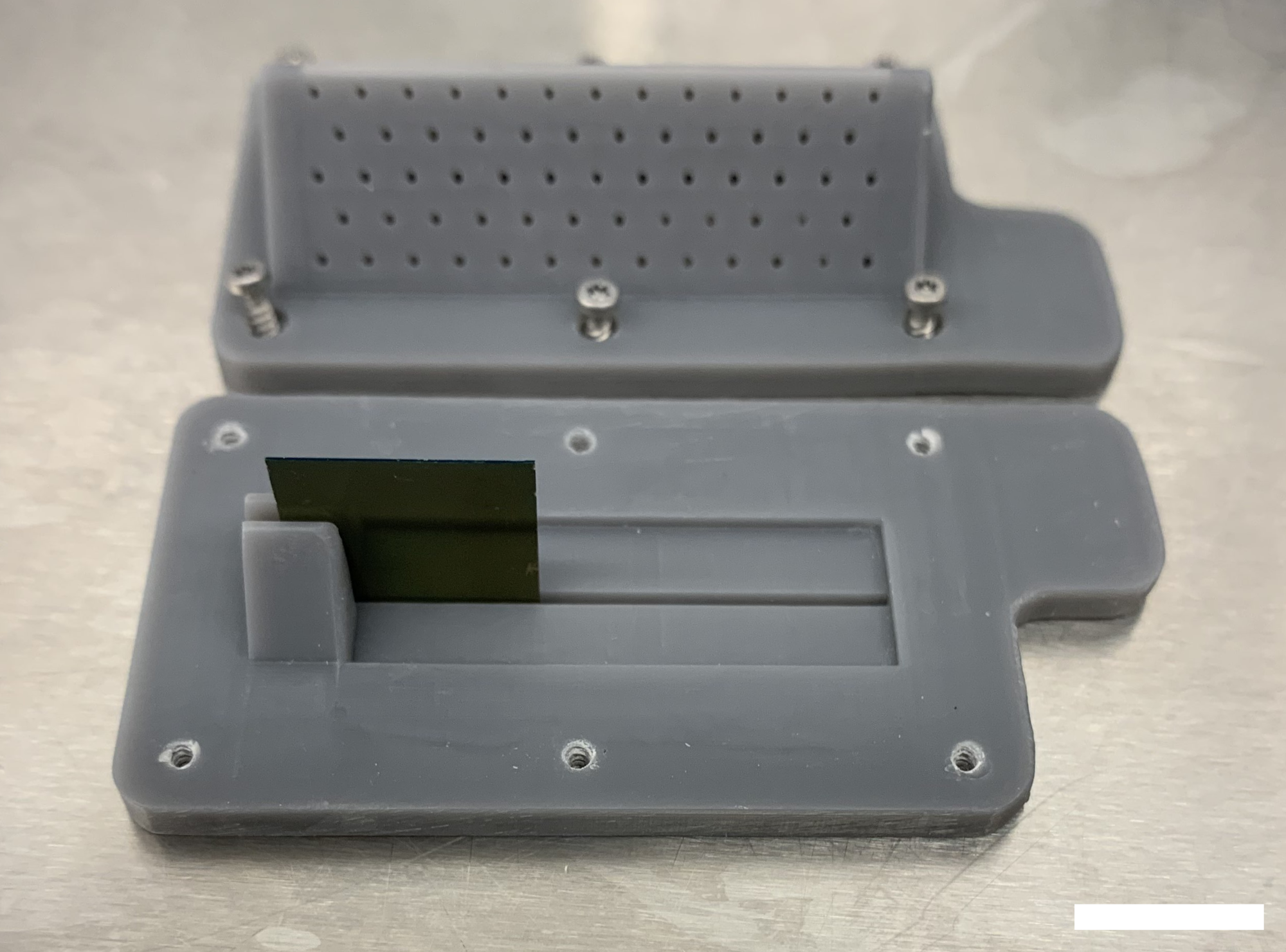}%
\caption{\textbf{\textbar~ Preparation for gaseous etching.} Photograph shows fixture used to hold chips during \ch{XeF2} etching. The vertical orientation of the chip ensures that gas encounters both of its sides at equal pressures. Scale bar: 10~\si{\milli\meter}.}%
\label{F:fixtureForXeF2}%
\end{figure*}

Here we present supplementary details concerning our film fabrication process.  Figure~\ref{F:fabDiagRibs} summarizes the fabrication steps for the protruding ribs film configuration (see also Figure~\ref{F:compositeChar}(a) in the main article). Figure~\ref{F:photosOfCompositeFilms} shows photographs of films on \ch{Si} chips. Figure~\ref{F:moreImagesBigSuspended} provides more images of the fully suspended prototype (indented trenches configuration) shown in Figure~\ref{F:overview}(b) in the main article. Figure~\ref{F:filmImagesOnSi} provides micrographs of the corrugated film shown in Figure~\ref{F:photosOfCompositeFilms}(b). The focal point for these images was the tops of the hexagons rather than the bottoms of the trenches.  Figure~\ref{F:laserDrilledHoles} shows a silicon chip with three 500~\si{\micro\meter} laser-drilled holes in its back side. Finally, Figure~\ref{F:fixtureForXeF2} shows the fixture that we used during \ch{XeF2} etching to promote equal gas pressure on both sides of the chips, protecting the films against excessive force when venting the etching chamber.

\section{Additional information on film characterization}\label{S:filmChar}

\begin{figure*}
\centering
\includegraphics[width=\figWidthFull]{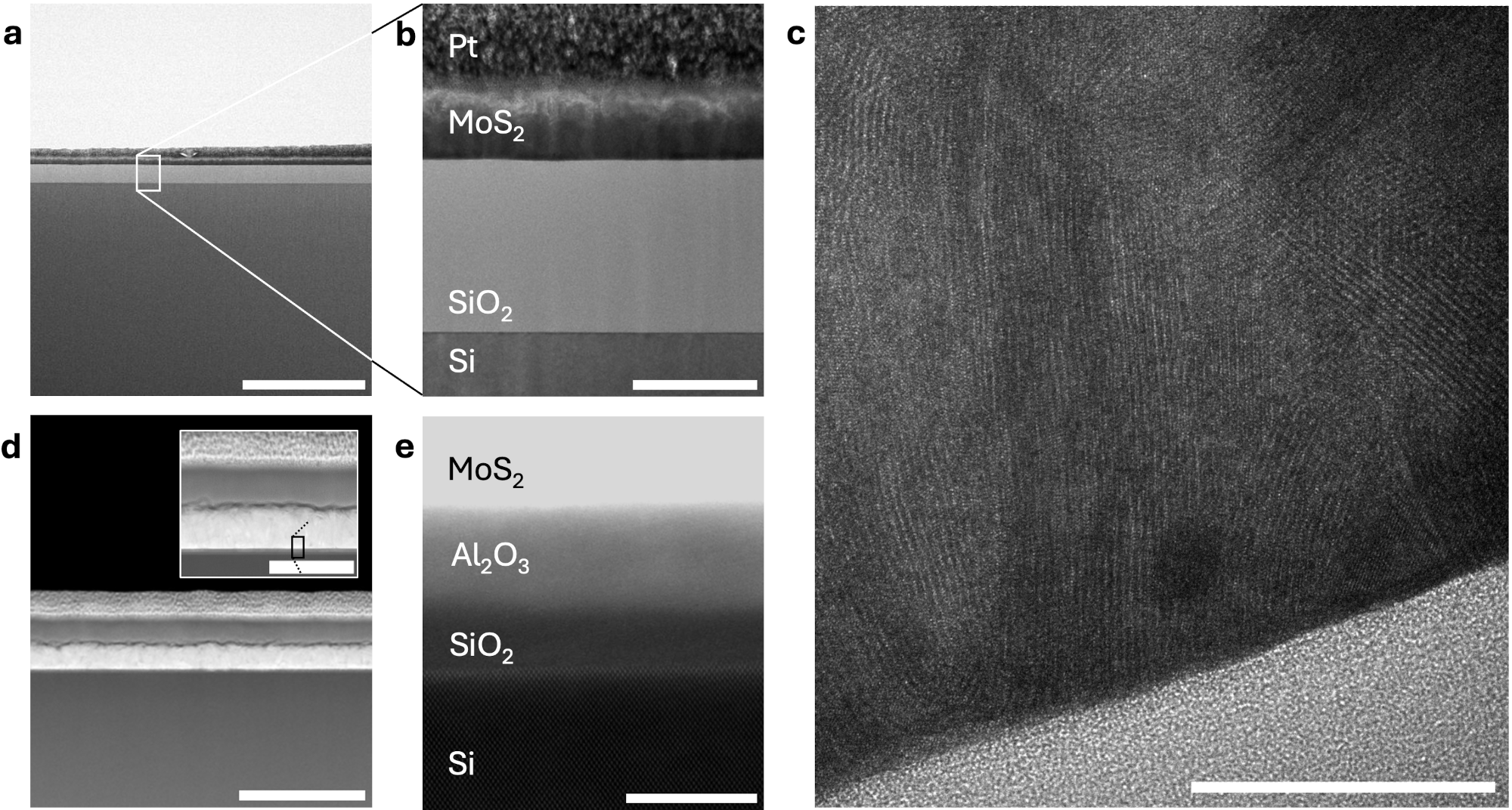}%
\caption{\textbf{\textbar~ Cross-sectional transmission electron microscopy (TEM) images of a \ch{MoS2} film.} Film thickness is about 75-\si{\nano\meter}. (\textbf{a}) Micrograph of \ch{MoS2} on \ch{SiO2}/\ch{Si} substrate. (\textbf{b}) Enlargement of panel~(\textbf{a}) with layers labeled.  The \ch{Pt} is used as a protective layer during the microscopy process. (\textbf{c}) Micrograph showing layered structure of \ch{MoS2}. (\textbf{d}) Micrograph obtained by imaging \ch{MoS2} directly on \ch{Al2O3} ribs of hexagonal mold pattern.  Inset shows enlargement of important layers.  (\textbf{e}) Further enlargement from the inset of panel (\textbf{d}) as indicated by black rectangle; native oxide (\ch{SiO2}) and \ch{Si} mold substrate can be seen. Scale bars: (\textbf{a}) 2~\si{\micro\meter}, (\textbf{b}) 200~\si{\nano\meter}, (\textbf{c}) 20~\si{\nano\meter}, (\textbf{d}) 500~\si{\nano\meter}, inset of (d): 200~\si{\nano\meter}, (\textbf{e}) 10~\si{\nano\meter}.}%
\label{F:SI_TEM_fig}%
\end{figure*}

Figure~\ref{F:SI_TEM_fig} presents several cross-sectional transmission electron microscopy (TEM) images of sample \ch{MoS2} films. For these images, we prepared samples using the \ch{Xe} plasma focused ion beam approach~\cite{Vitale2022-646} and subsequently transferred them to half-grids using the \insitu\xspace liftoff technique~\cite{Kim2023-1044}. The orientation of the grains is determined by various factors that include strain, transformation rate, and defects present during the sulfurization process~\cite{Fei2016-12206, Kumar2017-5068, Altvater2024-2400463}. 

\begin{figure*}
\centering
\includegraphics[width=\figWidthCol]{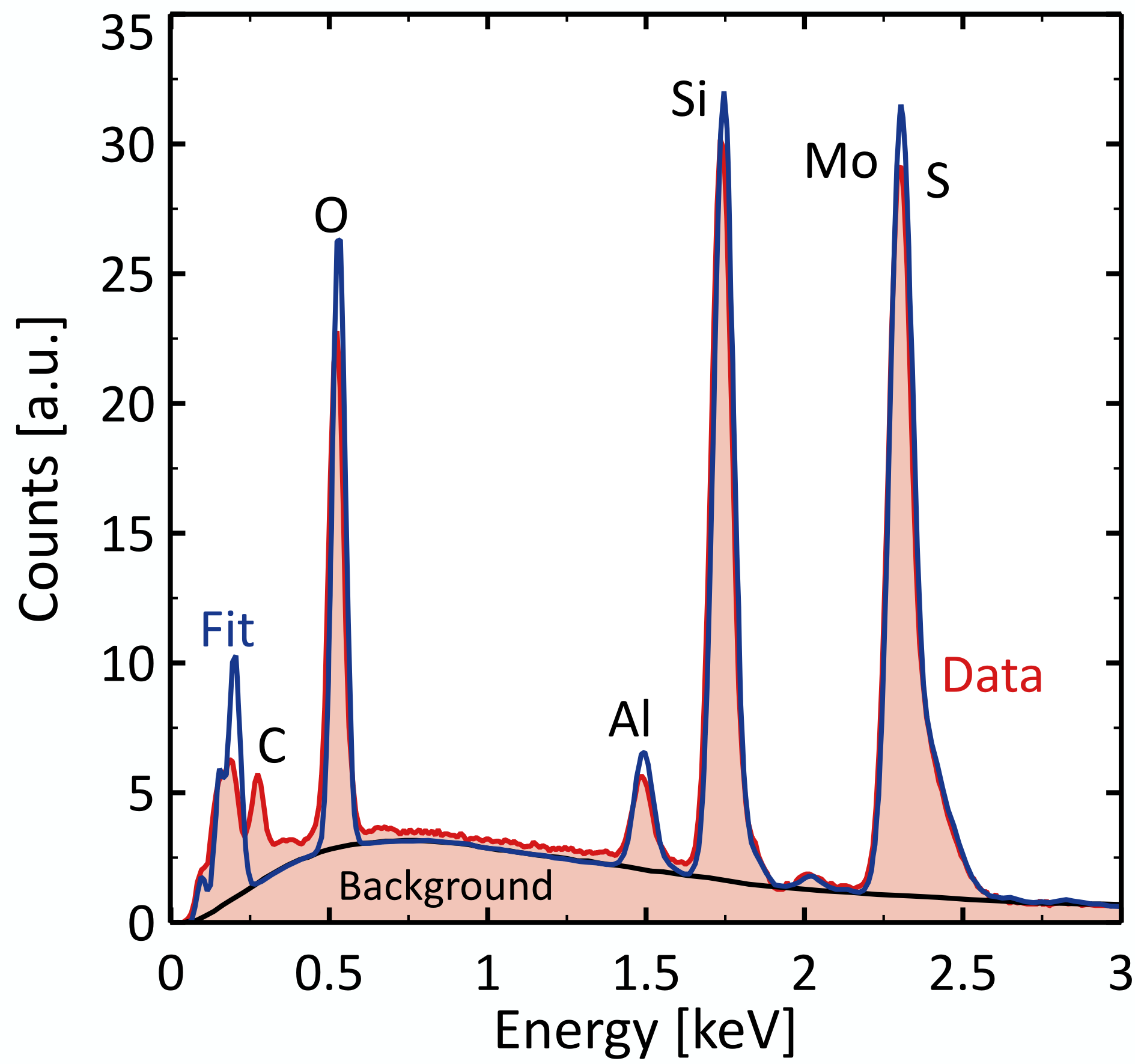}%
\caption{\textbf{\textbar~ Energy dispersive X-ray spectroscopy (EDS) spectrum.} Spectrum corresponds to Figure~\ref{F:compositeChar}(b) in the main article.}%
\label{F:edsFitPlotNormalSingle}%
\end{figure*}

\begin{figure*}
\centering
\includegraphics[width=\figWidthFull]{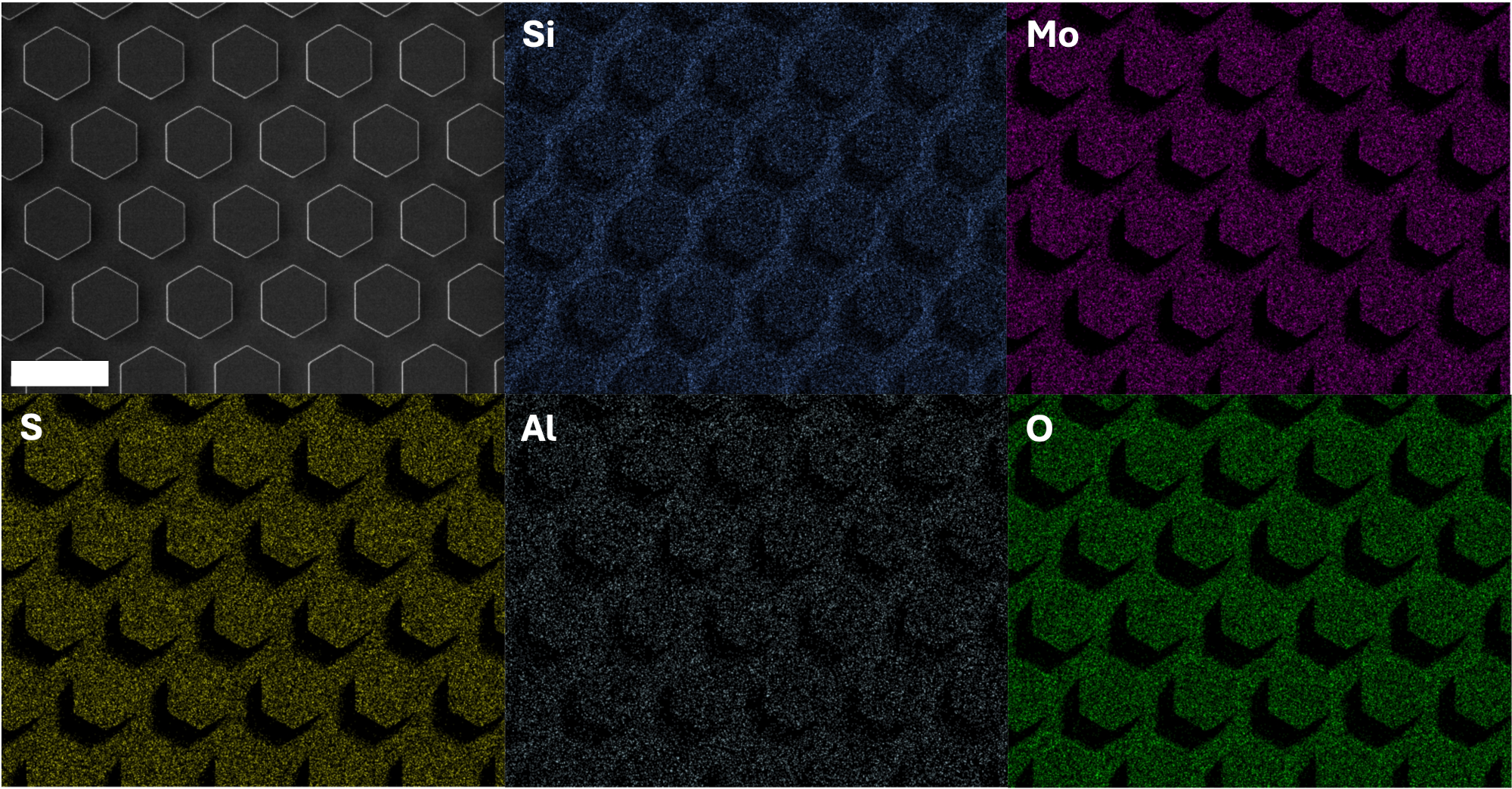}%
\caption{\textbf{\textbar~ SEM and EDS images obtained normal to the film surface}. Depicted is a corrugated nanolaminate film on \ch{Si} substrate etched in the indented trench configuration, showing elemental composition in several unit cells.  See also Figure~\ref{F:eds45several}. Scale bar: 50~\si{\micro\meter}.}%
\label{F:edsNormalSeveral}%
\end{figure*}

\begin{figure*}
\centering
\includegraphics[width=\figWidthFull]{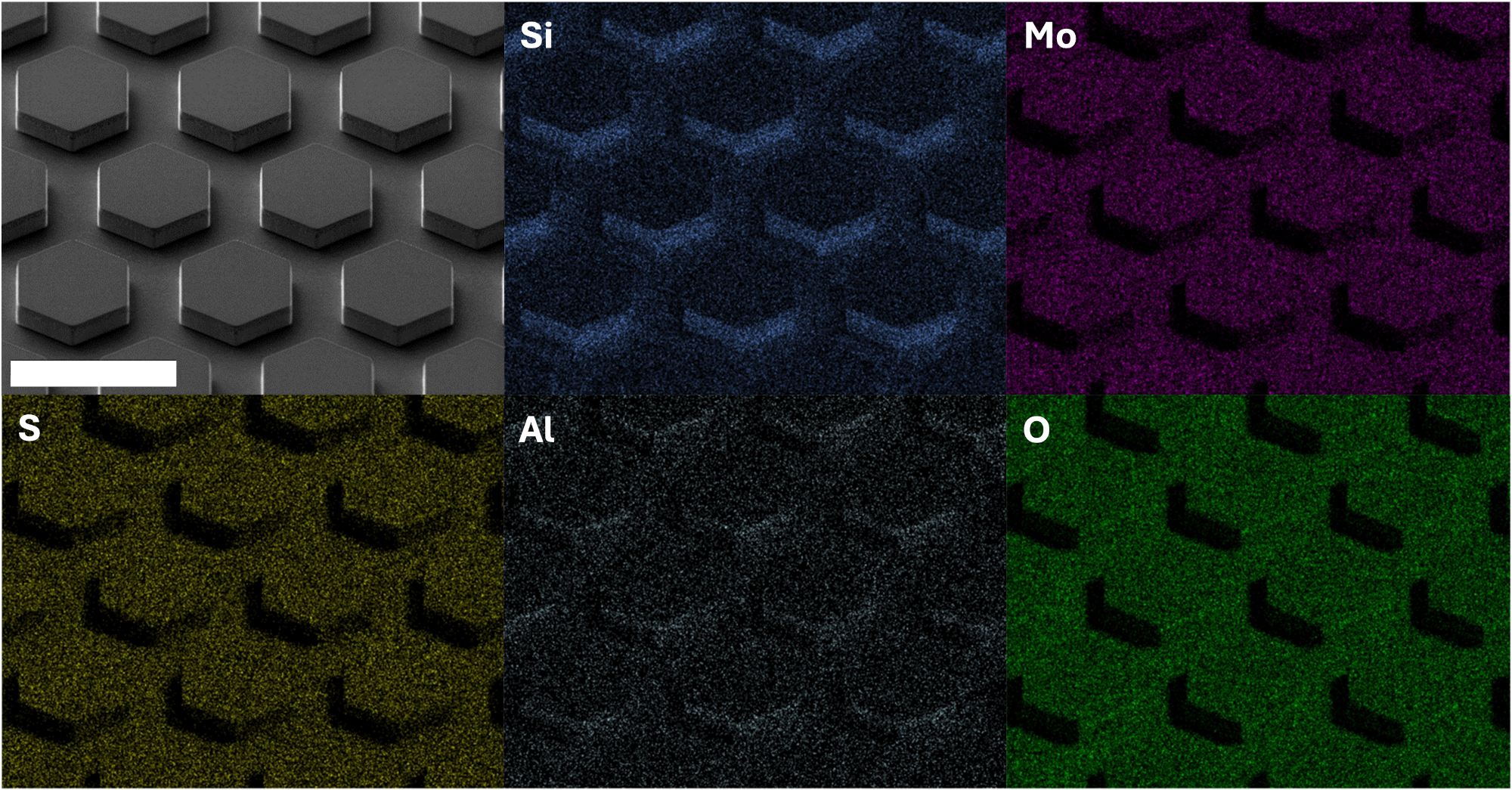}%
\caption{\textbf{\textbar~ SEM and EDS images obtained at 45\textdegree\xspace to the film surface.} See also Figure~\ref{F:edsNormalSeveral}. Scale bar: 50~\si{\micro\meter}.}%
\label{F:eds45several}%
\end{figure*}

Figure~\ref{F:edsFitPlotNormalSingle} provides the spectrum obtained in the energy dispersive X-ray spectroscopy (EDS) imaging corresponding to Figure~\ref{F:compositeChar}(b) in the main article.  The elemental \ch{C} peak is likely due to hydrocarbons adsorbed to the sample surface and is common to most EDS spectra. Figures~\ref{F:edsNormalSeveral} and~\ref{F:eds45several} are similar to Figure~\ref{F:compositeChar}(b) in the main article, except they provide normal and 45\textdegree\xspace perspective views of several unit cells, respectively. 

\section{Mechanics of corrugated films}\label{S:mechCorrFilms}

Here we derive the areal density of a composite hexagonally-corrugated film and explain how we calculate the enhanced bending stiffness and reduced tensile stiffness that hexagonal corrugation brings to films.  Figures~\ref{F:hexDimsTrenches} and~\ref{F:hexDimsRibs} show the film dimensions relevant to this work in the indented trenches and protruding ribs configurations, respectively.  The important dimensions are the hexagon diameter $d_h$, trench width $w_t$ (or rib width $w_r$), trench height $h_t$ (or rib height $h_r$),  and film thicknesses $t_{A,b}$ (bottom), $t_M$ (middle), and $t_{A,t}$ (top). For the purposes of this derivation, we measure $d_h$, $w_t$ (and $w_r$), and $h_t$ (and $h_r$) based on the \ch{Si} mold's dimensions.  Also, the side length of the hexagons $a_h$ can be calculated by
\begin{linenomath}\begin{equation}
	a_h = \frac{d_h}{\sqrt{3}}
	\label{E:aHex}
\end{equation}\end{linenomath}

We calculate the areal density for non-corrugated (flat, planar) films $\rho_{a,p}$ using
\begin{linenomath}\begin{equation}
	\rho_{a,p} = \rho_A \left(t_{A,b}+t_{A,t}\right) + \rho_M t_M
	\label{E:arealdensitynoncorr}
\end{equation}\end{linenomath}
Here, $\rho_A=3200$~\si{\kilo\gram\per\meter\cubed} is the density of \ch{Al2O3}~\cite{Ilic2010-044317} and $\rho_M=5060$~\si{\kilo\gram\per\meter\cubed} is the density of \ch{MoS2}~\cite{Graczykowski2017-7647} (see Table~\ref{T:materialProps}). Calculating the areal density for corrugated films, $\rho_{a,c,t}$ or $\rho_{a,c,r}$, is more involved.  We obtain these values by finding the ratio of the mass of a unit cell $M_{uc}$ to its planar area $A_{uc}$. For the indented trench configuration this is written
\begin{linenomath}\begin{equation}
	\rho_{a,c,t} = \frac{M_{uc,t}}{A_{uc,t}}
	\label{E:arealdensitycorrT}
\end{equation}\end{linenomath}
and for the protruding ribs films this is written 
\begin{linenomath}\begin{equation}
	\rho_{a,c,r} = \frac{M_{uc,r}}{A_{uc,r}}.
	\label{E:arealdensitycorrR}
\end{equation}\end{linenomath}
In the case of indented trenches, the unit cell mass is given by
\scriptsize
\begin{linenomath}\begin{align}
	M_{uc} &= \frac{\sqrt{3}}{2} \rho_A \Bigg( \Bigg. t_{A,b}\left(d_h+w_t\right)^2 \nonumber\\
                 & \;\;\;\;\;\; + h_t\left(\left(d_h+2t_{A,b}\right)^2 - d_h^2\right) \Bigg. \Bigg)  \nonumber\\
           &+ \frac{\sqrt{3}}{2} \rho_M \Bigg( \Bigg. t_M\left(d_h+w_t\right)^2 \nonumber\\
                 & \;\;\;\;\;\; + h_t\left(\left(d_h+2\left(t_{A,b}+t_M\right)\right)^2 - \left(d_h+2t_{A,b}\right)^2\right) \Bigg. \Bigg) \nonumber\\
           &+ \frac{\sqrt{3}}{2} \rho_A \Bigg( \Bigg. t_{A,t}\left(d_h+w_t\right)^2 \nonumber\\
                 & \;\;\;\;\;\; + h_t \Big( \Big. \left(d_h+2\left(t_{A,b}+t_M+t_{A,t}\right)\right)^2 \nonumber\\
                 & \;\;\;\;\;\; \;\;\;\;\;\;- \left(d_h+2\left(t_{A,b}+t_M\right)\right)^2 \Big. \Big) \Bigg. \Bigg) \label{E:Muct}
\end{align}\end{linenomath}
\normalsize
and in the case of protruding ribs, the unit cell mass is
\scriptsize
\begin{linenomath}\begin{align}
	M_{uc} &= \frac{\sqrt{3}}{2} \rho_A \Bigg( \Bigg. t_{A,b}\left(d_h+w_r\right)^2 \nonumber\\
                 & \;\;\;\;\;\; + h_r\left(d_h^2 - \left(d_h-2t_{A,b}\right)^2\right) \Bigg. \Bigg)  \nonumber\\
           &+ \frac{\sqrt{3}}{2} \rho_M \Bigg( \Bigg. t_M\left(d_h+w_r\right)^2 \nonumber\\
                 & \;\;\;\;\;\; + h_r\left(\left(d_h-2t_{A,b}\right)^2 - \left(d_h-2\left(t_{A,b}+t_M\right)\right)^2\right) \Bigg. \Bigg) \nonumber\\
           &+ \frac{\sqrt{3}}{2} \rho_A \Bigg( \Bigg. t_{A,t}\left(d_h+w_r\right)^2 \nonumber\\
                 & \;\;\;\;\;\; + h_r \Big( \Big. \left(d_h-2\left(t_{A,b}+t_M\right)\right)^2 \nonumber\\
                 & \;\;\;\;\;\; \;\;\;\;\;\;- \left(d_h-2\left(t_{A,b}+t_M+t_{A,t}\right)\right)^2 \Big. \Big) \Bigg. \Bigg) \label{E:Mucr}
\end{align}\end{linenomath}
\normalsize
In Equations~\ref{E:Muct} and~\ref{E:Mucr}, the first term in each parenthetical group (multiplied by the film thickness) computes the mass associated with the hexagon bases and bottoms of the trenches (or tops of the ribs), and the second term (multiplied by the rib or trench height) determines the mass of the vertical sections of the trenches (ribs). The area of the unit cell is given by 
\begin{linenomath}\begin{equation}
	A_{uc,t} = \frac{\sqrt{3}}{2} \left(d_h + w_t\right)^2
	\label{E:Auct}
\end{equation}\end{linenomath}
and
\begin{linenomath}\begin{equation}
	A_{uc,r} = \frac{\sqrt{3}}{2} \left(d_h + w_r\right)^2
	\label{E:Aucr}
\end{equation}\end{linenomath}
for the trench and rib configurations, respectively. We will henceforth write the equations in terms of the indented trench configuration using $w_t$ and $h_t$, but these two variables can be substituted for $w_r$ and $h_r$, respectively, for the protruding ribs configuration in the remaining expressions.

\begin{figure*}
\centering
\begin{tikzpicture}

    \pgfmathsetlengthmacro{\dh}{3cm}; 
    \pgfmathsetlengthmacro{\wt}{1.4cm}; 
    \pgfmathsetlengthmacro{\htt}{1cm}; 
    \pgfmathsetlengthmacro{\tfC}{0.1cm}; 
    \pgfmathsetlengthmacro{\tfB}{0.2cm}; 
    \pgfmathsetlengthmacro{\tfA}{0.05cm}; 
    \pgfmathsetlengthmacro{\ah}{\dh/sqrt(3)}; 
    \pgfmathsetlengthmacro{\dhe}{\dh+2*\tfA+2*\tfB+2*\tfC}; 
    \pgfmathsetlengthmacro{\ahe}{\dhe/sqrt(3)}; 
    \pgfmathsetlengthmacro{\wte}{\wt-2*\tfA-2*\tfB-2*\tfC}; 
    \coordinate (tA) at (0,\dhe);
    \path (tA) +(0,\dhe*1.6) node{\small Top view}; 
    \foreach \angle in {30,90,...,330} {\path (tA) +(\angle:\ahe) coordinate (a\angle); }
    \foreach \angle in {30,90,...,330} {
        \path (a\angle) +({(\angle-30)}:\wte) coordinate (b\angle);
        \path (b\angle) +(\angle:\ahe) coordinate (c\angle);
        \path (c\angle) +({(\angle+90)}:\wte) coordinate (d\angle);
        \path (a\angle) +({(\angle+30)}:\wte) coordinate (e\angle);
        \ifthenelse{\angle < 330}{\pgfmathsetlengthmacro{\temp}{\angle+60}}{\pgfmathsetlengthmacro{\temp}{30}};
        \path (a\temp) +({(\angle+30)}:\wte) coordinate (f\angle);
        };
    \draw[fill=gray!10] (b30) \foreach \angle in {30,90,...,330}{-- (c\angle) decorate[decoration=snake]{-- (d\angle)} -- (e\angle) -- (f\angle)} -- cycle;
    \draw[gray!30] (d30) decorate[decoration=snake]{-- (c90)};
    \draw[gray!30] (d90) decorate[decoration=snake]{-- (c150)};
    \draw[gray!30] (d150) decorate[decoration=snake]{-- (c210)};
    \draw[gray!30] (d210) decorate[decoration=snake]{-- (c270)};
    \draw[gray!30] (d270) decorate[decoration=snake]{-- (c330)};
    \draw[gray!30] (d330) decorate[decoration=snake]{-- (c30)};
    \draw[fill=white] (a30) \foreach \angle in {30,90,...,330}{-- (a\angle)} -- cycle;
    \draw[dashed] (d150 |- tA) node[above right]{A} -- node[above]{Section} (c30 |- tA) node[above left]{A}; 
    
    \coordinate (bA) at (0,-\dhe*1.1);
    \path (bA) +(0,\htt) node{\small Section A-A}; 
    \path (bA) +(0,\tfA) coordinate (O1); 
    \path (bA) +(0,\tfA+\tfB) coordinate (O2); 
    \path (bA) +(0,\tfA+\tfB+\tfC) coordinate (O3); 
    \pgfmathsetlengthmacro{\L}{\dhe+\wte}; 
    \pgfmathsetlengthmacro{\MIC}{\dhe/2}; 
    \pgfmathsetlengthmacro{\MOC}{\dhe/2+\wte}; 
    \pgfmathsetlengthmacro{\MIB}{\dhe/2-\tfC}; 
    \pgfmathsetlengthmacro{\MOB}{\dhe/2+\wte+\tfC}; 
    \pgfmathsetlengthmacro{\MIA}{\dhe/2-\tfC-\tfB}; 
    \pgfmathsetlengthmacro{\MOA}{\dhe/2+\wte+\tfC+\tfB}; 
    \pgfmathsetlengthmacro{\MIO}{\dhe/2-\tfC-\tfB-\tfA}; 
    \pgfmathsetlengthmacro{\MOO}{\dhe/2+\wte+\tfC+\tfB+\tfA}; 
    \path (O3) +(-\L,0) coordinate (L3);
    \path (O3) +(-\MOC,0) coordinate (LMO3);
    \path (O3) +(-\MIC,0) coordinate (LMI3);
    \path (O3) +(\L,0) coordinate (R3);
    \path (O3) +(\MOC,0) coordinate (RMO3);
    \path (O3) +(\MIC,0) coordinate (RMI3);
    \path (RMI3) +(0,-\htt) coordinate (H3);
    \path (O2) +(-\L,0) coordinate (L2);
    \path (O2) +(-\MOB,0) coordinate (LMO2);
    \path (O2) +(-\MIB,0) coordinate (LMI2);
    \path (O2) +(\L,0) coordinate (R2);
    \path (O2) +(\MOB,0) coordinate (RMO2);
    \path (O2) +(\MIB,0) coordinate (RMI2);
    \path (RMI2) +(0,-\htt) coordinate (H2);
    \path (O1) +(-\L,0) coordinate (L1);
    \path (O1) +(-\MOA,0) coordinate (LMO1);
    \path (O1) +(-\MIA,0) coordinate (LMI1);
    \path (O1) +(\L,0) coordinate (R1);
    \path (O1) +(\MOA,0) coordinate (RMO1);
    \path (O1) +(\MIA,0) coordinate (RMI1);
    \path (RMI1) +(0,-\htt) coordinate (H1);
    \path (bA) +(-\L,0) coordinate (L0);
    \path (bA) +(-\MOO,0) coordinate (LMO0);
    \path (bA) +(-\MIO,0) coordinate (LMI0);
    \path (bA) +(\L,0) coordinate (R0);
    \path (bA) +(\MOO,0) coordinate (RMO0);
    \path (bA) +(\MIO,0) coordinate (RMI0);
    \path (RMI0) +(0,-\htt) coordinate (H0);
    \draw[fill=gray!30] (O3) -- (RMI3) -- (H3) -- (RMO3 |- H3) -- (RMO3) -- (R3) -- (R2) -- (RMO2) -- (RMO2 |- H2) -- (H2) -- (RMI2) -- (O2) -- (LMI2) -- (LMI2 |- H2) -- (LMO2 |- H2) -- (LMO2) -- (L2) -- (L3) -- (LMO3) -- (LMO3 |- H3) -- (LMI3 |- H3) -- (LMI3) -- cycle;
    \draw[fill=gray!50] (O2) -- (RMI2) -- (H2) -- (RMO2 |- H2) -- (RMO2) -- (R2) -- (R1) -- (RMO1) -- (RMO1 |- H1) -- (H1) -- (RMI1) -- (O1) -- (LMI1) -- (LMI1 |- H1) -- (LMO1 |- H1) -- (LMO1) -- (L1) -- (L2) -- (LMO2) -- (LMO2 |- H2) -- (LMI2 |- H2) -- (LMI2) -- cycle;
    \draw[fill=gray!30] (O1) -- (RMI1) -- (H1) -- (RMO1 |- H1) -- (RMO1) -- (R1) -- (R0) -- (RMO0) -- (RMO0 |- H0) -- (H0) -- (RMI0) -- (bA) -- (LMI0) -- (LMI0 |- H0) -- (LMO0 |- H0) -- (LMO0) -- (L0) -- (L1) -- (LMO1) -- (LMO1 |- H1) -- (LMI1 |- H1) -- (LMI1) -- cycle;
    \path (bA) +(-\MIO,-\htt*5/4) coordinate (Larrow1);
    \path (bA) +(\MIO,-\htt*5/4) coordinate (Rarrow1);
    \path (bA) +(-\MOO,-\htt*5/4) coordinate (Larrow2);
    \path (H3) +(\wt/8,0) coordinate (Darrow1);
    \path (H3) +(\wt/8,+\htt) coordinate (Uarrow1);
    \draw[stealth-stealth] (Larrow1) -- node[below]{$\mathsf{d_h}$} (Rarrow1); 
    \draw[stealth-stealth] (Larrow2) -- node[below]{$\mathsf{w_t}$} (Larrow1); 
    \draw[stealth-stealth] (Uarrow1) -- node[right]{$\mathsf{h_t}$} (Darrow1); 
    \path (RMO3) +(\tfC/2,-\tfC/2) coordinate (H3arrowTip);
    \path (RMO3) +(\tfC/2+\wt/4,-\tfC-\tfB/2) coordinate (H2arrowTip);
    \path (RMO3) +(\tfC/2+\wt/2,-\tfC-\tfB-\tfA/2) coordinate (H1arrowTip);
    \path (bA) +(0,\htt*1.5) coordinate (H3arrowTopY);
    \path (bA) +(0,\htt*1) coordinate (H2arrowTopY);
    \path (bA) +(0,\htt*0.5) coordinate (H1arrowTopY);
    \draw[Stealth-] (H3arrowTip) -- (H3arrowTip |- H3arrowTopY) node[above]{$\mathsf{t_{A,t}}$}; 
    \draw[Stealth-] (H2arrowTip) -- (H2arrowTip |- H2arrowTopY) node[above]{$\mathsf{t_M}$}; 
    \draw[Stealth-] (H1arrowTip) -- (H1arrowTip |- H1arrowTopY) node[above]{$\mathsf{t_{A,b}}$}; 

 \end{tikzpicture}
\caption{\textbf{\textbar~ Schematic diagram of a hexagonally-corrugated film in the indented trench configuration.} Upper graphic presents a top view; lower graphic provides a section view showing the film thicknesses $t_{A,b}$ (bottom), $t_M$ (middle), and $t_{A,t}$ (top), with thickness ratios 1:4:2. We have exaggerated the dimensions to provide clarity, and, in the top view, shaded the lower trench regions. The hexagon diameter $d_h$, trench height $h_t$, and trench width $w_t$ are defined based on the dimensions of the \ch{Si} mold. }%
\label{F:hexDimsTrenches}%
\end{figure*}

\begin{figure*}
\centering
\begin{tikzpicture}

    \pgfmathsetlengthmacro{\dh}{3.75cm}; 
    \pgfmathsetlengthmacro{\wr}{0.75cm}; 
    \pgfmathsetlengthmacro{\hr}{1cm}; 
    \pgfmathsetlengthmacro{\tfC}{0.1cm}; 
    \pgfmathsetlengthmacro{\tfB}{0.2cm}; 
    \pgfmathsetlengthmacro{\tfA}{0.05cm}; 
    \pgfmathsetlengthmacro{\ah}{\dh/sqrt(3)}; 
    \pgfmathsetlengthmacro{\dhe}{\dh-2*\tfA-2*\tfB-2*\tfC}; 
    \pgfmathsetlengthmacro{\ahe}{\dhe/sqrt(3)}; 
    \pgfmathsetlengthmacro{\wre}{\wr+2*\tfA+2*\tfB+2*\tfC}; 
    \coordinate (tA) at (0,\dhe);
    \path (tA) +(0,\dhe*1.95) node{\small Top view}; 
    \foreach \angle in {30,90,...,330} {\path (tA) +(\angle:\ahe) coordinate (a\angle); }
    \foreach \angle in {30,90,...,330} {
        \path (a\angle) +({(\angle-30)}:\wre) coordinate (b\angle);
        \path (b\angle) +(\angle:\ahe) coordinate (c\angle);
        \path (c\angle) +({(\angle+90)}:\wre) coordinate (d\angle);
        \path (a\angle) +({(\angle+30)}:\wre) coordinate (e\angle);
        \ifthenelse{\angle < 330}{\pgfmathsetlengthmacro{\temp}{\angle+60}}{\pgfmathsetlengthmacro{\temp}{30}};
        \path (a\temp) +({(\angle+30)}:\wre) coordinate (f\angle);
        };
    \draw[fill=white] (b30) \foreach \angle in {30,90,...,330}{-- (c\angle) decorate[decoration=snake]{-- (d\angle)} -- (e\angle) -- (f\angle)} -- cycle;
    \draw[fill=gray!10] (d30) decorate[decoration=snake]{-- (c90)} -- (f30) -- (e30) -- cycle;
    \draw[gray!30] (d30) decorate[decoration=snake]{-- (c90)};
    \draw[fill=gray!10] (d90) decorate[decoration=snake]{-- (c150)} -- (f90) -- (e90) -- cycle;
    \draw[gray!30] (d90) decorate[decoration=snake]{-- (c150)};
    \draw[fill=gray!10] (d150) decorate[decoration=snake]{-- (c210)} -- (f150) -- (e150) -- cycle;
    \draw[gray!30] (d150) decorate[decoration=snake]{-- (c210)};
    \draw[fill=gray!10] (d210) decorate[decoration=snake]{-- (c270)} -- (f210) -- (e210) -- cycle;
    \draw[gray!30] (d210) decorate[decoration=snake]{-- (c270)};
    \draw[fill=gray!10] (d270) decorate[decoration=snake]{-- (c330)} -- (f270) -- (e270) -- cycle;
    \draw[gray!30] (d270) decorate[decoration=snake]{-- (c330)};
    \draw[fill=gray!10] (d330) decorate[decoration=snake]{-- (c30)} -- (f330) -- (e330) -- cycle;
    \draw[gray!30] (d330) decorate[decoration=snake]{-- (c30)};
    \draw[fill=gray!10] (a30) \foreach \angle in {30,90,...,330}{-- (a\angle)} -- cycle;
    \draw[dashed] (d150 |- tA) node[above right]{A} -- node[above]{Section} (c30 |- tA) node[above left]{A}; 
    
    \coordinate (bA) at (0,-\dhe*1.8);
    \path (bA) +(0,\hr*1.7) node{\small Section A-A}; 
    \path (bA) +(0,\tfA) coordinate (O1); 
    \path (bA) +(0,\tfA+\tfB) coordinate (O2); 
    \path (bA) +(0,\tfA+\tfB+\tfC) coordinate (O3); 
    \pgfmathsetlengthmacro{\L}{\dhe+\wre}; 
    \pgfmathsetlengthmacro{\MIC}{\dhe/2}; 
    \pgfmathsetlengthmacro{\MOC}{\dhe/2+\wre}; 
    \pgfmathsetlengthmacro{\MIB}{\dhe/2+\tfC}; 
    \pgfmathsetlengthmacro{\MOB}{\dhe/2+\wre-\tfC}; 
    \pgfmathsetlengthmacro{\MIA}{\dhe/2+\tfC+\tfB}; 
    \pgfmathsetlengthmacro{\MOA}{\dhe/2+\wre-\tfC-\tfB}; 
    \pgfmathsetlengthmacro{\MIO}{\dhe/2+\tfC+\tfB+\tfA}; 
    \pgfmathsetlengthmacro{\MOO}{\dhe/2+\wre-\tfC-\tfB-\tfA}; 
    \path (O3) +(-\L,0) coordinate (L3);
    \path (O3) +(-\MOC,0) coordinate (LMO3);
    \path (O3) +(-\MIC,0) coordinate (LMI3);
    \path (O3) +(\L,0) coordinate (R3);
    \path (O3) +(\MOC,0) coordinate (RMO3);
    \path (O3) +(\MIC,0) coordinate (RMI3);
    \path (RMI3) +(0,\hr) coordinate (H3);
    \path (O2) +(-\L,0) coordinate (L2);
    \path (O2) +(-\MOB,0) coordinate (LMO2);
    \path (O2) +(-\MIB,0) coordinate (LMI2);
    \path (O2) +(\L,0) coordinate (R2);
    \path (O2) +(\MOB,0) coordinate (RMO2);
    \path (O2) +(\MIB,0) coordinate (RMI2);
    \path (RMI2) +(0,\hr) coordinate (H2);
    \path (O1) +(-\L,0) coordinate (L1);
    \path (O1) +(-\MOA,0) coordinate (LMO1);
    \path (O1) +(-\MIA,0) coordinate (LMI1);
    \path (O1) +(\L,0) coordinate (R1);
    \path (O1) +(\MOA,0) coordinate (RMO1);
    \path (O1) +(\MIA,0) coordinate (RMI1);
    \path (RMI1) +(0,\hr) coordinate (H1);
    \path (bA) +(-\L,0) coordinate (L0);
    \path (bA) +(-\MOO,0) coordinate (LMO0);
    \path (bA) +(-\MIO,0) coordinate (LMI0);
    \path (bA) +(\L,0) coordinate (R0);
    \path (bA) +(\MOO,0) coordinate (RMO0);
    \path (bA) +(\MIO,0) coordinate (RMI0);
    \path (RMI0) +(0,\hr) coordinate (H0);
    \draw[fill=gray!30] (O3) -- (RMI3) -- (H3) -- (RMO3 |- H3) -- (RMO3) -- (R3) -- (R2) -- (RMO2) -- (RMO2 |- H2) -- (H2) -- (RMI2) -- (O2) -- (LMI2) -- (LMI2 |- H2) -- (LMO2 |- H2) -- (LMO2) -- (L2) -- (L3) -- (LMO3) -- (LMO3 |- H3) -- (LMI3 |- H3) -- (LMI3) -- cycle;
    \draw[fill=gray!50] (O2) -- (RMI2) -- (H2) -- (RMO2 |- H2) -- (RMO2) -- (R2) -- (R1) -- (RMO1) -- (RMO1 |- H1) -- (H1) -- (RMI1) -- (O1) -- (LMI1) -- (LMI1 |- H1) -- (LMO1 |- H1) -- (LMO1) -- (L1) -- (L2) -- (LMO2) -- (LMO2 |- H2) -- (LMI2 |- H2) -- (LMI2) -- cycle;
    \draw[fill=gray!30] (O1) -- (RMI1) -- (H1) -- (RMO1 |- H1) -- (RMO1) -- (R1) -- (R0) -- (RMO0) -- (RMO0 |- H0) -- (H0) -- (RMI0) -- (bA) -- (LMI0) -- (LMI0 |- H0) -- (LMO0 |- H0) -- (LMO0) -- (L0) -- (L1) -- (LMO1) -- (LMO1 |- H1) -- (LMI1 |- H1) -- (LMI1) -- cycle;
    \path (bA) +(-\MIO,-\hr/4) coordinate (Larrow1);
    \path (bA) +(\MIO,-\hr/4) coordinate (Rarrow1);
    \path (bA) +(-\MOO,-\hr/4) coordinate (Larrow2);
    \path (H0) +(\wr/4,0) coordinate (Uarrow1);
    \path (H0) +(\wr/4,-\hr) coordinate (Darrow1);
    \draw[stealth-stealth] (Larrow1) -- node[below]{$\mathsf{d_h}$} (Rarrow1); 
    \draw[stealth-stealth] (Larrow2) -- node[below]{$\mathsf{w_r}$} (Larrow1); 
    \draw[stealth-stealth] (Uarrow1) -- node[right]{$\mathsf{h_r}$} (Darrow1); 
    \path (H0) +(0,\tfA+\tfB+\tfC/2) coordinate (H3arrowTip);
    \path (H0) +(\wr/2,\tfA+\tfB/2) coordinate (H2arrowTip);
    \path (H0) +(\wr,\tfA/2) coordinate (H1arrowTip);
    \path (bA) +(0,\hr*2.5) coordinate (H3arrowTopY);
    \path (bA) +(0,\hr*2) coordinate (H2arrowTopY);
    \path (bA) +(0,\hr*1.5) coordinate (H1arrowTopY);
    \draw[Stealth-] (H3arrowTip) -- (H3arrowTip |- H3arrowTopY) node[above]{$\mathsf{t_{A,t}}$}; 
    \draw[Stealth-] (H2arrowTip) -- (H2arrowTip |- H2arrowTopY) node[above]{$\mathsf{t_M}$}; 
    \draw[Stealth-] (H1arrowTip) -- (H1arrowTip |- H1arrowTopY) node[above]{$\mathsf{t_{A,b}}$}; 
    
 \end{tikzpicture}
\caption{\textbf{\textbar~ Schematic diagram of a hexagonally-corrugated film in the protruding rib configuration.} Upper graphic presents a top view; lower graphic provides a section view showing the film thicknesses $t_{A,b}$ (bottom), $t_M$ (middle), and $t_{A,t}$ (top), with thickness ratios 1:4:2. We have exaggerated the dimensions to provide clarity, and, in the top view, shaded the lower hexagon regions. The hexagon diameter $d_h$, rib height $h_r$, and rib width $w_r$ are defined based on the dimensions of the \ch{Si} mold.}%
\label{F:hexDimsRibs}%
\end{figure*}

The indented trench or protruding rib areas increase the bending stiffness of the film relative to its planar (non-corrugated) counterpart.  Davami~\etal~\cite{Davami2015-10019} quantified this performance using the bending stiffness enhancement factor $\mathbb{B}$, defined here as the ratio of the bending stiffness for a corrugated film $K_c$ to that of a planar film, $K_p$. To be explicit, corrugated films with $\mathbb{B}>1$ are more stiff in bending than their planar versions. 
\begin{linenomath}\begin{equation}
	\mathbb{B} = \frac{K_c}{K_p}
	\label{E:BSEF}
\end{equation}\end{linenomath}

The original derivation provided by Davami~\etal~\cite{Davami2015-10019} concerned a single-layer film, whereas our films are three-layer composites. A rigorous treatment of the bending properties would ideally account for the impact of the three layers independently~\cite{Foral1979-200, Roy1992-479}. However, the purpose of this analysis is to illustrate the general improvement that hexagonal corrugation provides to light sail films, rather than to obtain the numerically exact performance enhancement.  Therefore, instead of re-deriving the Davami~\etal\xspace equations for composite films, we instead for expediency calculate here the equivalent single-layer film properties for our three-layer films, and then proceed to use the Davami~\etal\xspace expressions as-is with these single-layer metrics. The required effective unified film properties are the thickness, Young's modulus, and Poisson's ratio, and we obtain these using standard mixing rules~\cite{Liu2009-2198, You2017-682, Raju2018-607}. 

The unified thickness $t_u$ of the composite film is sum of the thicknesses of the individual layers. 
\begin{linenomath}\begin{equation}
	t_u = t_{A,b} + t_M + t_{A,t}
	\label{E:tu}
\end{equation}\end{linenomath}
We calculate the unified Young's modulus $E_u$ using a slight modification of the traditional Voigt equation that accounts for the Poisson effect~\cite{Liu2009-2198}.  
\begin{linenomath}\begin{align}
	E_u &= \left(f_A E_A + f_M E_M\right) \nonumber \\
        &+ \frac{f_A f_M E_A E_M \left(\nu_A - \nu_M\right)^2}{f_A E_A\left(1-\nu_M^2\right) + f_M E_M\left(1-\nu_A^2\right)} 
        \label{E:Eu}
\end{align}\end{linenomath}
Here, $\nu_A=0.24$ and $\nu_M=0.25$ are the Poisson's ratios for \ch{Al2O3}~\cite{Miller2010-58, Ylivaara2014-124} and \ch{MoS2}~\cite{Cooper2013-035423, Peng2013-19427, Woo2016-075420, Graczykowski2017-7647}, respectively, and $E_A=170$~\si{\giga\pascal}~\cite{Tripp2006-419, Ilic2010-044317, Ylivaara2014-124} and $E_M=15$~\si{\giga\pascal}~\cite{Graczykowski2017-7647} are the corresponding Young's moduli (see Table~\ref{T:materialProps}). The terms $f_A$ and $f_M$ are the fractional thicknesses of the \ch{Al2O3} and \ch{MoS2} materials, respectively, defined as 
\begin{linenomath}\begin{equation}
	f_A = \frac{t_{A,b}+t_{A,t}}{t_u}
	\label{E:FA}
\end{equation}\end{linenomath}
and
\begin{linenomath}\begin{equation}
	f_M = \frac{t_M}{t_u}.
	\label{E:FM}
\end{equation}\end{linenomath}
Similarly, we calculate the unified Poisson's ratio $\nu_u$ value using~\cite{Raju2018-607}
\begin{linenomath}\begin{equation}
	\nu_u = f_A \nu_A + f_M \nu_M.
	\label{E:nuu}
\end{equation}\end{linenomath}

Returning to Equation~\ref{E:BSEF}, the $K_p$ term corresponds to the bending stiffness of a planar (non-corrugated) plate with the same thickness, Young's modulus, and Poisson's ratio as the composite, namely $t_u$, $E_u$, and $\nu_u$, respectively.  This can be calculated by
\begin{linenomath}\begin{equation}
	K_p = \frac{E_u t_u^3}{12\left(1-\nu_u^2\right)}
	\label{E:Kp}
\end{equation}\end{linenomath}
The effective corrugated bending stiffness can be obtained by taking into account the relative area and stiffness of the trench (rib) intersections and the main region of the hexagonal unit cell (see Davami~\etal~\cite{Davami2015-10019}), according to
\begin{linenomath}\begin{equation}
	K_c = \frac{A_{uc,t}}{\frac{A_m}{K_m}+\frac{A_i}{K_p}}. 
	\label{E:Kc}
\end{equation}\end{linenomath}
In this expression, the area of the unit cell $A_{uc,t}$ is given in Equation~\ref{E:Auct}, $A_i$ is the area of the rib intersections, given by 
\begin{linenomath}\begin{equation}
	A_i = \frac{\sqrt{3}}{2} w_t^2, 
	\label{E:Ai}
\end{equation}\end{linenomath}
and $A_m$ is the area of the main region of the unit cell, obtained by subtraction: 
\begin{linenomath}\begin{equation}
	A_m = A_{uc,t}-A_i. 
	\label{E:Am}
\end{equation}\end{linenomath}
The parameter $K_p$ in Equation~\ref{E:Kc} corresponds to the stiffness of the intersections of the trenches (ribs). Since these areas do not contain any vertical components, the associated stiffness is identical to that of a flat plate (Equation~\ref{E:Kp}). The parameter $K_m$ in Equation~\ref{E:Kc} corresponds to the stiffness of the main area of the unit cell, which is also said to contain the vertical trench (rib) walls.  These greatly enhance the stiffness of the main area, as can be seen by comparing Equation~\ref{E:Kp} to Equation~\ref{E:Km}, below. 
\begin{linenomath}\begin{equation}
	K_m = \frac{E_u}{12\left(1-\nu_u^2\right)} \left(t_u^3 + 3 t_u h_t^2 + \frac{2t_u h_t^3}{a_h+w_t} \right).
	\label{E:Km}
\end{equation}\end{linenomath}

\begin{figure*}
\centering
\includegraphics[width=\figWidthFull]{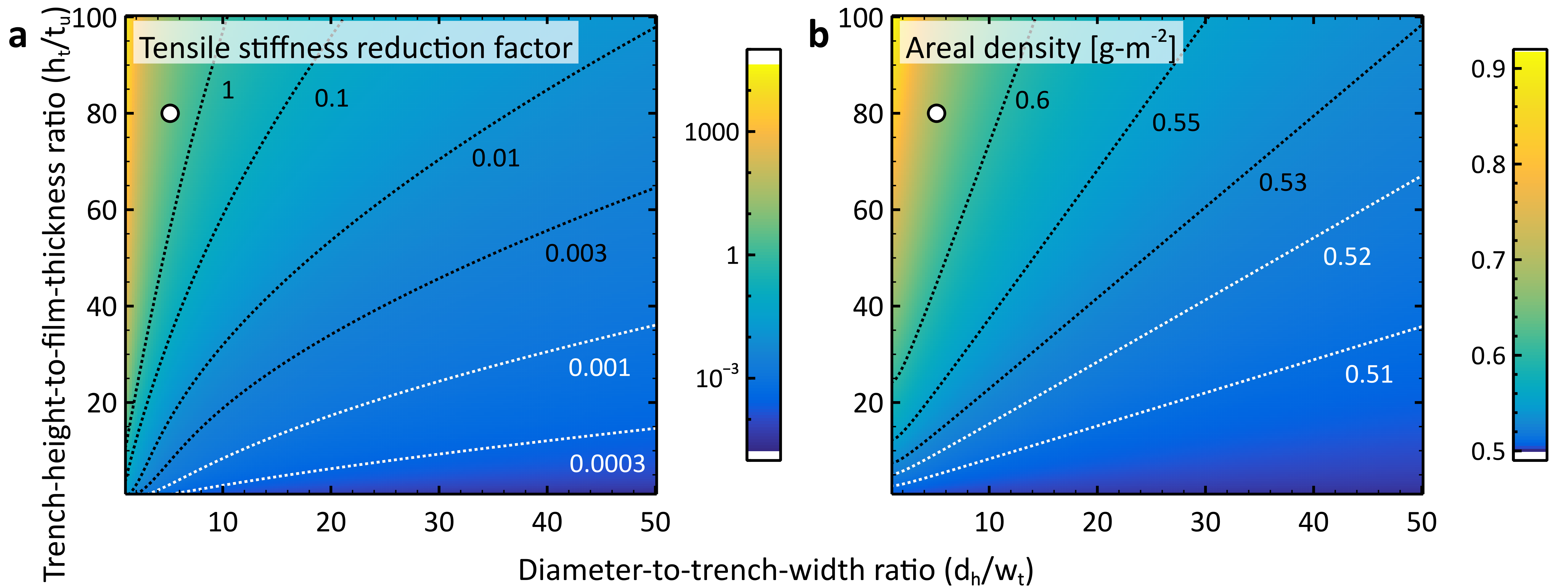}%
\caption{\textbf{\textbar~ Tensile stiffness and areal density calculations.} Panel~(\textbf{a}) shows the tensile stiffness reduction factor $\mathbb{T}$ and panel~(\textbf{b}) shows the corrugated areal density $\rho_{a,c,t}$ for hexagonally corrugated films (indented trench configuration) with $t_{A,b}\approx 21~\si{\nano\meter}$ (bottom \ch{Al2O3} thickness), $t_{M}\approx 53~\si{\nano\meter}$ (\ch{MoS2}), and $t_{A,t}\approx 51~\si{\nano\meter}$ (top \ch{Al2O3}). The fabricated prototype design, featuring  $d_h\approx77~\si{\micro\meter}$, $w_t\approx15~\si{\micro\meter}$, $h_t\approx10~\si{\micro\meter}$, $\rho_{a,c,t}\approx0.7$~\si{\gram\per\meter\squared}, $\mathbb{B}\approx38$, and $\mathbb{T}\approx2.4$, is shown with the white circle. In this case, the value $\mathbb{T}>1$ indicates no reduction in the tensile stiffness relative to a planar (non-corrugated) film. See also Figure~\ref{F:tsrfAdImp}}%
\label{F:tsrfAdProto}%
\end{figure*}

\begin{figure*}
\centering
\includegraphics[width=\figWidthFull]{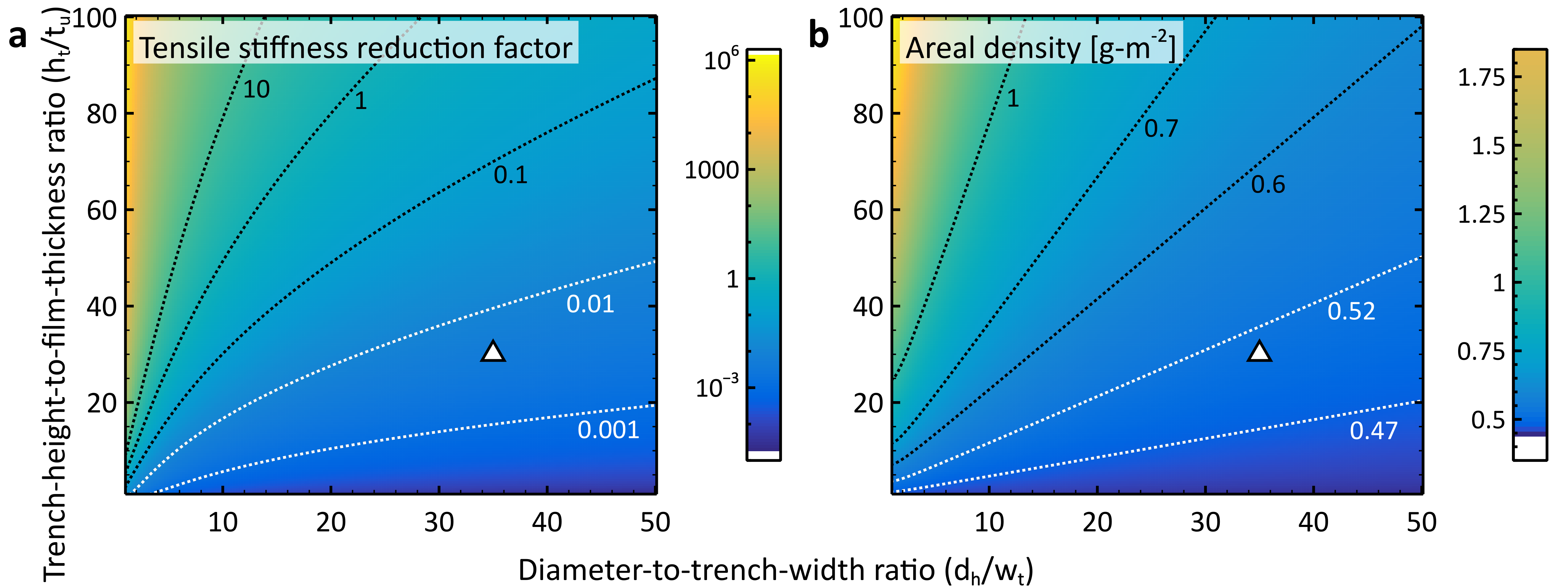}%
\caption{\textbf{\textbar~ Tensile stiffness and areal density calculations.} Panel~(\textbf{a}) shows the tensile stiffness reduction factor $\mathbb{T}$ and panel~(\textbf{b}) shows the corrugated areal density $\rho_{a,c,t}$ for hexagonally corrugated films (indented trench configuration) with $t_{A,b}=t_{A,t}\approx 19~\si{\nano\meter}$ and $t_{M}\approx 63~\si{\nano\meter}$. The proposed optimized design, featuring  $d_h=70~\si{\micro\meter}$, $w_t=2~\si{\micro\meter}$, $h_t=3~\si{\micro\meter}$, $\rho_{a,c,t}\approx0.5$~\si{\gram\per\meter\squared}, $\mathbb{B}\approx890$, and $\mathbb{T}\approx0.0033$, is shown with the white triangle. See also Figure~\ref{F:tsrfAdProto}.}%
\label{F:tsrfAdImp}%
\end{figure*}

Jiao~\etal~\cite{Jiao2020-100599} later demonstrated that hexagonal corrugation reduces the tensile stiffness, or increases the ``stretchiness,'' of films relative to their non-corrugated counterparts.  In this work, we define the ratio of the tensile stiffness of a corrugated plate $S_c$ to that of a planar (non-corrugated) plate $S_p$ as the tensile stiffness reduction factor, $\mathbb{T}$. Specifically, corrugated films with $\mathbb{T}<1$ have lower tensile stiffness, or are more stretchy, than their planar relatives. As with $\mathbb{B}$ (Equation~\ref{E:BSEF}), we employ the unified thickness and mechanical properties derived \via mixing rules from our composite sail film in the following equations. 
\begin{linenomath}\begin{equation}
	\mathbb{T} =  \frac{S_c}{S_p}
	\label{E:TSRF}
\end{equation}\end{linenomath}
According to the formulation of Jiao~\etal~\cite{Jiao2020-100599}, the tensile stiffness of a planar plate can be calculated by
\begin{linenomath}\begin{equation}
	S_p = \frac{E_u t_u}{1-\nu_u^2},
	\label{E:Sp}
\end{equation}\end{linenomath}
where $E_u$ is obtained in Equation~\ref{E:Eu}. The tensile stiffness of a hexagonally-corrugated plate can be obtained through
\begin{linenomath}\begin{equation}
	S_c = \frac{E_e h_t}{1-\nu_u^2},
	\label{E:Sc}
\end{equation}\end{linenomath}
where $E_e$ is the effective Young's modulus for the corrugated film. We obtain the effective Young's modulus according to
\begin{linenomath}\begin{equation}
	E_e = \frac{E_u}{1-\nu_e^2} \left( \alpha_c \frac{I_c}{t_u d_h^3} + \beta_c 
 \frac{h_t^4 w_t^4}{t_u^2 d_h^5 \left(2 h_t + w_t\right)}\right)
	\label{E:Eeff}
\end{equation}\end{linenomath}
wherein $I_c$ is the moment of inertia of the U-shaped trenches (ribs), $\nu_e$ is the effective Poisson's ratio of the corrugated film, and $\alpha_c=0.234$ and $\beta_c=0.062$ are empirical parameters derived from numerical plate stretching simulations~\cite{Jiao2020-100599}. 
\begin{linenomath}\begin{equation}
	I_c = \frac{1}{12}t_u w_t^3 + \frac{1}{6}h_t t_u^3  + \frac{1}{2}h_t t_u w_t^2
	\label{E:Ic}
\end{equation}\end{linenomath}
\begin{linenomath}\begin{equation}
	\nu_e = \frac{\Lambda_c - 36 I_c}{\Lambda_c + 108 I_c}
	\label{E:nue}
\end{equation}\end{linenomath}
In Equation~\ref{E:nue}, $\Lambda_c$ is given by
\begin{linenomath}\begin{equation}
	\Lambda_c = t_u \left(2 h_t + w_t\right) \left( d_h^2 + 3\left(2.4+1.5\nu_u\right)w_t^2\right).
	\label{E:lambdaM}
\end{equation}\end{linenomath}
We used these equations to predict $\rho_{a,c,t}$, $\mathbb{B}$, and $\mathbb{T}$ as a function of $\frac{d_h}{w_t}$ and $\frac{h_t}{t_u}$ for hexagonally-corrugated \ch{Al2O3}-\ch{MoS2}-\ch{Al2O3} films. The results for $\mathbb{B}$ are shown in Figure~\ref{F:compositeChar}(e) in the main article, and those of $\rho_{a,c,t}$ and $\mathbb{T}$ are shown in Figures~\ref{F:tsrfAdProto} and~\ref{F:tsrfAdImp} for design families corresponding to our fabricated prototype film and for our proposed optimized film, respectively (separate plots are required to display $\rho_{a,c,t}$ and $\mathbb{T}$ for the two films because of their different compositions and thicknesses; however, in the case of $\mathbb{B}$ in Figure~\ref{F:compositeChar}(e) of the main article, we could combine the plots for the fabricated prototype and proposed optimized films since the Young's modulus cancels out in Equation~\ref{E:BSEF} and the Poisson ratios of \ch{Al2O3} and \ch{MoS2} are very similar). 

Examining Figure~\ref{F:compositeChar}(e) in the main article, we may observe that, for sufficiently high $\frac{h_t}{t_u}$, $\mathbb{B}$ can be increased by increasing $\frac{d_h}{w_t}$.  Likewise, for high $\frac{d_h}{w_t}$, $\mathbb{B}$ can be increased by increasing $\frac{h_t}{t_u}$. For hexagonal patterns with $\frac{h_t}{t_u} \approx \frac{d_h}{w_t}$, $\mathbb{B}$ can be increased by increasing both ratios. Our proposed optimized design features $\mathbb{B}\approx900$, which we achieved by increasing $\frac{d_h}{w_t}$ relative to our fabricated prototype film. Simultaneously, our proposed optimized design features smaller trench heights $h_t$, which, due to the shape of the $\mathbb{B}$ contours, maintains a high bending stiffness while decreasing the film's areal density (see Figure~\ref{F:tsrfAdImp}). 

Viewing Figures~\ref{F:tsrfAdProto}(a) and \ref{F:tsrfAdImp}(a), we observe that ``stretchier'' designs (lower $\mathbb{T}$), which are less likely to tear, feature high $\frac{d_h}{w_t}$ and moderate $\frac{h_t}{t_u}$. Such designs also correspond to lower areal density values (Figures~\ref{F:tsrfAdProto}(b) and \ref{F:tsrfAdImp}(b)), which generally lead to faster light sail acceleration. When $\frac{h_t}{t_u}$ ratios become very low, however, $\mathbb{B}$ values decline (Figure~\ref{F:compositeChar}(e) in the main article), making the sail more less robust.  Our proposed optimized sail corrugation pattern navigates this trade-off. 

It can be useful to estimate the equivalent film thickness and material Young's modulus required to achieve the bending and tensile stiffness values in a non-corrugated (flat) plate that are equivalent to those of a corrugated plate. We refer to these as $t_{eq}$ and $E_{eq}$, and obtain them using the corrugated plate bending and tensile stiffness values, $K_c$ and $S_c$, respectively, in Equations~\ref{E:Kp} and~\ref{E:Sp}, respectively.
\begin{linenomath}\begin{equation}
	t_{eq} = \sqrt{\frac{12 K_c}{S_c}}
	\label{E:teq}
\end{equation}\end{linenomath}
\begin{linenomath}\begin{equation}
    E_{eq} = \sqrt{\frac{S_c^3 \left(1-\nu^2\right)^2}{12 K_c}}
	\label{E:Eeq}
\end{equation}\end{linenomath}
In Equation~\ref{E:Eeq}, the Poisson's ratio $\nu$ refers to that of the equivalent flat film, which we assume to be similar to that of the corrugated film $\nu_e$. For both our fabricated prototype and proposed optimized hexagonal pattern designs, $t_{eq} > t_u$ and $E_{eq} < E_u$, reflecting the benefits of corrugation in terms of mass savings and material selection. In other words, a non-corrugated (flat) film with the same bending stiffness and tensile ``stretchiness'' as our corrugated films would need to be thicker and be made from materials with lower Young's moduli. 

Finally, to avoid the possibility of folds or creases in hexagonally corrugated films, we mention the so-called ``no-straight-line rule,''~\cite{Jiao2019-034055}.  This guideline states that kinks are less likely when it is impossible to draw a straight line through the pattern without contacting vertical sidewalls. The mathematical interpretation of this statement is that $d_h \ge 3 w_t$.  Notice that the film in Figure~\ref{F:overview}(c) features $d_h\approx36$~\si{\micro\meter} and $w_t\approx15$~\si{\micro\meter}, which allowed us to bend the film to show its corrugation (\ie, $d_h < 3w_t$, breaking the rule).  

\section{Explanation of mechanical properties}\label{S:explainMechProps}

Here we explain the values of the mechanical properties that used in our calculations. Whenever possible, we obtained values directly from published works. However, in some cases where data were not available, we derived or inferred these values.  

All density values shown in Table~\ref{T:materialProps} are for microfabricated thin films, with the exception of \ch{Si}, for which we report the crystalline density. 

We obtained the tensile yield stress for \ch{Al2O3} using 
\begin{linenomath}\begin{equation}
	\sigma_y = E \epsilon_y, 
	\label{E:hooke}
\end{equation}\end{linenomath}
where $\epsilon_y$ is the tensile yield strain.  Jen, Bertrand, and George reported critical tensile strains for atomic layer deposition-based \ch{Al2O3} films ranging from 0.0052 to 0.024 for thicknesses ranging from 80~\si{\nano\meter} to 5~\si{\nano\meter} (respectively)~\cite{Jen2011-084305}. Selecting an intermediate yield strain, together with a Young's modulus of $E=170$~\si{\giga\pascal}, obtained for thin atomic layer deposition-based films~\cite{Ilic2010-044317, Tripp2006-419, Ylivaara2014-124}, provided us with a conservative tensile yield stress of $\sigma_y=2$~\si{\giga\pascal}. We used the melting point of alumina as its thermal limit, $T_{max}=2320$~\si{\kelvin}, because of its high vacuum heating stability~\cite{Schneider1967-317}.

We also derived the tensile yield stress for \ch{MoS2} using Equation~\ref{E:hooke}, with $E=15$~\si{\giga\pascal}~\cite{Graczykowski2017-7647} and $\epsilon_y=0.05$~\cite{Sledzinska2020-1169}. These values are taken for polycrystalline \ch{MoS2} fabricated using a process similar to that which we used. However, as discussed in Sections~\ref{S:optimizingThickness} and~\ref{SS:optimizedFilmCalc}, for the proposed optimized design, we assumed that future fabrication advances would allow \ch{MoS2} to be deposited in thick crystalline layers that would absorb minimally in the laser band. For the associated acceleration calculations, we conservatively assumed that these thicker layers could achieve 10\% of the perfect monolayer crystalline tensile yield strength~\cite{Bertolazzi2011-9703}, giving a value of $\sigma_y=2.3$~\si{\giga\pascal}. Recent measurements of the thickness-dependence of the Young's modulus of \ch{MoSe2}, which is similar to \ch{MoS2}, suggest this is a reasonable assumption~\cite{Babacic2021-2008614}. Note that the Young's modulus cancels out of the bending and tensile stiffness enhancement factor ratios (Equations~\ref{E:BSEF} and~\ref{E:TSRF}), thus making these calculations agnostic to our selection of polycrystalline or crystalline properties for \ch{MoS2}. For the thermal limit of \ch{MoS2} we chose $T_{max}=1000$~\si{\kelvin}, which is the point at which it begins to sublimate in a vacuum~\cite{Cui2018-44}.

For both crystalline and polycrystalline \ch{Si}, tensile yield stresses of about 1-3~\si{\giga\pascal} have been reported~\cite{Sato1998-148, Sharpe1999-162, Tsuchiya2005-665, Tsuchiya2010-1}. We selected $\sigma_y=2$~\si{\giga\pascal} as an intermediate value. Although the melting point of ~\ch{Si} is 1690~\si{\kelvin}~\cite{Gayler1938-478}, it begins to sublime at around 1470~\si{\kelvin}~\cite{Nannichi1963-586}, which we thus selected for its thermal limit. Note that recent studies have suggested that, due to its increasing absorption coefficients with temperature, a limiting temperature for \ch{Si} for laser propulsion applications could be as low as 500~\si{\kelvin}~\cite{Holdman2022-2102835}. However, since similar information is not presently available for the other materials considered in this study, we have adopted limits based on melting/sublimation/decomposition points.  

For \ch{SiO2}, tensile yield stress values in the range 0.6-1.9~\si{\giga\pascal} have been reported; we selected $\sigma_y=1.5$~\si{\giga\pascal}~\cite{Tsuchiya2000-286, Yoshioka2000-291}. \ch{SiO2} melts at 1980~\si{\kelvin}~\cite{CRCbook2024} but begins to chemically break down around 1450~\si{\kelvin}~\cite{Liehr1987-1559}, so we used the latter as our thermal limit. 

For \ch{Si3N4}, we used Equation~\ref{E:hooke} and the results of Yoshioka~\etal~\cite{Yoshioka2000-291} to derive $\sigma_y=14$~\si{\giga\pascal}. In terms of the thermal limit, \ch{Si3N4} melts at about 2170~\si{\kelvin}~\cite{CRCbook2024} but decomposes at 1670~\si{\kelvin}~\cite{Batha1973-365}, so we used the latter as its limiting temperature. 

Finally, for \ch{TiO2}, we were unable to find a tensile yield stress in the literature. However, using Equation~\ref{E:hooke} with a Young's modulus of 151~\si{\giga\pascal}~\cite{Borgese2012-2459} and a yield strain of 0.0075~\cite{Tavares2008-1434}, we obtained $\sigma_y=1.1$~\si{\giga\pascal}. \ch{TiO2} is known to be unstable when heated under vacuum conditions; although its melting point is 2110~\si{\kelvin}~\cite{StPierre1952-188}, it begins to decompose around 670~\si{\kelvin}~\cite{Mizuno2002-1716}.  We used the latter temperature as its thermal limit. 

\begin{table*}
  \caption{\textbf{\textbar~ Select properties for the materials examined in this study.}  The yield stress values are specific to tensile loading. The properties of \ch{MoS2} listed here are for its thin-film polycrystalline form, which is applicable to our fabricated prototype films.  For our optimized design, we assumed that a tensile yield stress of 10\% of the monolayer crystalline value will be achievable using future fabrication advances, \ie, $\sigma_y=2.3$~\si{\giga\pascal}~\cite{Bertolazzi2011-9703}. } 
  \begin{tabular}{ l l l l l l }
    \hline
    Material      & Density       & Yield stress          & Maximum temperature & Young's modulus    & Poisson's ratio \\
                  & $\rho$        & $\sigma_y$            & $T_{max}$           & $E$                & $\nu$           \\
                  & \si{\kilo\gram\per\meter\cubed} & \si{\giga\pascal} & \si{\kelvin} & \si{\giga\pascal} &           \\
    \hline
    \ch{Al2O3}    & 3200~\cite{Groner2004-639, Ilic2010-044317} & 2~\cite{Miller2010-58, Jen2011-084305} & 2320~\cite{Schneider1967-317} & 170~\cite{Ilic2010-044317, Tripp2006-419, Ylivaara2014-124} & 0.24~\cite{Miller2010-58, Ylivaara2014-124} \\
    \ch{MoS2}     & 5060~\cite{Graczykowski2017-7647} & 0.75~\cite{Graczykowski2017-7647, Sledzinska2020-1169} & 1000~\cite{Cui2018-44} & 15~\cite{Graczykowski2017-7647} & 0.25~\cite{Graczykowski2017-7647, Cooper2013-035423, Peng2013-19427, Woo2016-075420} \\
    \ch{Si}       & 2330~\cite{Petersen1982-420} & 2~\cite{Sato1998-148, Sharpe1999-162, Tsuchiya2005-665, Tsuchiya2010-1} & 1470~\cite{Nannichi1963-586} & - & - \\
    \ch{Si3N4}    & 3200~\cite{Yen2003-1895} & 14~\cite{Yoshioka2000-291} & 1670~\cite{Batha1973-365} & - & - \\
    \ch{SiO2}     & 2030~\cite{Kawase2009-101401} & 1.5~\cite{Tsuchiya2000-286, Yoshioka2000-291} & 1450~\cite{Liehr1987-1559} & - & - \\
    \ch{TiO2}     & 3700~\cite{Saari2022-15357} & 1.1~\cite{Borgese2012-2459, Tavares2008-1434} & 670~\cite{Mizuno2002-1716} & - & - \\
    \hline
  \end{tabular}
  \label{T:materialProps}
\end{table*}

\section{Relativistic acceleration of light sails}\label{S:relAcc}

Here we outline the equations used to determine the acceleration, distance, time, and laser output energy throughout the laser-illumination phase of a relativistic light sail's journey.  These equations are largely obtained from the Supporting Information for Campbell~\etal~\cite{Campbell2022-90}, where a full derivation can be found. They are similar to those found in Parkin~\cite{Parkin2018-370}, Kulkarni, Lubin, and Zhang~\cite{Kulkarni2016-43}, Ilic, Went, and Atwater~\cite{Ilic2018-5583}, F\H{u}zfa, Dhelonga-Biarufu, and Welcomme~\mbox{\cite{Fuzfa2020-043186}}, and Pegoraro, Livi, and Macchi~\mbox{\cite{Pegoraro2021-485}}.

\subsection{Acceleration}\label{SS:accel}

Laser photons arriving in the accelerating sail's frame of reference will have wavelengths $\lambda$ that are Doppler-shifted relative to their initial values, according to 
\begin{linenomath}\begin{equation}
	\lambda_s=\lambda_l \gamma \left(1+\beta\right),
	\label{E:lambdaS}
\end{equation}\end{linenomath}
where 
\begin{linenomath}\begin{equation}
	\gamma=\frac{1}{\sqrt{1-\beta^2}}
	\label{E:gamma}
\end{equation}\end{linenomath}
is the Lorentz factor,  
\begin{linenomath}\begin{equation}
	\beta=\frac{v}{c}
	\label{E:beta}
\end{equation}\end{linenomath}
is the relative velocity, $v$ is the sail velocity in the laser's reference frame,  $c$ is the speed of light, and subscripts $s$ and $l$ denote the sail and laser reference frames, respectively. 

It is more straightforward to describe the acceleration of the sail in terms of $\beta$ rather than in terms of time, because the reference frames of the sail and laser must be related through relativity. The acceleration of the sail in the Earth-bound laser array's frame of reference can be written 
\begin{linenomath}\begin{equation}
	a_{s,l} = \frac{2 \varrho_{\beta,a} \Phi_{l,\beta} }{m_{tot} c \gamma^3}\!\left(\frac{1-\beta}{1 + \beta} \right),
	\label{E:asl}
\end{equation}\end{linenomath}
and in the sail's frame of reference (the proper acceleration) is
\begin{linenomath}\begin{equation}
		a_{s,s} = \frac{2 \varrho_{\beta,a} \Phi_{l,\beta} }{m_{tot} c}\!\left(\frac{1-\beta}{1 + \beta} \right).
	\label{E:ass}
\end{equation}\end{linenomath}
In these expressions, $\varrho_{\beta,a}$ is the average reflectivity of the sail film when the sail's relative velocity is $\beta$ (an average is required if the sail has a curved shape, rather than a flat shape~\cite{Campbell2022-90}, see Section~\ref{SS:avgReflectSailShape}), $\Phi_{l,\beta}$ is the output power of the laser (subscript $\beta$ indicates that we could choose to allow this output power to change as the sail accelerates), and $m_{tot}$ is the total rest (\ie, proper, invariant, intrinsic) mass of the sailcraft (including sail, payload, and any connecting tethers).  We emphasize that, because of the Doppler shift of the wavelength of the incident photons, the reflectivity of the sail changes as it accelerates. Since $\gamma>1$ for $\beta>0$, $a_{s,s}>a_{s,l}$, \ie, the sail locally experiences a stronger acceleration than what the laser observes from its fixed vantage point. 

\subsection{Power}\label{SS:power}

The laser power experienced by the sail $\Phi_{s,\beta}$ is related to that outputted by the laser according to 
\begin{linenomath}\begin{equation}
    \Phi_{s,\beta} = \Phi_{l,\beta} \! \left(\frac{1-\beta}{1+\beta}\right).
	\label{E:Phisbeta}
\end{equation}\end{linenomath}
Thus, for $\beta>0$, the power experienced by the sail is less than that outputted by the laser. This decrease is associated with the red-shift of the photons in the sail's frame of reference and with the decrease in photon flux as the sail accelerates away from the incoming laser light. Note that $\Phi_{l,\beta}$ is \textit{not} the laser output power at the instant in time that the sail's relative velocity is $\beta$; rather, it is the power required to have been outputted by the laser at some previous time (see Equation~\ref{E:tpSimple}) such that those photons will reach the sail when its relative velocity is $\beta$. Also, for the sake of clarity, $\Phi_{l,\beta}$ is a constant in the case of constant laser output power. 

\subsection{Distance}\label{SS:dist}

As explained in the Supporting Information for Campbell~\etal~\cite{Campbell2022-90}, the distance traveled by the sail when its relative velocity has reached $\beta$ can be expressed as~\cite{Atwater2018-861, Ilic2018-5583}
\begin{linenomath}\begin{equation}
	D = \frac{m_{tot} c^3}{2} \! \int_0^{\beta} \! \frac{\beta\gamma}{\varrho_{\beta,a} \Phi_{l,\beta} (1-\beta)^2} d\beta.
	\label{E:D}
\end{equation}\end{linenomath}
We define the acceleration length $L$ as the distance traveled by the sail while being illuminated by laser-generated photons as it accelerates to a final velocity of $v_f=\frac{1}{5}c$:
\begin{linenomath}\begin{equation}
    L = D|_{\beta=\beta_f=0.2}
    \label{E:L}
\end{equation}\end{linenomath}

\subsection{Time}\label{SS:time}

The time $t$ required for the sail to achieve a relative velocity of $\beta$, as counted in the laser's frame of reference, is given by 
\begin{linenomath}\begin{equation}
	t = \frac{m_{tot} c^2}{2} \! \int_0^{\beta} \! \frac{\gamma}{\varrho_{\beta,a} \Phi_{l,\beta} (1-\beta)^2} d\beta. 
	\label{E:t}
\end{equation}\end{linenomath}
Thus, we define the acceleration time of the sail as 
\begin{linenomath}\begin{equation}
    t_a = t|_{\beta=\beta_f=0.2}.
    \label{E:ta}
\end{equation}\end{linenomath}
However, photons require a finite amount of time to travel from the laser to the accelerating sail.  Thus, in order for a photon to reach the sail at time $t$, it must be released at an earlier time~\cite{Parkin2018-370, Kulkarni2016-43, Kulkarni2018-155} 
\begin{linenomath}\begin{equation}
	t_p = t - \frac{D}{c}. 
	\label{E:tpSimple}
\end{equation}\end{linenomath}
Notice that $t_p < t$ for $\beta>0$. The photon release time can be expressed in terms of $\beta$ in the form 
\begin{linenomath}\begin{equation}
	t_p = \frac{m_{tot} c^2}{2} \! \int_0^{\beta} \! \frac{\gamma}{\varrho_{\beta,a} \Phi_{l,\beta} (1-\beta)} d\beta, 
	\label{E:tp}
\end{equation}\end{linenomath}
and the total time duration for which the laser array must emit photons $t_l$ can be expressed as
\begin{linenomath}\begin{equation}
    t_l = t_p|_{\beta=\beta_f=0.2}.
    \label{E:tl}
\end{equation}\end{linenomath}

\subsection{Laser array size}\label{SS:laserSize}

It is useful to estimate the required areal size of the Earth-bound laser array.  Optimal light sail acceleration occurs when the laser spot diameter $d_{l,s}$ is centered and focused on, and does not exceed, the light sail diameter $d_s$ until the sail has reached its final relative velocity of $\beta=\beta_f$~\cite{Kulkarni2016-43}. The spot size is limited by diffraction to be approximately
\begin{linenomath}\begin{equation}
	d_{l,s} = 2\lambda_l \frac{D}{d_{l,E}},
\end{equation}\end{linenomath}
where $d_{l,E}$ is the laser array diameter on Earth. Thus, for $\beta=\beta_f$ at the optimal condition $d_{l,s}=d_s$, 
\begin{linenomath}\begin{equation}
	d_{l,E} = 2\lambda_l \frac{L}{d_s}.
	\label{E:dlE}
\end{equation}\end{linenomath}

\subsection{Pressure}\label{SS:pressure}

The incident laser light will cause radiation pressure on the sail~\cite{Nichols1903-315}. The pressure will vary with position on the sail, but for circular curved sails will likely have its maximum at the sail's radial center~\cite{Campbell2022-90}. This maximum pressure is given by 
\begin{linenomath}\begin{equation}
    P = \frac{8 \varrho_{\beta,\perp} \Phi_{l,\beta} }{ \pi c d_s^2}\!\left(\frac{1-\beta}{1 + \beta} \right). 
    \label{E:Psail}
\end{equation}\end{linenomath}
Here, $\varrho_{\beta,\perp}$ is the relative-velocity-dependent perpendicular (normal-to-surface) reflectivity of the sail and $d_s$ is the sail's circular diameter. For a flat (uncurved) two-dimensional-isotropic sail perfectly aligned to the incident photons, $\varrho_{\beta,\perp}$ is the same everywhere on the surface. For a curved sail, $\varrho_{\beta,\perp}$ is the normal-to-surface reflectivity at the sail's radial center. We can observe that, if the laser output power and reflectivity are held constant, the pressure experienced by the sail will decrease throughout the acceleration phase (as $\beta$ increases) due the Doppler shift and photon flux effects mentioned above. 

\subsection{Thermal balance}\label{SS:thermBal}

The light sail will absorb a small fraction of the incident laser photons, and will thus adopt an elevated temperature at which the energy it absorbs equals its radiant exitance. The peak temperature, likely occurring at the point where the sail film is perpendicular to the incident photons (\ie, at the radial center of a curved sail), can be determined implicitly through an energy balance of the form~\cite{Campbell2022-90, Ilic2018-5583, Salary2020-1900311, Brewer2022-594} 
\begin{linenomath}\begin{equation}
    \frac{\alpha_{\beta,\perp} \Phi_{l,\beta}}{A_\perp} \! \left(\frac{1-\beta}{1+\beta}\right)  = \int_{\lambda_1}^{\lambda_2} \frac{4\pi h c^2 \varepsilon_\lambda}{\lambda^5 \! \left(\exp{\!\left(\frac{h c}{\lambda k_B T}\right)}-1\right)} d\lambda.
    \label{E:implicitT}
\end{equation}\end{linenomath}
Here, $\alpha_{\beta,\perp}$ is the perpendicular-to-sail (normal) absorptivity of the sail film when the sail's relative velocity is $\beta$, $\varepsilon_\lambda$ is the hemispherical spectral emissivity for the sail film at wavelength $\lambda$, $h$ is Planck's constant, $k_B$ is Boltzmann's constant, $T$ is the sail's temperature at its radial center, and 
\begin{linenomath}\begin{equation}
    A_\perp = \pi \left(\frac{d_s}{2}\right)^2
    \label{E:Aperp}
\end{equation}\end{linenomath}
is the perpendicular-to-laser beam area of the sail. The integral is shown to occur between two wavelengths $\langle\lambda_1,\lambda_2\rangle$ for which spectral information for the sail film is available and a majority of the emission is likely to take place according to Wien's displacement law.

We note that, in principle, the optical coefficients $\alpha_{\beta,\perp}$ and $\varepsilon_\lambda$ are temperature-dependent. Due to a lack of experimental data over a sufficiently wide wavelength range at elevated temperatures, we have opted to adopt only room-temperature ($T=300$~\si{\kelvin}) optical information (see Section~\ref{S:spectroData}). We consider that this simplification is useful and reasonable given the other large uncertainties present in these simulations. However, future studies would greatly benefit from high-temperature optical properties (see Section~\ref{S:impactOfTemperature}). 

\section{Transfer-matrix method}\label{S:transMatMeth}

\emph{The following section is copied, with minor editorial changes, from the Supporting Information for Campbell~\etal~\cite{Campbell2022-90}.}

\begin{figure*}
\centering
\begin{tikzpicture}[decoration={markings, mark= between positions 0.1 and 0.9 step 5mm with {\arrow{stealth}}}] 
    \pgfmathsetlengthmacro{\lx}{4cm};
    \pgfmathsetlengthmacro{\ly}{0.5cm}; 
    \coordinate (OO) at (0,0);
    \coordinate (A) at (-\lx/2,0);
    \coordinate (B) at (\lx/2,0);
    \path (OO) +(0,-\ly) coordinate (C);
    \path (OO) +(0,-2*\ly) coordinate (D);
    \path (OO) +(0,-3*\ly) coordinate (E);
    \path (OO) +(0,-4*\ly) coordinate (F);
    \path (OO) +(0,-5*\ly) coordinate (G);
    \draw[fill=gray!30] (A) -- (B) decorate[decoration=snake]{-- (B |- C)} -- (A |- C) decorate[decoration=snake]{-- (A)};
    \draw[fill=gray!50] (A |- C) -- (B |- C) decorate[decoration=snake]{-- (B |- D)} -- (A |- D) decorate[decoration=snake]{-- (A |- C)};
    \draw[fill=gray!40] (A |- D) -- (B |- D) decorate[decoration=snake]{-- (B |- E)} -- (A |- E) decorate[decoration=snake]{-- (A |- D)};
    \draw[fill=gray!60] (A |- E) -- (B |- E) decorate[decoration=snake]{-- (B |- F)} -- (A |- F) decorate[decoration=snake]{-- (A |- E)};
    \draw[fill=gray!80] (A |- F) -- (B |- F) decorate[decoration=snake]{-- (B |- G)} -- (A |- G) decorate[decoration=snake]{-- (A |- F)};
    \draw[dashed] (OO) -- +(0,4*\ly); 
    \path (OO) +({90-25}:4*\ly) coordinate (H); 
    \draw[gray!90,postaction={decorate}] (H) -- (OO); 
    \path (OO) -- +(0,2*\ly) coordinate (I); 
    \draw[-stealth] (I) arc (90:90-25:2*\ly); 
    \path (OO) +({90-25/2}:\ly*5/2) node{$\mathsf{\theta_1}$}; 
    \node (MO) at (-\lx/4,\ly/2) {$\mathsf{m=1}$};
    \node (MC) at (-\lx/4,-\ly/2) {$\mathsf{m=2}$};
    \node (MD) at (-\lx/4,-\ly*3/2) {$\mathsf{m=3}$};
    \node (ME) at (-\lx/4,-\ly*4.65/2) {$\vdots$};
    \node (MF) at (-\lx/4,-\ly*7/2) {$\mathsf{m=\psi-2}$};
    \node (MG) at (-\lx/4,-\ly*9/2) {$\mathsf{m=\psi-1}$};
    \node (MGE) at (-\lx/4,-\ly*11/2) {$\mathsf{m=\psi}$};
\end{tikzpicture}
\caption{\textbf{\textbar~ Film layers in the transfer-matrix method.} 
The film is shown with $\psi$ layers, including the space on both sides of the structure.  Each $m$\textsuperscript{th} layer has a complex index of refraction $\mathfrak{n}_m=n_m+\ji\kappa_m$, a thickness $t_m$, and an angle of incidence (the polar angle) relative to the normal direction $\theta_m$. In the case of a sail in the vacuum of space, the index of refraction of the first and $\psi$\textsuperscript{th} layers is $\mathfrak{n}_1=n_1+\ji\kappa_1 = \mathfrak{n}_\psi = n_\psi+\ji\kappa_\psi = 1 + 0 \ji$ and the thicknesses $t_1$ and $t_\psi$ are semi-infinite. The azimuthal angle $\phi$ (not shown) quantifies the rotation about the vertical dashed line. }%
\label{F:transMatMeth}%
\end{figure*}
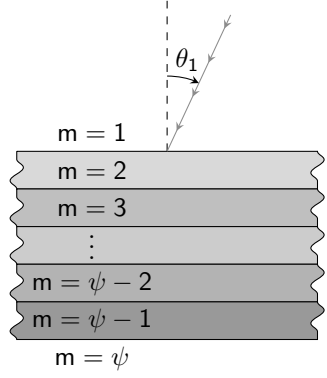

Similarly to others~\cite{Ilic2018-5583, Islam2021-2000180}, we used the transfer-matrix method~\cite{Ohta1990-2466, Macleod2017-book, Campbell2022-90} to obtain optical properties (reflectivity and absorptivity/emissivity) for the thin sail films used in this work. Consider the structure shown in Figure~\ref{F:transMatMeth}, which consists of $\psi$ layers (including the space on both sides), where each $m$\textsuperscript{th} layer has a thickness $t_m$, an angle of incidence relative to the normal direction $\theta_m$, and a complex index of refraction $\mathfrak{n}_m = n_m+\ji\kappa_m$ (here $n$ is the real component, $\kappa$ is the extinction coefficient, $\ji=\sqrt{-1}$, and $n$ and $\kappa$ depend on the light wavelength $\lambda$). In the case of a light sail in the vacuum of space, the index of refraction of the first and $\psi$\textsuperscript{th} layers is $\mathfrak{n}_1 = n_1+\ji\kappa_1 = \mathfrak{n}_\psi = n_\psi+\ji\kappa_\psi = 1 + 0 \ji$ and the thicknesses $t_1$ and $t_\psi$ are semi-infinite. The value of $\theta_1$ is equal to the angle of incidence upon the (possibly curved) sail, and the remaining complex angles $(2\le m \le \psi)$~\cite{Boyajian1923-155, Macleod2017-book} can be calculated according to
\begin{linenomath}\begin{align}
    \theta_m &= \Re{\left(\arcsin{\!\left(\frac{\mathfrak{n}_{m-1}}{\mathfrak{n}_m}\sin{(\theta_{m-1})}\right)}\right)} \nonumber\\
             & \;\;\;\;- \ji\bigg|\Im{\left(\arcsin{\!\left(\frac{\mathfrak{n}_{m-1}}{\mathfrak{n}_m}\sin{(\theta_{m-1})}\right)}\right)}\bigg|,
    \label{E:tmmTheta}
\end{align}\end{linenomath}
where $\Re{}$ and $\Im{}$ denote the real and imaginary components of a complex number, respectively, and the vertical bars $\|$ signify the absolute value. Since, according to Kirchhoff's law of radiation, emissivity is equal to absorptivity at a given wavelength for a body in thermal equilibrium, for emission calculations the angle of exitance can be thought of as the angle of incidence for absorption.  

The Fresnel equations for perpendicular $(\mathtt{s})$ and parallel $(\mathtt{p})$ polarizations for reflection $(r)$ and transmission $(t)$ can be used to obtain the reflection and transmission coefficients in each layer:
\begin{linenomath}\begin{align}
    r_{\mathtt{s},m} &= \frac{\mathfrak{n}_m\cos{(\theta_m)}-\mathfrak{n}_{m+1}\cos{(\theta_{m+1})}}{\mathfrak{n}_m\cos{(\theta_m)}+\mathfrak{n}_{m+1}\cos{(\theta_{m+1})}} \label{E:tmmRsm} \\
    t_{\mathtt{s},m} &= \frac{2\mathfrak{n}_m\cos{(\theta_m)}}{\mathfrak{n}_m\cos{(\theta_m)}+\mathfrak{n}_{m+1}\cos{(\theta_{m+1})}} \label{E:tmmTsm} \\
    r_{\mathtt{p},m} &= \frac{\mathfrak{n}_m\cos{(\theta_{m+1})}-\mathfrak{n}_{m+1}\cos{(\theta_m)}}{\mathfrak{n}_m\cos{(\theta_{m+1})}+\mathfrak{n}_{m+1}\cos{(\theta_m)}} \label{E:tmmRpm} \\
    t_{\mathtt{p},m} &= \frac{2\mathfrak{n}_m\cos{(\theta_m)}}{\mathfrak{n}_m\cos{(\theta_{m+1})}+\mathfrak{n}_{m+1}\cos{(\theta_m)}}. \label{E:tmmTpm}
\end{align}\end{linenomath}
In Equations~\ref{E:tmmRsm}-\ref{E:tmmTpm}, $1\le m \le \psi-1$. The phase shift factors $\Delta_m$ can be calculated using 
\begin{linenomath}\begin{equation}
    \Delta_m = 2 \pi \frac{t_m}{\lambda} \mathfrak{n}_m \cos{(\theta_m)}
    \label{E:tmmPsf}
\end{equation}\end{linenomath}
for $2\le m \le \psi-1$. Transfer matrices for each polarization and layer can be formed according to
\begin{linenomath}\begin{align}
    \mathbf{T}_{\mathtt{s},m} &= \begin{bmatrix}
        \frac{1}{t_{\mathtt{s},m}}\exp{(-\ji\Delta_m)} & \frac{r_{\mathtt{s},m}}{t_{\mathtt{s},m}}\exp{(\ji\Delta_m)} \\
        \frac{r_{\mathtt{s},m}}{t_{\mathtt{s},m}}\exp{(-\ji\Delta_m)} & \frac{1}{t_{\mathtt{s},m}}\exp{(\ji\Delta_m)} 
                \end{bmatrix} \label{E:tmmTmsm} \\
    \mathbf{T}_{\mathtt{p},m} &= \begin{bmatrix}
        \frac{1}{t_{\mathtt{p},m}}\exp{(-\ji\Delta_m)} & \frac{r_{\mathtt{p},m}}{t_{\mathtt{p},m}}\exp{(\ji\Delta_m)} \\
        \frac{r_{\mathtt{p},m}}{t_{\mathtt{p},m}}\exp{(-\ji\Delta_m)} & \frac{1}{t_{\mathtt{p},m}}\exp{(\ji\Delta_m)} 
                \end{bmatrix} \label{E:tmmTmpm}
\end{align}\end{linenomath}
for $2\le m \le \psi-1$, and 
\begin{linenomath}\begin{align}
    \mathbf{T}_{\mathtt{s},\psi} &= \begin{bmatrix}
        \frac{1}{t_{\mathtt{s},\psi}} & \frac{r_{\mathtt{s},\psi}}{t_{\mathtt{s},\psi}} \\
        \frac{r_{\mathtt{s},\psi}}{t_{\mathtt{s},\psi}} & \frac{1}{t_{\mathtt{s},\psi}} 
                \end{bmatrix} \label{E:tmmTmspsi} \\
    \mathbf{T}_{\mathtt{p},\psi} &= \begin{bmatrix}
        \frac{1}{t_{\mathtt{p},\psi}} & \frac{r_{\mathtt{p},\psi}}{t_{\mathtt{p},\psi}} \\
        \frac{r_{\mathtt{p},\psi}}{t_{\mathtt{p},\psi}} & \frac{1}{t_{\mathtt{p},\psi}} 
                \end{bmatrix} \label{E:tmmTmppsi}
\end{align}\end{linenomath}
for $m=\psi$. These layer-wise transfer matrices can be combined into a single transfer-matrix for the entire film stack according to 
\begin{linenomath}\begin{align}
    \mathbf{T}_{\mathtt{s},\theta_1} &= \prod_{m=2}^{\psi} \mathbf{T}_{\mathtt{s},m} \label{E:tmmTtotals} \\
    \mathbf{T}_{\mathtt{p},\theta_1} &= \prod_{m=2}^{\psi} \mathbf{T}_{\mathtt{p},m} \label{E:tmmTtotalp}
\end{align}\end{linenomath}
where the product $\prod$ implies right two-dimensional matrix multiplication
\begin{linenomath}\begin{align}
    \mathbf{M}_{m} \mathbf{M}_{m+1} 
    &= \begin{bmatrix}
        a_m & b_m \\
        c_m & d_m
      \end{bmatrix} 
      \begin{bmatrix}
        a_{m+1} & b_{m+1} \\
        c_{m+1} & d_{m+1}
      \end{bmatrix} \nonumber \\
    &= \begin{bmatrix}
        a_m a_{m+1}+b_m c_{m+1} & a_m b_{m+1}+b_m d_{m+1} \\
        c_m a_{m+1}+d_m c_{m+1} & c_m b_{m+1}+d_m d_{m+1}
      \end{bmatrix} 
    \label{E:matMult}
\end{align}\end{linenomath}
and the polar angle subscript $\theta_1$ indicates that these are calculated at the specified angle of incidence. 

The total Fresnel coefficients for reflection and transmission are then
\begin{linenomath}\begin{align}
    r_{\mathtt{s},\theta_1} &= \frac{\mathbf{T}_{\mathtt{s},\langle 2,1 \rangle}}{\mathbf{T}_{\mathtt{s},\langle 1,1 \rangle}} \label{E:tmmrtotals} \\
    t_{\mathtt{s},\theta_1} &= \frac{1}{\mathbf{T}_{\mathtt{s},\langle 1,1 \rangle}} \label{E:tmmttotals} \\
    r_{\mathtt{p},\theta_1} &= \frac{\mathbf{T}_{\mathtt{p},\langle 2,1 \rangle}}{\mathbf{T}_{\mathtt{p},\langle 1,1 \rangle}} \label{E:tmmrtotalp} \\
    t_{\mathtt{p},\theta_1} &= \frac{1}{\mathbf{T}_{\mathtt{p},\langle 1,1 \rangle}} \label{E:tmmttotalp},
\end{align}\end{linenomath}
where the bracketed subscript indices $\langle i,j \rangle$ indicate the row and column in the matrix, respectively. Finally, the total angular spectral reflectivity $\varrho$, transmissivity $\tau$, and absorptivity $\alpha$ in intensity can be obtained using
\begin{linenomath}\begin{align}
    \varrho_{\mathtt{s},\lambda,\theta,\phi} &= \abs{r_\mathtt{s}}^2 \label{E:rslt} \\
    \tau_{\mathtt{s},\lambda,\theta,\phi} &= \abs{t_\mathtt{s}}^2 \frac{\Re{\left(\mathfrak{n}_\psi \cos{(\theta_\psi)}\right)}}{\Re{\left(\mathfrak{n}_1 \cos{(\theta_1)}\right)}} \label{E:tslt} \\
    \alpha_{\mathtt{s},\lambda,\theta,\phi} &= 1-\varrho_{\mathtt{s},\lambda,\theta}-\tau_{\mathtt{s},\lambda,\theta} \label{E:aslt} \\
    \varrho_{\mathtt{p},\lambda,\theta,\phi} &= \abs{r_\mathtt{p}}^2 \label{E:rplt} \\
    \tau_{\mathtt{p},\lambda,\theta,\phi} &= \abs{t_\mathtt{p}}^2 \frac{\Re{\left(\mathfrak{n}_\psi \cos{(\theta_\psi)}\right)}}{\Re{\left(\mathfrak{n}_1 \cos{(\theta_1)}\right)}} \label{E:tplt} \\
    \alpha_{\mathtt{p},\lambda,\theta,\phi} &= 1-\varrho_{\mathtt{p},\lambda,\theta}-\tau_{\mathtt{p},\lambda,\theta}, \label{E:aplt} 
\end{align}\end{linenomath}
where here we have added the aziumuthal angle $\phi$ as a subscript to indicate that these are spectral-specular values. In practice, the data produced by the transfer-matrix method in planar-isotropic films are azimuthally symmetric. 

\section{Optical property calculations}\label{S:optPropCalcs}

\subsection{Polarization}\label{SS:polarization}

The transfer-matrix method yields both perpendicular $(\mathtt{s})$ and parallel $(\mathtt{p})$ polarizations for the reflectivity $\varrho$, transmissivity $\tau$, and absorptivity $\alpha$.  We assume that the incoming laser light is unpolarized, and as such, we average the $\mathtt{s}$ and $\mathtt{p}$ data to obtain single values for reflection and absorption.  
\begin{linenomath}\begin{equation}
    \varrho_{\beta,\theta,\phi} = \frac{\varrho_{\mathtt{s},\lambda_s,\theta,\phi}+\varrho_{\mathtt{p},\lambda_s,\theta,\phi}}{2}
    \label{E:reflectivityThetaBeta}
\end{equation}\end{linenomath}
\begin{linenomath}\begin{equation}
    \alpha_{\beta,\theta,\phi} = \frac{\alpha_{\mathtt{s},\lambda_s,\theta,\phi}+\alpha_{\mathtt{p},\lambda_s,\theta,\phi}}{2}
    \label{E:absorptivityBetaEdit}
\end{equation}\end{linenomath}

The emitted light is also unpolarized, allowing us likewise to average the $\mathtt{s}$ and $\mathtt{p}$ components.  In addition, according to Kirchhoff's law of radiation, we equate the film's emissivity and absorptivity by assuming the sail is in thermal equilibrium.
\begin{linenomath}\begin{equation}
    \varepsilon_{\lambda,\theta,\phi} = \alpha_{\lambda,\theta,\phi} = \frac{\alpha_{\mathtt{s},\lambda,\theta,\phi}+\alpha_{\mathtt{p},\lambda,\theta,\phi}}{2} 
    \label{E:epsilonLambdaTheta}
\end{equation}\end{linenomath}

\subsection{Reflectivity}\label{SS:avgReflectSailShape}

For a flat two-dimensional isotropic sail oriented perpendicular to the incoming photons, the average reflectivity is equal to the normal reflectivity at any point on the sail: 
\begin{equation}
    \varrho_{\beta,a}^{flat} = \varrho_{\beta,\perp}.
    \label{E:avgRefFlat}
\end{equation}

The perpendicular-to-sail reflectivity is obtained using $\theta=0$~\si{\radian} in Equation~\ref{E:reflectivityThetaBeta} (for $\theta=0$~\si{\radian}, the $\phi$-component is eliminated):
\begin{linenomath}\begin{equation}
    \varrho_{\beta,\perp} = \varrho_{\beta,\theta=0,\phi}.
    \label{E:varrhoPerp}
\end{equation}\end{linenomath}

In contrast, the reflectivity for curved sails, which have better laser-beam-riding stability and feature lower mechanical stresses than their flat counterparts~\cite{Popova2016-1346, Gieseler2021-21562, Campbell2022-90, Gao2024-4203}, is impacted by their off-normal-axis reflectivity values and angle-dependent photon reflection directions~\cite{Campbell2022-90}.  We can account for these two factors using 
\begin{linenomath}\begin{equation}
	\varrho_{\beta,a} = \frac{4}{\pi} \left(\frac{s_s}{d_s}\right)^2 \! \int_0^{2\pi} \! \int_0^{\theta_{max}} \! \varrho_{\beta,\theta,\phi} \cos^3(\theta) \sin(\theta) d\theta d\phi, 
	\label{E:varrhoA}
\end{equation}\end{linenomath}
where $\theta$ is the angular polar coordinate (measured relative to the
incident laser light), $\phi$ is the aziumuthal angle, $\varrho_{\beta,\theta,\phi}$ is the angle-dependent reflectivity of the
surface when the sail’s relative velocity is $\beta$, and 
\begin{linenomath}\begin{equation}
    \theta_{max} = \arcsin \! \left(\frac{d_s}{2 s_s}\right). 
    \label{E:thetaMax}
\end{equation}\end{linenomath}

It is expedient to compare the different sail films by examining their average normal (perpendicular-to-surface) reflectivity within the applicable Doppler-shifted wavelength range corresponding to their acceleration from $\beta=0$ to $\beta=0.2$. We calculate this value according to
\begin{linenomath}\begin{equation}
    \overbar{\varrho_\perp} = \frac{\int_{\lambda_l}^{\lambda_f} \varrho_{\lambda,\perp} d\lambda }{\lambda_f-\lambda_l},
    \label{E:barVarrho}
\end{equation}\end{linenomath}
where $\lambda_l$ is the laser output wavelength (\ie, that which hits the sail at $\beta=0$) and $\lambda_f$ is the wavelength of the photons hitting the sail when $\beta=0.2$ (Equation~\ref{E:lambdaS}).  Note that the integral in Equation~\ref{E:barVarrho} is within $\lambda$ space, whereas other integrals in this work (\eg, Equation~\ref{E:D}) are in $\beta$ space. The two formulations produce slightly different average values due to differences in the wavelength spacing between evenly-spaced $\beta$ and $\lambda$ points. Campbell~\etal~\cite{Campbell2022-90} selected the $\beta$-basis, whereas here we use the $\lambda$-basis. 

\subsection{Absorptivity}\label{SS:absorpt}

For the thermal energy balance calculation, we used the normal (perpendicular-to-sail-surface) absorptivity in order to conservatively estimate the sail temperature (the absorptivity in the normal direction is usually the largest). These values are simply the $\theta=0$ absorptivities (there is no $\phi$-contribution for $\theta=0$) provided by the transfer matrix method.
\begin{linenomath}\begin{equation}
    \alpha_{\beta,\perp} = \alpha_{\beta,\theta=0,\phi} 
    \label{E:absorptivityBetaTMM}
\end{equation}\end{linenomath}

Estimating the absorptivity of the sails in the literature against which we compared our designs was complicated by the fact that several contained patterned geometric elements not captured well by the transfer matrix method.  Therefore, for the sail comparison of Figure~\ref{F:benchmarking} of the main article, we estimated the absorptivity of the designs using 
\begin{linenomath}\begin{align}
    \alpha_{\lambda,\perp} &= 1 - \prod_{i=1}^{n} \! \left( 1 - \alpha_{\lambda,\perp,i}\right) \nonumber \\
    &\approx \sum_{i=1}^{n} \! \alpha_{\lambda,\perp,i} \nonumber \\
    &= \sum_{i=1}^{n} \! \frac{4 \pi \kappa_{\lambda,i} t_{f,i} F_i}{\lambda} ,
    \label{E:absorptivityBetaEstimate}
\end{align}\end{linenomath}
where $\alpha_{\lambda,\perp,i}$ is the absorptivity of the $i$\textsuperscript{th} layer ($n$ total layers) at wavelength $\lambda$, $\kappa_{\lambda,i}$ is the extinction coefficient of the $i$\textsuperscript{th} layer at wavelength $\lambda$, $t_{f,i}$ is the $i$\textsuperscript{th} layer's thickness, and $F_i$ is the layer's material fill factor, estimated using the given geometric pattern of the sail (see Section~\ref{S:compareFilms}).

We calculated the laser-band average absorptivities by integrating in $\lambda$ space:
\begin{linenomath}\begin{equation}
    \overbar{\alpha_\perp} = \frac{\int_{\lambda_l}^{\lambda_f} \alpha_{\lambda,\perp} d\lambda }{\lambda_f-\lambda_l},
    \label{E:barAlpha}
\end{equation}\end{linenomath}

\subsection{Emissivity}\label{SS:emis}

We obtain the spectral hemispherical emissivity according to
\begin{linenomath}\begin{equation}
    \varepsilon_\lambda = \frac{ \int_0^{2\pi} \! \int_0^{\frac{\pi}{2}} \! \varepsilon_{\lambda,\theta,\phi} \sin{(\theta)} \cos{(\theta)} d\theta d\phi }{\pi},
    \label{E:epsilonLambda}
\end{equation}\end{linenomath}
where of course $\int_0^{2\pi}d\phi = 2\pi$. This calculation involves integrating emissivity (\ie, absorptivity) data calculated using the transfer-matrix method over the polar angle $\theta$ and the azimuthal angle $\phi$. A full derivation for this equation and other emissivity equations is presented in the Supporting Information for Campbell~\etal~\cite{Campbell2022-90}.  

It is useful to compare the emissivity values of different sail designs, and for this purpose we calculate the effective emissivity:
\begin{linenomath}\begin{equation}
    \varepsilon_e = \frac{\int_{\lambda_1}^{\lambda_2} \varepsilon_\lambda I_{b,\lambda} d\lambda}{\int_{\lambda_1}^{\lambda_2} I_{b,\lambda} d\lambda}. 
    \label{E:epsilonEff}
\end{equation}\end{linenomath}
The integration is performed over the wavelength range for which optical information is available $\langle\lambda_1,\lambda_2\rangle$, which ideally should include a majority of the emission (this can be estimated using Wien's displacement law). The term $I_{b,\lambda}$ is the black-body emission intensity, calculated according to 
\begin{linenomath}\begin{equation}
    I_{b,\lambda} = \frac{2 h c^2}{\lambda^5 \! \left(\exp{\!\left(\frac{h c}{\lambda k_B T}\right)}-1\right)},
    \label{E:Ib}
\end{equation}\end{linenomath}
where $h$ is Planck's constant, $k_B$ is Boltzmann's constant, and $T$ is the sail's temperature.  For the purpose of comparing the effective emissivity values of several sail films, we used a consistent temperature of $T=1000$~\si{\kelvin} (but, due to the limited amount of data available, we used refractive index information at roughly 300~\si{\kelvin}). The exitance of the sail in its radial center $\mathcal{E}_s$ can be determined through
\begin{linenomath}\begin{equation}
    \mathcal{E}_s = 2 \int_{\lambda_1}^{\lambda_2} \pi \varepsilon_\lambda I_{b,\lambda} d\lambda,
    \label{E:exitance}
\end{equation}\end{linenomath}
where the factor of two denotes that emission can occur from both sides of the sail. For sail designs that are not symmetric across their thicknesses, the emissivity values of the front (facing the laser photons, $\varepsilon_{f}$) and back (facing away from the laser photons, $\varepsilon_{b}$) sides are different.  In this case, we simply calculate the average hemispherical emissivity spectrum
\begin{linenomath}\begin{equation}
    \varepsilon_{\lambda,a} = \frac{\varepsilon_{\lambda,f}+\varepsilon_{\lambda,b}}{2}
    \label{E:epsilonLambdaAvg}
\end{equation}\end{linenomath}
and average effective emissivity 
\begin{linenomath}\begin{equation}
    \varepsilon_{e,a} = \frac{\varepsilon_{e,f}+\varepsilon_{e,b}}{2}
    \label{E:epsilonEffAvg}
\end{equation}\end{linenomath}
and use these in place of $\varepsilon_\lambda$ and $\varepsilon_e$, respectively. 

\section{Spectroscopic data}\label{S:spectroData}

\begin{figure*}
\centering
\includegraphics[width=\figWidthCol]{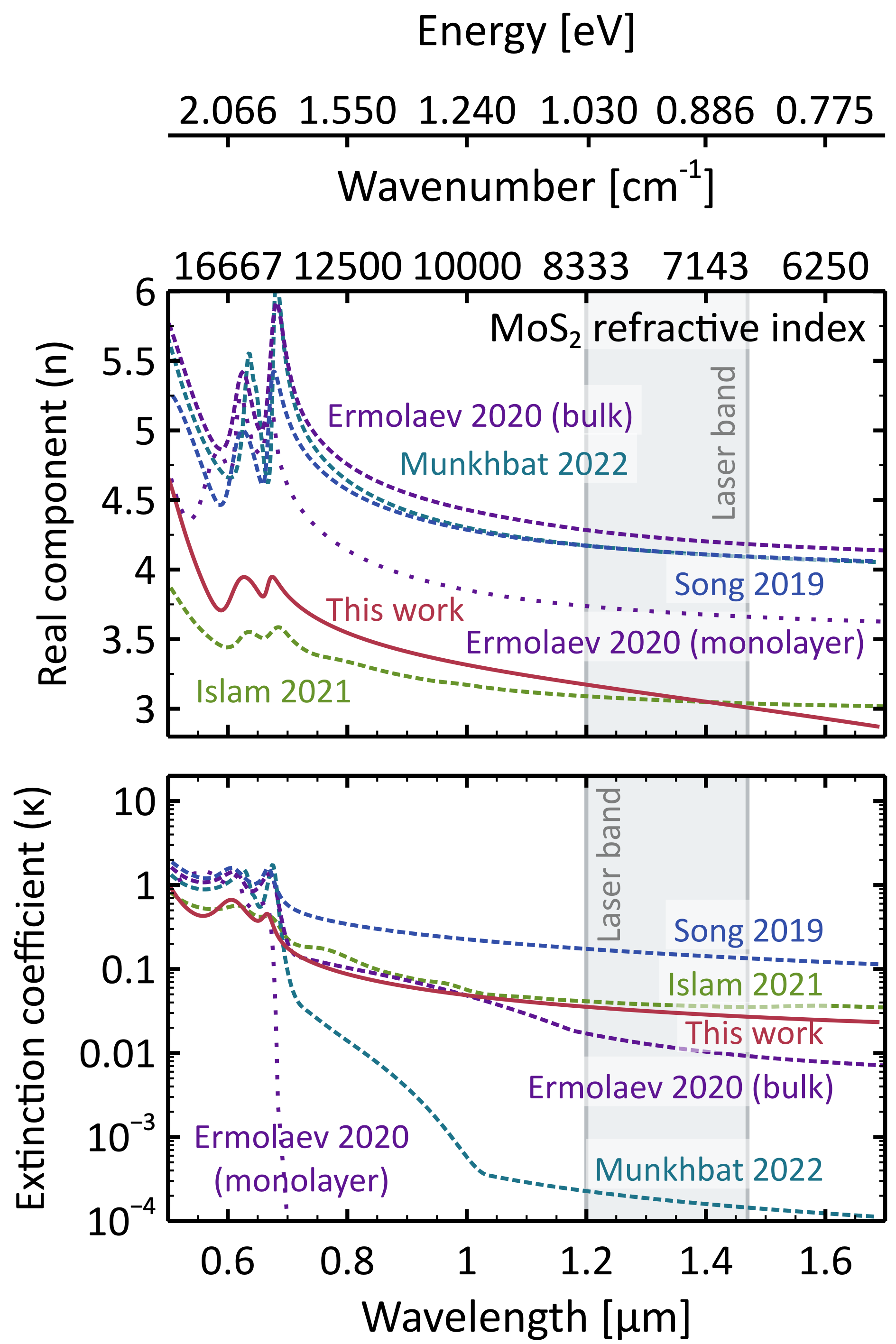}%
\caption{\textbf{\textbar~ Index of refraction measurements for \ch{MoS2} films in the literature.} Comparison is provided of the measured index of refraction $\mathfrak{n} = n+i\kappa$ for our \ch{MoS2} film with those reported by Song~\etal~\cite{Song2019-1801250},  Ermolaev~\etal~\cite{Ermolaev2020-21}, Islam~\etal~\cite{Islam2021-2000180}, and Munkhbat~\etal~\cite{Munkhbat2022-2398}.  The bulk sample of Ermolaev~\etal~\cite{Ermolaev2020-21} was roughly 1~\si{\milli\meter} thick. Here and elsewhere, the shaded gray range denotes the Doppler-shifted wavelength range corresponding to a final relative velocity of $\beta_f=0.2$: $\lambda = \langle 1.2 , 1.4697 \rangle$~\si{\micro\meter}. }%
\label{F:nkMoS2compare}%
\end{figure*}

\begin{figure*}
\centering
\includegraphics[width=\figWidthCol]{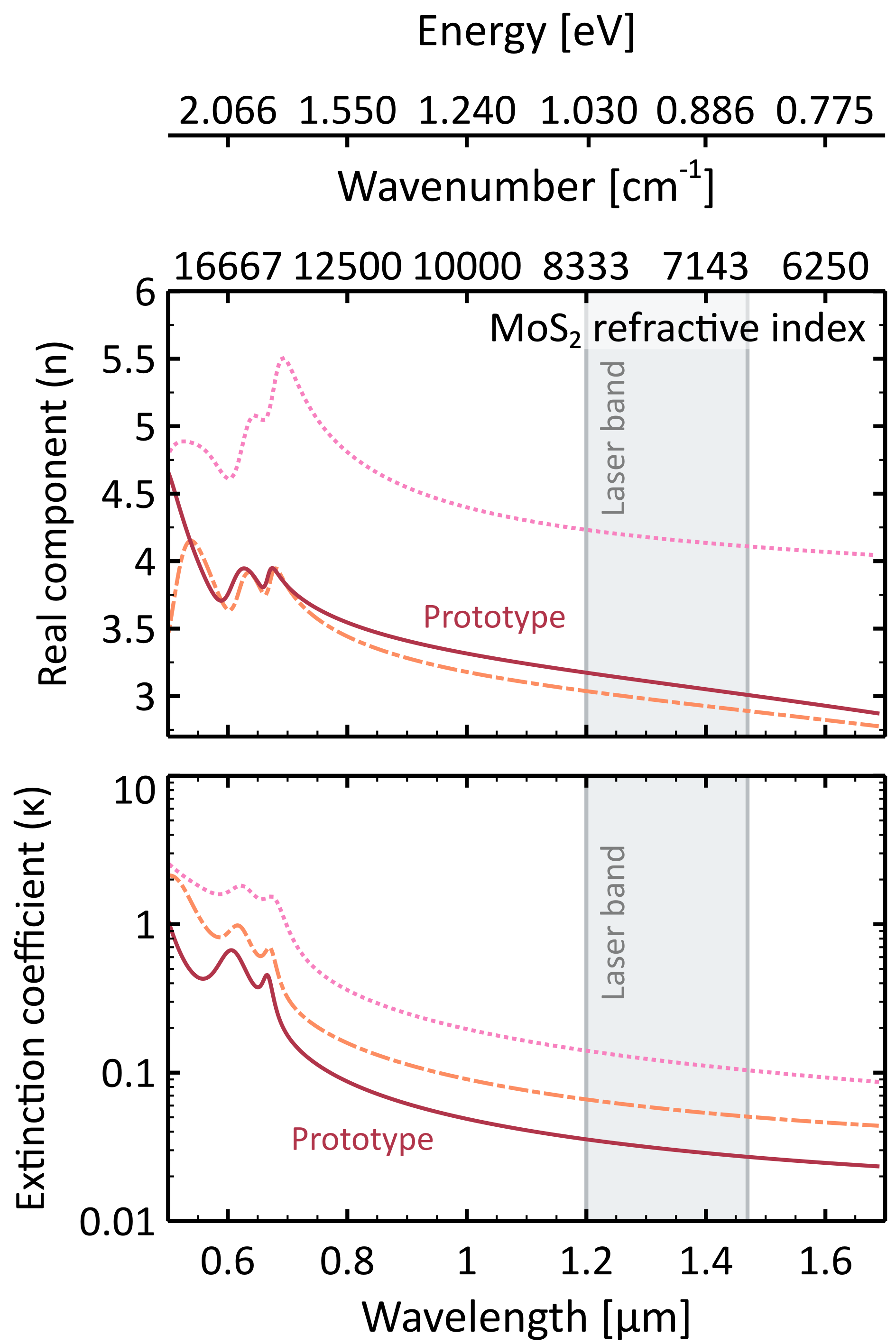}%
\caption{\textbf{\textbar~ Index of refraction measurements for three \ch{MoS2} films fabricated in-house.} The record labeled ``Prototype'' is also shown in Figure~\ref{F:nkMoS2compare}. }%
\label{F:nkMoS2other}%
\end{figure*}

\begin{figure*}
\centering
\includegraphics[width=\figWidthFull]{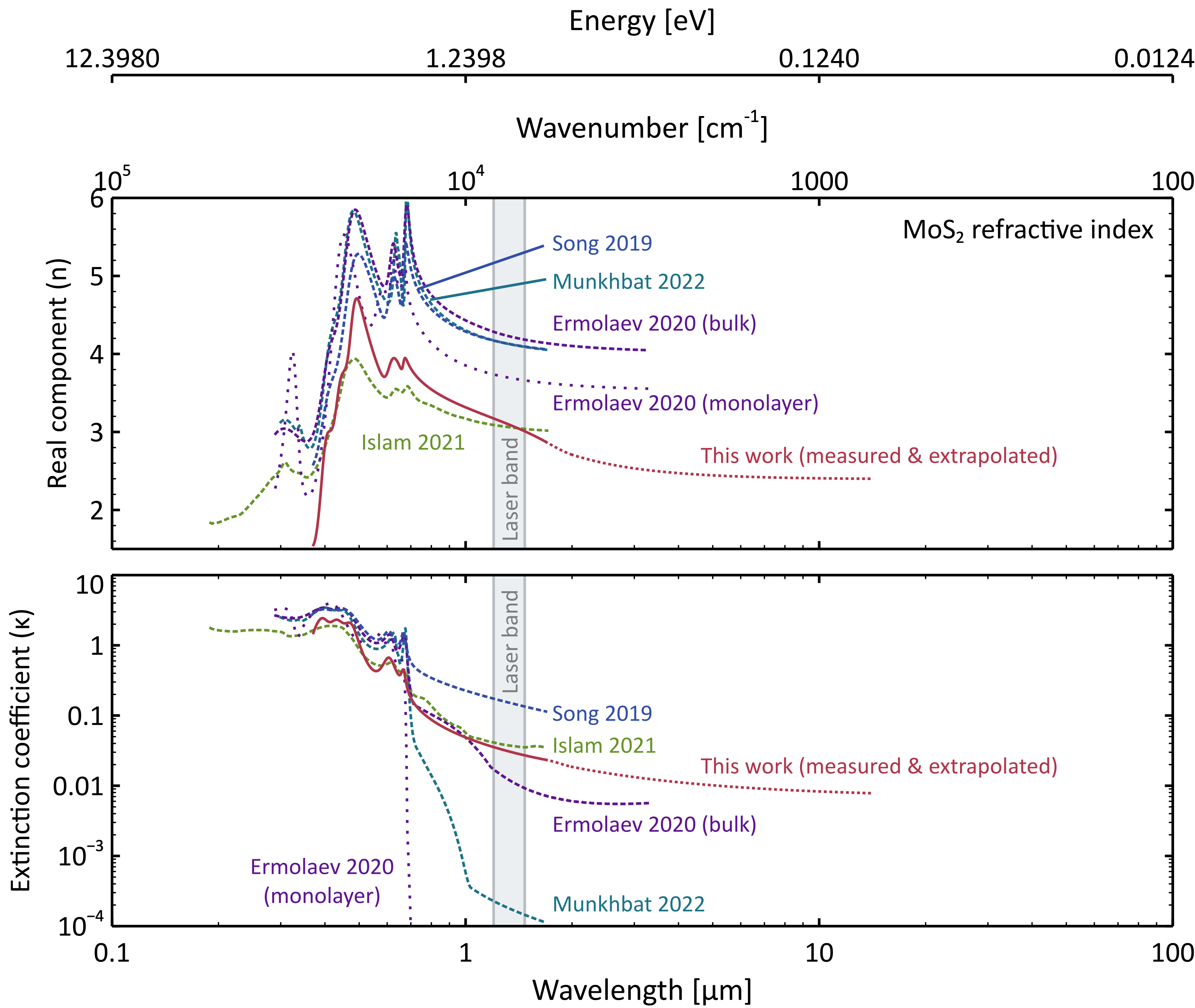}%
\caption{\textbf{\textbar~ Index of refraction of \ch{MoS2} films over an extended wavelength range.} Data are obtained from Song~\etal~\cite{Song2019-1801250},  Ermolaev~\etal~\cite{Ermolaev2020-21}, Islam~\etal~\cite{Islam2021-2000180}, and Munkhbat~\etal~\cite{Munkhbat2022-2398}.
The extrapolation of our measured data is shown in fine dots. }%
\label{F:nkMoS2extrap}%
\end{figure*}

Figure~\ref{F:nkMoS2compare} provides a comparison of the index of refraction $\mathfrak{n} = n+\ji\kappa$ that we measured for a \ch{MoS2} film that we fabricated with others reported in the literature~\cite{Song2019-1801250, Ermolaev2020-21, Islam2021-2000180, Munkhbat2022-2398}. This film in particular is that which was part of the fabricated prototype whose reflectivity and transmissivity are shown in Figure~\ref{F:compositeChar}(c) of the main article. Peaks in the $n$ data are evident near wavelengths of $\lambda=675$~\si{\nano\meter} and $625$~\si{\nano\meter} in all four records, indicating the presence of strong A and B excitons, respectively. Figure~\ref{F:nkMoS2other} shows the indices of refraction measured for two other films that we fabricated. We were able to grow large-area \ch{MoS2} with a refractive index as large as $n=4.25$ at $\lambda = 1.5$~\si{\micro\meter}, which is comparable to the value observed in exfoliated crystals~\cite{Ermolaev2020-21, Munkhbat2022-2398}. This demonstrates the potential for high-quality, wafer-scale growth of \ch{MoS2}. The reduced refractive index in the record corresponding to our prototype is most likely caused by an increased number of voids within the \ch{MoS2} layer, since both $n$ and $\kappa$ are reduced, whereas $\kappa$ would increase below the band gap (1.2-1.8~\si{\eV}) if the variation was the result of imperfections in the crystal~\cite{Jan2023-2911}. In order to calculate the emissivity of our fabricated prototype, we used a Cauchy model (with the first two terms only) and an Urbach-tail extension to extrapolate our measured $n$ and $\kappa$ data, respectively, to longer wavelengths~\cite{Fujiwara2007-book}, as shown in Figure~\ref{F:nkMoS2extrap} (see Section~\ref{S:compareFilms}). To estimate the emissivity of our proposed optimized design, we used \ch{MoS2} data from Munkhbat~\etal~\cite{Munkhbat2022-2398}, similarly extrapolated. Other data for \ch{MoS2} are also available in the literature, and generally agree with the results displayed here~\cite{Beal1979-881, Roxio1987-555, Yim2014-103114, Hsu2019-1900239, Liu2020-15282, Ermolaev2021-854, Zotev2023-2200957, Polyanskiy2024-94}. 

\begin{figure*}
\centering
\includegraphics[width=\figWidthCol]{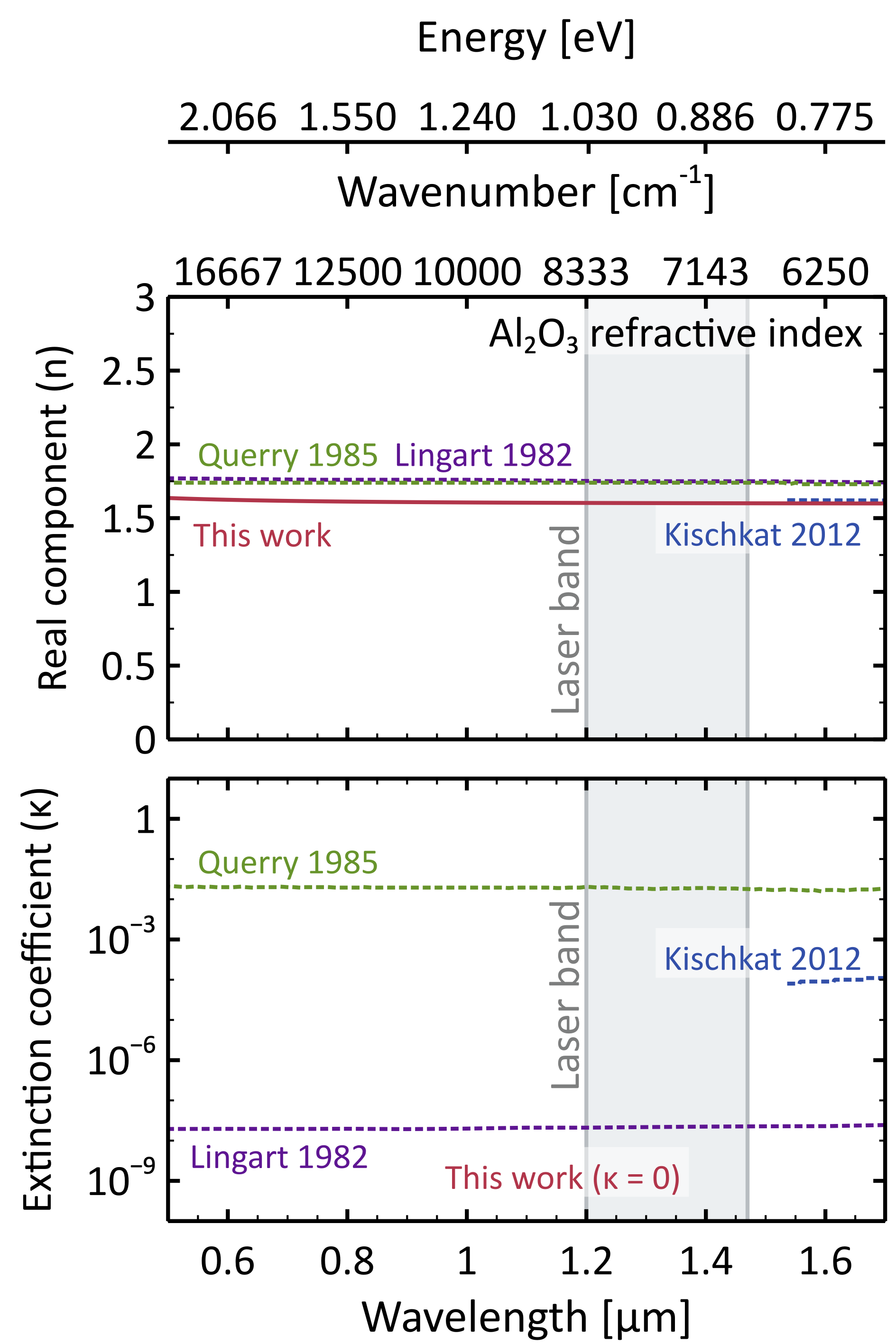}%
\caption{\textbf{\textbar~ Index of refraction measurements for \ch{Al2O3} films in the literature.} Comparison is provided of the measured index of refraction $\mathfrak{n} = n+i\kappa$ for our \ch{Al2O3} film with those reported by Lingart, Petrov, and Tikhonova~\cite{Lingart1982-706} at ($T=300$~\si{\kelvin}), Querry~\cite{Querry1985-report}, and Kischkat~\etal~\cite{Kischkat2012-6789}. 
Our measurements did not reveal any extinction from the alumina ($\kappa = 0)$. }%
\label{F:Al2O3refIndShort}%
\end{figure*}

Figure~\ref{F:Al2O3refIndShort} shows our measured index of refraction information for \ch{Al2O3} in comparison to those reported by Lingart, Petrov, and Tikhonova~\cite{Lingart1982-706} at ($T=300$~\si{\kelvin}), Querry~\cite{Querry1985-report}, and Kischkat~\etal~\cite{Kischkat2012-6789} at wavelengths near the laser band ($\lambda = \langle 1.2 , 1.4697 \rangle$~\si{\micro\meter}). Both the $n$ and $\kappa$ data are relatively constant in this range, although there are significant differences between the $\kappa$ values from different sources. For simulating the optimized film's optical properties for the comparison in Figure~\ref{F:benchmarking} in the main article, we extrapolated the data of Kischkat~\etal~\cite{Kischkat2012-6789} back to the laser band; we selected this source because it is more recent than the prior sources, its extinction coefficients fall within the reported range of literature values, and its $n$ values closely match those that we measured. 

The \ch{Al2O3} data of Lingart, Petrov, and Tikhonova~\cite{Lingart1982-706} is compiled from several sources, including References~\cite{Lee1960-594, Oppenheim1962-1078, Gillespie1965-1488, Gryvnak1965-625, Billard1974-943, Billard1976-117}. Though it contains information up to $T = 2300$~\si{\kelvin}, we used only the $T=300$~\si{\kelvin} data because elevated-temperature information was not also available at wavelengths longer than 7~\si{\micro\meter}, nor were high-temperature spectroscopic properties available for \ch{MoS2}. The absorption data in this paper are provided as Naperian attenuation coefficients $\varpi_\lambda$, which we converted to extinction coefficients $\kappa_\lambda$ through
\begin{linenomath}\begin{equation}
    \kappa_\lambda = \frac{\varpi_\lambda \lambda}{4 \pi},
    \label{E:attenToExt}
\end{equation}\end{linenomath}
where $\lambda$ is the light wavelength.  We attach the subscript $\lambda$ to emphasize that the index of refraction parameters are wavelength-dependent. 

The report by Querry~\cite{Querry1985-report} presents both ordinary $(\mathtt{o})$ and extraordinary $(\mathtt{e})$ parts of the refractive index for \ch{Al2O3}, which we averaged according to
\begin{linenomath}\begin{equation}
    n_\lambda = \frac{n_{\mathtt{o},\lambda}+n_{\mathtt{e},\lambda}}{2}
    \label{E:averageOEn}
\end{equation}\end{linenomath}
and
\begin{linenomath}\begin{equation}
    \kappa_\lambda = \frac{\kappa_{\mathtt{o},\lambda}+\kappa_{\mathtt{e},\lambda}}{2}.
    \label{E:averageOEkappa}
\end{equation}\end{linenomath}
Other references for \ch{Al2O3} exist and generally agree with the works we have included in Figure~\ref{F:Al2O3refIndShort}, although we have not attempted to conduct an exhaustive comparison here~\cite{Malitson1962-1377, Hagemann1975-742, Billard1980-641, Cabannes1987-97, Sarou-Kanian2005-1263, Lee2011-1448, Kischkat2012-6789, Franta2015-96281U, Kalman2015-74, Yang2016-111, Boidin2016-1177}. See also Section~\ref{S:impactOfTemperature} for limited temperature-dependent optical property information. 

\begin{figure*}
\centering
\includegraphics[width=170mm]{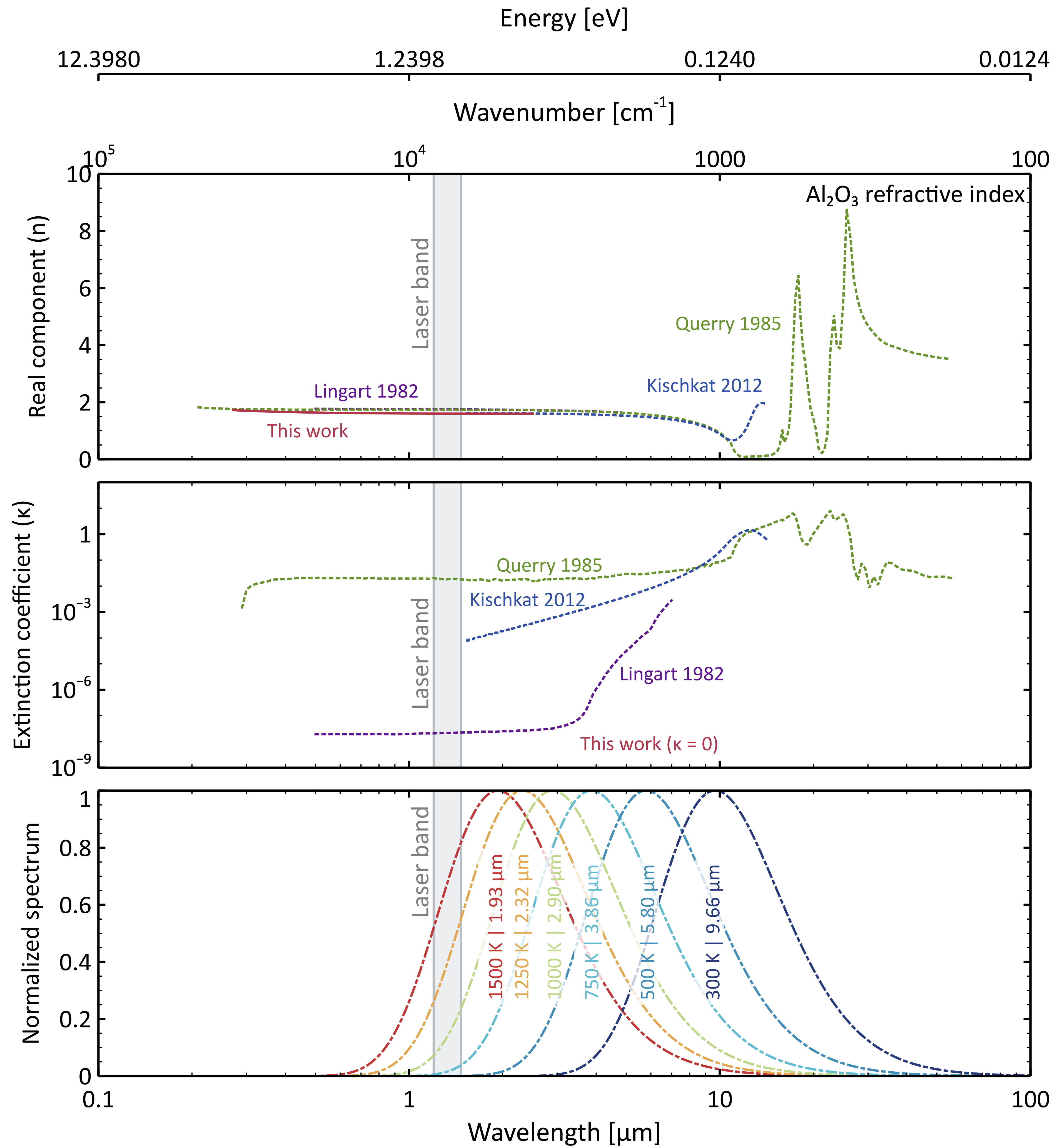}%
\caption{\textbf{\textbar~ Index of refraction of \ch{Al2O3} films over an extended wavelength range.} Top and middle: Index of refraction. Comparison is given between our measurements and those reported by Lingart, Petrov, and Tikhonova~\cite{Lingart1982-706}, Querry~\cite{Querry1985-report}, and Kischkat~\etal~\cite{Kischkat2012-6789}. Our measurements did not reveal any extinction from the alumina ($\kappa = 0)$. Bottom: Normalized Planck black-body emission spectrum parametrized by wavelength (Equation~\ref{E:Ib}) at several temperatures. The peak wavelength in each distribution according to Wien's displacement law is provided as well. Note that the peaks would have different positions (at longer wavelengths) if the distribution was parameterized by frequency~\cite{Marr2012-399}. Though the peaks at higher temperatures correspond to wavelengths where the extinction coefficient of \ch{Al2O3} is not as large, the overall exitance can still be significant because the peak of the black-body radiation curve increases dramatically in magnitude (curves in this plot are normalized to unity). }%
\label{F:Al2O3refIndLong}%
\end{figure*}

Figure~\ref{F:Al2O3refIndLong} shows longer-wavelength index of refraction information for $\ch{Al2O3}$, which is particularly useful in judging its performance as a thermal emitter.  We have included several unity-normalized Planck black-body curves, calculated according to Equation~\ref{E:Ib}, to show how the emission shifts with temperature.  

To simulate the optical properties of the other films reported in Figure~\ref{F:benchmarking} of the main article, we drew upon indices of refraction reported in the literature (see Section~\ref{S:compareFilms}): \ch{MoS2}~\cite{Munkhbat2022-2398}, \ch{Si}~\cite{Poruba2000-148, Franta2017-405}, \ch{SiO2}~\cite{Kischkat2012-6789, RodriguezdeMarcos2016-3622}, \ch{Si3N4}~\cite{Kischkat2012-6789}, and \ch{TiO2}~\cite{Kischkat2012-6789}. 

\section{Laser reflection and transmission measurements}\label{S:laserMeas}

\begin{figure*}
\centering
\includegraphics[width=\figWidthFull]{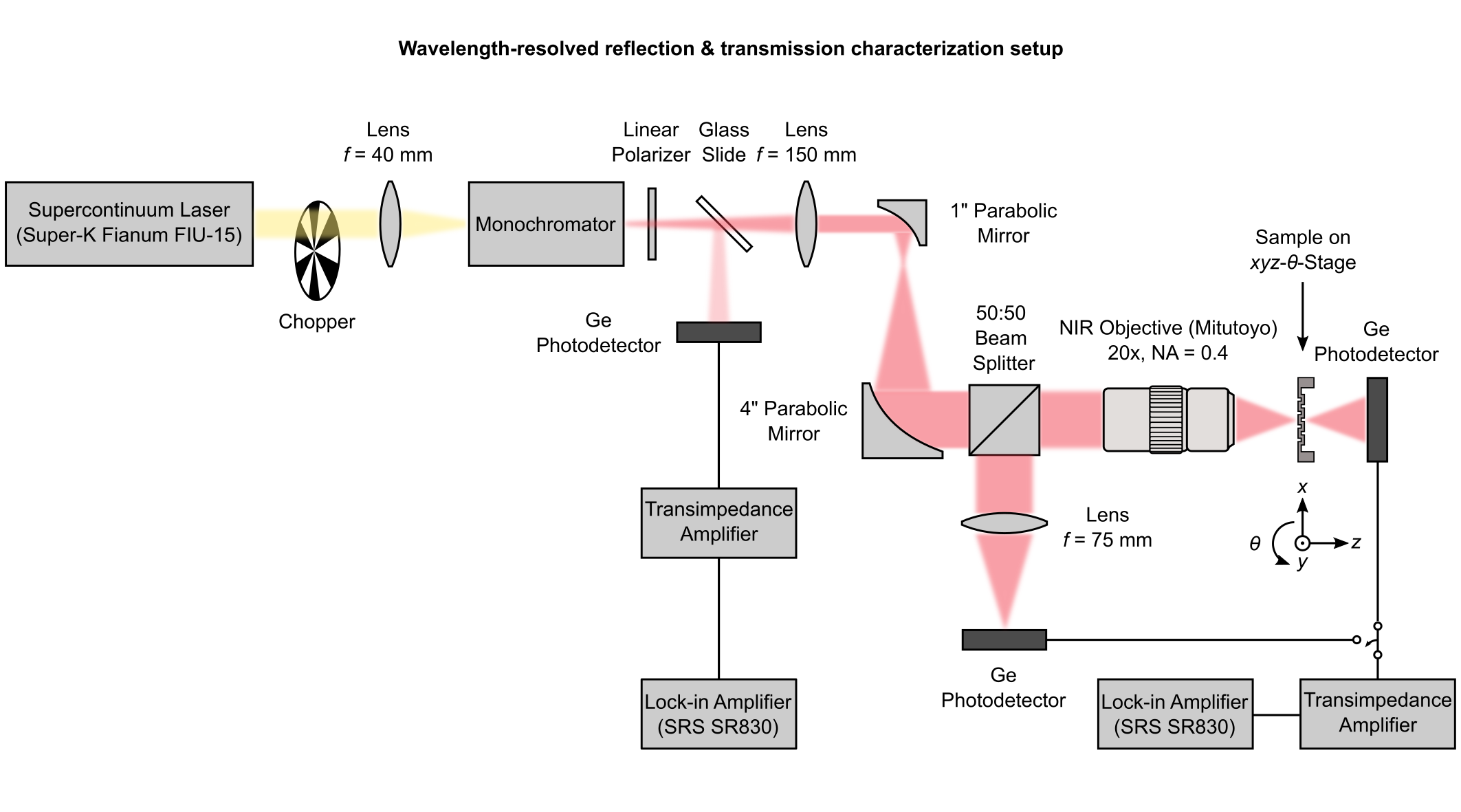}%
\caption{\textbf{\textbar~ Schematic diagram showing experimental laser setup for reflection and transmission measurements.}}%
\label{F:laserSetupCaltech}%
\end{figure*}

\begin{figure*}
\centering
\includegraphics[width=\figWidthFull]{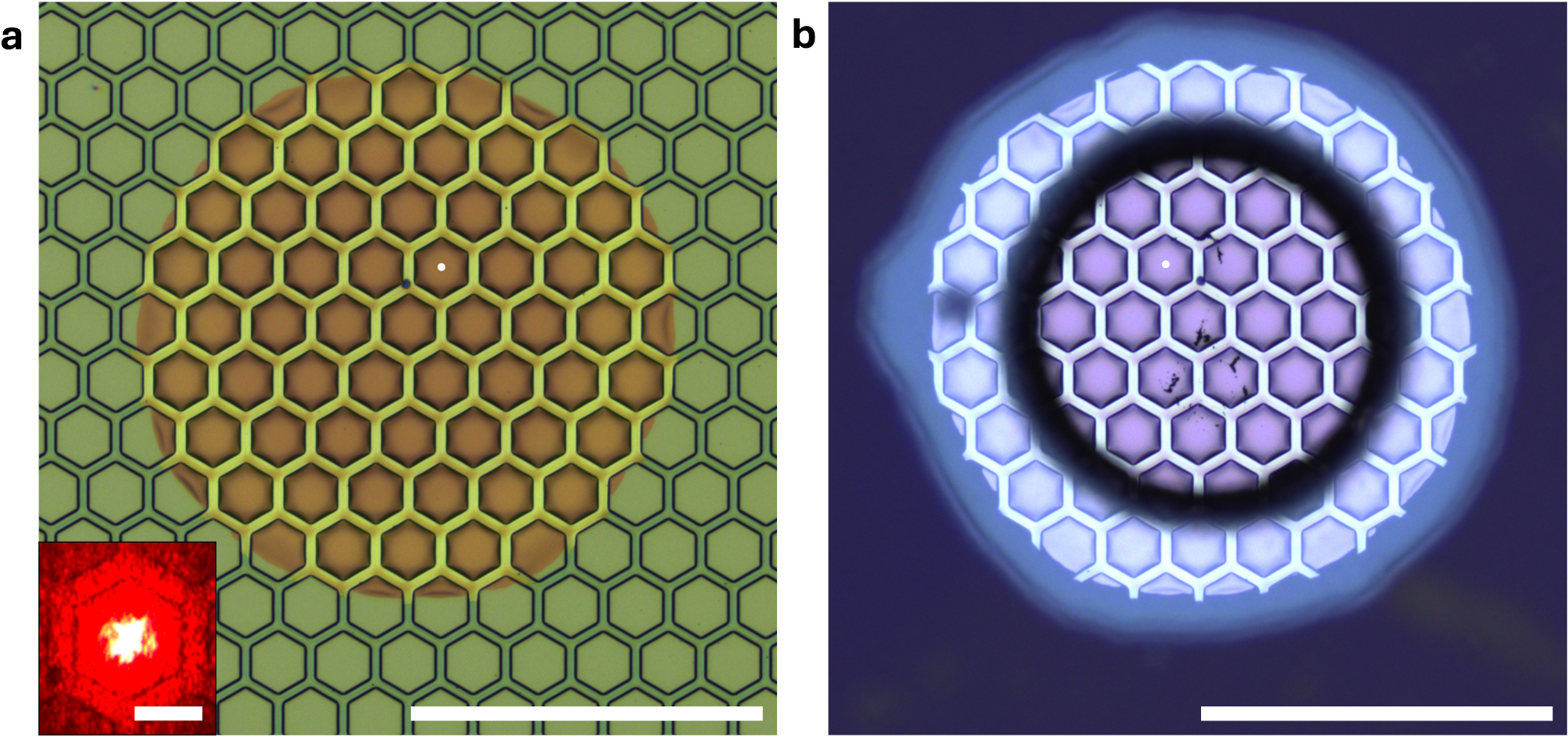}%
\caption{\textbf{\textbar~ Micrographs showing suspended film used for laser measurements.}  (\textbf{a}) Front/top side, showing indented trenches. Inset: laser spot, focused on the center of a single hexagon. (\textbf{b}) Back/bottom side. The small white dots show the single hexagon that was illuminated with the laser spot. The dark residue on the backside is likely residual \ch{SiO2} debris created in the laser etching fabrication step. The backside image shows evidence of lateral etching that occurred during the \ch{XeF2} release fabrication step, in that the original laser-drilled hole in the backside had a diameter of roughly 500~\si{\micro\meter}, but the final suspended diameter is closer to 770~\si{\micro\meter}. Fabricated prototype film dimensions: $d_h\approx77~\si{\micro\meter}$, $w_t\approx15~\si{\micro\meter}$, $h_t\approx10~\si{\micro\meter}$, $t_{A,b}\approx 21~\si{\nano\meter}$ (bottom \ch{Al2O3} thickness), $t_{M}\approx 53~\si{\nano\meter}$ (\ch{MoS2}), $t_{A,t}\approx 51~\si{\nano\meter}$ (top \ch{Al2O3}), $\rho_{a,c,t}\approx0.7$~\si{\gram\per\meter\squared} (corrugated film areal density). Scale bars: (\textbf{a}) 500~\si{\micro\meter}, inset of (a): 50~\si{\micro\meter}, (\textbf{b}) 500~\si{\micro\meter}. }%
\label{F:spotForLaserMeasurement}%
\end{figure*}

Here we present additional details about the laser reflection and transmission measurements.  All laser characterization measurements were conducted at the California Institute of Technology in Pasadena, CA; we designed special fixtures to ship chips with fully suspended films to there from the University of Pennsylvania in Philadelphia, PA. In our experiments, we coupled the output of a supercontinuum white light laser source (SuperK Fianium FIU-15) to a grating monochromator to produce wavelength-tunable, monochromatic light to illuminate our suspended test films in the wavelength range $\lambda = \langle 1000 , 1600 \rangle$~\si{\nano\meter} (Figure~\ref{F:laserSetupCaltech}). We used an optical chopper at a frequency of 417~\si{\hertz} to modulate the light, enabling us to make phase-sensitive measurements and reject stray radiation. Moreover, we used a wire-grid linear polarizer to polarize the laser beam in the perpendicular $(\mathtt{s})$ direction. Note that, for light normally incident on a surface, the perpendicular $(\mathtt{s})$ and parallel $(\mathtt{p})$ transmissivity values are equal, and the $(\mathtt{s})$ and $(\mathtt{p})$ reflectivity values are equal. After the monochromator, our beam path included lenses, parabolic mirrors, and apertures to collimate and expand the monochromatic laser beam, filling the objective's rear aperture ($20\times$ M Plan APO NIR, numerical aperture $NA = 0.4$, Mitutoyo) for a diffraction-limited laser spot on the sample (roughly 10-\si{\micro\meter} diameter; see Figure~\ref{F:spotForLaserMeasurement}). We clamped the chip containing the suspended film vertically on a stage for $X-Y$ translation and single-degree-of-freedom rotation (incidence angle). We included a beam splitter prior to the objective lens to allow us to collect the reflected light for imaging onto a camera or for detecting the reflected intensity using a \ch{Ge} photodetector. Importantly, our use of this beamsplitter allowed us to measure the normal-incidence (\ie, $0$\textdegree) reflectivity. We normalized the reflection measurements using the reflection from a template-stripped, flat \ch{Au} sample on \ch{Si}, whose wavelength-dependent reflectivity we calculated using the transfer-matrix method based on literature-obtained indices of refraction~\cite{Olmon2012-235147, Schinke2015-067168}. For transmission measurements, we placed the \ch{Ge} photodetector behind the sample.  We obtained the dark current measurements (no light) by moving the sample out of the beam path with motorized stages.  During all measurements, we took a reference measurement of the laser beam source simultaneously in order to normalize out intensity fluctuations.  We amplified and biased the detected signals (reference, reflection, and transmission) using transimpedance amplifiers and ultimately measured the signals using lock-in amplifiers. We measured the reflectivity and transmissivity at three spots on the film (henceforth denoted $A$, $B$, and $C$) and averaged the results. 

We analyzed the laser data as follows. We calculated the wavelength-dependent transmissivity using 
\begin{linenomath}\begin{equation}
    \tau_\lambda = \frac{T_{\lambda,f}}{T_{\lambda,100}},
    \label{E:transmissivity}
\end{equation}\end{linenomath}
where $T_{\lambda,f}$ is the normalized transmission detector signal with the film in place,
\begin{linenomath}\begin{equation}
    T_{\lambda,f} = \frac{J_{\lambda,f}}{J_{\lambda,f,r}}, 
    \label{E:Tlambdaf}
\end{equation}\end{linenomath}
and $T_{\lambda,100}$ is the normalized transmission detector signal with the film shifted out of the beam path,
\begin{linenomath}\begin{equation}
    T_{\lambda,100} = \frac{J_{\lambda,100}}{J_{\lambda,100,r}}.
    \label{E:Tlambda100}
\end{equation}\end{linenomath}
Here $J$ indicates the current produced by the photodetector and subscript $r$ denotes the reference detector measurement used for normalization of the laser intensity fluctuations. 

We calculated the wavelength-dependent reflectivity according to 
\begin{linenomath}\begin{equation}
    \varrho_\lambda = \frac{R_{\lambda,f}-R_{\lambda,0}}{R_{\lambda,100}-R_{\lambda,0}}. 
    \label{E:reflectivity}
\end{equation}\end{linenomath}
Here, $R_{\lambda,f}$ is the normalized reflection detector signal with the film in place
\begin{linenomath}\begin{equation}
    R_{\lambda,f} = \frac{J_{\lambda,f}}{J_{\lambda,f,r}},
    \label{E:RlambdaF}
\end{equation}\end{linenomath}
$R_{\lambda,0}$ is the normalized reflection detector signal with the film shifted out of the beam path
\begin{linenomath}\begin{equation}
    R_{\lambda,0} = \frac{J_{\lambda,0}}{J_{\lambda,0,r}},
    \label{E:Rlambda0}
\end{equation}\end{linenomath}
and $R_{\lambda,100}$ is the normalized reflection detector signal with the \ch{Au} reference chip in place of the film, normalized by the calculated reflectivity for the chip $\varrho_{\lambda,c}$
\begin{linenomath}\begin{equation}
    R_{\lambda,100} = \left(\frac{J_{\lambda,100}}{J_{\lambda,100,r}}\right)\!\left(\frac{1}{\varrho_{\lambda,c}}\right).
    \label{E:Rlambda100}
\end{equation}\end{linenomath}

We estimated a $\pm3\%$ uncertainty for the reflectivity and transmissivity values, which we applied to each of the three spots that we measured.  To obtain the error bars shown in the main article, we determined the maximum and minimum values among the three datasets, \eg, $\tau_\lambda^{max} = \text{max}\left(\tau_{\lambda,A}^+,\tau_{\lambda,B}^+,\tau_{\lambda,C}^+\right)$ or $\varrho_\lambda^{min} = \text{min}\left(\varrho_{\lambda,A}^-,\varrho_{\lambda,B}^-,\varrho_{\lambda,C}^-\right)$, where here the $+$ and $-$ superscripts indicate the $+3\%$ and $-3\%$ uncertainty values, respectively.  We compounded these errors for the absorptivity uncertainty values, \eg, $\alpha_\lambda^{max} = 1 -\tau_\lambda^{min} - \varrho_\lambda^{min}$.

\begin{figure*}
\centering
\includegraphics[width=\figWidthFull]{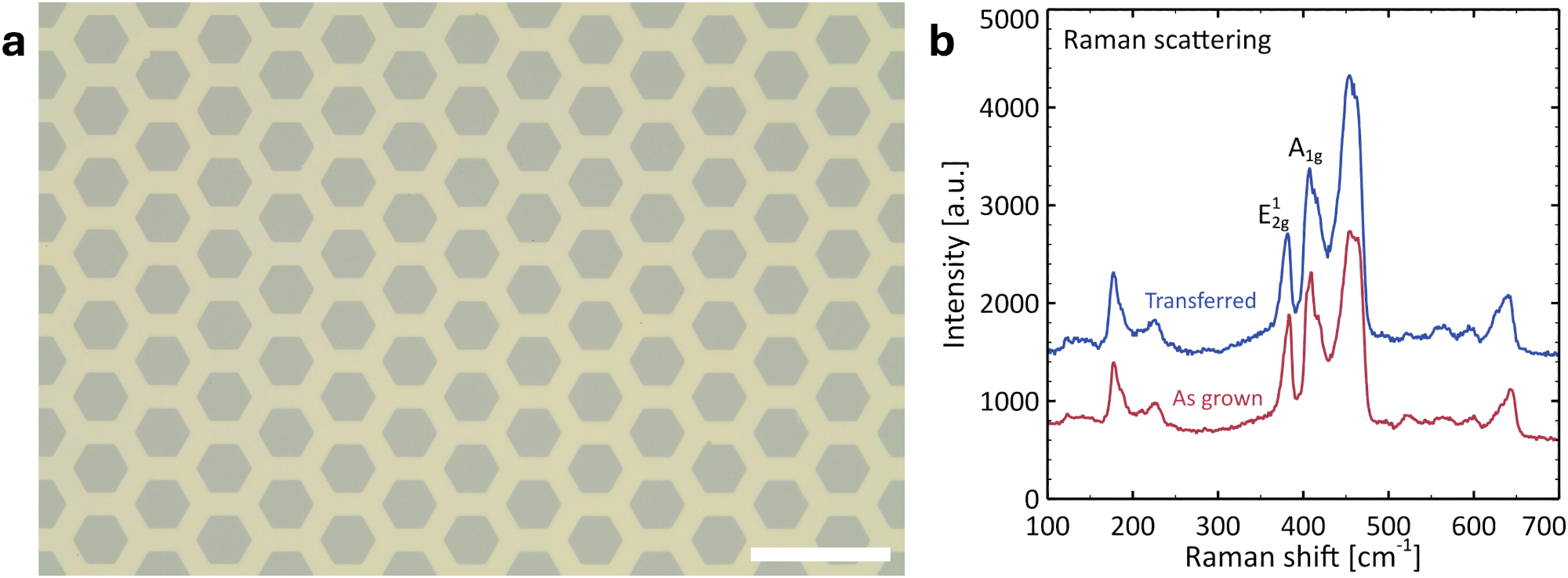}
\caption{\textbf{\textbar~ Visualization and characterization of wet chemical transfer prototype film.} (\textbf{a}) Optical micrograph showing prototype film on \ch{Si} substrate (indented trench configuration) with \ch{MoS2} film (originally grown on a flat substrate) wet chemical transferred on top~\cite{Kumar2022-182}. (\textbf{b}) Raman scattering measurements of a \ch{MoS2} film on a \ch{Si} substrate and after transfer to a \ch{Al2O3}-coated corrugated \ch{Si} substrate. Sample dimensions: $d_h\approx36~\si{\micro\meter}$, $w_t\approx15~\si{\micro\meter}$, $h_t\approx10~\si{\micro\meter}$. Scale bar: (\textbf{a}) 100~\si{\micro\meter}.  }%
\label{F:MoS2onTrenches}%
\end{figure*}

Finally, we note that each measurement was made with the laser spot focused in the center of one of the corrugated hexagon areas, rather than on the perimeter of a hexagon near the vertical walls. This was a practical decision on our part in order to simplify the experiments. In a practical light sail, laser light would naturally be incident upon these vertical components, which present a longer path length for absorption than the planar film areas. Here we offer several comments on this. First, any practical sail film must exhibit  virtually zero absorption in the Doppler-shifted wavelength range. Though the absorption path length would be roughly two orders of magnitude larger in the vertical components than the planar regions, this difference might be inconsequential compared to extinction coefficients close to zero for both \ch{Al2O3} and \ch{MoS2}. Second, for the more immediately achievable goal of intra-solar-system light sail travel in which lower laser powers were required, incident photon fluxes on the sail would conceivably not be high enough to cause damage, even in the vertical corrugated components. Third, future research may reveal sail configurations that do not require corrugation to prevent wrinkles; in such cases, a non-corrugated embodiment of our \ch{Al2O3}-\ch{MoS2}-\ch{Al2O3} sail would still be advantageous due to its low areal density and high reflectivity.  Fourth, we have developed an alternative corrugated architecture in which planar \ch{MoS2} films are wet chemical transferred onto the tops of \ch{Al2O3}-coated trenches (Figure~\ref{F:MoS2onTrenches})~\cite{Kumar2022-182}. This implementation reduces the amount of absorbing material in the vertical components, thereby mitigating the risk of excessive absorption there.  Finally, detailed simulations should be done to specifically account for the presence of the vertical walls and hexagons (including any diffractive effects) at the small scales employed in this work, which we leave to future work.  

\section{Estimate of optical requirements and thermally-limited power}\label{S:estimateOptical}

Here we estimate the required ratio of infrared emissivity to laser band absorptivity for the sail film.  We assume to first order that at all points throughout its acceleration, the sail will be in thermal equilibrium such that any energy it absorbs (per unit area) it also must radiate out. We perform this estimate on a per-unit-area basis because the thermal conductivity of nanometer-thick sails will likely be so low that, for curved sails, the center region (perpendicular to the incident laser photons) may be significantly hotter than the perimeter areas~\cite{Gao2024-4203}. 
\begin{linenomath}\begin{equation}
    e_{absorb} = e_{radiate}
\label{E:EconsSail}
\end{equation}\end{linenomath}
The energy absorbed from the incident laser photons per unit area can be estimated by 
\begin{linenomath}\begin{equation}
    e_{absorb} = \frac{\overbar{\alpha_\perp} \Phi_s}{A_\perp}
\label{E:EabsorbSail}
\end{equation}\end{linenomath}
where $\overbar{\alpha_\perp}$ is the laser band-average absorptivity (Equation~\ref{E:barAlpha}), $\Phi_s$ is the instantaneous incident photon power hitting the sail, and $A_\perp$ is the area of the sail perpendicular to the incident laser beam (Equation~\ref{E:Aperp}). The energy radiated per unit area can be estimated by
\begin{linenomath}\begin{equation}
    e_{radiate} = 2 \varepsilon_e \sigma T^4
\label{E:EradiateSail}
\end{equation}\end{linenomath}
where $\varepsilon_e$ is the hemispherical effective emissivity of the sail at $T=1000$~\si{\kelvin} (Equation~\ref{E:epsilonEff}), $\sigma$ is the Stefan-Boltzmann constant, $T$ is the sail temperature, and the factor of two indicates that radiation can occur from both sides of the sail film. Equating Equations~\ref{E:EabsorbSail} and~\ref{E:EradiateSail} and solving for the ratio $\frac{\varepsilon_e}{\overbar{\alpha_\perp}}$ gives
\begin{linenomath}\begin{equation}
    \frac{\varepsilon_e}{\overbar{\alpha_\perp}} = \frac{\Phi_s}{ 2 A_\perp \sigma T^4} . 
\label{E:epsAlphaRatio}
\end{equation}\end{linenomath}
At a laser power of $\Phi_s=100$~\si{\giga\watt} (when the $\beta=0$, the power incident on the sail is equal to the laser output power; see Equation~\ref{E:Phisbeta}), a sail perpendicular-to-laser area of $A_\perp=1$~\si{\meter\squared}, and a reasonable maximum sail temperature of $T=1000$~\si{\kelvin} (at which point the sail would likely begin to sublime or melt~\cite{Brewer2022-594}), the ratio becomes roughly $\frac{\varepsilon_e}{\overbar{\alpha_\perp}}\approx10^6$. Since the emissivity of films this thin is likely to be $\varepsilon_e \sim 10^{-3}$, it is essential that the laser band absorptivity be $\overbar{\alpha_\perp} \sim 10^{-9}$, or virtually zero. 

Similarly, we can estimate the maximum power that a sail can tolerate, given an emissivity-absorptivity ratio $\frac{\varepsilon_e}{\overbar{\alpha_\perp}}$, a maximum material temperature $T_{max}$, and a perpendicular-to-laser area $A_\perp$.
\begin{linenomath}\begin{equation}
    \Phi_{s,max} = 2 \frac{\varepsilon_e}{\overbar{\alpha_\perp}} A_\perp \sigma T_{max}^4 
\label{E:maxPowerCalc}
\end{equation}\end{linenomath}
In our calculations, we have used a consistent temperature of $T=1000$~\si{\kelvin} to calculate the effective emissivity (Equation~\ref{E:epsilonEff}), although some sails could tolerate higher maximum temperatures before melting or vaporizing. This simplification is reasonable given other uncertainties, such as the temperature dependence of the indices of refraction of the sail materials (see Section~\ref{S:impactOfTemperature}). 

According to Equation~\ref{E:maxPowerCalc}, sails with higher emissivity-absorptivity ratios can tolerate higher laser power values. In addition, the fourth-order dependence on the material temperature suggests that sails that have higher thermal limits will realize substantial accelerative improvements. Finally, sail films that have lower areal densities can be made to be larger for the same total sail mass, reducing the optical intensity (power per area) at any point on the sail and allowing higher overall laser powers to be used. 

\section{Strength analysis}\label{S:strength}

The photon pressure (Equation~\ref{E:Psail}) experienced by the sail will cause stress and strain in the sail film.  For a thin spherically-curved sail with thickness $t_f$ and radius of curvature $s_s$ experiencing a pressure $P$, the membrane stress $\sigma_s$ can be approximated by~\cite{Campbell2022-90, Gordon1978-book, Leff2002-792, Sakamoto2007-514, Timoshenko1959-book}
\begin{linenomath}\begin{equation}
    \sigma_s = \frac{P s_s}{2 t_f} 
    \label{E:sigmaSail} 
\end{equation}\end{linenomath}
\begin{figure*}
\centering
\begin{tikzpicture}[decoration={markings, mark= between positions 0.1 and 0.9 step 5mm with {\arrow{stealth}}}] 
    \pgfmathsetlengthmacro{\ds}{4cm};
    \pgfmathsetlengthmacro{\ss}{\ds}; 
    \pgfmathsetlengthmacro{\thetaMax}{asin(\ds/(2*\ss))}; 
    \pgfmathsetlengthmacro{\sRing}{(\ds/2)/2}; 
    \coordinate (B) at (0,0);
    \node at (B) [circle, fill, inner sep=1.5pt]{}; 
    \path (B) +(0:\ss) coordinate (A); 
    \draw[dashed] (A) arc (0:180:\ss); 
    \draw[dashed] (B) -- +({90-\thetaMax}:\ss) coordinate (C); 
    \draw[dashed] (B) -- +({90+\thetaMax}:\ss) coordinate (Q);
    \draw[dashed] (B) -- +(90:\ss*1) coordinate (D);
    \draw[pattern=north west lines, pattern color=gray] (C) arc ({90-\thetaMax}:{90+\thetaMax}:\ss) -- cycle; 
    \path (B) -- +(90:\ss*1.1) coordinate (R);
    \draw[stealth-stealth] (Q |- R) -- node[above]{$\mathsf{d_s}$} (C |- R);
    \draw[stealth-stealth] (B) -- node[below]{$\mathsf{s_s}$} +(25:\ss);
    \draw[gray!90,-{Triangle[width=30pt,length=15pt]}, line width=10pt](0,-\ds*5/10)  node[black,below]{\small Incident laser light} -- (0, -\ds*1/10); 
\end{tikzpicture}
\caption{\textbf{\textbar~ Schematic diagram of a spherically curved circular sail.} The perspective provided is a side view, and the sail is shaded.  The sail's diameter is equal to its spherical radius of curvature: $d_s = s_s$. }%
\label{F:geometrySail}%
\end{figure*}
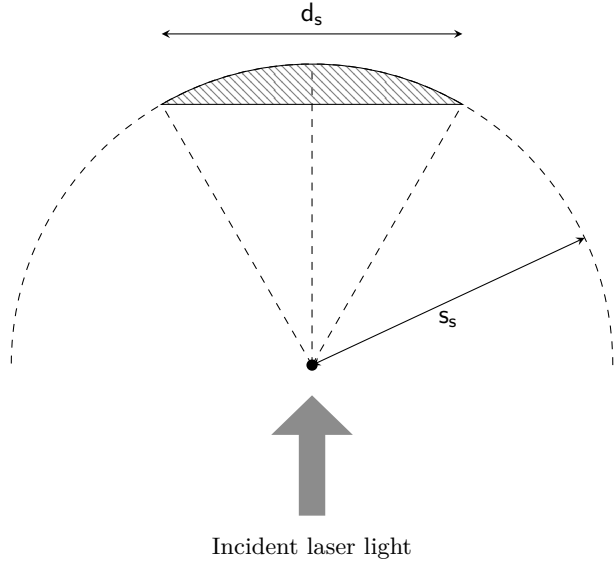
(See Figure~\ref{F:geometrySail} for an explanation of the sail diameter and radius of curvature.) More complicated analyses can be done for the stress in multilayer nanolaminate films~\cite{Foral1979-200, Roy1992-479}, but the simple form above is expedient for analyses in which the uncertainty of many other variables (\eg, the laser power, the sail temperature, the sail's optical properties, \etc) is relatively large. Rearranging Equation~\ref{E:sigmaSail} produces 
\begin{linenomath}\begin{equation}
    \sigma_s t_f = \frac{P s_s}{2},
    \label{E:varRobustness} 
\end{equation}\end{linenomath}
where we define the right side of this expression as the \emph{photon-induced tension}, 
\begin{linenomath}\begin{equation}
    \Upsilon = \frac{P s_s}{2}.
    \label{E:photonInducedTension} 
\end{equation}\end{linenomath}
This quantity, with dimensions force per unit length, can be thought of as the force acting to create a tear of some length in the sail. At the verge of failure (sail tearing), the membrane stress will be equal to the tensile yield stress $\sigma_y$ of the material, \ie, $\sigma_s = \sigma_y$.  Making this substitution in Equation~\ref{E:varRobustness} gives 
\begin{linenomath}\begin{equation}
    \sigma_s t_f = \sigma_y t_f = \Upsilon.
    \label{E:varRobustnessFailure} 
\end{equation}\end{linenomath}
We can extend the left side of Equation~\ref{E:varRobustnessFailure} to be applicable to composite materials using the rule of mixtures~\cite{Liu2009-2198, You2017-682, Raju2018-607}. The rule of mixtures states that the yield stress of a composite film $\sigma_{y,c}$ composed of $n$ layers can be estimated by
\begin{linenomath}\begin{equation}
    \sigma_{y,c} = \sum_{i=1}^{n} f_i \sigma_{y,i}, 
    \label{E:sigmaYieldComp} 
\end{equation}\end{linenomath}
where, if we define the total composite film thickness to be 
\begin{linenomath}\begin{equation}
    t_c = \sum_{i=1}^{n} t_{f,i}
    \label{E:tc} 
\end{equation}\end{linenomath}
(see also Equation~\ref{E:tu}), the fractional composition of the $i$\textsuperscript{th} component can be obtained by 
\begin{linenomath}\begin{equation}
    f_i = \frac{t_{f,i}}{t_c}
    \label{E:fi} 
\end{equation}\end{linenomath}
(see also Equations~\ref{E:FA} and~\ref{E:FM}). Multiplying Equation~\ref{E:sigmaYieldComp} by $t_c$ and simplifying, we obtain
\begin{linenomath}\begin{align}
    t_c \sigma_{y,c} &= t_c \sum_{i=1}^{n} f_i \sigma_{y,i} \nonumber \\
                     &= t_c \sum_{i=1}^{n} \frac{t_{f,i}}{t_c} \sigma_{y,i} \nonumber \\ 
                     &= \sum_{i=1}^{n} t_{f,i} \sigma_{y,i}. \label{E:sigmaYieldComp1} 
\end{align}\end{linenomath}
Equation~\ref{E:sigmaYieldComp1} allows us to calculate the product of the unified composite thickness and its unified tensile yield strength from the thicknesses and yield strengths of its individual layers and materials. We define the right side of this equation as the \emph{membrane mechanical robustness}, 
\begin{linenomath}\begin{equation}
    \wp = \sum_{i=1}^{n} G_i t_{f,i} \sigma_{y,i},
    \label{E:memMechRobust} 
\end{equation}\end{linenomath}
where $G_i$ are geometrical parameters to account for the presence of holes or other architected features in the film layers, discussed below. Using this result in Equation~\ref{E:varRobustnessFailure}, in our simulations, we defined mechanical failure of the film at the point where
\begin{linenomath}\begin{equation}
    \wp = \Upsilon.
    \label{E:memMechRobFailure} 
\end{equation}\end{linenomath}
The geometrical parameters can be calculated using Equation~\ref{E:GiCalc}. 
For films that are planar or have repeating sections that are planar, $G_i$ are taken as unity. For periodic films with holes of diameter $d_{hole}$ and period $a_{pattern}$, $G_i$ are the fractional connected lengths.
\begin{linenomath}\begin{equation}
    G_i =
        \begin{cases}
          1                                            &  \mathrm{planar} \\
          \frac{a_{pattern}-d_{hole}}{a_{pattern}}     &  \mathrm{periodic~holes}
        \end{cases}
    \label{E:GiCalc}
\end{equation}\end{linenomath}
Finally, for reference, we calculated the membrane mechanical robustness for monolayer graphene to be about 44~\si{\newton\per\meter}~\cite{Lee2008-385}, and that for a sheet of standard letter paper to be roughly 3~\si{\kilo\newton\per\meter}~\cite{Yokoyama2007-s68}. 

\section{Calculation of maximum relative velocity}\label{S:calcMaxRelVel}

Here we explain our procedure for calculating the maximum relative velocity achievable by a given film design, $\beta_{max}$. We will provide specific details about the inputs to this method, such as areal density and laser output wavelength, for each sail film that we evaluated~\cite{Ilic2018-5583, Lien2022-3032, Taghavi2022-20034, Salary2020-1900311, Brewer2022-594, Santi2022-16, Chang2024-6689}, in Section~\ref{S:compareFilms}. 

For a given sail film, we began by using its material thickness, architecture, and yield stress information to calculate its areal density (simply denoted $\rho_a$ here) and membrane mechanical robustness $\wp$ (Equation~\ref{E:memMechRobust}). We then used this areal density to calculate the surface area of one side of a spherically curved sail with a mass of $m_s=1$~\si{\gram}:
\begin{linenomath}\begin{equation}
    A_s = \frac{m_s}{\rho_a}
\label{E:sailArea}
\end{equation}\end{linenomath}
Next, we used this area to calculate the diameter $d_s$ of a spherically-curved circular sail that has a spherical radius of curvature equal to its diameter, $s_s = d_s$. See Figure~\ref{F:geometrySail} for an explanation of the sail diameter and radius of curvature.
\begin{linenomath}\begin{equation}
    d_s = s_s = \sqrt{\frac{A_s}{\pi\left(2-\sqrt{3}\right)}}
\label{E:sailDiameterFromAreaForDsEqSs}
\end{equation}\end{linenomath}
Equation~\ref{E:sailDiameterFromAreaForDsEqSs} can be derived from Equation~S44 in the Supporting Information for Campbell~\etal~\cite{Campbell2022-90} by setting $d_s=s_s$. The condition $s_s \approx d_s$ is optimal for curved light sails~\cite{Campbell2022-90}. We next used this sail diameter to estimate the maximum distance $L_{max}$ over which the sail could be accelerated with the laser perfectly focused on the full sail area for an Earth-bound laser array diameter $d_{l,E} = 30$~\si{\kilo\meter} and a constant laser output wavelength $\lambda_l$~\cite{Kulkarni2016-43}: 
\begin{linenomath}\begin{equation}
    L_{max} = \frac{d_{l,E} d_s}{2 \lambda_l} .
\label{E:LmaxFromDs}
\end{equation}\end{linenomath}
We obtained this expression from Equation~\ref{E:dlE} in this document. Notice that, given the mass constraint, the sail diameter is inversely proportional to the areal density, which implies sails with higher areal densities have shorter acceleration ``runways'' $L$. This is one constraining factor on $\beta_{max}$. 

We next iteratively determined the maximum constant laser output power $\Phi_l$ that the sail could sustain without tearing as it accelerated out to a distance $L$, up to a maximum power of $\Phi_l = 100$~\si{\giga\watt}. We defined the tearing failure mode according to Equation~\ref{E:memMechRobFailure}, \ie, at the point where the photon-induced tension $\Upsilon$ (Equation~\ref{E:photonInducedTension}) would be equal to the film's membrane mechanical robustness $\wp$ (Equation~\ref{E:memMechRobust}).
\begin{linenomath}\begin{equation}
    \wp = \Upsilon.
    \tag{\ref{E:memMechRobFailure}}
\end{equation}\end{linenomath}
In this calculation we assumed a total rest sailcraft mass (including sail, payload, and any connecting tethers) of $m_{tot}=2$~\si{\gram} (\ie, twice the sail mass); having approximately equal sail and payload masses is optimal for acceleration~\cite{Lubin2016-40, Ilic2018-5583, Kulkarni2016-43, Campbell2022-90}. 

It is instructive to obtain an expression for the photon-induced tension in terms of the laser power and other sail design parameters. We achieved this by substituting Equation~\ref{E:Psail} into Equation~\ref{E:photonInducedTension} and simplifying knowing that we chose $d_s = s_s$ for this analysis. In the following, we also substituted $\Phi_{l,\beta} = \Phi_l$, which states that the output power produced by the laser is constant. 
\begin{linenomath}\begin{equation}
    \frac{P s_s}{2} = \Upsilon = \frac{4 \varrho_{\beta,\perp} \Phi_l }{ \pi c d_s}\!\left(\frac{1-\beta}{1 + \beta} \right). 
    \label{E:photonInducedTensionV2}
\end{equation}\end{linenomath}
The photon-induced tension is proportional to the film's laser band reflectivity $\varrho_{\beta,\perp}$ and the constant laser output power $\Phi_l$, and inversely proportional to the sail diameter $d_s$. It also decreases as $\beta$ increases. Since $\varrho_{\beta,\perp}$ and $d_s$ are prescribed by the sail design, some films required that the laser power be throttled back to a value lower than 100~\si{\giga\watt}. For other films, a power greater than this could be tolerated, in which case we simply applied the maximum power of 100~\si{\giga\watt}.  

Finally, given the maximum achievable laser power, we stepped through Equation~\ref{E:D} by incrementally increasing the relative velocity $\beta$ until the distance the sail has traveled is equal to $L_{max}$. We defined the corresponding relative velocity as $\beta_{max}$. 

One may observe that Equation~\ref{E:photonInducedTensionV2}, which we use to calculate the maximum power that the sail can sustain without tearing, depends on $\beta$. Thus, the power iteration step required calculating over a range of $\beta$ values, the upper limit of which was initially an unknown. In practice, we set a high upper limit for $\beta$ and later checked to ensure that, if the laser output power was required to be limited at some $\beta=\beta_{limit}$, that limit occurred for $\beta_{limit}\leq\beta_{max}$. 

Lastly, note that Equation~\ref{E:photonInducedTensionV2} is based on the perpendicular reflectivity of the sail ($\varrho_{\beta,\perp}$), whereas Equation~\ref{E:D} uses the spatially-averaged reflectivity that accounts for the impact of the sail curvature ($\varrho_{\beta,a}$; see Section~\ref{SS:avgReflectSailShape}). In in our $\beta_{max}$ calculations, we used the perpendicular reflectivity in place of the average reflectivity. This was because, for many of the sails in Figure~\ref{F:benchmarking} of the main article, angle-dependent reflectivity information, necessary for calculating the spatially-averaged reflectivity, was not available. We expect this approximation to slightly inflate the $\beta_{max}$ values calculated for some designs.

\section{Information on film thickness optimization}\label{S:optimizingThickness}

We determined the \ch{Al2O3} and \ch{MoS2} film thicknesses of our proposed optimized design using the following method. We first selected hexagonal corrugation dimensions that would increase the film's bending stiffness enhancement factor $\mathbb{B}$ (thereby making the film more resistant to wrinkling) and decrease its tensile stiffness reduction factor $\mathbb{T}$ (thereby making the film more stretchy), as described in Section~\ref{S:mechCorrFilms} and illustrated in Figure~\ref{F:compositeChar}(e) in the main article and Figure~\ref{F:tsrfAdImp}(a). We chose a hexagon diameter $d_h=70~\si{\micro\meter}$, a trench width $w_t=2~\si{\micro\meter}$, and a trench height  $h_t=3~\si{\micro\meter}$. Also, for simplicity, we added the constraint that the top and bottom \ch{Al2O3} thicknesses should be equal. For optical property information, to determine the best sail accelerative performance within state-of-the-art fabrication capabilities, we used \ch{Al2O3} index of refraction data from Kischkat~\etal~\cite{Kischkat2012-6789} and \ch{MoS2} index of refraction data from Munkhbat~\etal~\cite{Munkhbat2022-2398}. In doing so, we assumed that future process improvements would allow multilayer \ch{MoS2} films to be fabricated with optical quality equivalent to that of bulk \ch{MoS2} crystals. Furthermore, we set the sail mass to $m_s=1$~\si{\gram} and the total sailcraft mass (sail with payload chip and any tethers) to be $m_{tot}=2$~\si{\gram}. 

Next, we examined multiple film design combinations spanning \ch{Al2O3} film thicknesses $\langle 1 \le t_{A,b}=t_{A,t} \le 100 \rangle$~\si{\nano\meter} and \ch{MoS2} film thicknesses $\langle 1 \le t_M \le 200 \rangle$~\si{\nano\meter}. For each, we calculated the areal density $\rho_{a,c,t}$ (Equations~\ref{E:arealdensitycorrT}, \ref{E:Muct}, and~\ref{E:Auct}, with $\rho_A=3200$~\si{\kilo\gram\per\meter\cubed}~\cite{Ilic2010-044317} and $\rho_M=5060$~\si{\kilo\gram\per\meter\cubed}~\cite{Graczykowski2017-7647}), membrane mechanical robustness $\wp$ (Equation~\ref{E:memMechRobust} with $G_i=1$ and tensile yield stress values $\sigma_{y,\ch{Al2O3}}=2$~\si{\giga\pascal}~\cite{Miller2010-58, Jen2011-084305} and $\sigma_{y,\ch{MoS2}}=2.3$~\si{\giga\pascal}~\cite{Bertolazzi2011-9703}), wavelength-dependent reflectivity within the Doppler-shifted laser band $\varrho_{\beta,\perp}$ (Equation~\ref{E:varrhoPerp}), and maximum relative velocity $\beta_{max}$ for a laser array diameter $d_{l,E}=30$~\si{\kilo\meter} and a constant maximum laser output power of $\Phi_l=100$~\si{\giga\watt}. As discussed in Section~\ref{S:explainMechProps}, the value we used for the \ch{MoS2} tensile yield stress assumes that future fabrication improvements will allow thick and strong crystalline films to be formed; our estimate is reasonable given comparable data for a similar material, \ch{MoSe2}~\cite{Babacic2021-2008614}. 

We considered the optimized design to be that which achieved the highest $\beta_{max}$ value. Our model predicted that this design, with \ch{Al2O3} thickness $t_{A,b}=t_{A,t} \approx 19$~\si{\nano\meter} and \ch{MoS2} thickness $t_{M}\approx 63$~\si{\nano\meter}), will achieve $\beta_{max}=0.26$. 
Our proposed optimized film has a corrugated areal density $\rho_{a,c,t} = 0.51$~\si{\gram\per\meter\squared}, a bending stiffness enhancement factor of $\mathbb{B} = 890$, and a tensile stiffness reduction factor of $\mathbb{T} = 0.0033$. We note that the optimization procedure outlined here does not account for thermal constraints. Depending on the extinction coefficients of \ch{Al2O3} and \ch{MoS2}, thermal considerations might require reducing the laser power to prevent the sail from sublimating, which would consequently decrease $\beta_{max}$. Such an optimization would be more relevant if reliable spectroscopic data at elevated temperatures for \ch{Al2O3} and \ch{MoS2} was available (see Section~\ref{S:impactOfTemperature}). 

\begin{figure*}
\centering
\includegraphics[width=\figWidthFull]{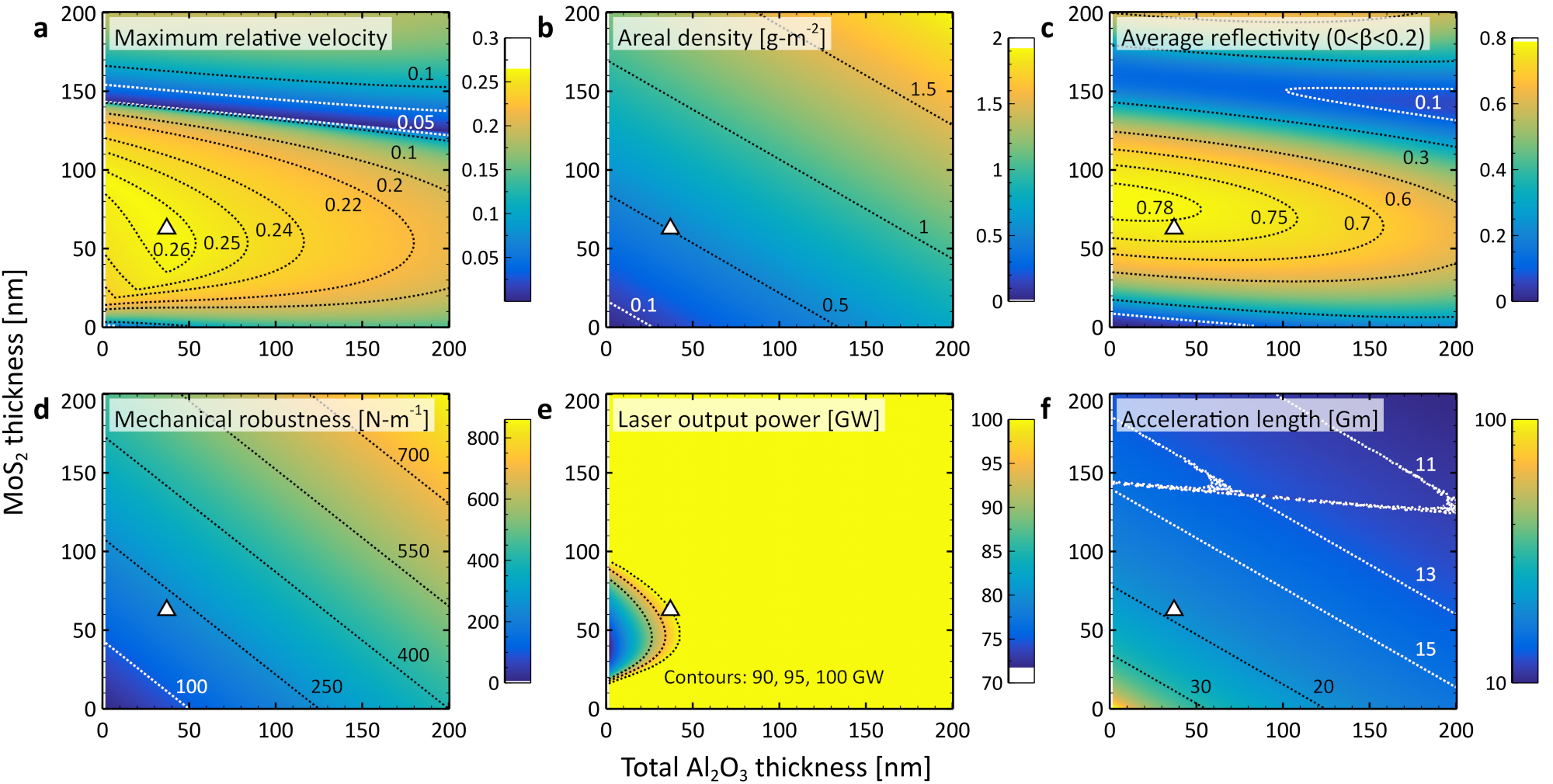}
\caption{\textbf{\textbar~ Contour plots showing thickness optimization process for the proposed optimized sail design.} Film is designed with the indented trench corrugation using parameters $d_h=70~\si{\micro\meter}$, $w_t=2~\si{\micro\meter}$, and $h_t=3~\si{\micro\meter}$.  Note that the abscissa is given in terms of the \emph{total} \ch{Al2O3} thickness, \ie, $t_{A,b}+t_{A,t}$, and that we have constrained the optimization to having equal top and bottom \ch{Al2O3} thickness, \ie, $t_{A,b}=t_{A,t}$. (\textbf{a}) Maximum achievable relative velocity $(\beta_{max})$. (\textbf{b}) Corrugated areal density $(\rho_{a,c,t})$. (\textbf{c}) laser band average ($\beta=0$ to $\beta=0.2$) reflectivity $\overbar{\varrho_\perp}$. Note that, for designs achieving $\beta>0.2$, reflectivity values corresponding to $\beta>0.2$ are not included in this average. (\textbf{d}) Membrane mechanical robustness $\wp$. (\textbf{e}) Maximum sustainable constant laser output power $\Phi_l$. (\textbf{f}) Acceleration length $L$. In all panels, the proposed optimized design, featuring \ch{Al2O3} thickness $t_{A,b}=t_{A,t} \approx 19$~\si{\nano\meter} and \ch{MoS2} thickness $t_{M}\approx 63$~\si{\nano\meter}), is shown with the white triangle. }%
\label{F:contourOptimizeThickness}%
\end{figure*}

Contour plots showing the results of the film thickness optimization are provided in Figure~\ref{F:contourOptimizeThickness}. The optimal design (with the highest $\beta_{max}$ value) does not achieve the highest reflectivity within the design space, in part because it decreases its areal density by reducing its \ch{MoS2} thickness. Notice that, according to Equation~\ref{E:LmaxFromDs}, the acceleration length $L_{max}$ scales with the sail diameter $d_s$, which depends on the sail area $A_s$ (Equation~\ref{E:sailDiameterFromAreaForDsEqSs}), which in turn depends on the areal density $\rho_a$ (Equation~\ref{E:sailArea}).  This explains why the shapes of the contours in Figure~\ref{F:contourOptimizeThickness}(f) mirror those of the areal density in panel~(b). 

\section{Comparison of films in the literature}\label{S:compareFilms}

\subsection{Excluded designs}

Figure~\ref{F:benchmarking} of the main article compares several light sail films introduced in the literature~\cite{Ilic2018-5583, Salary2020-1900311, Brewer2022-594, Lien2022-3032, Santi2022-16, Taghavi2022-20034, Chang2024-6689}.  The data from this figure are presented in Table~\ref{T:dataForFilmComparison}. Other sail designs have been suggested, as well, which we excluded for a variety of reasons. Some were theoretical proposals~\cite{Gieseler2021-21562, Tung2022-1108, Campbell2022-90, Gao2024-4203}; others were non-continuous designs where a membrane mechanical robustness $\wp$ was not definable~\cite{Jin2020-2350, Kudyshev2022-190}; one was a preliminary design outside the feasible atmospheric transmission wavelength range~\cite{Gao2022-1965}; and others only reflected strongly at a single wavelength rather than across the entire Doppler-shifted laser wavelength band~\cite{Siegel2019-2032, Myilswamy2020-8223}. A noteworthy design by Norder~\etal~\cite{Norder2025-2753} presented experimentally-measured reflectivity data, but the wavelength range provided ($\lambda = \langle1.53, 1.62\rangle$~\si{\micro\meter}) was too small to allow for a full $\beta_{max}$ simulation ($\lambda = \langle1.55, 1.8984\rangle$~\si{\micro\meter}). Finally, a recent design by Whittam~\etal~\cite{Whittam2025-345} employed a polydimethylsiloxane (PDMS) connecting structure that had a low thermal limit (around 600~\si{\kelvin}~\cite{Camino2001-2395}) and a low tensile yield strength (around 5~\si{\mega\pascal}~\cite{Ariati2021-4258}), which limited it to low laser powers.  

\end{multicols}

\begin{sidewaystable}
  \caption{\textbf{\textbar~ Tabulated data of Figure~\ref{F:benchmarking} in the main article.} Data included are information on the laser band average ($\beta=0$ to $\beta=0.2$) reflectivity $\overbar{\varrho_\perp}$, laser band average absorptivity $\overbar{\alpha_\perp}$, infrared effective emissivity $\varepsilon_e$ (at $T=1000$~\si{\kelvin}, $\lambda=2-14$~\si{\micro\meter}), emissivity-to-absorptivity ratio, thermally-limited power $\Phi_{s,max}$, maximum relative velocity $\beta_{max}$, maximum power used in the $\beta_{max}$ simulation, membrane mechanical robustness $\wp$, areal density $\rho_a$, laser output wavelength $\lambda_l$, and perpendicular-to-laser sail area $A_\perp$. Experimentally-measured values are noted with an asterisk symbol $(\ast)$. Note that the units of the thermally limited power $\Phi_{s,max}$ are [\si{\mega\watt}] whereas those of the power used for the $\beta_{max}$ simulations are [\si{\giga\watt}].}
  \tiny
  \begin{tabular}{ l l l l l l l l l l l l l }
    \hline
    Citation & Description & $\overbar{\varrho_\perp}$ & $\overbar{\alpha_\perp}$ & $\varepsilon_e$ & $\frac{\varepsilon_e}{\overbar{\alpha_\perp}}$ & $\Phi_{s,max}$ (thermal limit) & $\beta_{max}$ & $\Phi_l$ for $\beta_{max}$ & $\wp$ & $\rho_a$ & $\lambda_l$ & $A_\perp$  \\
    ~ & ~ & ~ & ~ & ~ & ~ & [\si{\mega\watt}] & ~ & [\si{\giga\watt}] & [\si{\newton\per\meter}] & [\si{\gram\per\meter\squared}] & [\si{\micro\meter}] & [\si{\meter\squared}] \\
    \hline
    Ilic, Went, and Atwater (2018)~\cite{Ilic2018-5583} & \ch{SiO2} layers with gaps                   & 0.79 & 0.0053   & 0.047   & 8.9   & 2.8 & 0.22 & 100 & 1077 & 1.46 & 1.2   & 0.64 \\
    Salary and Mosallaei (2020)~\cite{Salary2020-1900311} & \ch{Si} disks on \ch{Si}-\ch{SiO2} bilayer & 0.76 & 0.00038  & 0.0030  & 8.0   & 6.9 & 0.25 & 100 & 268  & 0.54 & 1.3   & 1.74 \\
    Brewer~\etal~(2022)~\cite{Brewer2022-594} & \ch{Si3N4}-\ch{MoS2}-\ch{Si3N4}; holes                 & 0.81 & 0.000041 & 0.00063  & 15    & 12  & 0.18 & 18  & 21   & 0.13 & 1.2   & 7.2 \\
    Lien~\etal~(2022)~\cite{Lien2022-3032} & \ch{Si3N4}; holes                                         & 0.44$\ast$  & 0.000077 & 0.063   & 823   & 348 & 0.17 & 100 & 5892 & 1.94 & 1.064 & 0.48 \\
    Santi~\etal~(2022)~\cite{Santi2022-16} & \ch{TiO2}-\ch{SiO2}-\ch{TiO2}                             & 0.78 & 0.0020   & 0.032   & 16    & 0.26 & 0.23 & 100 & 571 & 1.31 & 1.064 & 0.71 \\
    Taghavi and Mosallaei (2022)~\cite{Taghavi2022-20034} & \ch{Si} disks on \ch{Si}-\ch{SiO2} bilayer & 0.76 & 0.00038  & 0.0047  & 12    & 4.9 & 0.19   & 79 & 296 & 1.18 & 1.3 & 0.79 \\
    Chang~\etal~(2024)~\cite{Chang2024-6689} & \ch{Si3N4} with holes on \ch{Si}                        & 0.66$\ast$  & 0.076$\ast$  & 0.017    & 0.23    & 0.074 & 0.19 & 100 & 2353 & 1.54 & 1.3 & 0.60 \\
    Fabricated prototype film (this study) & Corrugated \ch{Al2O3}-\ch{MoS2}-\ch{Al2O3}                & 0.51$\ast$  & 0.029$\ast$  & 0.023    & 0.78    & 0.12 & 0.21 & 100 & 184  & 0.68 & 1.2 & 1.37 \\
    Proposed optimized film (this study) & Corrugated \ch{Al2O3}-\ch{MoS2}-\ch{Al2O3}                  & 0.77 & 0.00014  & 0.0038   & 27   & 5.6  & 0.26 & 100 & 218  & 0.51 & 1.2 & 1.84 \\
    \hline
  \end{tabular}
  \label{T:dataForFilmComparison}
\end{sidewaystable}

\begin{multicols}{2}

\subsection[Ilic, Went, and Atwater (2018)]{Ilic, Went, and Atwater (2018)~\cite{Ilic2018-5583}}\label{S:calcIlic}

The paper by Ilic, Went, and Atwater~\cite{Ilic2018-5583} contained several silica-based designs and used a laser output wavelength of $\lambda_l=1.2$~\si{\micro\meter}. For comparison, we selected what the paper refers to as design ``A11,'' which consisted of \ch{SiO2} layers separated by air/vacuum gaps, due to its high reflectivity and high membrane mechanical robustness (see Table~S1 in the Supplementary Information for that paper). We used the design's layer thicknesses and index of refraction information for \ch{SiO2} from Rodriguez-de Marcos~\etal~\cite{RodriguezdeMarcos2016-3622} to calculate the wavelength-dependent reflectivity for this design using the transfer-matrix method (Section~\ref{S:transMatMeth}), which we then averaged to determine the laser band ($\beta=0$ to $\beta=0.2$) average reflectivity $\overbar{\varrho_\perp}$  (Equation~\ref{E:barVarrho}). We used the same data in Equations~\ref{E:absorptivityBetaEstimate} and~\ref{E:barAlpha} to estimate the laser-band-average absorptivity, with fill factors for the layers $F_i=1$ since the layers in this design were planar.  

We used spectroscopic information from Kischkat~\etal~\cite{Kischkat2012-6789} and the transfer-matrix method to determine the sail's spectral directional emissivity in the wavelength range $\lambda = \langle 2 , 14 \rangle$~\si{\micro\meter}, and subsequently integrated this data at a temperature of $T=1000$~\si{\kelvin} to find the film's effective emissivity $\varepsilon_e$ (Equation~\ref{E:epsilonEff}). The wavelength range $\lambda = \langle 2 , 14 \rangle$~\si{\micro\meter} includes a majority of the emissive power at $T=1000$~\si{\kelvin} and conveniently allowed us to make use of the Kischkat~\etal~\cite{Kischkat2012-6789} datasets for  \ch{Al2O3}, \ch{Si3N4}, \ch{SiO2}, and \ch{TiO2} without extrapolation. 

We calculated the membrane mechanical robustness $\wp$ using Equation~\ref{E:memMechRobust}, with geometrical parameters $G_i=1$ and yield stress $\sigma_{y,\ch{SiO2}}=1.5$~\si{\giga\pascal}~\cite{Tsuchiya2000-286, Yoshioka2000-291}. We calculated the areal density $\rho_a$ using 
\begin{linenomath}\begin{equation}
	\rho_a = \sum_{i=1}^n \rho_i t_{f,i}
	\label{E:arealdensitygeneral}
\end{equation}\end{linenomath}
where $\rho_i$ and $t_{f,i}$ are the density and thickness of each layer, respectively; in this case $\rho_{\ch{SiO2}}=2030$~\si{\kilo\gram\per\meter\cubed}~\cite{Kawase2009-101401} and $\rho_{gap}=0$~\si{\kilo\gram\per\meter\cubed} (see Table~\ref{T:materialProps}). We obtained the maximum relative velocity achievable for this design, $\beta_{max}$, using the method outlined in Section~\ref{S:calcMaxRelVel} with the reflectivity profile that we calculated above. Notably, the relatively high membrane mechanical robustness value of this design allowed it to sustain the maximum laser power of $\Phi_l=100$~\si{\giga\watt} in this simulation. Lastly, we estimated the thermally-limited power $\Phi_{s,max}$ using Equation~\ref{E:maxPowerCalc}, with $A_\perp$ obtained \via Equation~\ref{E:Aperp} (using the sail diameter $d_s$ derived from the sail's areal density and a mass of $m_s=1$~\si{\gram}, Equations~\ref{E:sailArea} and~\ref{E:sailDiameterFromAreaForDsEqSs}) and $T_{max}=1450$~\si{\kelvin} (limited by \ch{SiO2}). 

\subsection[Salary and Mosallaei (2020)]{Salary and Mosallaei (2020)~\cite{Salary2020-1900311}}

The paper by Salary and Mosallaei~\cite{Salary2020-1900311} proposed designs that featured protruding \ch{Si} disks of differing diameters on top of a continuous \ch{Si}-\ch{SiO2} backbone film, and used a laser output wavelength of $\lambda_l=1.3$~\si{\micro\meter}. For simplicity, we used the reflectivity information contained in Figure~2(b) of their paper~\cite{Salary2020-1900311} with a constant disk diameter of 300~\si{\nano\meter}, which we averaged to estimate the mean laser band reflectivity $\overbar{\varrho_\perp}$ (Equation~\ref{E:barVarrho}).

We estimated the absorptivity using Equations~\ref{E:absorptivityBetaEstimate} and~\ref{E:barAlpha} with optical data from Poruba~\etal~\cite{Poruba2000-148} and Rodriguez-de Marcos~\etal~\cite{RodriguezdeMarcos2016-3622} for \ch{Si} and \ch{SiO2}, respectively. We accounted for the protruding \ch{Si} disks using fill factor 
\begin{linenomath}\begin{equation}
	F_i = \frac{\pi\left(\frac{d_{disk}}{2}\right)^2}{a^2},
	\label{E:FillFactorSalary}
\end{equation}\end{linenomath}
where $d_{disk}$ is the protruding disk diameter and $a$ is the period of the pattern, and used a fill factor $F_i=1$ for the connecting (continuous) 89-\si{\nano\meter}-thick \ch{Si} and 60-\si{\nano\meter}-thick \ch{SiO2} layers below. 

We calculated the sail's emissivity as in Section~\ref{S:calcIlic}, using optical data from Franta~\etal~\cite{Franta2017-405} and Kischkat~\etal~\cite{Kischkat2012-6789} for \ch{Si} and \ch{SiO2}, respectively, along with an effective medium approximation~\cite{El-Haija2003-2590}. We calculated the membrane mechanical robustness using Equation~\ref{E:memMechRobust} with $G_i=1$, accounting only for the strength of the backbone \ch{Si}-\ch{SiO2} film (not the \ch{Si} disks), with yield stress values $\sigma_{y,\ch{Si}}=2$~\si{\giga\pascal}~\cite{Sato1998-148, Sharpe1999-162, Tsuchiya2005-665, Tsuchiya2010-1} and $\sigma_{y,\ch{SiO2}}=1.5$~\si{\giga\pascal}~\cite{Tsuchiya2000-286, Yoshioka2000-291}. We estimated the sail's areal density by averaging the two extreme areal density values given in the paper~\cite{Salary2020-1900311} for the largest and smallest protruding \ch{Si} disks. We calculated the sail's maximum relative velocity using the method outlined in Section~\ref{S:calcMaxRelVel} with the reflectivity profile obtained from Figure~2(b) of their paper~\cite{Salary2020-1900311} as discussed above. Lastly, we estimated the thermally-limited power $\Phi_{s,max}$ as in Section~\ref{S:calcIlic}, using $T_{max}=1450$~\si{\kelvin} (limited by \ch{SiO2}). 

\subsection[Brewer~\etal~(2022)]{Brewer~\etal~(2022)~\cite{Brewer2022-594}}

Brewer~\etal~\cite{Brewer2022-594} proposed a three-layer \ch{Si3N4}-\ch{MoS2}-\ch{Si3N4} composite film (thicknesses $t_{\ch{Si3N4},t}=5$~\si{\nano\meter}, $t_{\ch{MoS2}}=90$~\si{\nano\meter}, and $t_{\ch{Si3N4},b}=5$~\si{\nano\meter}) with patterned holes (diameter $d_{hole}=1044$~\si{\nano\meter}, $x$-period $a_x=2009$~\si{\nano\meter}, and $y$-period $a_y=1160$~\si{\nano\meter}).  We averaged the reflectivity information provided in Figure~2(c) of their paper~\cite{Brewer2022-594} to find the mean laser band reflectivity $\overbar{\varrho_\perp}$ (Equation~\ref{E:barVarrho}), and used Equations~\ref{E:absorptivityBetaEstimate} and~\ref{E:barAlpha} with spectroscopic data from Kischkat~\etal~\cite{Kischkat2012-6789} and Munkhbat~\etal~\cite{Munkhbat2022-2398} for \ch{Si3N4} and \ch{MoS2}, respectively, to estimate the average absorptivity. We accounted for the holes in the sail film in Equation~\ref{E:absorptivityBetaEstimate} using fill factor 
\begin{linenomath}\begin{equation}
	F_i = \frac{a_x a_y - 2\pi\left(\frac{d_{hole}}{2}\right)^2}{a_x a_y}, 
	\label{E:FillFactorBrewer}
\end{equation}\end{linenomath}
where the factor of two in front of $\pi$ is included because each unit cell in their design contained the equivalent of two holes. We obtained the effective emissivity as in Section~\ref{S:calcIlic} using spectroscopic data from Kischkat~\etal~\cite{Kischkat2012-6789} and Munkhbat~\etal~\cite{Munkhbat2022-2398} for \ch{Si3N4} and \ch{MoS2}, respectively.  

We derived the sail's membrane mechanical robustness using Equation~\ref{E:memMechRobust} with
\begin{linenomath}\begin{equation}
	G_i=\frac{a_y-d_{hole}}{a_y}
	\label{E:GiBrewer}
\end{equation}\end{linenomath}
(see Equation~\ref{E:GiCalc}) and yield stress values $\sigma_{y,\ch{Si3N4}}=14$~\si{\giga\pascal}~\cite{Yoshioka2000-291} and $\sigma_{y,\ch{MoS2}}=0.75$~\si{\giga\pascal}~\cite{Graczykowski2017-7647, Sledzinska2020-1169}. We calculated the areal density of the film using the ratio of the unit cell mass to the unit cell planar area, \ie,
\begin{linenomath}\begin{equation}
	\rho_a = \frac{M_{uc}}{A_{uc}},
	\label{E:arealdensityholes}
\end{equation}\end{linenomath}
wherein
\begin{linenomath}\begin{align}
	M_{uc} &= \left(\rho_{\ch{Si3N4}}\left(t_{\ch{Si3N4},t}+t_{\ch{Si3N4},b}\right)+\rho_{\ch{MoS2}}t_{\ch{MoS2}}\right) \nonumber \\
       & \;\;\;\;\;\; \times \left(a_x a_y - 2\pi\left(\frac{d_{hole}}{2}\right)^2\right)
	\label{E:arealdensityholesMuc}
\end{align}\end{linenomath}
and
\begin{linenomath}\begin{equation}
	A_{uc} = a_y a_x .
	\label{E:arealdensityholesAuc}
\end{equation}\end{linenomath}
Here, $\rho_{\ch{Si3N4}}=3200$~\si{\kilo\gram\per\meter\cubed}~\cite{Yen2003-1895} and  $\rho_{\ch{MoS2}}=5060$~\si{\kilo\gram\per\meter\cubed}~\cite{Graczykowski2017-7647}. As above, the factor of two in front of $\pi$ is included because each unit cell in their design contained the equivalent of two holes. We computed the sail's maximum relative velocity using the method outlined in Section~\ref{S:calcMaxRelVel} with the reflectivity profile shown in Figure~2(c) of their paper~\cite{Brewer2022-594}. In this simulation, we reduced the laser power because the mechanical robustness of the sail was low. Finally, we  estimated the thermally-limited power $\Phi_{s,max}$ as in Section~\ref{S:calcIlic}, using $T_{max}=1000$~\si{\kelvin} (limited by \ch{MoS2}). 

\subsection[Lien~\etal~(2022)]{Lien~\etal~(2022)~\cite{Lien2022-3032}}

Lien~\etal~\cite{Lien2022-3032} fabricated a silicon nitride photonic crystal film by patterning a $t_{\ch{Si3N4}}=690$~\si{\nano\meter} thick \ch{Si3N4} layer with $d_{hole}=415$~\si{\nano\meter} holes in a square grid with spacing $a=1064$~\si{\nano\meter}. They optimized their design for a laser wavelength of $\lambda_l=1.064$~\si{\micro\meter} and measured the transmissivity of their prototype; assuming negligible absorption, they estimated the reflectivity by subtracting the transmissivity from unity (see Figure~2(c) of their paper~\cite{Lien2022-3032}). We used this definition of reflectivity to estimate the average laser band reflectivity $\overbar{\varrho_\perp}$ for their design (Equation~\ref{E:barVarrho}). 

We used optical data from Kischkat~\etal~\cite{Kischkat2012-6789} in Equations~\ref{E:absorptivityBetaEstimate} and~\ref{E:barAlpha} to calculate the average laser band absorptivity, and used the method of Section~\ref{S:calcIlic} to obtain the effective emissivity. We calculated the appropriate fill factor for the absorptivity calculation using 
\begin{linenomath}\begin{equation}
	F_i = \frac{a_x a_y - \pi\left(\frac{d_{hole}}{2}\right)^2}{a_x a_y}, 
	\label{E:FillFactorLien}
\end{equation}\end{linenomath}
where unlike Equation~\ref{E:arealdensityholesMuc}, the pattern in the design contained only one hole per unit cell.

We derived the sail's membrane mechanical robustness using 
\begin{linenomath}\begin{equation}
	G_i=\frac{a-d_{hole}}{a}
	\label{E:GiLien}
\end{equation}\end{linenomath}
(see Equation~\ref{E:GiCalc}; for this single-layer film $i=1$ only) and yield stress $\sigma_{y,\ch{Si3N4}}=14$~\si{\giga\pascal}~\cite{Yoshioka2000-291} in Equation~\ref{E:memMechRobust}. We calculated the areal density using Equation~\ref{E:arealdensityholes}, with 
\begin{linenomath}\begin{equation}
	M_{uc} = \rho_{\ch{Si3N4}} t_{\ch{Si3N4}} \! \left(a^2- \pi \left(\frac{d_{hole}}{2}\right)^2\right)
	\label{E:arealdensityholesMuc2}
\end{equation}\end{linenomath}
and
\begin{linenomath}\begin{equation}
	A_{uc} = a^2 .
	\label{E:arealdensityholesAuc2}
\end{equation}\end{linenomath}
Here $\rho_{\ch{Si3N4}}=3200$~\si{\kilo\gram\per\meter\cubed}~\cite{Yen2003-1895} and as mentioned above the pattern in the design contained only one hole per unit cell. We computed the sail's maximum relative velocity using the method outlined in Section~\ref{S:calcMaxRelVel} with the reflectivity profile reported in Figure~2(c) of their paper~\cite{Lien2022-3032}. Finally, we  estimated the thermally-limited power $\Phi_{s,max}$ as in Section~\ref{S:calcIlic}, using $T_{max}=1670$~\si{\kelvin} (limited by \ch{Si3N4}). 

\subsection[Santi~\etal~(2022)]{Santi~\etal~(2022)~\cite{Santi2022-16}}

Santi~\etal~\cite{Santi2022-16} numerically investigated a family of light sails composed of one, two, three, or four layers of various materials. Their optimization resulted in a three-layer \ch{TiO2}-\ch{SiO2}-\ch{TiO2} composite film with thicknesses $t_{\ch{TiO2},t}=121$~\si{\nano\meter}, $t_{\ch{SiO2}}=203$~\si{\nano\meter}, and $t_{\ch{TiO2},b}=121$~\si{\nano\meter}. We used these layer thicknesses and index of refraction information for \ch{TiO2} and \ch{SiO2} from Kischkat~\etal~\cite{Kischkat2012-6789} and Rodriguez-de Marcos~\etal~\cite{RodriguezdeMarcos2016-3622}, respectively, to calculate the wavelength-dependent reflectivity for this design using the transfer-matrix method (Section~\ref{S:transMatMeth}), which we then averaged to determine the laser band average reflectivity $\overbar{\varrho_\perp}$  (Equation~\ref{E:barVarrho}). We also used this optical data in Equations Equations~\ref{E:absorptivityBetaEstimate} and~\ref{E:barAlpha} with $F_i=1$ to obtain the laser-band average absorptivity $\overbar{\alpha_\perp}$. 
We used optical data from Kischkat~\etal~\cite{Kischkat2012-6789} for both \ch{TiO2} and \ch{SiO2} to estimate the effective emissivity of the film, as in Section~\ref{S:calcIlic}. 

We obtained the film's membrane mechanical robustness using Equation~\ref{E:memMechRobust} with yield stresses $\sigma_{y,\ch{TiO2}}=1.1$~\si{\giga\pascal}~\cite{Borgese2012-2459, Tavares2008-1434} and $\sigma_{y,\ch{SiO2}}=1.5$~\si{\giga\pascal}~\cite{Tsuchiya2000-286, Yoshioka2000-291}. We derived the areal density \via Equation~\ref{E:arealdensitygeneral}, with $\rho_{\ch{TiO2}}=3700$~\si{\kilo\gram\per\meter\cubed}~\cite{Saari2022-15357} $\rho_{\ch{SiO2}}=2030$~\si{\kilo\gram\per\meter\cubed}~\cite{Kawase2009-101401}. Note that the areal density we calculated is slightly lower than that listed in their paper~\cite{Santi2022-16} because we used different material density values. We computed the sail's maximum relative velocity using the method outlined in Section~\ref{S:calcMaxRelVel} with the reflectivity profile that we calculated above. Finally, we  estimated the thermally-limited power $\Phi_{s,max}$ as in Section~\ref{S:calcIlic}, using $T_{max}=670$~\si{\kelvin} (limited by \ch{TiO2}). 

\subsection[Taghavi and Mosallaei (2022)]{Taghavi and Mosallaei (2022)~\cite{Taghavi2022-20034}}

The paper by Taghavi and Mosallaei~\cite{Taghavi2022-20034} proposed a similar design and methodology to that of Salary and Mosallaei~\cite{Salary2020-1900311}. We averaged the reflectivity information provided in Figure~2(b) of their paper~\cite{Taghavi2022-20034} at a constant disk diameter of 300~\si{\nano\meter} to obtain the laser band average reflectivity $\overbar{\varrho_\perp}$ (Equation~\ref{E:barVarrho}). We estimated the absorptivity using Equations~\ref{E:absorptivityBetaEstimate} and~\ref{E:barAlpha} with optical data from Poruba~\etal~\cite{Poruba2000-148} and Rodriguez-de Marcos~\etal~\cite{RodriguezdeMarcos2016-3622} for \ch{Si} and \ch{SiO2}, respectively, with a disk fill factor calculated using Equation~\ref{E:FillFactorSalary} and $F_i=1$ for the continuous 103-\si{\nano\meter}-thick \ch{Si} and 60-\si{\nano\meter}-thick \ch{SiO2} layers underneath. We obtained the sail's emissivity as in Section~\ref{S:calcIlic}, using optical information  from Franta~\etal~\cite{Franta2017-405} and Kischkat~\etal~\cite{Kischkat2012-6789} for \ch{Si} and \ch{SiO2}, respectively. 

We calculated the membrane mechanical robustness using Equation~\ref{E:memMechRobust} with $G_i=1$, accounting only for the strength of the backbone \ch{Si}-\ch{SiO2} film (not the \ch{Si} disks), with yield stress values $\sigma_{y,\ch{Si}}=2$~\si{\giga\pascal}~\cite{Sato1998-148, Sharpe1999-162, Tsuchiya2005-665, Tsuchiya2010-1} and $\sigma_{y,\ch{SiO2}}=1.5$~\si{\giga\pascal}~\cite{Tsuchiya2000-286, Yoshioka2000-291}. We obtained the sail's areal density using the ratio of the total sail mass ($m_s=4.7$~\si{\gram}) to the sail area ($A_s=4$~\si{\meter\squared}; see Equation~\ref{E:sailArea}). We calculated the sail's maximum relative velocity according to the method outlined in Section~\ref{S:calcMaxRelVel} using the reflectivity profile that we obtained from Figure~2(b) of their paper~\cite{Taghavi2022-20034}, as discussed above. Lastly, we  estimated the thermally-limited power $\Phi_{s,max}$ as in Section~\ref{S:calcIlic}, using $T_{max}=1450$~\si{\kelvin} (limited by \ch{SiO2}). 

\subsection[Chang~\etal~(2024)]{Chang~\etal~(2024)~\cite{Chang2024-6689}}

Chang~\etal~\cite{Chang2024-6689} fabricated a bilayer \ch{Si3N4}-\ch{Si} sail with thicknesses $t_{\ch{Si3N4}}=400$~\si{\nano\meter} and $t_{\ch{Si}}=321$~\si{\nano\meter}. They patterned the \ch{Si3N4} layer with a square grid of holes ($d_{hole}=415$~\si{\nano\meter} and $a=1064$~\si{\nano\meter}) but left the \ch{Si} intact. They measured both the reflectivity $\varrho_\lambda$ and the transmissivity $\tau_\lambda$ of their prototype near their proposed laser wavelength of $\lambda_l = 1.3$~\si{\micro\meter}, and corrected their reflectivity for system losses using a gold mirror standard. We estimated the absorptivity spectrum of their prototype by subtracting their corrected reflectivity and transmissivity spectra from unity, and subsequently calculated the laser band reflectivity $\overbar{\varrho_\perp}$ and absorptivity $\overbar{\alpha_\perp}$ (Equations~\ref{E:barVarrho} and~\ref{E:barAlpha}, respectively). We derived the effective emissivity using the process outlined in Section~\ref{S:calcIlic}, with optical data from Franta~\etal~\cite{Franta2017-405} and Kischkat~\etal~\cite{Kischkat2012-6789} for \ch{Si} and \ch{Si3N4}, respectively. We calculated the membrane mechanical robustness using Equation~\ref{E:memMechRobust} with $G_i$ given by Equation~\ref{E:GiLien} for the $(i=1)$ \ch{Si3N4} layer and $G_i=1$ for the $(i=2)$ \ch{Si} layer, with $\sigma_{y,\ch{Si3N4}}=14$~\si{\giga\pascal}~\cite{Yoshioka2000-291} and $\sigma_{y,\ch{Si}}=2$~\si{\giga\pascal}~\cite{Sato1998-148, Sharpe1999-162, Tsuchiya2005-665, Tsuchiya2010-1}. We derived the areal density using Equation~\ref{E:arealdensityholes} with 
\begin{linenomath}\begin{equation}
	M_{uc} = \left(\rho_{\ch{Si3N4}} t_{\ch{Si3N4}}\right) \! \left(a^2 - \pi\left(\frac{d_{hole}}{2}\right)^2\right) + \rho_{\ch{Si}} t_{\ch{Si}} a^2
	\label{E:arealdensityholesMuc3}
\end{equation}\end{linenomath}
and $A_{uc}$ from Equation~\ref{E:arealdensityholesAuc2}, with $\rho_{\ch{Si3N4}}=3200$~\si{\kilo\gram\per\meter\cubed}~\cite{Yen2003-1895} and $\rho_{\ch{Si}}=2330$~\si{\kilo\gram\per\meter\cubed}~\cite{Petersen1982-420}. We calculated the sail's maximum relative velocity using the method outlined in Section~\ref{S:calcMaxRelVel} using the study's measured reflectivity data. Lastly, we  estimated the thermally-limited power $\Phi_{s,max}$ as in Section~\ref{S:calcIlic}, using $T_{max}=1470$~\si{\kelvin} (limited by \ch{Si}).

\subsection{Fabricated prototype film (this study)}\label{SS:protoFilm}

The fabricated prototype film characterized in Figure~\ref{F:compositeChar}(c) of the main article consists of a \ch{Al2O3}-\ch{MoS2}-\ch{Al2O3} composite ($t_{A,b}\approx 21~\si{\nano\meter}$ (bottom \ch{Al2O3} thickness), $t_{M}\approx 53~\si{\nano\meter}$ (\ch{MoS2}), $t_{A,t}\approx 51~\si{\nano\meter}$ (top \ch{Al2O3})) with a hexagonally corrugated indented trench structure ($d_h\approx77~\si{\micro\meter}$, $w_t\approx15~\si{\micro\meter}$, $h_t\approx10~\si{\micro\meter}$; see Figure~\ref{F:hexDimsTrenches}). We averaged our measured reflectivity and absorptivity data $(\alpha_{\lambda,\perp}=1-\varrho_{\lambda,\perp}-\tau_{\lambda,\perp})$ to obtain the laser band average values ($\beta=0$ to $\beta=0.2$, or equivalently for our laser wavelength, $\lambda = \langle 1.2 , 1.4697 \rangle$~\si{\micro\meter}) for reflectivity $\overbar{\varrho_\perp}$ and absorptivity $\overbar{\alpha_\perp}$ (Equations~\ref{E:barVarrho} and~\ref{E:barAlpha}, respectively). We used spectroscopic information from Kischkat~\etal~\cite{Kischkat2012-6789} for \ch{Al2O3} and a two-term Cauchy expansion of our experimentally-measured \ch{MoS2} index of refraction (see Figure~\ref{F:nkMoS2extrap}) to estimate the film's effective emissivity, as in Section~\ref{S:calcIlic}. We calculated the membrane mechanical robustness of our design using Equation~\ref{E:memMechRobust} with $G_i=1$ and yield stress values $\sigma_{y,\ch{Al2O3}}=2$~\si{\giga\pascal}~\cite{Miller2010-58, Jen2011-084305} and $\sigma_{y,\ch{MoS2}}=0.75$~\si{\giga\pascal}~\cite{Graczykowski2017-7647, Sledzinska2020-1169}. Our choice of $G=1$ is conservative and allows us to examine the robustness of the film in a planar (non-corrugated) state.  However, the hexagonal structure can actually be designed to increase compliance to in-plane stresses, thereby increasing the practical membrane mechanical robustness~\cite{Jiao2020-100599}. We derived the areal density using Equations~\ref{E:arealdensitycorrT}, \ref{E:Muct}, and~\ref{E:Auct}, with $\rho_A=3200$~\si{\kilo\gram\per\meter\cubed}~\cite{Ilic2010-044317} and $\rho_M=5060$~\si{\kilo\gram\per\meter\cubed}~\cite{Graczykowski2017-7647}. We calculated the film's maximum relative velocity using the method outlined in Section~\ref{S:calcMaxRelVel} using our measured reflectivity data. Finally, we  estimated the thermally-limited power $\Phi_{s,max}$ as in Section~\ref{S:calcIlic}, using $T_{max}=1000$~\si{\kelvin} (limited by \ch{MoS2}). 

\subsection{Optimized film (this study)}\label{SS:optimizedFilmCalc}

The optimized film that we propose consists of a \ch{Al2O3}-\ch{MoS2}-\ch{Al2O3} composite ($t_{A,b}\approx 19~\si{\nano\meter}$ (bottom \ch{Al2O3} thickness), $t_{M}\approx 63~\si{\nano\meter}$ (\ch{MoS2}), $t_{A,t}\approx 19~\si{\nano\meter}$ (top \ch{Al2O3})) with a hexagonally corrugated indented trench structure ($d_h=70~\si{\micro\meter}$, $w_t=2~\si{\micro\meter}$, $h_t=3~\si{\micro\meter}$; see Figure~\ref{F:hexDimsTrenches}). We used index of refraction information for \ch{Al2O3} from Kischkat~\etal~\cite{Kischkat2012-6789} and for \ch{MoS2} from Munkhbat~\etal~\cite{Munkhbat2022-2398} to determine the laser band reflectivity spectrum according to the transfer-matrix method, and averaged it to determine $\overbar{\varrho_\perp}$ (Equation~\ref{E:barVarrho}). We used the same optical constants to derive the absorptivity using Equations~\ref{E:absorptivityBetaEstimate} and~\ref{E:barAlpha}, and also to obtain the effective emissivity using the method of Section~\ref{S:calcIlic}. 

We calculated the membrane mechanical robustness using $G_i=1$ and tensile yield stress values $\sigma_{y,\ch{Al2O3}}=2$~\si{\giga\pascal}~\cite{Miller2010-58, Jen2011-084305} and $\sigma_{y,\ch{MoS2}}=2.3$~\si{\giga\pascal}~\cite{Bertolazzi2011-9703} (see Equation~\ref{E:memMechRobust} and Sections~\ref{S:explainMechProps} and~\ref{S:optimizingThickness}), and the areal density using $\rho_A=3200$~\si{\kilo\gram\per\meter\cubed}~\cite{Ilic2010-044317} and $\rho_M=5060$~\si{\kilo\gram\per\meter\cubed}~\cite{Graczykowski2017-7647}. We calculated the maximum relative velocity as outlined in Section~\ref{S:calcMaxRelVel} using the reflectivity data that we calculated above. Finally, we  estimated the thermally-limited power $\Phi_{s,max}$ as in Section~\ref{S:calcIlic}, using $T_{max}=1000$~\si{\kelvin} (limited by \ch{MoS2}). 

\section{Energy analysis of accelerating sail}\label{S:energyAnalysis}

Here we describe our calculation of the equilibrium temperature during the acceleration of our optimized design. We performed this simulation with similar constraints to the others in this work: a sail mass of $m_s=1$~\si{\gram}, a total sailcraft mass (sail, chip, and tethers) of $m_{tot}=2$~\si{\gram}, a circular sail with a spherical radius of curvature equal to its diameter $(d_s=s_s)$, and a maximum constant laser output power of $\Phi_l=100$~\si{\giga\watt}. Our approach was to divide the acceleration interval ($\beta=0$ to $\beta=\beta_f=0.2$) into discrete $\beta$ steps and solve Equation~\ref{E:implicitT} to determine the sail temperature at each relative velocity value. We also checked the mechanical integrity of the sail (photon-induced tension) during its acceleration, using the same material properties as in Section~\ref{S:optimizingThickness}. 

For the optical properties of \ch{MoS2} we used the data of Munkhbat~\etal~\cite{Munkhbat2022-2398}, extrapolating the $n$ values using a Cauchy expansion and setting the $\kappa$ values equal to $10^{-8}$ at wavelengths equal to and longer than the laser wavelength of $\lambda_l=1.2$~\si{\micro\meter}. For the $n$ optical coefficients of \ch{Al2O3} we used the data of  Querry~\cite{Querry1985-report} at all wavelengths.  For the $\kappa$ data of \ch{Al2O3}, at wavelengths equal to and shorter than $\lambda_A=7$~\si{\micro\meter} we used the values of Lingart, Petrov, and Tikhonova~\cite{Lingart1982-706}; at wavelengths equal to and greater than $\lambda_B=10$~\si{\micro\meter} we used the values of Querry~\cite{Querry1985-report}; and in between we used a power-law interpolation of the form
\begin{linenomath}\begin{equation}
    \kappa = a\lambda^b
    \label{E:kappaInterpolation}
\end{equation}\end{linenomath}
where 
\begin{linenomath}\begin{equation}
    b = \frac{\ln\left(\frac{\kappa_A}{\kappa_B}\right)}{\ln\left(\frac{\lambda_A}{\lambda_B}\right)}
    \label{E:kappaInterpolation_b}
\end{equation}\end{linenomath}
and
\begin{linenomath}\begin{equation}
    a = \frac{\kappa_A}{\lambda_A^b}. 
    \label{E:kappaInterpolation_a}
\end{equation}\end{linenomath}
Here, $\kappa_A$ and $\kappa_B$ correspond to the extinction coefficients provided by Lingart, Petrov, and Tikhonova~\cite{Lingart1982-706} and Querry~\cite{Querry1985-report} at $\lambda_A$ and $\lambda_B$, respectively. We used these optical coefficients and the transfer-matrix method to calculate the reflectivity and absorptivity at each $\beta$-step throughout the acceleration, and also used the transfer-matrix method to obtain the spectral hemispherical emissivity in the wavelength range $\lambda = \langle 0.301 , 20 \rangle$~\si{\micro\meter} (Equation~\ref{E:epsilonLambda}). Also, in order to provide a more realistic portrait of the sail's accelerative performance, we accounted for the impact of the spherically-curved sail's shape on its reflectivity using Equation~\ref{E:varrhoA}, which required calculating angle-dependent reflectivity values. 

\begin{figure*}
\centering
\includegraphics[width=\figWidthFullExtra]{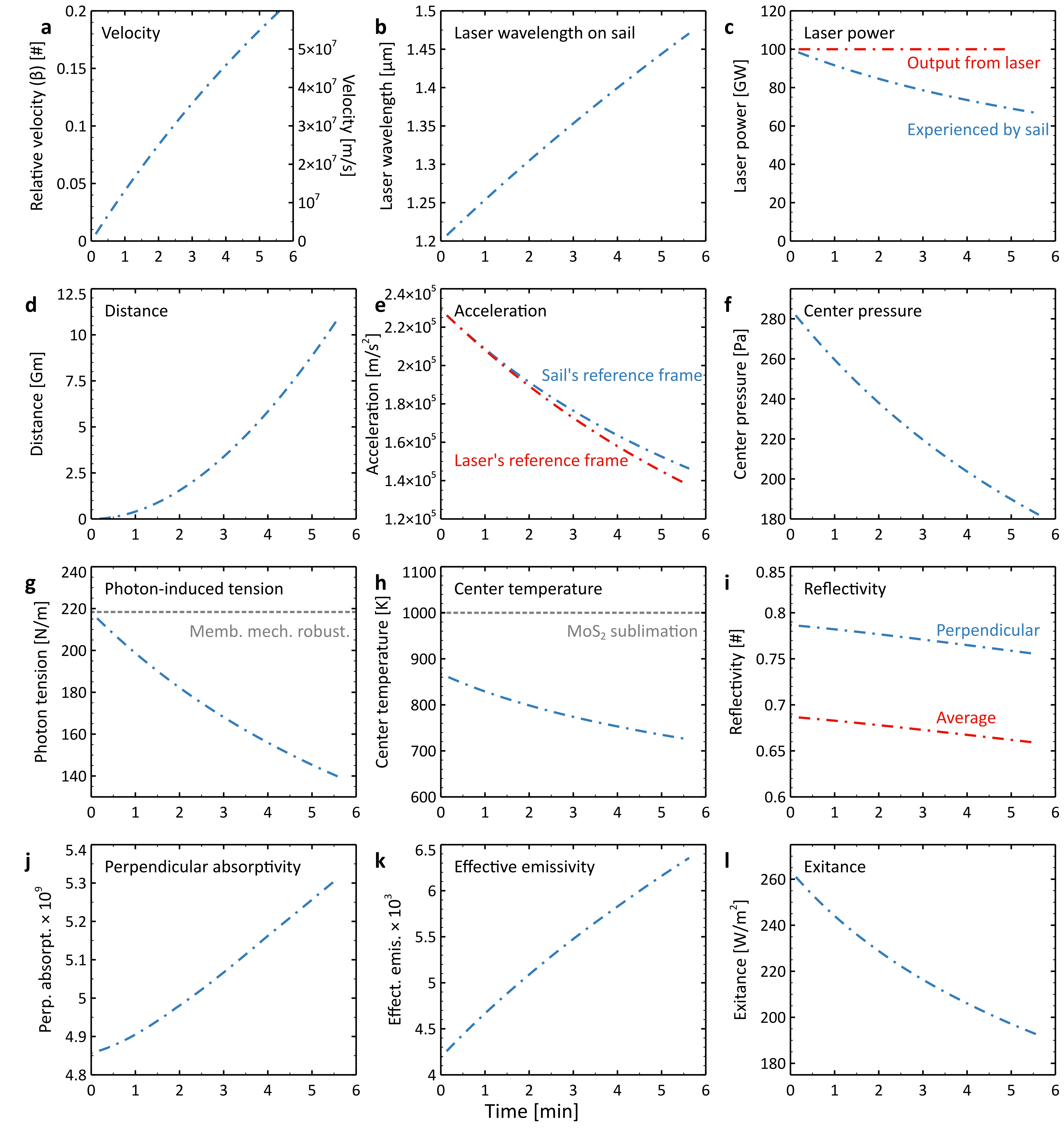}
\caption{\textbf{\textbar~ Calculations for the acceleration of our proposed optimized sail design.} The simulation is for a spherically-curved circular sail configuration with total sailcraft mass $m_{tot}=2$~\si{\gram}, sail mass $m_s=1$~\si{\gram}, equal spherical radius of curvature and diameter $(d_s=s_s)$, and optimal optical coefficients for minimal absorptivity. Panels provide show the
(\textbf{a}) the sailcraft's relative velocity $\beta$ and laser reference frame velocity $v$, 
(\textbf{b}) Doppler-shifted laser wavelength $\lambda_s$, 
(\textbf{c}) laser output power $\Phi_{l}$ (constant) and power incident on the sail $\Phi_{s,\beta}$, 
(\textbf{d}) distance of the sailcraft from Earth $D$, 
(\textbf{e}) acceleration in the laser's reference frame $a_{s,l}$ and in the sail's reference frame $a_{s,s}$, 
(\textbf{f}) center-of-sail photon pressure $P$, 
(\textbf{g}) photon-induced tension $\Upsilon$, 
(\textbf{h}) center-of-sail temperature $T$, 
(\textbf{i}) average reflectivity $\varrho_{\beta,a}$ and perpendicular reflectivity $\varrho_{\beta,\perp}$, 
(\textbf{j}) perpendicular (normal) absorptivity $\alpha_{\beta,\perp}$, 
(\textbf{k}) effective emissivity $\varepsilon_e$, and 
(\textbf{l}) exitance $\mathcal{E}_s$.
The calculation yields an acceleration distance of $L\approx11$~\si{\giga\meter}, a maximum sail temperature of $T_s\approx868$~\si{\kelvin}, an acceleration time of $t_a=5.6$~\si{\minute}, and a laser-on time of $t_l=5.0$~\si{\minute}. 
The proposed optimized film consists of a \ch{Al2O3}-\ch{MoS2}-\ch{Al2O3} composite ($t_{A,b}\approx 19~\si{\nano\meter}$ (bottom \ch{Al2O3} thickness), $t_{M}\approx 63~\si{\nano\meter}$ (\ch{MoS2}), $t_{A,t}\approx 19~\si{\nano\meter}$ (top \ch{Al2O3})) with a hexagonally corrugated indented trench structure ($d_h=70~\si{\micro\meter}$, $w_t=2~\si{\micro\meter}$, $h_t=3~\si{\micro\meter}$; see Figure~\ref{F:hexDimsTrenches}).}%
\label{F:accelThermalBalance}%
\end{figure*}

The results of our calculations are provided in Figure~\ref{F:accelThermalBalance}. Panel~(h) shows that the peak temperature occurred at the beginning of the acceleration phase, and that, even with a laser output power of $\Phi_l=100$~\si{\giga\watt}, the sail stayed cooler than than the \ch{MoS2} vacuum sublimation point of about 1000~\si{\kelvin}~\cite{Cui2018-44}. This result is predicated on our selection of $\kappa=10^{-8}$ for \ch{MoS2} in the laser band; higher extinction coefficients would require reducing the laser output power to avoid overheating the sail. In this particular simulation, the constraining parameter for this design was the photon-induced tension (panel (g)), which was the factor for which we optimized this design (see Section~\ref{S:optimizingThickness}).

A few more aspects of these acceleration calculations are worth discussing. Though we held the laser output power, shown in Figure~\ref{F:accelThermalBalance}(c), constant, the power incident on the sail decreased with time due to the red shift of the photons and the reduced photon flux relative to the accelerating sail.  Notice that the time duration during which the laser needed to produce photons, $t_l=5.0$~\si{\minute}, was less than the sail's acceleration time, $t_a=5.6$~\si{\minute}, because the last photon leaving the laser took about 0.6~\si{\minute} to reach the sail. The acceleration experienced by the sail (in its own reference frame) was also higher than that in the laser's reference frame due to the relativistic increase in the sailcraft's mass from the laser's perspective (panel (e)). The average reflectivity of the spherically curved sail (panel (i)) was smaller than that of the perpendicular component (at the sail center); both decreased with time as the laser wavelength reaching the sail (panel (b)) was redshifted. 
Ultimately, our calculations indicated that our proposed optimized film could accelerate from rest to $\beta=0.2$ over a distance of $L\approx11$~\si{\giga\meter}, and would require a laser array diameter on Earth of $d_{l,E}\approx17$~\si{\kilo\meter}.

Our use of an extinction coefficient of $\kappa=10^{-8}$ for \ch{MoS2} is optimistic but not unreasonable. For instance,  Ermolaev~\etal~\cite{Ermolaev2020-21}, Hsu~\etal~\cite{Hsu2019-1900239}, and Liu~\etal~\cite{Liu2020-15282} reported optical coefficients of $\kappa=0$ for \ch{MoS2} at or near $\lambda=1.2$~\si{\micro\meter}. We anticipate that future fabrication improvements will allow perfect \ch{MoS2} sheets to be conformally fabricated over large areas.  Our use of the extinction coefficients of Lingart, Petrov, and Tikhonova~\cite{Lingart1982-706} in the laser wavelength range is done knowing that, if the \ch{Al2O3} was amorphous as deposited using atomic layer deposition (ALD), it would need to be annealed so it would adopt a crystalline form. Recent work has shown that annealing ALD \ch{Al2O3} films for just 30~\si{\minute} in a \ch{H2} or \ch{N2} atmosphere at a pressure of $10^5$~\si{\pascal} and a temperature of around 1000~\si{\kelvin} is sufficient to accomplish this~\cite{Broas2017-3390, Broas2019-147} (but note that another study suggested a temperature of around 1400~\si{\kelvin}~\cite{Zhang2007-3707}). (We also note that, while annealing greatly reduces the extinction coefficient of \ch{Al2O3}, it has only a small impact its real component of the index of refraction~\cite{Wang2015-46}.) Annealing the alumina may impact its mechanical properties; the yield stress of bulk sapphire is roughly an order of magnitude lower than that for thin ALD alumina films.  We are unaware of studies on the mechanical properties of thin sapphire films, but there is some evidence that annealing \ch{Al2O3} films at temperatures up to 1000~\si{\kelvin} does not adversely impact them~\cite{Ylivaara2022-062424}. Annealing of the face alumina films would need to take place in the presence of the inner \ch{MoS2} film. As we noted earlier, \ch{MoS2} is known to sublimate in vacuum environments at temperatures exceeding 1000~\si{\kelvin}. However, recent work suggests that it can be heated to 1500~\si{\kelvin} without evaporating if it is pressurized to $>100$~\si{\pascal} (gaseous environment)~\cite{Cui2018-44}. Even if sublimation were to occur, it may be relatively slow; a study by Lu~\etal\xspace found that at 923~\si{\kelvin} and 1300~\si{\pascal} it took 1~\si{\hour} to remove one layer (about 0.65~\si{\nano\meter}) of a \ch{MoS2} flake~\cite{Lu2013-8904}. Together, these factors suggest that future \ch{Al2O3}-\ch{MoS2}-\ch{Al2O3} light sails could feature alumina and \ch{MoS2} extinction coefficients as low as or even lower than $\kappa=10^{-7}$ and $\kappa=10^{-8}$, respectively. 

\section{Impact of temperature}\label{S:impactOfTemperature}

Elevated temperatures will impact the mechanical and optical properties of the sail. While a dedicated study of these effects for thin films will ultimately be required (but is now beyond the scope of this work), we can make some observations from related studies in the literature~\cite{Holdman2022-2102835}.  

The mechanical properties of \ch{Al2O3} and \ch{MoS2} will degrade with temperature. The fracture stress of sapphire has been found to decrease by roughly 15\% from room temperature to 1000~\si{\kelvin}~\cite{Brenner1962-33}, and the yield stress of polycrystalline alumina was found to decrease by as much as 62\% over the same range~\cite{Sanchez-Gonzalez2007-3345}. Likewise, molecular dynamics simulations of monolayer \ch{MoS2} flakes suggest that their ultimate stress decreases by roughly 23\% from room temperature to 600~\si{\kelvin}~\cite{Pham2022-7777}. This degradation could be offset by using lower laser powers (decreasing the photon flux and hence the photon-induced tension) or thicker film layers (increasing the sail's membrane mechanical robustness). 

\begin{figure*}
\centering
\includegraphics[width=\figWidthCol]{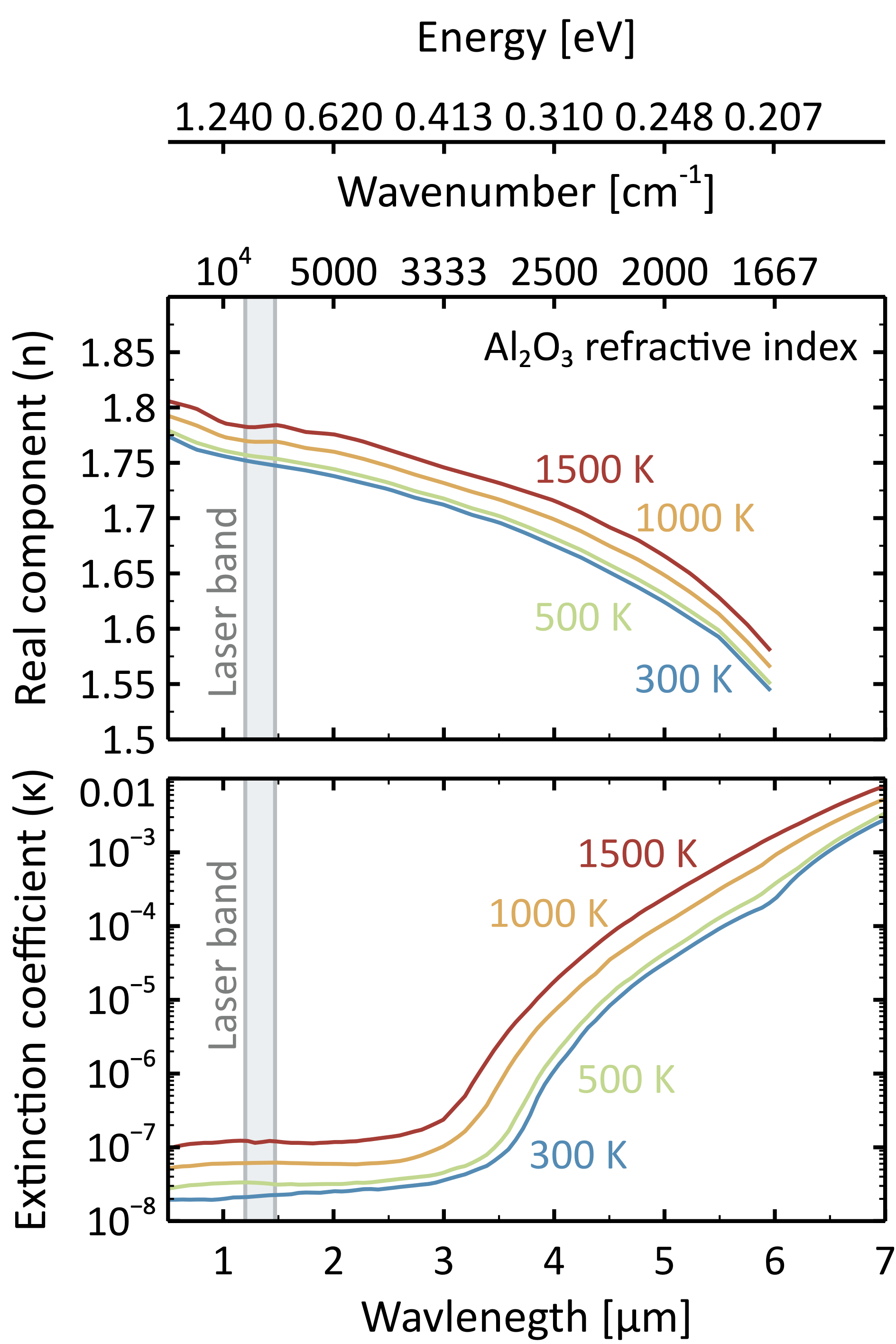}
\caption{\textbf{\textbar~ Index of refraction of sapphire (\ch{Al2O3}) as a function of temperature.} Data are obtained from Lingart, Petrov, and Tikhonova~\cite{Lingart1982-706}. }%
\label{F:lingartAl2O3highTopticalData}%
\end{figure*}

\begin{figure*}
\centering
\includegraphics[width=\figWidthCol]{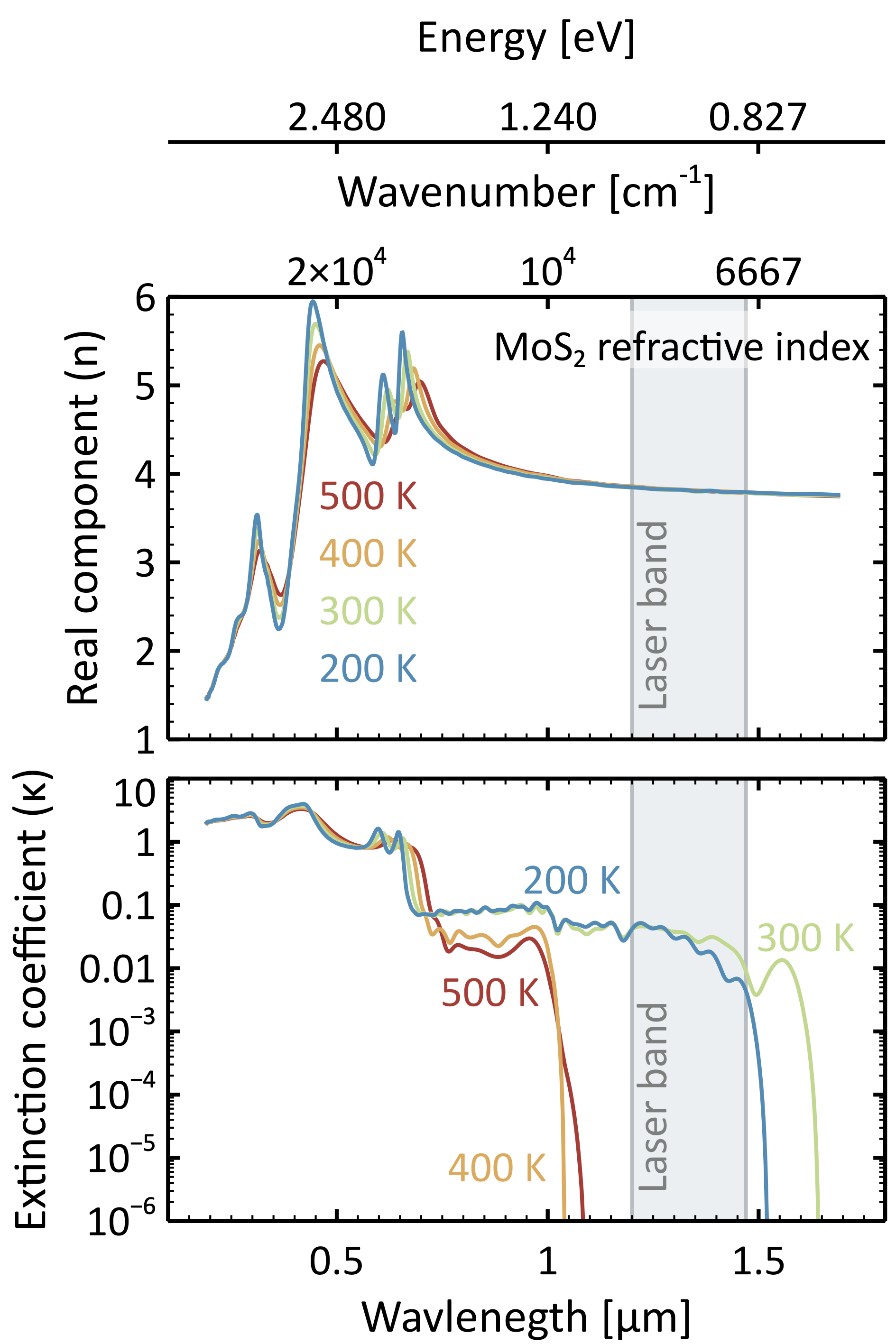}
\caption{\textbf{\textbar~ Index of refraction of \ch{MoS2} as a function of temperature.} Data are obtained from Liu~\etal~\cite{Liu2020-15282}. }%
\label{F:Liu2020MoS2TopticalData}%
\end{figure*}

The extinction coefficients for \ch{Al2O3} and \ch{MoS2} will likely increase with temperature; unfortunately, high-temperature optical data within the laser band for these materials is scarce. Lingart, Petrov, and Tikhonova~\cite{Lingart1982-706} compiled information for sapphire at elevated temperatures, which suggests that its extinction coefficient in the laser band roughly triples (190\% increase) from room temperature to 1000~\si{\kelvin} (Figure~\ref{F:lingartAl2O3highTopticalData}). However, this is accompanied by a similar increase at longer wavelengths, suggesting that the sail will be able to emit more radiation to slightly offset the increased photon absorption. The real component of the index of refraction of sapphire increases negligibly (1\%) over the same temperature range. Data reported by Liu~\etal\xspace for \ch{MoS2} suggest that its extinction coefficient in the laser band does not change significantly from 4.5~\si{\kelvin} to 300~\si{\kelvin}, and that it may decrease at higher temperatures~\cite{Liu2020-15282} (Figure~\ref{F:Liu2020MoS2TopticalData}). Also, according to their results, the real component of the index of refraction of \ch{MoS2} changes by less than 1\% from room temperature to 500~\si{\kelvin}. 

\begin{figure*}
\centering
\includegraphics[width=\figWidthCol]{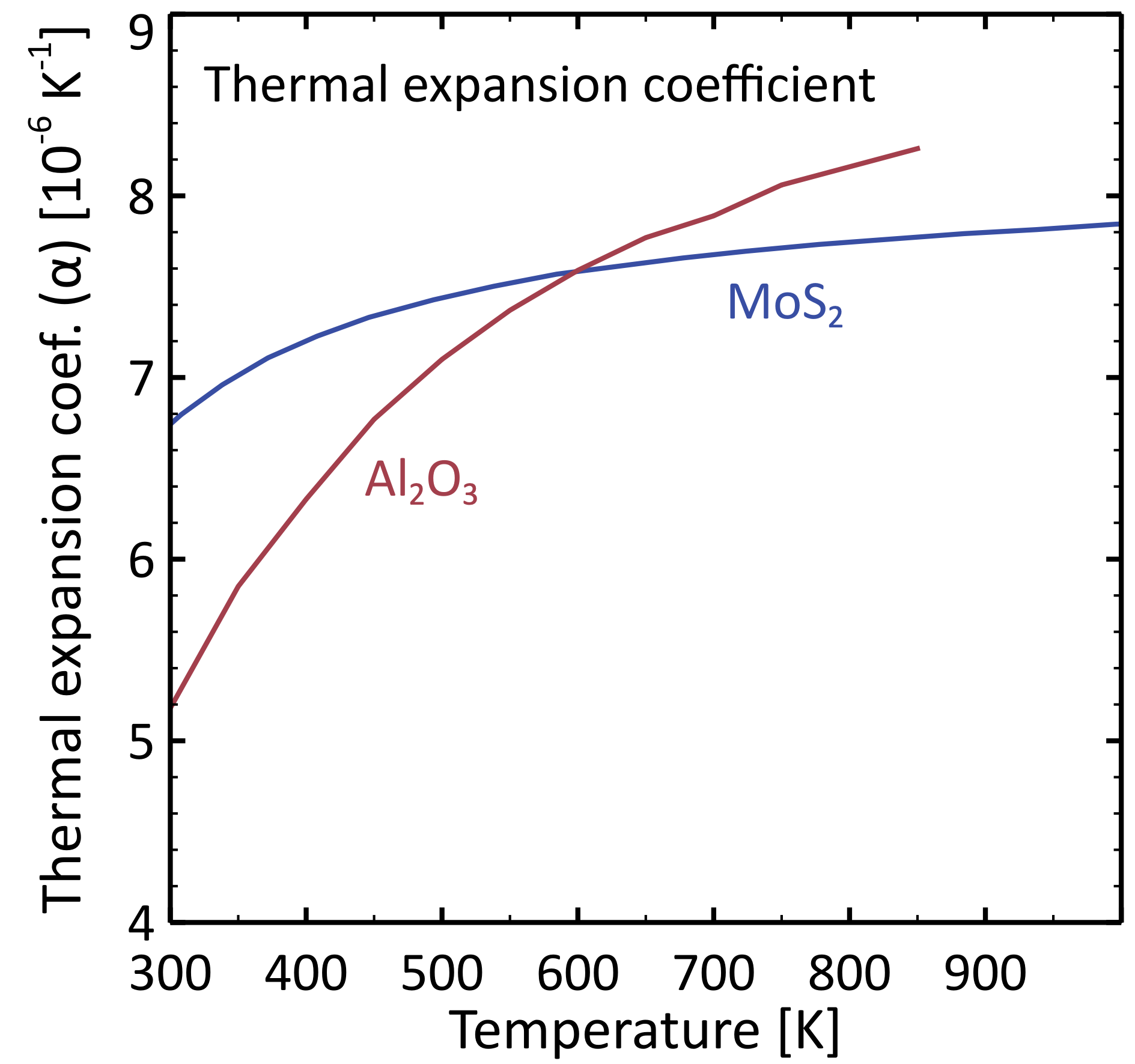}
\caption{\textbf{\textbar~ Thermal expansion coefficients of \ch{Al2O3} and \ch{MoS2} as a function of temperature.} Data are obtained from Hayashi, Watanabe, and Inaba~\cite{Hayashi2000-77} (\ch{Al2O3}) and Huang, Gong, and Zeng~\cite{Huang2014-045409} (\ch{MoS2}). }%
\label{F:plotThermalExpansCoef}%
\end{figure*}

Finally, thermal expansion effects will impact the sail. As shown in Figure~\ref{F:plotThermalExpansCoef}, \ch{Al2O3} and \ch{MoS2} have well-matched thermal expansion coefficients, $\alpha_A$ and $\alpha_M$, respectively - within about 5\% in the range 500-800~\si{\kelvin} and within 10-15\% up to 1000~\si{\kelvin}~\cite{Despont1996-219, Hayashi2000-77, Huang2014-045409, Touloukian1977-report}. We can estimate the strain $\epsilon_{T}$ that this thermal expansion induces using 
\begin{linenomath}\begin{equation}
    \epsilon_{T} = (T-T_0)\left|\alpha_A-\alpha_M\right|
	\label{E:thermalStress}
\end{equation}\end{linenomath}
where $T$ is the temperature of interest and $T_0$ is a reference temperature at which the strain is negligible. This equation indicates that the thermal strain will be less than a tenth of a percent from room temperature up to 1000~\si{\kelvin}, and therefore suggests that thermal strain is not likely to be a failure mode for the sail. 




\end{multicols}

\end{document}